\ifdefined\pdfoutput
\pdfoutput=1 
\fi
\documentclass{article}

\usepackage[letterpaper,margin=1in]{geometry}
\usepackage{graphicx}
\usepackage{lmodern}
\usepackage{array}
\usepackage[export]{adjustbox}
\usepackage{booktabs}
\usepackage{rotating}
\usepackage{tabularx}
\usepackage{multirow}
\usepackage{subcaption}
\usepackage{algorithm}
\usepackage{algorithmic}
\usepackage{enumitem}
\usepackage{longtable}
\usepackage{pdflscape}

\usepackage{wrapfig}
\usepackage{url}
\usepackage{amsmath, amsfonts, amssymb, amsthm}
\usepackage[table]{xcolor}
\definecolor{okGreen}{HTML}{009E73}
\definecolor{okOrange}{HTML}{E69F00}
\definecolor{okRed}{HTML}{D55E00}
\newcommand{\clsID}{\cellcolor{okGreen!25}Identifiable}
\newcommand{\clsIDlow}{\cellcolor{okGreen!8}Low PII, low coverage}
\newcommand{\clsWI}{\cellcolor{okOrange!30}Weakly id.}
\newcommand{\clsNI}{\cellcolor{okRed!35}Non-id.}
\usepackage[sort,nocompress]{cite}
\usepackage{authblk}
\usepackage{hyperref}
\usepackage{tikz}
\usetikzlibrary{arrows.meta, positioning}

\theoremstyle{remark}
\newtheorem{remark}{Remark}


\makeatletter
\providecommand{\insert@pcolumn}{\insert@column}
\newcommand*\SM@save[1]{%
  \expandafter\let\csname SM@saved@\string#1\endcsname#1}
\newcommand*\SM@restore[1]{%
  \expandafter\let\expandafter#1\csname SM@saved@\string#1\endcsname}
\SM@save\title \SM@save\author \SM@save\date \SM@save\and \SM@save\thanks
\SM@save\@title \SM@save\@author \SM@save\@date \SM@save\@thanks
\SM@save\@maketitle
\SM@save\AB@authlist \SM@save\AB@affillist \SM@save\AB@authors
\SM@save\AB@author \SM@save\AB@affilsep \SM@save\AB@blk@and
\expandafter\SM@save\csname @sep1\endcsname
\expandafter\SM@save\csname @sep2\endcsname
\edef\SM@authblkcounters{%
  \noexpand\setcounter{authors}{\the\c@authors}%
  \noexpand\setcounter{affil}{\the\c@affil}%
  \noexpand\setcounter{Maxaffil}{\the\c@Maxaffil}%
  \noexpand\setlength\noexpand\@affilsep{\the\@affilsep}}
\AtBeginDocument{\SM@save\maketitle}
\DeclareRobustCommand\SM@cite{%
  \@ifnextchar[{\@tempswatrue\SM@citex}{\@tempswafalse\SM@citex[]}}
\def\SM@citex[#1]#2{\leavevmode
  \let\@citea\@empty
  \SM@@cite{\@for\@citeb:=#2\do
    {\@citea\def\@citea{,\penalty\@m\ }%
     \edef\@citeb{supp:\expandafter\@firstofone\@citeb\@empty}%
     \@ifundefined{b@\@citeb}{\hbox{\reset@font\bfseries ?}%
       \G@refundefinedtrue
       \@latex@warning{Citation `\@citeb' on page \thepage\space undefined}}%
      {\@cite@ofmt{\csname b@\@citeb\endcsname}}}}{#1}}
\def\SM@@cite#1#2{[{#1\if@tempswa , #2\fi}]}
\def\SM@bibitem{\@ifnextchar[\SM@lbibitem\SM@nbibitem}
\def\SM@lbibitem[#1]#2{\@lbibitem[#1]{supp:#2}}
\def\SM@nbibitem#1{\@bibitem{supp:#1}}
\newcommand*\beginsupplement{%
  \clearpage
  \SM@restore\maketitle \SM@restore\@maketitle
  \SM@restore\title \SM@restore\author \SM@restore\date \SM@restore\and
  \SM@restore\thanks \SM@restore\@title \SM@restore\@author
  \SM@restore\@date \SM@restore\@thanks
  \SM@restore\AB@authlist \SM@restore\AB@affillist \SM@restore\AB@authors
  \SM@restore\AB@author \SM@restore\AB@affilsep \SM@restore\AB@blk@and
  \expandafter\SM@restore\csname @sep1\endcsname
  \expandafter\SM@restore\csname @sep2\endcsname
  \newaffiltrue \SM@authblkcounters
  \setcounter{section}{0}\setcounter{subsection}{0}%
  \setcounter{subsubsection}{0}\setcounter{paragraph}{0}%
  \setcounter{subparagraph}{0}\setcounter{figure}{0}%
  \setcounter{table}{0}\setcounter{equation}{0}\setcounter{footnote}{0}%
  \def\theHsection{supp.\arabic{section}}%
  \def\theHfigure{supp.\arabic{figure}}%
  \def\theHtable{supp.\arabic{table}}%
  \let\cite\SM@cite
  \let\bibitem\SM@bibitem
  \pdfbookmark[0]{Supplementary Material}{SupplementaryMaterial}%
}
\makeatother

\title{Parameter uncertainty in dynamical models: a practical identifiability index}
\author[1,3]{Hamed Karami}
\author[1]{Alexandra Smirnova}
\author[2]{Sunmi Lee}
\author[2,3,*]{Gerardo Chowell}

\affil[1]{Department of Mathematics \& Statistics, Georgia State University, Atlanta, GA 30303, USA}
\affil[2]{Department of Applied Mathematics, Kyung Hee University, Yongin 17104, Korea}
\affil[3]{Department of Population Health Sciences, Georgia State University, Atlanta, GA 30303, USA}
\affil[*]{Corresponding author: \href{mailto:gchowell@gsu.edu}{gchowell@gsu.edu}}

\date{}
\begin{document}

\maketitle
\begingroup
\setlength{\parindent}{0pt}
\begin{abstract}
\noindent
Ordinary differential equation models are widely used to describe how complex
systems change over time, but reliable conclusions depend on how well their
parameters can be estimated from limited, noisy data. Structural identifiability
asks whether parameters can be uniquely determined from ideal data, whereas
practical identifiability asks how well they can be estimated from the data
actually available. Common measures of uncertainty, such as confidence intervals
and coefficients of variation, are useful, but they do not provide a single,
easy-to-interpret scale for comparing uncertainty across parameters, models,
error structures, and data-collection designs.

\noindent
We introduce the Practical Identifiability Index (PII), a unit-free,
scale-invariant measure of uncertainty for individual positive-valued parameters.
PII is defined as the base-10 logarithm of the ratio between the upper and lower
bounds of a confidence interval. This gives PII a simple interpretation: a value
of 1 corresponds to a tenfold ratio between the interval bounds. Because PII is
based on this ratio, parameter uncertainty can be compared on the same scale
across different parameters, models, and data conditions.

\noindent
We study PII using parametric bootstrap experiments in growth and compartmental
epidemic models. In general, PII decreases as longer data windows are used for
model fitting and tends to increase with observation noise and when several
parameters must be estimated together, although the effect of noise is not
uniform across all settings. PII also remains larger for parameters associated
with latent or indirectly observed processes, while adding more measured
quantities substantially improves estimation of progression and recovery
parameters. We then apply PII to daily influenza incidence from the 1918 San
Francisco epidemic to show how parameter uncertainty changes as more data become
available. As a benchmark, we compare PII with normalized confidence-interval
width (NCIW) using equivalent thresholds. The two measures give the same
classification in 134 of 135
model--scenario--parameter--error-structure combinations (99.3\% agreement).
Thus, PII gives nearly the same classification while allowing uncertainty to be
interpreted directly as a multiplicative range, without normalization by the
parameter estimate.

\noindent
PII is meant to complement, rather than replace, other identifiability and
uncertainty analyses. A narrow confidence interval does not necessarily mean that
the interval contains the true parameter value at the expected rate, so PII
should be considered together with empirical coverage and other diagnostic
measures. Overall, PII provides a simple way to report and compare uncertainty
in individual parameters and to assess how the amount and type of available data
affect practical identifiability.
\end{abstract}

\textbf{Keywords:} practical identifiability; parameter uncertainty; dynamical systems; ordinary differential equations; epidemic models; confidence intervals; parametric bootstrap; observation design.

\section{Introduction}

Mathematical models based on systems of ordinary differential equations (ODEs)
are widely used to understand, forecast, and control processes that change over
time across disciplines, including population biology, ecology,
pharmacokinetics, and infectious disease epidemiology~
\cite{chowell2024investigating,hoang2025differential,munday2025forecasting}.
Growth and compartmental epidemic models are especially important for studying
underlying mechanisms, estimating key parameters, and informing policy
decisions~
\cite{strogatz2024nonlinear,edelstein2005mathematical,brauer2019mathematical,
huang2025infectious,liyanage2025structural}.
Despite their widespread use, an ongoing challenge in ODE modeling is
determining whether model parameters can be reliably estimated from data,
since limited information about parameter values can make inference and
prediction less reliable even when models fit the observed data well~
\cite{pawitan2001all,gutenkunst2007universally,raue2009structural,
dankwa2022structural,roosa2019assessing}.
This problem occurs in many biological and biophysical systems fitted to noisy,
incomplete, or partially observed data, including systems biology, transport
models, and processes in heterogeneous populations~
\cite{villaverde2017dynamical,eisenberg2013identifiability,
balsa2025quantifying,villaverde2016identifiability}.

Identifiability is commonly divided into structural and practical components.
Structural identifiability asks whether parameters can, in principle, be
uniquely determined from ideal, noise-free observations given the model
structure and observables~
\cite{bellman1970structural,raue2009structural,cobelli1980parameter,wieland2021structural}.
Many studies have developed algebraic, differential, and computational methods
for nonlinear ODE systems~
\cite{dong2023differential,bellman1970structural,walter1997identification,
ljung1994global,saccomani2003parameter,chis2011structural,
villaverde2017dynamical,tuncer2018structural},
including applications to epidemic models where parameters may be
non-identifiable even in simple compartmental structures~
\cite{tuncer2018structural},
as well as tools for structural identifiability analysis~
\cite{rey2023benchmarking,diaz2023strike,heinrich2025structural}.
Benchmarking studies further evaluate the performance, scalability, and
reliability of these tools~\cite{rey2023benchmarking}.
However, structural identifiability is not sufficient for reliable inference
from real data: observations are finite, noisy, partially observed, and often
limited in time~\cite{simpson2020practical}, so structurally identifiable
parameters may still have large uncertainty, strong correlations, or
sensitivity to small changes in the data~
\cite{gevertz2024minimally,gallo2022lack,daly2018inference}.
This problem, known as practical identifiability, has long been recognized in
systems biology and epidemiological modeling~
\cite{raue2009structural,gutenkunst2007universally,
eisenberg2013identifiability,transtrum2015perspective,
heinrich2025different,wang2025systematic,tuncer2025structural}.
In practice, parameter constraint depends on how informative the data are~
\cite{preston2025think,porthiyas2024practical},
as well as data quality, measurement timing, and the analysis workflow used for
calibration and uncertainty quantification~
\cite{preston2025think,liu2024parameter,murphy2024implementing}.

Practical identifiability is commonly evaluated using profile likelihoods,
confidence intervals, sensitivity analysis, diagnostics based on Fisher
information, or Bayesian posterior summaries~
\cite{venzon1988method,pawitan2001all,raue2009structural,raue2013joining,tonsing2018profile}.
These approaches are useful, but they often require interpretation based on
plots, rules of thumb, or specific parameters, such as deciding whether a
likelihood profile is ``flat'' or whether a confidence interval is ``too wide''~
\cite{heinrich2025different,raue2009structural}.
Quantitative measures based on confidence-interval width, relative uncertainty,
or coefficients of variation have also been proposed, but they lack a common
standard across models and can be difficult to compare for parameters with
different units, scales, or magnitudes~
\cite{ashyraliyev2009systems,raue2009structural,kreutz2012likelihood,
tuncer2025structural}.
These limitations matter in epidemic modeling, where parameter estimates inform
projections, evaluations of interventions, and public health responses~
\cite{chowell2017fitting,kao2018practical}.
Common practical identifiability challenges include trade-offs between
transmission and recovery rates, confounding between initial conditions and
parameters, and sensitivity to data aggregation and reporting noise~
\cite{capaldi2012parameter,king2008inapparent,tuncer2018structural,
roosa2019assessing}.
These challenges were especially clear in real-time COVID-19 analyses~
\cite{roda2020difficult}, and similar issues arise in growth models~
\cite{liyanage2025structural,chowell2016characterizing} and other mechanistic systems
where limited information in the data, observation design, or measurement noise
can hinder estimation~
\cite{villaverde2019observability,simpson2020practical}.
These findings motivate clear quantitative summaries of how strongly the
available data constrain model parameters~\cite{gallo2022lack}.

In this study, we introduce the Practical Identifiability Index (PII), a
single-number summary of marginal parameter uncertainty based on the logarithmic
span of confidence intervals. Because it depends only on the confidence-interval
endpoints, the PII does not need to be scaled by the parameter estimate.
Instead of a yes-or-no test, the PII expresses uncertainty on an
order-of-magnitude scale, allowing comparisons across positive-valued
parameters, models, and data conditions. A PII value of 1 corresponds exactly
to a factor of ten between the confidence-interval endpoints, that is, one
order of magnitude, when $\theta_{j,L}>0$. Because the PII is computed from
uncertainty intervals for individual parameters, it should be interpreted as a
marginal interval-width diagnostic rather than a full identifiability analysis.
It can be used in frequentist or Bayesian workflows whenever confidence or
credible intervals are available~\cite{chowell2024parameter,karami2025bayesianfitforecast}.

We present the PII as a standardized, scale-invariant reporting approach rather
than a new method for inference. It allows marginal parameter uncertainty to be
compared across parameters, models, error structures, and observation designs.
We also compare PII with normalized confidence-interval width (NCIW) to determine
whether the two interval-based summaries give consistent classifications of
parameter constraint. The PII provides a brief and clear summary that supports
consistent comparison and synthesis of parameter-uncertainty patterns across
models and data conditions.
In this way, the PII connects identifiability theory with applied model
calibration through a diagnostic that is easy to compute, interpret, and
compare.
We demonstrate the index across increasingly complex growth and epidemic
models. Growth models provide a controlled setting for studying how uncertainty
changes with calibration-window length, while epidemic models illustrate the
effects of latent compartments, parameter coupling, observation noise, and the
amount of usable data. Through these analyses, we show that the PII provides a
compact way to compare marginal parameter uncertainty, identify weakly
constrained parameters, and assess how additional observables improve parameter constraint, a question raised in previous identifiability studies~\cite{sauer2021identifiability,kiss2023parameter}.
\section{Models\label{sec:models}}

In this section, we describe the ODE models used to study the proposed
Practical Identifiability Index. We consider two broad families of widely used
models: (1) growth models, which describe overall epidemic trajectories without
explicit assumptions about transmission mechanisms, and (2) compartmental
epidemic models, which divide the population into epidemiological states and
define transition rates based on the natural course of the disease. These two
model classes are widely used in epidemic modeling and provide complementary
settings for studying identifiability challenges. For each model, we present
the model equations, define the observation operator linked to the observed
data, and summarize the parameters to be estimated.

\subsection{Growth Models}

Growth models describe cumulative case counts $C(t)$ without detailed
assumptions about disease transmission~\cite{turner1976theory}. These models
are especially useful early in an outbreak, when limited data make parameter
estimation harder. Because they involve relatively few parameters and simple
model structures, they provide a useful setting for studying how parameter
constraint changes as more data become available. They also provide a natural
setting for practical identifiability analysis because their parameters can
show different uncertainty patterns even when the fitted trajectory closely
matches the observations~\cite{liyanage2025structural}. We consider three
nested growth models with increasing flexibility
(Figure~\ref{fig:growth_models}).\\
The exponential growth model (EXP) is given by
\begin{equation}
\frac{dC}{dt}=rC,
\label{eq:exp}
\end{equation}
where $r>0$ denotes the growth rate. The generalized growth model (GGM)
\cite{viboud2016generalized} builds on this formulation as
\begin{equation}
\frac{dC}{dt}=rC^{p},
\label{eq:ggm}
\end{equation}
where $p\in[0,1]$ controls the growth, with $p=1$ corresponding to
exponential growth and $0<p<1$ allowing sub-exponential growth. The
generalized logistic growth model (GLM) \cite{chowell2016using} further
modifies the growth through
\begin{equation}
\frac{dC}{dt}=rC^{p}\left(1-\frac{C}{K}\right),
\label{eq:glm}
\end{equation}
where $K>0$ denotes the final epidemic size.

Within the ODE framework, all growth models use the same continuous-time
notation for new cases: $y(t)$ is given by the rate of change of cumulative
cases, i.e., $y(t)=\frac{dC}{dt}$. In the fitting model described below,
discrete observations are evaluated using one-step changes in the cumulative
state. The initial condition $C(0)=C_0$, representing the initial case count,
is fixed to the first observed value.
\begin{figure}[H]
\centering
\begin{tikzpicture}[>=Latex]
\tikzstyle{comp}=[circle, draw, minimum size=12mm, inner sep=0pt, thick]
\tikzstyle{lab}=[font=\small, align=center]

\node[lab] (EXP_Title) at (0, 2.2) {\textbf{EXP}};
\node[comp] (EXP_C) at (0,0) {$C$};
\draw[->, very thick] (EXP_C) to[out=60, in=120, looseness=8] (EXP_C);
\node[lab] at (-0.9, 1.2) {$rC$};

\node[lab] (GGM_Title) at (4, 2.2) {\textbf{GGM}};
\node[comp] (GGM_C) at (4,0) {$C$};
\draw[->, very thick] (GGM_C) to[out=60, in=120, looseness=8] (GGM_C);
\node[lab] at (3, 1.2) {$rC^p$};

\node[lab] (GLM_Title) at (8, 2.2) {\textbf{GLM}};
\node[comp] (GLM_C) at (8,0) {$C$};
\draw[->, very thick] (GLM_C) to[out=60, in=120, looseness=8] (GLM_C);
\node[lab] at (6.5, 1.2) {$rC^p(1-\frac{C}{K})$};

\node[draw, rounded corners=3pt, fill=gray!10, minimum width=55mm, minimum height=8mm] (Obs) at (4,-2.2) {Observation: $y(t) = \dfrac{dC}{dt}$};
\draw[dashed, ->, shorten >=2pt] (EXP_C.south) -- (Obs.north west);
\draw[dashed, ->, shorten >=2pt] (GGM_C.south) -- (Obs.north);
\draw[dashed, ->, shorten >=2pt] (GLM_C.south) -- (Obs.north east);
\end{tikzpicture}
\caption{Diagrams of the three growth models. Circles represent the cumulative
case count $C(t)$. Self-loops show the growth dynamics using model-specific
functions. Dashed arrows show that all models use the same observation source,
with new cases $y(t)$ represented by the derivative of cumulative cases with
respect to time.}
\label{fig:growth_models}
\end{figure}

\subsection{Compartmental Epidemic Models}

Compartmental models divide the population into distinct epidemiological
states and define transition rates based on assumptions about how the disease
spreads and progresses. We consider increasingly complex models, from the
standard SIR model to extensions that include exposed periods, unreported
infections, asymptomatic transmission, and disease-induced mortality.

\subsubsection{SIR Model}

The Susceptible--Infectious--Recovered (SIR) model is a standard framework for
infectious disease dynamics
\cite{kermack1927contribution,hethcote2000mathematics}. The population of size
$N$ is divided into susceptible ($S$), infectious ($I$), and recovered ($R$)
compartments. We also track cumulative infections $C$ to relate the model to
observed cases:
\begin{equation}
\begin{aligned}
\frac{dS}{dt} = -\frac{\beta S I}{N}, \quad
\frac{dI}{dt} = \frac{\beta S I}{N} - \gamma I, \quad
\frac{dR}{dt} = \gamma I, \quad
\frac{dC}{dt} = \frac{\beta S I}{N},
\end{aligned}
\label{eq:sir}
\end{equation}
where $\beta$ is the transmission rate and $\gamma$ is the recovery rate.

\begin{figure}[H]
\centering
\begin{tikzpicture}[>=Latex]
\tikzstyle{comp}=[circle, draw, minimum size=12mm, inner sep=0pt, font=\large]
\tikzstyle{lab}=[font=\small, align=center]

\node[comp] (S) at (0,0) {$S$};
\node[comp] (I) at (4,0) {$I$};
\node[comp] (R) at (8,0) {$R$};

\draw[->, thick] (S) -- (I) node[lab, pos=0.5, above] {$\dfrac{\beta S I}{N}$};
\draw[->, thick] (I) -- (R) node[lab, pos=0.5, above] {$\gamma I$};

\node[draw, rounded corners=3pt, fill=gray!10, minimum width=35mm, minimum height=8mm, anchor=north] (obs) at (4,-1.8) {Observation: $y(t) = \dfrac{\beta S I}{N}$};
\draw[dashed, ->, shorten >=2pt] (I.south) -- (obs.north);
\end{tikzpicture}
\caption{SIR compartment diagram. Solid arrows show transitions between compartments; the dashed arrow shows the observation source (new infections).}
\label{fig:sir}
\end{figure}

\subsubsection{SEIR Model}

The SEIR model extends the SIR model by including an exposed compartment ($E$)
representing individuals who have been infected but are not yet infectious~\cite{brauer2019mathematical}:
\begin{equation}
\begin{aligned}
\frac{dS}{dt} = -\frac{\beta S I}{N}, \quad
\frac{dE}{dt} = \frac{\beta S I}{N} - \kappa E,\quad
\frac{dI}{dt} = \kappa E - \gamma I, \quad
\frac{dR}{dt} = \gamma I, \quad
\frac{dC}{dt} = \kappa E,
\end{aligned}
\label{eq:seir}
\end{equation}
where $\kappa$ is the transition rate from the exposed to the infectious state.

\begin{figure}[H]
\centering
\begin{tikzpicture}[>=Latex]
\tikzstyle{comp}=[circle, draw, minimum size=12mm, inner sep=0pt, font=\large]
\tikzstyle{lab}=[font=\small, align=center]

\node[comp] (S) at (0,0) {$S$};
\node[comp] (E) at (3,0) {$E$};
\node[comp] (I) at (6,0) {$I$};
\node[comp] (R) at (9,0) {$R$};

\draw[->, thick] (S) -- (E) node[lab, pos=0.5, above] {$\dfrac{\beta S I}{N}$};
\draw[->, thick] (E) -- (I) node[lab, pos=0.5, above] {$\kappa E$};
\draw[->, thick] (I) -- (R) node[lab, pos=0.5, above] {$\gamma I$};

\node[draw, rounded corners=3pt, fill=gray!10, minimum width=35mm, minimum height=8mm, anchor=north] (obs) at (4.5,-1.8) {Observation: $y(t) = \kappa E$};
\draw[dashed, ->, shorten >=2pt] (E.south) -- ++(0,-0.5) -| (obs.north);
\end{tikzpicture}
\caption{SEIR compartment diagram. The exposed compartment represents the
latent period before individuals become infectious.}
\label{fig:seir}
\end{figure}

\begin{remark}
In addition to this standard single-observable configuration
($y = \kappa E$), we also examine multi-observable configurations in which
infectious individuals $I$ and recovered individuals $R$ are available as
additional observables for calibration. These configurations are denoted
SEIRMO.
\end{remark}

\subsubsection{SEIR Model with Unreported Infections (SEIR--UR)}

This extension accounts for incomplete case reporting by introducing a
reporting fraction $\rho \in (0,1]$~\cite{karami2026comparative}:
\begin{equation}
\begin{aligned}
\frac{dS}{dt} = -\frac{\beta S I}{N}, \quad
\frac{dE}{dt} = \frac{\beta S I}{N} - \kappa E, \quad
\frac{dI}{dt} = \kappa E - \gamma I, \quad
\frac{dR}{dt} = \gamma I, \quad
\frac{dC}{dt} = \kappa \rho E,
\end{aligned}
\label{eq:seir_unrep}
\end{equation}
where only a fraction $\rho$ of infections are reported and contribute to the
observed case count. The reporting fraction $\rho$ can create identifiability
challenges when estimated together with transmission parameters.

\begin{figure}[H]
\centering
\begin{tikzpicture}[>=Latex]
\tikzstyle{comp}=[circle, draw, minimum size=12mm, inner sep=0pt, font=\large]
\tikzstyle{lab}=[font=\small, align=center]

\node[comp] (S) at (0,0) {$S$};
\node[comp] (E) at (3,0) {$E$};
\node[comp] (I) at (6,0) {$I$};
\node[comp] (R) at (9,0) {$R$};

\draw[->, thick] (S) -- (E) node[lab, pos=0.5, above] {$\dfrac{\beta S I}{N}$};
\draw[->, thick] (E) -- (I) node[lab, pos=0.5, above] {$\kappa E$};
\draw[->, thick] (I) -- (R) node[lab, pos=0.5, above] {$\gamma I$};

\node[draw, rounded corners=3pt, fill=gray!10, minimum width=40mm, minimum height=8mm, anchor=north] (obs) at (4.5,-1.8) {Observation: $y(t) = \rho \kappa E$};
\draw[dashed, ->, shorten >=2pt] (E.south) -- ++(0,-0.5) -| (obs.north);
\end{tikzpicture}
\caption{SEIR--UR compartment diagram. Only a fraction $\rho$ of new infections are reported.}
\label{fig:seir_unrep}
\end{figure}

\subsubsection{SEIR Model with Asymptomatic Transmission (SEIAR)}

This model treats symptomatic ($I$) and asymptomatic ($A$) infectious
individuals separately, with potentially different transmission rates~
\cite{chowell2023structural}:
\begin{equation}
\begin{gathered}
\frac{dS}{dt} = -\frac{(\beta_0 I + \beta_1 A) S}{N},\quad
\frac{dE}{dt} = \frac{(\beta_0 I + \beta_1 A) S}{N} - \kappa E, \quad
\frac{dI}{dt} = \kappa \rho E - \gamma I,\\
\frac{dA}{dt} = \kappa (1-\rho) E - \gamma A,\quad
\frac{dR}{dt} = \gamma (I + A),\quad
\frac{dC}{dt} = \kappa \rho E,
\end{gathered}
\label{eq:seir_asymp}
\end{equation}
where $\beta_0$ and $\beta_1$ are the transmission rates from symptomatic and
asymptomatic individuals, respectively, and $\rho$ is the fraction of
infections that become symptomatic. Only symptomatic cases contribute to the
observed new cases. This structure represents the role of asymptomatic
transmission in many respiratory infections.

\begin{figure}[H]
\centering
\begin{tikzpicture}[>=Latex]
\tikzstyle{comp}=[circle, draw, minimum size=12mm, inner sep=0pt, font=\large]
\tikzstyle{lab}=[font=\small, align=center]

\node[comp] (S) at (0,0) {$S$};
\node[comp] (E) at (3,0) {$E$};
\node[comp] (I) at (6,1) {$I$};
\node[comp] (A) at (6,-1) {$A$};
\node[comp] (R) at (9,0) {$R$};

\draw[->, thick] (S) -- (E) node[lab, pos=0.5, above] {$\dfrac{(\beta_0 I + \beta_1 A)S}{N}$};
\draw[->, thick] (E) -- (I) node[lab, pos=0.5, above left] {$\kappa \rho E$};
\draw[->, thick] (E) -- (A) node[lab, pos=0.5, below left] {$\kappa(1-\rho)E$};
\draw[->, thick] (I) -- (R) node[lab, pos=0.5, above right] {$\gamma I$};
\draw[->, thick] (A) -- (R) node[lab, pos=0.5, below right] {$\gamma A$};

\node[draw, rounded corners=3pt, fill=gray!10, minimum width=40mm, minimum height=8mm, anchor=north] (obs) at (4.5,-2.8) {Observation: $y(t) = \kappa \rho E$};
\draw[dashed, ->, shorten >=2pt] (I.south) -- ++(0,-0.8) -| (obs.north);
\end{tikzpicture}
\caption{SEIAR compartment diagram. Exposed individuals progress to either
symptomatic ($I$) or asymptomatic ($A$) states; only symptomatic cases are
observed.}
\label{fig:seir_asymp}
\end{figure}

\subsubsection{SEIR Model with Disease-Induced Mortality (SEIRD)}

The SEIRD model extends the SEIR model by adding a death compartment ($D$) for
disease-induced mortality~\cite{karami2026comparative}:
\begin{equation}
\begin{aligned}
\frac{dS}{dt} = -\frac{\beta S I}{N},\quad
\frac{dE}{dt} = \frac{\beta S I}{N} - \kappa E,\quad
\frac{dI}{dt} = \kappa E - \gamma I, \quad
\frac{dR}{dt} = (1-\rho) \gamma I, \quad
\frac{dD}{dt} = \rho \gamma I,
\end{aligned}
\label{eq:seird}
\end{equation}
where $\gamma$ is the removal rate from the infectious compartment and $\rho$
is the infection fatality fraction, representing the fraction of infected
individuals who die from the disease. Thus, $(1-\rho)\gamma I$ and
$\rho\gamma I$ represent the flows to recovery and disease-induced death,
respectively. When death data are used for calibration, the observation is the
number of new deaths in each observation interval.

\begin{figure}[H]
\centering
\begin{tikzpicture}[>=Latex]
\tikzstyle{comp}=[circle, draw, minimum size=12mm, inner sep=0pt, font=\large]
\tikzstyle{lab}=[font=\small, align=center]

\node[comp] (S) at (0,0) {$S$};
\node[comp] (E) at (3,0) {$E$};
\node[comp] (I) at (6,0) {$I$};
\node[comp] (R) at (9,1) {$R$};
\node[comp] (D) at (9,-1) {$D$};

\draw[->, thick] (S) -- (E) node[lab, pos=0.5, above] {$\dfrac{\beta S I}{N}$};
\draw[->, thick] (E) -- (I) node[lab, pos=0.5, above] {$\kappa E$};
\draw[->, thick] (I) -- (R) node[lab, pos=0.5, above] {$(1-\rho)\gamma I$};
\draw[->, thick] (I) -- (D) node[lab, pos=0.5, below] {$\rho \gamma I$};

\node[draw, rounded corners=3pt, fill=gray!10, minimum width=40mm, minimum height=8mm, anchor=north] (obs) at (4.5,-2.5) {Observation: $y(t) = \rho \gamma I$};
\draw[dashed, ->, shorten >=2pt] (D.south) -- ++(0,-0.3) -| (obs.east);
\end{tikzpicture}
\caption{SEIRD compartment diagram. Infectious individuals either recover or
die; new deaths are observed.}
\label{fig:seird}
\end{figure}
\section{Methodology}\label{sec:methodology}

This section describes the computational framework used to study practical
identifiability across ODE models, calibration-window lengths
(Table~\ref{tab:calibration_windows}), estimation scenarios
(Table~\ref{tab:scenarios}), and error structures, as described below.
The analysis combines ODE simulations, likelihood-based model fitting,
parametric bootstrap for uncertainty quantification, and results from repeated
runs to evaluate practical parameter constraint.

For each configuration defined by the model, estimation scenario, error
structure, and calibration-window length $T$, we: (i) generate synthetic data
from the ODE model, (ii) fit the model to the first $T$ observations,
(iii) quantify parameter uncertainty using parametric bootstrap, and
(iv) compute the PII from the spans of the bootstrap confidence intervals.
This procedure is repeated over $500$ independent replicates, each generated
by adding random noise to the noise-free synthetic data.

\subsection{Mathematical Framework\label{subsec:frame}}

Consider an ODE model with state vector
$\mathbf{x}(t)\in\mathbb{R}^{s}$ and parameter vector
$\boldsymbol{\theta}\in\Theta\subset\mathbb{R}^{p}$.
Given an initial condition $\mathbf{x}_0$, let
$\mathbf{x}(t;\boldsymbol{\theta})$ denote the solution of
\begin{equation}
  \frac{d\mathbf{x}}{dt}=\mathbf{g}(\mathbf{x},\boldsymbol{\theta}),
  \qquad \mathbf{x}(0)=\mathbf{x}_0.
  \label{eq:ivp}
\end{equation}

Assume that data are observed at discrete times $t=1,\dots,T$, where $T$
denotes the calibration-window length. As described in the previous section,
each observed quantity corresponds either to a state variable or to a quantity
derived from a cumulative state variable, such as its first-order finite
difference. Although the model diagrams and descriptions use continuous-time
notation such as $\frac{dC}{dt}$ or $\frac{dD}{dt}$ to represent new cases or
new deaths, the likelihood is evaluated using discrete one-step changes in the
cumulative variables. Specifically, fitted observations are based on increments
such as
$C(t;\boldsymbol{\theta})-C(t-1;\boldsymbol{\theta})$ or
$D(t;\boldsymbol{\theta})-D(t-1;\boldsymbol{\theta})$, representing the number
of new cases or new deaths over the observation interval.

Let $Q$ denote the number of observed quantities. For each $q=1,\ldots,Q$,
let $m_q$ denote the index of the associated state variable.
Let $\mathcal{J}\subset\{1,\ldots,Q\}$ be the set of indices for which
observations are given directly by state variables
(e.g., $I$ and $R$ in the multi-observable SEIR model), and let
$\mathcal{J}^c$ denote its complement, corresponding to first-order
finite-difference observations (e.g., new cases from $C$ in the growth, SIR,
SEIR, SEIR-UR, and SEIAR models, and new deaths from $D$ in the SEIRD model).
For each time $t$ and observation index $q$, we define the model-predicted value
as

\begin{equation}
\mu_{t,q}(\boldsymbol{\theta})=
\begin{cases}
x_{m_q}(t;\boldsymbol{\theta}), & q\in\mathcal{J}, \\[4pt]
x_{m_q}(t;\boldsymbol{\theta})-x_{m_q}(t-1;\boldsymbol{\theta}), & q\in\mathcal{J}^c.
\end{cases}
\label{eq:obs_map}
\end{equation}

We collect all values $\mu_{t,q}(\boldsymbol{\theta})$ and the corresponding
observations $y_{t,q}$, for $t=1,\dots,T$ and $q=1,\dots,Q$, into the vectors
$\boldsymbol{\mu}(\boldsymbol{\theta})\in\mathbb{R}^n$ and
$\mathbf{y}\in\mathbb{R}^n$, where $n=TQ$.

Given the fitted values $\mu_{t,q}(\boldsymbol{\theta})$, we model the
observations $Y_{t,q}$ using a stochastic process with mean
$\mu_{t,q}(\boldsymbol{\theta})$. Given the model-predicted means and, where
applicable, the dispersion parameter, observations are assumed independent
across observation times and observed quantities. Under a Poisson observation
model, we assume that
\[
Y_{t,q}\mid \mu_{t,q}(\boldsymbol{\theta})
\sim \mathrm{Poisson}\bigl(\mu_{t,q}(\boldsymbol{\theta})\bigr),
\]
and the corresponding negative log-likelihood is
\begin{equation}
\mathcal{J}_{\mathrm{Pois}}(\boldsymbol{\theta})
=-\sum_{t=1}^{T}\sum_{q=1}^{Q}
\left[
y_{t,q}\log \mu_{t,q}(\boldsymbol{\theta})
-\mu_{t,q}(\boldsymbol{\theta})
\right].
\label{eq:pois_obj}
\end{equation}

Alternatively, to account for overdispersion, we consider a negative-binomial
observation model with variance
\begin{equation}
\mathrm{Var}(Y_{t,q}\mid\mu_{t,q}(\boldsymbol{\theta}))
=\mu_{t,q}(\boldsymbol{\theta})+\alpha\,\mu_{t,q}(\boldsymbol{\theta}),
\qquad \alpha>0,
\label{eq:nb_var}
\end{equation}
where $\alpha>0$ is a dispersion parameter. Negative-binomial synthetic
datasets were generated using fixed dispersion values. The corresponding
negative log-likelihood is
\begin{equation}
\mathcal{J}_{\mathrm{NB}}(\boldsymbol{\theta},\alpha)
=
-\sum_{t=1}^{T}\sum_{q=1}^{Q}
\left[
\sum_{k=0}^{y_{t,q}-1}
\log\left(k+\frac{\mu_{t,q}(\boldsymbol{\theta})}{\alpha}\right)
+
y_{t,q}\log\alpha
-
\left(y_{t,q}+\frac{\mu_{t,q}(\boldsymbol{\theta})}{\alpha}\right)\log(1+\alpha)
\right].
\label{eq:nb_obj}
\end{equation}
Terms that do not depend on the fitted parameters, such as factorial terms,
are omitted from the negative log-likelihood.

We next describe how the simulated data are generated. Observations are
generated from the model using the observation operator described earlier and
the true parameter vector $\boldsymbol{\theta}^*$, after which observation
noise is added according to the specified error structures. We consider three
error structures: \texttt{Poisson}, \texttt{Negbin5}, and \texttt{Negbin10}.
For the negative-binomial synthetic datasets, \texttt{Negbin5} and
\texttt{Negbin10} denote dispersion values $\alpha=5$ and $\alpha=10$ used
to generate the data, respectively. During fitting, however, $\alpha$ is
estimated as an additional dispersion parameter rather than fixed at these
data-generating values. Each scenario is analyzed under all three error
structures.

Given the synthetic observations $\mathbf{y}$, the inverse problem is to
reconstruct $\boldsymbol{\theta}$ from data generated at the true parameter
vector $\boldsymbol{\theta}^*$ under the scenarios described below.
The calibrated estimate is obtained by solving the box-constrained
optimization problem

\begin{equation}
\hat{\boldsymbol{\theta}}
=\operatorname*{arg\,min}_{\boldsymbol{\theta}\in
[\boldsymbol{\theta}_L,\,\boldsymbol{\theta}_U]}
\mathcal{J}(\boldsymbol{\theta}),
\label{eq:opt}
\end{equation}
where $\mathcal{J}$ denotes the objective function defined in
Eqs.~\eqref{eq:pois_obj} or~\eqref{eq:nb_obj}. For negative-binomial fits, the
numerical optimization includes both the ODE parameter vector
$\boldsymbol{\theta}$ and the nuisance dispersion parameter $\alpha$, while PII
summaries are reported only for the model parameters of interest. Each
evaluation of $\mathcal{J}$ requires solving the initial-value
problem~\eqref{eq:ivp}. The optimization bounds for estimated parameters and
the fixed initial conditions used in the simulation experiments are reported in
Supplementary Tables~\ref{tab:supp_bounds} and
\ref{tab:supp_initial_conditions}.

Because $\mathcal{J}$ is generally non-convex, we use a global--local
optimization strategy. First, 30 initial points are sampled using a
Latin-hypercube maximin design within the feasible parameter box
$[\boldsymbol{\theta}_L,\boldsymbol{\theta}_U]$. From each initial point, a
sequential quadratic programming (SQP) algorithm~
\cite{nocedal2006numerical} solves the local box-constrained subproblem.
The solution with the smallest objective value across all starting points is
returned as $\hat{\boldsymbol{\theta}}$. This multi-start procedure is
implemented using MATLAB's \texttt{fmincon} solver with the SQP algorithm,
coordinated by the \texttt{MultiStart} framework. The estimation and
bootstrapping pipeline builds on the \textsc{QuantDiffForecast} toolbox~
\cite{chowell2024parameter}, which we extend to support the practical
identifiability analysis described below.

Structural identifiability depends on the exact model equations, observation
map, fixed and estimated parameters, and treatment of initial conditions.
The PII cannot resolve structural confounding; it should therefore be
interpreted only as a marginal interval-width diagnostic given the specified
inferential configuration and a separate assessment of structural
identifiability. The structural identifiability of the estimated parameters
under the corresponding model and observation configurations is summarized in
Supplementary Table~\ref{tab:structural_identifiability}.

\subsection{Parametric Bootstrap and Practical Identifiability Index
\label{subsec:Parametric Bootstrap}}

After estimating the parameter vector $\hat{\boldsymbol{\theta}}$,
we quantify parameter uncertainty using a parametric bootstrap~\cite{chowell2017fitting,karami2025parameter}. The fitted mean
curve $\hat{\mu}_{t,q}=\mu_{t,q}(\hat{\boldsymbol{\theta}})$ and the assumed
observation model are used to generate $B$ bootstrap datasets:
\[
y^{(b)}_{t,q}\sim D\big(\hat{\mu}_{t,q},\hat{\alpha}\big),
\qquad t=1,\ldots,T,\quad q=1,\ldots,Q,\quad b=1,\ldots,B,
\]
where $D$ denotes the assumed observation model. For Poisson analyses, no
dispersion parameter is used. For negative-binomial analyses, bootstrap
datasets are generated using the fitted dispersion parameter
$\hat{\alpha}$ from the original fit. Each bootstrap dataset is then refit by
solving the optimization problem~\eqref{eq:opt}. In negative-binomial
analyses, this refit re-estimates both the ODE parameter vector
$\boldsymbol{\theta}$ and the dispersion parameter $\alpha$, yielding
$\{(\hat{\boldsymbol{\theta}}^{(b)},\hat{\alpha}^{(b)})\}_{b=1}^{B}$.
However, the PII is computed only from the bootstrap distribution
$\{\hat{\boldsymbol{\theta}}^{(b)}\}_{b=1}^{B}$ for the ODE parameters of
interest.

For each parameter $\theta_j$, a 95\% bootstrap percentile confidence
interval is constructed as
\begin{equation}
\bigl[\theta_{j,L},\;\theta_{j,U}\bigr]
=\bigl[Q_{0.025}\!\bigl(\hat{\theta}_j^{(1:B)}\bigr),\;
Q_{0.975}\!\bigl(\hat{\theta}_j^{(1:B)}\bigr)\bigr],
\label{eq:ci_boot}
\end{equation}
where $Q_\tau$ denotes the $\tau$-quantile of the bootstrap sample.

To assess how tightly each parameter is constrained, we define the PII as the
base-10 logarithmic span of its confidence interval:
\begin{equation}
\mathrm{PII}_j =
\begin{cases}
\displaystyle
\log_{10}\!\left(
\frac{\theta_{j,U}}{\theta_{j,L}}
\right),
& 0<\theta_{j,L}\leq \theta_{j,U},\\[6pt]
+\infty,
& \theta_{j,L}=0<\theta_{j,U}.
\end{cases}
\label{eq:pii}
\end{equation}

For $0<\theta_{j,L}\leq\theta_{j,U}$,
\[
\mathrm{PII}_j
=
\log_{10}(\theta_{j,U})
-
\log_{10}(\theta_{j,L}),
\]
so the PII is exactly the width of the confidence interval on the base-10 logarithmic parameter scale.
Equivalently, it is the base-10 logarithm of the multiplicative
confidence-interval ratio. Thus,
$\mathrm{PII}_j=0$ only for a degenerate interval with
$\theta_{j,U}=\theta_{j,L}>0$, whereas larger values indicate greater
multiplicative confidence-interval width. If
$\theta_{j,L}=0<\theta_{j,U}$, we define $\mathrm{PII}_j=+\infty$
because the multiplicative confidence-interval ratio is unbounded when the
lower confidence bound reaches zero. This boundary case is not a numerical
failure; it indicates an unbounded span on the multiplicative scale, which may
differ from the interval width on the original parameter scale.

For any positive unit-conversion factor $c$, replacing
$(\theta_{j,L},\theta_{j,U})$ by
$(c\theta_{j,L},c\theta_{j,U})$ leaves the index unchanged because
$\log_{10}(c\theta_{j,U}/c\theta_{j,L})
=\log_{10}(\theta_{j,U}/\theta_{j,L})$.
Therefore, the PII is unchanged by multiplicative changes in parameter units.
It is also invariant under reciprocal transformation when the confidence
interval is transformed accordingly: an interval $[L,U]$ for $\theta$
corresponds to $[1/U,1/L]$ for $1/\theta$ and yields the same PII.

For reference, because $U/L=10^{\mathrm{PII}}$, PII values of
$0.01$, $0.1$, and $1$ correspond to endpoint ratios
$U/L\approx1.023$, $1.259$, and $10$, respectively.
These are endpoint ratios rather than symmetric percentage errors around
the parameter estimate.

Table~\ref{tab:pii_interpretation} provides an interpretive scale. Because the
PII is based on one-dimensional interval width, it does not by itself describe
the full shape of the likelihood or posterior, including parameter
correlations, flat likelihood ridges, multimodality, structural
non-identifiability, or estimator bias.

\begin{table}[H]
\centering
\caption{Interpretation of Practical Identifiability Index (PII) values.}
\label{tab:pii_interpretation}
\begin{tabular}{cl}
\toprule
\textbf{PII Range} & \textbf{Interpretation} \\
\midrule
$\mathrm{PII} < 0.1$ &
Tightly constrained; uncertainty is tightly concentrated \\
$0.1 \leq \mathrm{PII} < 1$ &
Weakly constrained; uncertainty is wider but remains below one order of magnitude \\
$\mathrm{PII} \geq 1$ &
Poorly constrained; uncertainty spans at least one order of magnitude ($\geq 10\times$) \\
\bottomrule
\end{tabular}
\end{table}

When the median PII is below 0.1 but the empirical 95\% CI coverage of the true
value is below 90\%, we retain the PII-based classification as tightly
constrained and flag the result as \emph{tightly constrained with low coverage}.
The 90\% cutoff is used only as a threshold for descriptive purposes to flag
low coverage relative to the nominal 95\% level and is not intended as a formal
test of interval calibration. Thus, low empirical coverage is treated as an
additional flag rather than as a separate PII constraint category.

The threshold $\mathrm{PII}<0.1$ should be interpreted as indicating tightly
constrained marginal uncertainty rather than guaranteed full practical
identifiability. Similarly, the thresholds used throughout this study are
intended primarily to support comparisons across models, parameters, and
calibration settings rather than to provide universal identifiability criteria.

For each model, error structure, calibration-window length $T$, and
parameter-estimation scenario (i.e., which subset of model parameters is
estimated versus held fixed), the procedure is repeated independently over
$R=500$ replicates. In each replicate, one synthetic dataset is generated by
adding observation noise to the true mean curve, the model is calibrated to
that dataset, and the parametric bootstrap ($B=300$) yields a 95\% CI from
which $\mathrm{PII}_j$ is computed, where $j\in\{1,\dots,p\}$ represents the
estimated model parameters of interest. The nuisance dispersion parameter
$\alpha$ is not included in the PII summaries. The $\mathrm{PII}_j$ values
are then summarized across replicates by their median and central 95\% interval

\begin{equation}
\widetilde{\mathrm{PII}}_j
=Q_{0.5}\!\bigl\{\mathrm{PII}_{j,r}\bigr\}_{r=1}^{R},
\qquad
\mathrm{PII}_j^{L/U}
=Q_{0.025/0.975}\!\bigl\{\mathrm{PII}_{j,r}\bigr\}_{r=1}^{R}.
\label{eq:pii_summary}
\end{equation}

Values of $+\infty$ are retained in the PII values across replicates and in
the corresponding summaries. For visualization only, when a subplot contains
PII summary values at or above 3, those values are displayed at an upper
plotting limit of 3 (labeled $\geq 3$) after the medians and quantiles have
been calculated. This display convention does not affect classifications,
coverage, threshold-window calculations, tables, or scientific conclusions.


\subsection{Relationship between PII and Existing Identifiability Diagnostics}

Practical identifiability is commonly assessed using profile likelihood
analysis, Fisher information--based measures, Bayesian posterior summaries,
and sensitivity-based diagnostics~
\cite{raue2009structural,gutenkunst2007universally,tuncer2018structural}.
Recent work emphasizes that these approaches are complementary, with profile
likelihood methods, sensitivity analyses, and Bayesian--frequentist frameworks
capturing different aspects of parameter estimability~
\cite{raue2009structural,heinrich2025different,raue2013joining,mohammed2025comparing,simpson2023profile}.
Although informative, these approaches often require interpretation based on
plots, heuristics, or individual parameters and do not provide a standardized
scalar summary for comparing many parameters, models, or calibration settings.

Profile likelihoods can reveal flat directions and support interval
construction, but their interpretation often depends on visual inspection.
Bayesian analyses characterize uncertainty through posterior distributions,
but comparisons across parameters usually require summaries such as
credible-interval widths or coefficients of variation. Sensitivity and
Fisher-information methods quantify local influence or estimability, but often
rely on linear approximations or heuristic thresholds. As a result, combining
parameter-constraint information across high-dimensional dynamical systems can
be difficult.

Confidence-interval width (CIW), normalized confidence-interval width (NCIW),
and coefficient-of-variation (CV) summaries have all been used to describe
parameter uncertainty~
\cite{lam2024evolving,eisenberg2014determining,
yoganandan2015lower,banerjee2020uncertainty}.
CIW describes uncertainty on the original parameter scale, whereas NCIW and CV
provide relative uncertainty measures normalized by the parameter estimate.
Table~\ref{tab:uncertainty_comparison} provides a simple comparison of these
summaries with the PII.

\begin{table}[H]
\centering
\caption{Comparison of the Practical Identifiability Index (PII) with common
summaries of marginal parameter uncertainty.}
\label{tab:uncertainty_comparison}
\small
\begin{tabular}{lccc}
\hline
\textbf{Summary} &
\textbf{Unit-free?} &
\textbf{Needs estimate?} &
\textbf{CI endpoints only?} \\
\hline
CIW  & No  & No  & Yes \\
NCIW & Yes & Yes & No  \\
CV   & Yes & Yes & No  \\
PII  & Yes & No  & Yes \\
\hline
\end{tabular}
\end{table}

CIW measures uncertainty on the original parameter scale and therefore depends
on the parameter units. NCIW and CV are unit-free but require the parameter
estimate in their calculation. The PII is also unit-free, but it is computed
directly from the confidence-interval endpoints. Its logarithmic form expresses
interval width on an order-of-magnitude scale and is invariant to multiplicative
changes of parameter units, allowing comparison across positive-valued
parameters with different magnitudes. A zero lower confidence bound is
represented explicitly by $\mathrm{PII}=+\infty$, indicating an unbounded
multiplicative confidence-interval ratio rather than a numerical failure.

The PII complements rather than replaces established diagnostics. A low PII
indicates a narrow marginal interval given the model, data, and inference
procedure used, but it does not guarantee unbiased estimation or rule out
correlations, likelihood ridges, multimodality, or structural
non-identifiability. Conversely, a high PII indicates that a parameter may
require further investigation using profile likelihoods, posterior diagnostics,
sensitivity analysis, or structural identifiability methods. Reported together
with bootstrap intervals, empirical coverage, profile likelihoods, or Bayesian
posterior summaries, the PII provides a simple way to compare marginal
parameter uncertainty across models, calibration windows, and
observation-noise scenarios.
\subsection{Simulation Study Design\label{subsec:simulation}}

As described earlier, we assess practical parameter constraint using
synthetic-data experiments in which the true parameters
$\boldsymbol{\theta}^{*}$ are known. This allows us to evaluate directly how
well the inverse problem~\eqref{eq:opt} recovers the parameters under
controlled conditions. Four factors are systematically varied: model,
calibration-window length $T$, estimation scenario, and error structure
(\texttt{Poisson}, \texttt{Negbin5}, \texttt{Negbin10}). The
negative-binomial labels denote dispersion values used to generate the data,
rather than fixed values used during fitting.

For each configuration, the model solution at the true parameter vector
$\boldsymbol{\theta}^{*}$ is used to generate synthetic observations according
to~\eqref{eq:obs_map} and the specified error structure. For a calibration
window of length $T$, only the first $T$ observations are used for model
fitting and PII calculation.

This procedure is repeated independently over $R=500$ replicates. In each
replicate, a synthetic dataset is generated, the model is fitted, and a
parametric bootstrap with $B=300$ bootstrap datasets is used to construct a
95\% confidence interval for each estimated model parameter, from which
$\mathrm{PII}_j$ is computed. The resulting $\mathrm{PII}_j$ values are
summarized across replicates using their median and central 95\% interval.

As described earlier, the calibration-window lengths for each model class are
listed in Table~\ref{tab:calibration_windows}.

\begin{table}[H]
\centering
\caption{Calibration-window lengths $T$ (in time units) used for each
model class.}
\label{tab:calibration_windows}
\begin{tabular}{ll}
\toprule
\textbf{Model} & \textbf{Window lengths} $T$ \\
\midrule
EXP, GGM     & 20, 30, 40, 50 \\
GLM          & 20, 30, \ldots, 90 \\
SIR          & 10, 20, \ldots, 70 \\
SEIR, SEIRD, SEIR-UR, SEIAR & 20, 30, \ldots, 100 \\
SEIRMO       & 30, 35, \ldots, 70 \\
\bottomrule
\end{tabular}
\end{table}

\begin{figure}[H]
\centering
\includegraphics[width=\linewidth]{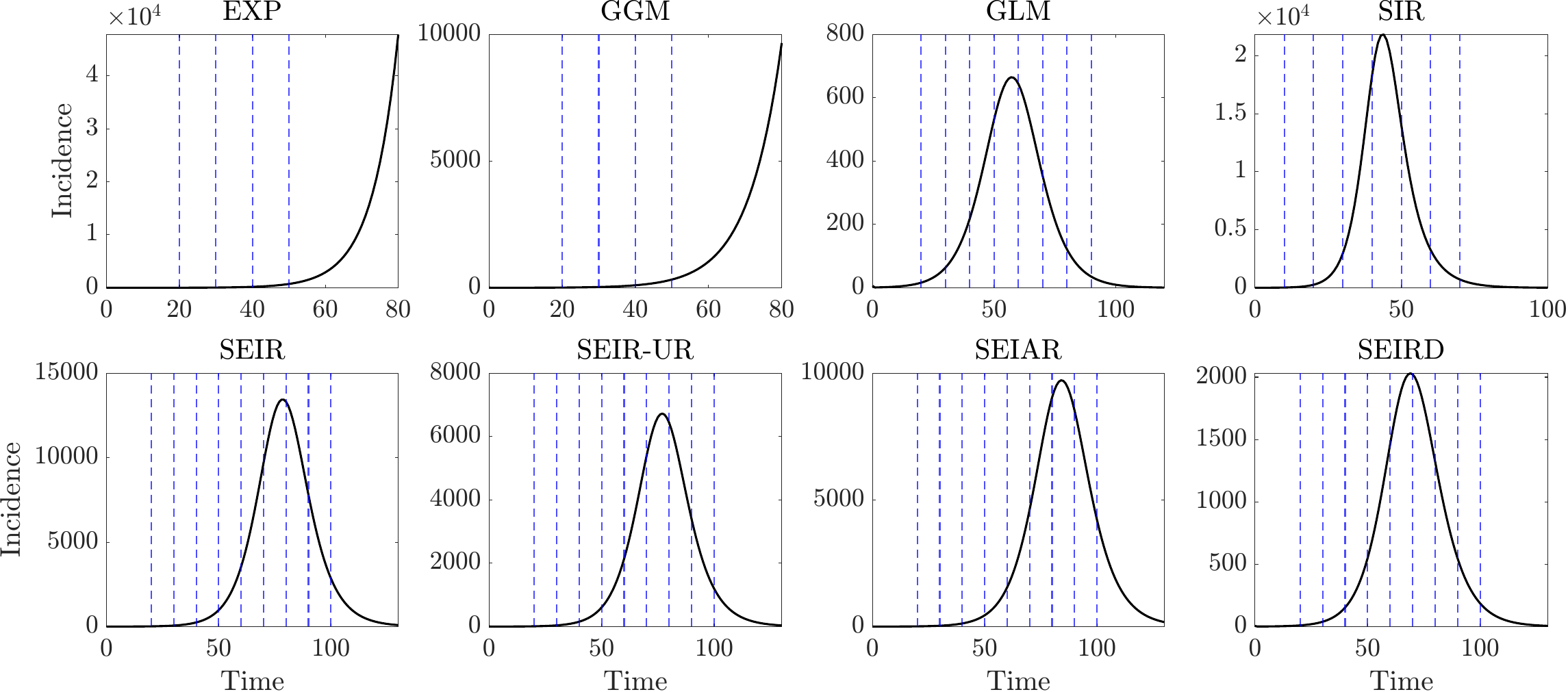}
\caption{
Noiseless simulated trajectories of new cases or new deaths for the models
considered: EXP, GGM, GLM, SIR, SEIR, SEIR-UR, SEIAR, and SEIRD. Vertical
blue dashed lines show the increasing calibration windows used for parameter
estimation. The SEIR multi-observable analysis (SEIRMO) uses the same SEIR
trajectory but varies the observation operator and is therefore summarized
separately in Figure~\ref{fig:seirmo_observability}.
}
\label{fig:simulated_data_all_models}
\end{figure}

Figure~\ref{fig:simulated_data_all_models} shows the noiseless trajectories
for each model together with the selected calibration windows (vertical dashed
lines). We use multiple calibration windows to assess how practical parameter
constraint depends on the amount of data available for model fitting. In
particular, we consider progressively longer windows to mimic realistic
settings in which only partial data are available. For models with a peak,
such as GLM and the compartmental models, these windows span different phases
of the epidemic trajectory, from the early growth phase to near-peak and
post-peak periods. For models with monotone growth, such as EXP and GGM, the
same sequence of window lengths covers only the early phase, because parameter
estimation becomes relatively straightforward once a sufficiently long segment
of the trajectory is observed.

For each model, we consider multiple scenarios in which progressively more
parameters are estimated while the remaining parameters are held fixed at
their true values (Table~\ref{tab:scenarios}). This design allows us to
examine how greater model flexibility and a larger number of estimated
parameters affect practical parameter constraint.

Table~\ref{tab:true_values} lists the true parameter values used to generate
the synthetic data for each model.

\begin{table}[H]
\centering
\caption{Estimated and fixed ODE parameters for each model and scenario. In
negative-binomial analyses, the dispersion parameter $\alpha$ is treated as a
nuisance parameter and estimated during fitting, but PII summaries are reported
only for the ODE parameters of interest. For SEIRMO observation labels,
$dC/dt$ is the continuous-time shorthand for new cases; fitting uses the
discrete increment of $C$ as in Eq.~\eqref{eq:obs_map}.}
\label{tab:scenarios}
\small
\begin{tabular}{llll}
\toprule
\textbf{Model} & \textbf{Scenario} & \textbf{Estimated} & \textbf{Fixed} \\
\midrule
EXP & 1 & $r$ & -- \\
\midrule
GGM & 1 & $r,\, p$ & -- \\
\midrule
GLM & 1 & $r,\, p,\, K$ & -- \\
\midrule
\multirow{2}{*}{SIR}
  & 1 & $\beta$ & $\gamma,\, N$ \\
  & 2 & $\beta,\, \gamma$ & $N$ \\
\midrule
\multirow{3}{*}{SEIR}
  & 1 & $\beta$ & $\kappa,\, \gamma,\, N$ \\
  & 2 & $\beta,\, \gamma$ & $\kappa,\, N$ \\
  & 3 & $\beta,\, \gamma,\, \kappa$ & $N$ \\
\midrule
\multirow{3}{*}{SEIR-UR}
  & 1 & $\beta$ & $\kappa,\, \rho,\, \gamma,\, N$ \\
  & 2 & $\beta,\, \rho$ & $\kappa,\, \gamma,\, N$ \\
  & 3 & $\beta,\, \rho,\, \gamma$ & $\kappa,\, N$ \\
\midrule
\multirow{3}{*}{SEIAR}
  & 1 & $\beta_0,\, \beta_1$ & $\kappa,\, \rho,\, \gamma,\, N$ \\
  & 2 & $\beta_0,\, \beta_1,\, \rho$ & $\kappa,\, \gamma,\, N$ \\
  & 3 & $\beta_0,\, \beta_1,\, \rho,\, \gamma$ & $\kappa,\, N$ \\
\midrule
\multirow{3}{*}{SEIRD}
  & 1 & $\beta$ & $\kappa,\, \rho,\, \gamma,\, N$ \\
  & 2 & $\beta,\, \rho$ & $\kappa,\, \gamma,\, N$ \\
  & 3 & $\beta,\, \rho,\, \gamma$ & $\kappa,\, N$ \\
\midrule
\multirow{3}{*}{SEIRMO}
  & 1 & $\beta,\, \kappa,\, \gamma$ \ (obs.\ $dC/dt$) & $N$ \\
  & 2 & $\beta,\, \kappa,\, \gamma$ \ (obs.\ $dC/dt,\, I$) & $N$ \\
  & 3 & $\beta,\, \kappa,\, \gamma$ \ (obs.\ $dC/dt,\, I,\, R$) & $N$ \\
\bottomrule
\end{tabular}
\end{table}

\begin{table}[H]
\centering
\caption{True parameter values used to generate synthetic data for each model.}
\label{tab:true_values}
\small
\begin{tabular}{ll}
\toprule
\textbf{Model} & \textbf{True parameter values} \\
\midrule
EXP     & $r=0.14$ \\
GGM     & $r=0.14,\; p=0.98$ \\
GLM     & $r=0.16,\; p=0.98,\; K=20{,}000$ \\
SIR     & $\beta=0.50,\; \gamma=0.25,\; N=500{,}000,\; I_0=4$ \\
SEIR    & $\beta=0.50,\; \kappa=0.50,\; \gamma=0.25,\; N=500{,}000,\; I_0=4$ \\
SEIR-UR & $\beta=0.50,\; \kappa=0.50,\; \rho=0.50,\; \gamma=0.25,\; N=500{,}000,\; I_0=5$ \\
SEIAR   & $\beta_0=0.50,\; \beta_1=0.40,\; \kappa=0.50,\; \rho=0.80,\; \gamma=0.25,\; N=500{,}000,\; I_0=2,\; A_0=2$ \\
SEIRD   & $\beta=0.50,\; \kappa=0.50,\; \rho=0.80,\; \gamma=0.25,\; N=100{,}000,\; I_0=5$ \\
SEIRMO  & $\beta=0.50,\; \kappa=0.50,\; \gamma=0.25,\; N=500{,}000,\; I_0=4$ \\
\bottomrule
\end{tabular}
\end{table}

We also examine whether additional observables improve practical parameter
constraint in SEIR Scenario 3, where $\beta$, $\kappa$, and $\gamma$ are
estimated together. In addition to the standard single-observable
configuration based on the continuous-time new-case rate
$y(t)=\kappa E$, we compare three observation configurations, where
$\frac{dC}{dt}$ denotes the continuous-time shorthand for new cases:
(i) new cases only, $\frac{dC}{dt}$;
(ii) new cases and prevalence, $\left(\frac{dC}{dt}, I\right)$; and
(iii) new cases, prevalence, and recovered counts,
$\left(\frac{dC}{dt}, I, R\right)$.
All other settings, including true parameter values, error structure, and
bootstrap settings, remain unchanged from the standard SEIR experiments.
The SEIRMO analysis uses calibration windows $T=30,35,\ldots,70$.

\subsection{Software Implementation}

The simulation, calibration, bootstrap, and plotting workflow is implemented
in MATLAB and builds on the \textsc{QuantDiffForecast} toolbox~
\cite{chowell2024parameter}. The code allows users to select the model,
estimation scenario, error structure, calibration-window lengths, and number
of simulation replicates. Both local and high-performance computing (HPC)
versions are provided. Details of the software workflow, including example
option files, model handles, and use of SLURM arrays, are provided in the
Supplementary Material and the public code repository.

\section{Results}\label{sec:results}

We evaluate how the PII summarizes marginal parameter uncertainty across
growth and compartmental epidemic models. In the simulation study, across
500 replicates, multiple calibration windows, and Poisson and
negative-binomial error structures, four main patterns appear repeatedly:
longer windows generally reduce PII, higher observation noise is associated
with larger PII in most but not all configurations, more complex models delay
parameter constraint through parameter coupling, and additional observables
improve constraint for latent or indirectly informed parameters. These trends
mainly serve as a check that the index behaves as expected.

\subsection{Growth Models (EXP, GGM, GLM)}

We first examine EXP, GGM, and GLM. Across these models, PII generally
decreases as the calibration-window length increases, with parameters moving
from $\mathrm{PII}\geq1$ to below the one-order-of-magnitude threshold
($\mathrm{PII}<1$), and reaching the more stringent tightly constrained range
only when $\mathrm{PII}<0.1$. This pattern is clearest for the exponential
growth rate $r$ (Figure~\ref{fig:EXP_S1_PII_parameters1}), whose median PII
declines steadily across error structures as more data are added; the
corresponding confidence intervals narrow around the true value
(Figure~\ref{fig:EXP_S1_CI_grid_param_1_1}).

The GGM shows parameter-specific uncertainty patterns. Across most windows
and error structures, the deceleration parameter $p$ has lower PII than the
growth rate $r$ (Supplementary Figure~\ref{fig:GGM_S1_PII_parameters}; see also
Supplementary Figures~\ref{fig:GGM_S1_CI_grid_param_1}--\ref{fig:GGM_S1_CI_grid_param_2}).
At the final window, median PII values are approximately $0.01$--$0.03$ for
$p$, compared with $0.07$--$0.14$ for $r$. The initial
\texttt{Negbin10} window is an exception, where $p$ has an infinite median PII.

In the GLM, the final epidemic size $K$ is the least well-constrained
parameter (Supplementary Figure~\ref{fig:GLM_S1_PII_parameters}; see also
Supplementary Figures~\ref{fig:GLM_S1_CI_grid_param_1}--\ref{fig:GLM_S1_CI_grid_param_3}).
Because $K$ governs saturation, pre-peak windows provide limited information
about its value. Under negative-binomial error structures, PII for $K$ can
increase as the trajectory begins to slow but still lacks enough post-peak
information; once post-peak observations are included, PII drops sharply.
Thus, parameters linked to early growth are constrained sooner than parameters
related to saturation, and observation noise generally increases uncertainty.

\begin{figure}[H]
\centering
\includegraphics[width=\linewidth]{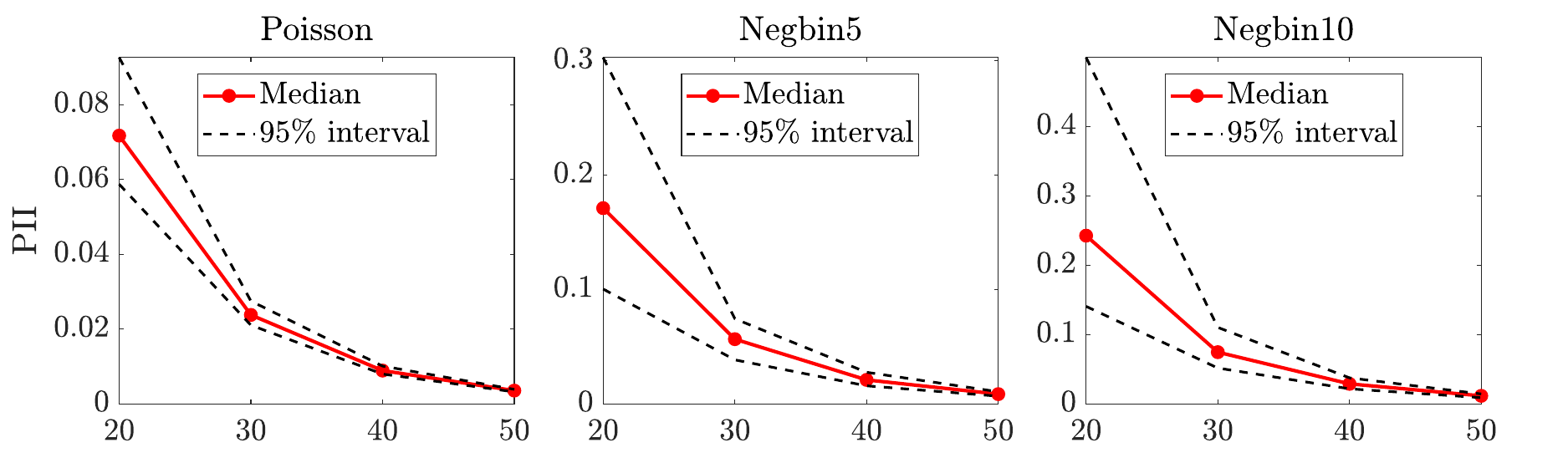}
\caption{Practical Identifiability Index (PII) for the exponential growth rate
$r$ in the EXP model across calibration-window lengths $T=20,30,40,50$ under
Poisson, Negbin5, and Negbin10 error structures. Red lines show the median PII
across replicates; dashed black curves show the central 95\% interval across
replicates.}
\label{fig:EXP_S1_PII_parameters1}
\end{figure}

\begin{figure}[H]
\centering
\includegraphics[width=0.9\linewidth]{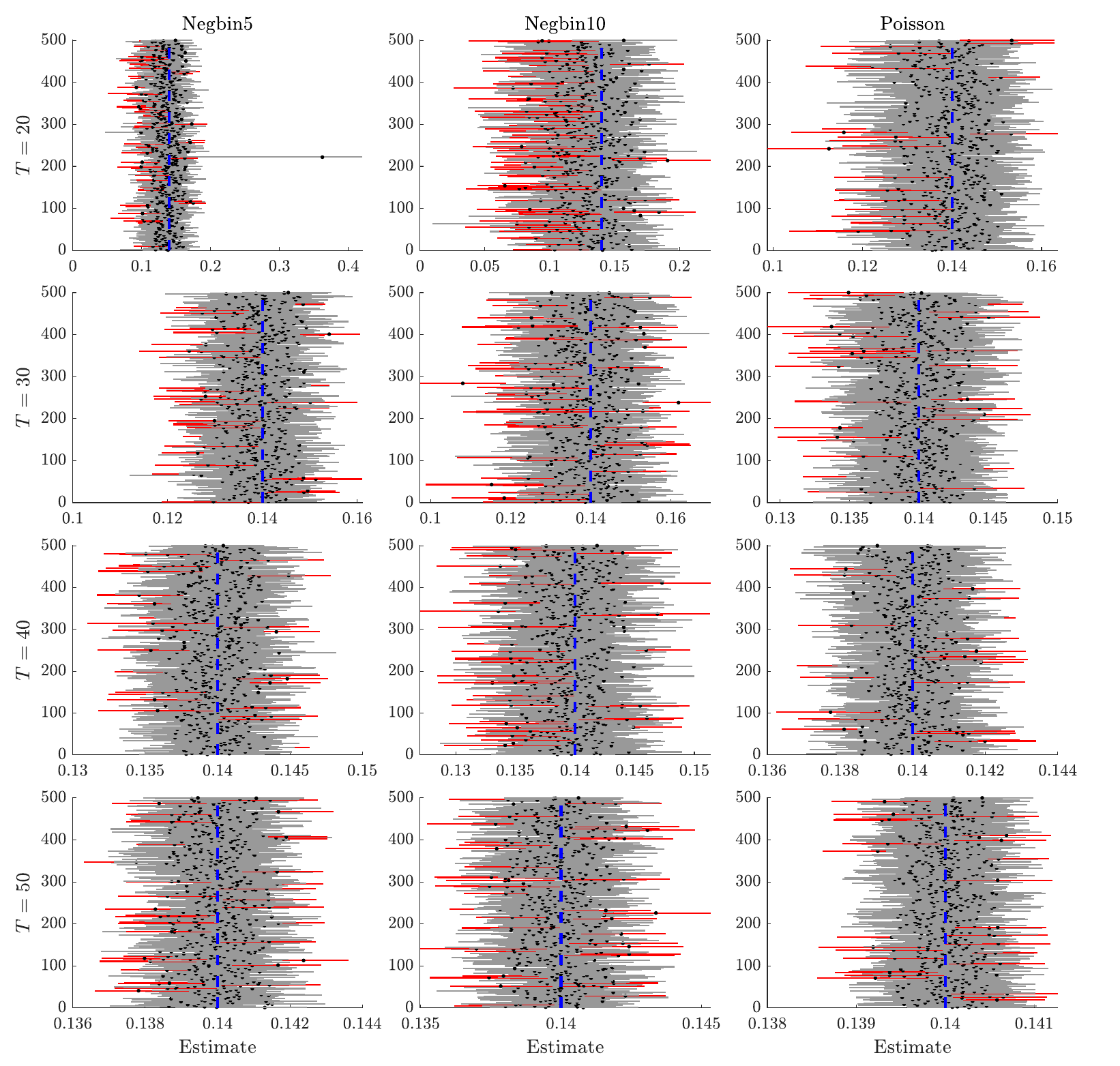}
\caption{Parameter estimates and 95\% bootstrap confidence intervals for the
exponential growth rate $r$ across 500 replicates and calibration windows
$T=20,30,40,50$. The vertical blue dashed line marks the true value $r=0.14$.
Columns show Negbin5, Negbin10, and Poisson error structures. Red intervals
fail to cover the true value, whereas gray intervals cover it. Red intervals
are more common under shorter windows and higher-noise error structures,
showing that empirical coverage should be considered alongside PII.}
\label{fig:EXP_S1_CI_grid_param_1_1}
\end{figure}

\subsection{SIR and SEIR Models}

We next consider the SIR and SEIR models. In the SIR model, the transmission
rate $\beta$ is well constrained by early epidemic growth when estimated
alone. When $\beta$ and $\gamma$ are estimated together, however, PII values
increase, especially for $\gamma$ and under negative-binomial error
structures, because early new-case data provide limited information for
distinguishing transmission from recovery.

The SEIR model shows the same pattern, but the latent exposed compartment
delays parameter constraint
(Figure~\ref{fig:SEIR_S1_PII_parameters1}; Supplementary
Figures~\ref{fig:SEIR_S2_PII_parameters} and~\ref{fig:SEIR_S3_PII_parameters}).
The incubation rate $\kappa$ and recovery rate $\gamma$ often remain at or
above $\mathrm{PII}=1$ for short-to-moderate windows and fall below this
threshold only when longer windows include later epidemic phases. They reach
the stricter $\mathrm{PII}<0.1$ range near the end of the simulated epidemic
(Supplementary Figures~\ref{fig:SEIR_S3_CI_grid_param_2}--\ref{fig:SEIR_S3_CI_grid_param_3}).

Across SIR and SEIR models, higher observation noise generally increases PII,
although this pattern is not uniform in jointly estimated SEIR Scenario~3 at
intermediate calibration windows.
Figure~\ref{fig:SEIR_S1_CI_grid_param_1_1} further shows that bootstrap
confidence intervals can fail to cover the true value under short windows and
higher-noise settings, even when the intervals are relatively narrow. Thus,
PII and empirical coverage provide complementary information: PII summarizes
marginal interval width, whereas coverage evaluates whether intervals contain
the true value across repeated simulations. Overall, parameters strongly
informed by early epidemic dynamics become constrained sooner, whereas
recovery, latency, and later-phase parameters require longer windows.

\subsection{Extended Models: SEIR-UR, SEIAR, SEIRD}

We next examine models with under-reporting, asymptomatic transmission, or
disease-induced mortality. In these models, parameters associated with latent,
partially observed, or indirectly observed processes tend to have larger PII
values than parameters directly linked to observed cases or deaths
(Supplementary Figures~\ref{fig:SEIRunrep_S1_PII_parameters}--\ref{fig:SEIRunrep_S3_PII_parameters}
for SEIR-UR; Supplementary
Figures~\ref{fig:SEIRasymp_S1_PII_parameters}--\ref{fig:SEIRasymp_S3_PII_parameters}
for SEIAR; Supplementary
Figures~\ref{fig:SEIRD_S1_PII_parameters}--\ref{fig:SEIRD_S3_PII_parameters}
for SEIRD).

Estimating more parameters jointly also increases uncertainty across
parameters. This effect is most apparent in scenarios with more estimated
parameters, where parameter coupling reduces the information available for
each parameter (see corresponding CI grid figures in the Supplementary
Material). The SEIAR model is the most challenging case: parameters governing
asymptomatic transmission are informed only indirectly by symptomatic case
observations and therefore remain weakly constrained for longer than parameters
directly linked to observed cases. These results show that practical parameter
constraint can worsen substantially as model complexity increases, even under
controlled synthetic-data conditions.

\begin{figure}[H]
\centering
\includegraphics[width=\linewidth]{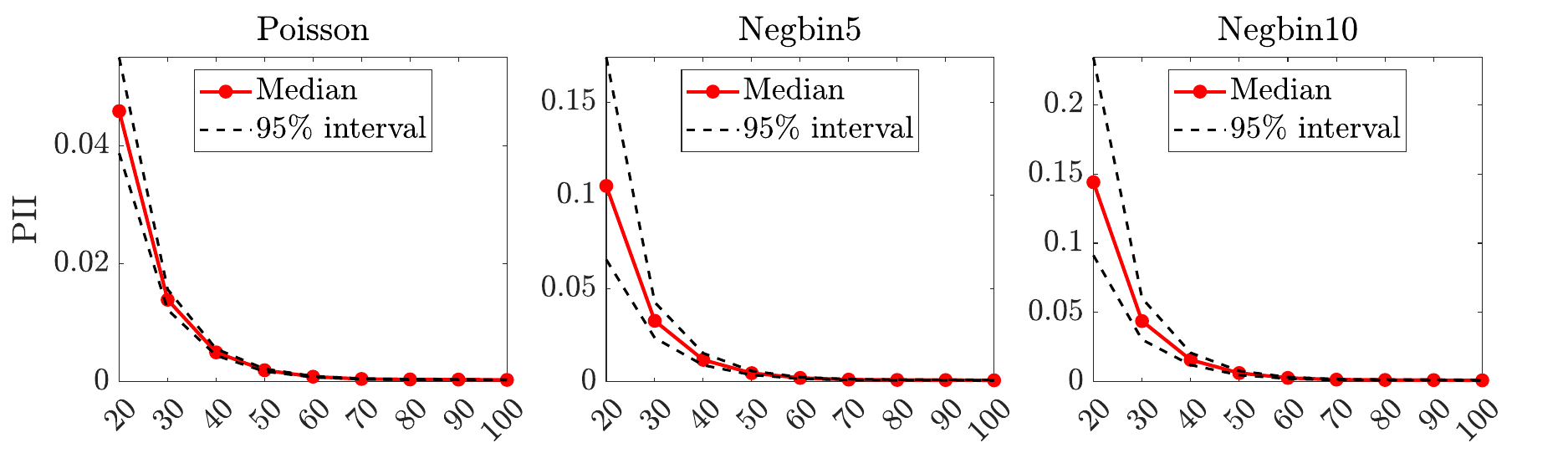}
\caption{Practical Identifiability Index (PII) for the transmission rate
$\beta$ in the SEIR model under Scenario~1 across calibration-window lengths
$T=20,30,\ldots,100$ under Poisson, Negbin5, and Negbin10 error structures.
Red lines show the median PII across replicates; dashed black curves show the
central 95\% interval across replicates.}
\label{fig:SEIR_S1_PII_parameters1}
\end{figure}

\begin{figure}[p]
\centering
\includegraphics[
width=\linewidth,
trim=0 13in 0 0,
clip
]{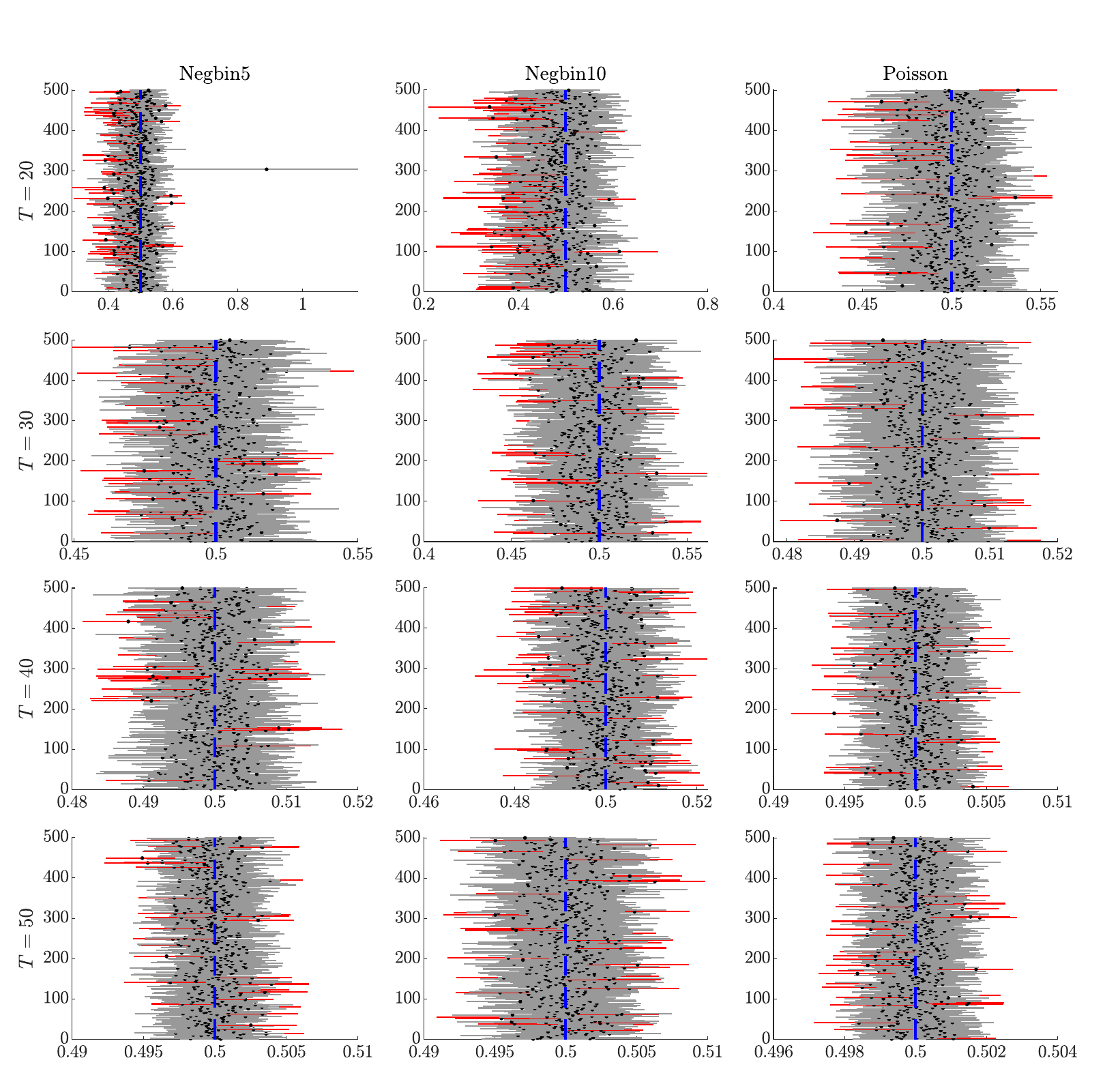}
\caption{Parameter estimates and 95\% bootstrap confidence intervals for the
transmission rate $\beta$ in SEIR Scenario~1 across 500 replicates and
calibration windows $T=20,\ldots,100$. The vertical blue dashed line marks the
true value $\beta=0.5$. Columns show Negbin5, Negbin10, and Poisson error
structures. Red intervals fail to cover the true value, whereas gray intervals
cover it. The figure shows that interval width and empirical coverage provide
complementary information.}
\label{fig:SEIR_S1_CI_grid_param_1_1}
\end{figure}

\begin{figure}[p]
\ContinuedFloat
\centering
\includegraphics[
width=\linewidth,
trim=0 0 0 10.4in,
clip
]{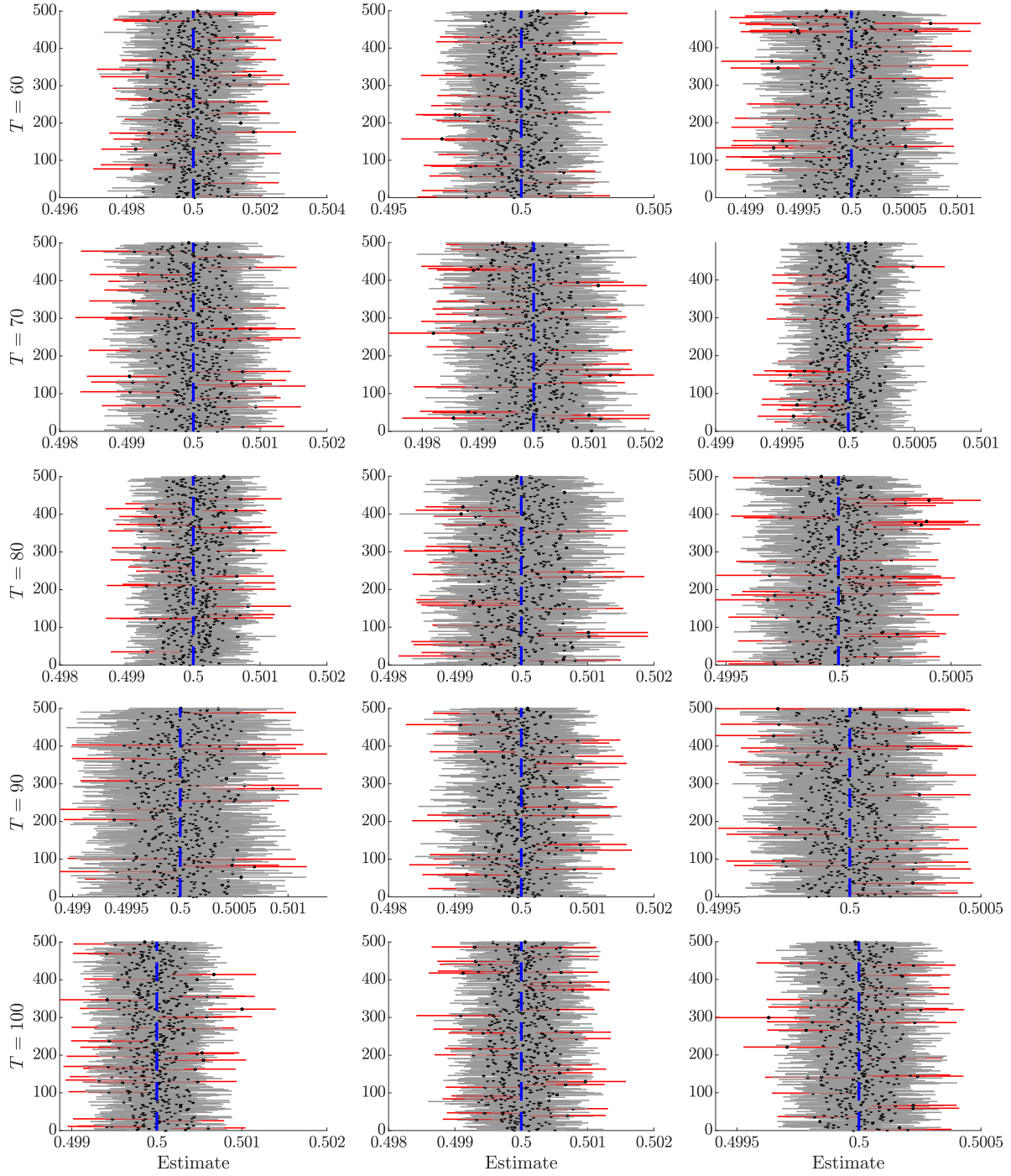}
\caption[]{Parameter estimates and 95\% bootstrap confidence intervals for the
transmission rate $\beta$ in the SEIR model under Scenario~1 (continued).}
\end{figure}

\subsection{Effect of Additional Observables}

We then assess how additional observables affect parameter constraint in the
SEIR model. The multi-observable analysis compares three observation
configurations based on new cases, infectious individuals, and recovered
counts (Figure~\ref{fig:seirmo_observability}). Adding observables substantially
reduces PII across parameters, often moving parameters from
$\mathrm{PII}\geq1$ under the single-observable configuration to below the
one-order-of-magnitude threshold in the multi-observable configurations
(Supplementary Figures~\ref{fig:SEIR_vars_dCdt_PII},
~\ref{fig:SEIR_vars_2vars_PII}, and~\ref{fig:SEIR_vars_3vars_PII}).

This effect is strongest for parameters associated with latent progression and
recovery. The incubation and recovery rates remain weakly constrained when
calibration uses only new cases but become more tightly constrained when
infectious and recovered counts are added
(compare Supplementary Figure~\ref{fig:SEIR_vars_dCdt_CI_param_2} with
Supplementary Figures~\ref{fig:SEIR_vars_2vars_CI_param_2}
and~\ref{fig:SEIR_vars_3vars_CI_param_2}). Thus, more informative observation
configurations can partly compensate for short calibration windows and
parameter coupling, and PII provides a direct way to quantify the added value
of additional observables.

\begin{figure}[H]
\centering
\includegraphics[width=\linewidth]{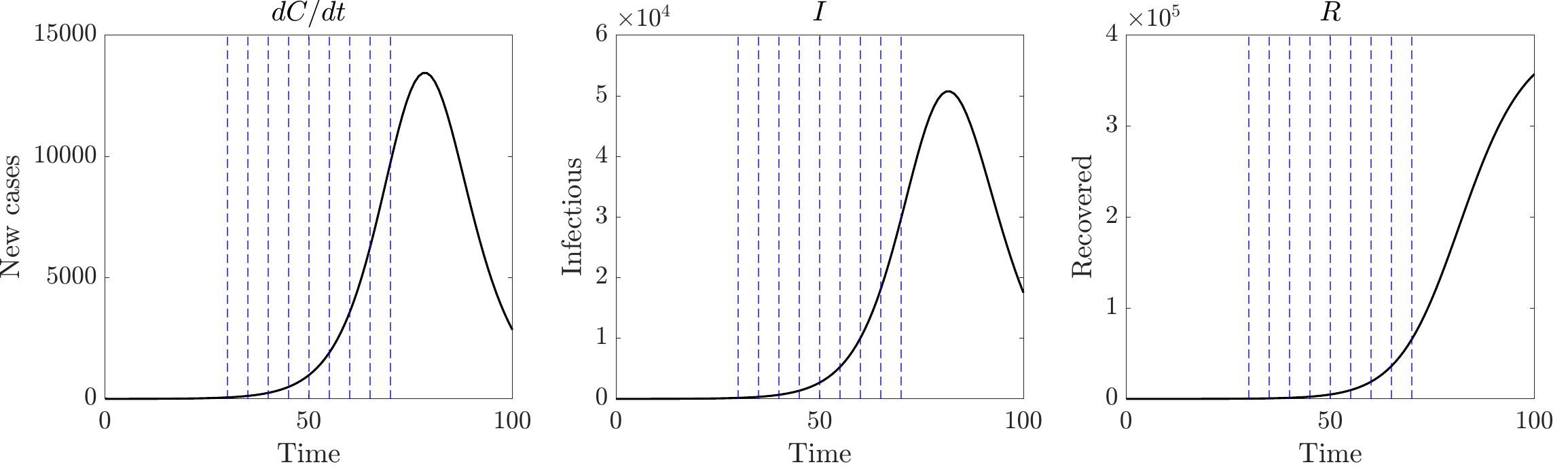}
\caption{SEIR multi-observable trajectories for calibration windows
$T=30,35,\ldots,70$. Panels show new cases, infectious individuals $I$, and
recovered counts $R$, which define the single- and multi-observable
configurations. Additional observables make the data more informative and
reduce PII for indirectly informed parameters.}
\label{fig:seirmo_observability}
\end{figure}

\subsection{Cross-Model Synthesis}

Figure~\ref{fig:pii_forest_summary} summarizes the distribution of
final-window PII values across models, while the full parameter-level summary
across models, including threshold windows, empirical coverage, final-window
PII values, and classifications, is provided in Supplementary
Table~\ref{tab:supp_cross_model_synthesis}. Each parameter is classified using
the median PII at the longest calibration window: \texttt{Tightly constrained}
if $\mathrm{PII}<0.1$, \texttt{Weakly constrained} if
$0.1\leq\mathrm{PII}<1$, and \texttt{Poorly constrained} if
$\mathrm{PII}\geq1$. These categories summarize marginal parameter constraint
for this study rather than serve as universal identifiability criteria. We
also flag combinations with $\mathrm{PII}<0.1$ and empirical coverage below
90\% as \emph{tightly constrained with low coverage}.

Of the 135 combinations of model, scenario, error structure, and parameter,
104 are classified as tightly constrained, 30 as weakly constrained, and one
as poorly constrained. Among the 104 tightly constrained combinations, 27 are
also flagged as \emph{tightly constrained with low coverage} because their
empirical coverage is below 90\%. The single poorly constrained case is SEIAR
Scenario~3, $\beta_1$ under Poisson error, with median PII
1.26 [0.81, 2.72]. Thus, most parameters eventually fall below
$\mathrm{PII}<1$, but many require longer calibration windows, and values in
$0.1\leq\mathrm{PII}<1$ remain weakly rather than tightly constrained.

Let $T^\star$ denote the first calibration window at which the median PII falls below 1. The coverage threshold is defined similarly as the first window at which empirical 95\% CI coverage reaches 90\% or higher. These are first-crossing thresholds and do not imply that the criteria remain satisfied at later windows. EXP, GGM, and
single-parameter SIR, SEIR, SEIR-UR, and SEIRD fits have
$T^{\star}\!\in\![10,30]$ time units. Multi-parameter SEIR-S3 and
SEIR-UR-S3 fits typically require $T^{\star}\!\in\![30,70]$. In SEIAR-S3,
$\beta_0$, $\beta_1$, and $\gamma$ generally require longer windows, whereas
$\rho$ crosses earlier ($T^\star=20$ under Poisson and $T^\star=30$ under
both negative-binomial settings); $\beta_1$ under Poisson error does not cross
within the windows studied. Higher observation noise increases final-window
PII values in most configurations, for example from $0.07$ to $0.14$ for GGM
$r$ and from $0.09$ to $0.26$ for SEIRD-S3 $\gamma$, although this pattern is
not universal. Overall, parameters tied to early, directly observed dynamics
reach lower final-window PII values, whereas parameters linked to latent or
partially observed processes require longer windows, additional observables,
or both.

\begin{figure}[H]
\centering
\includegraphics[width=\linewidth]{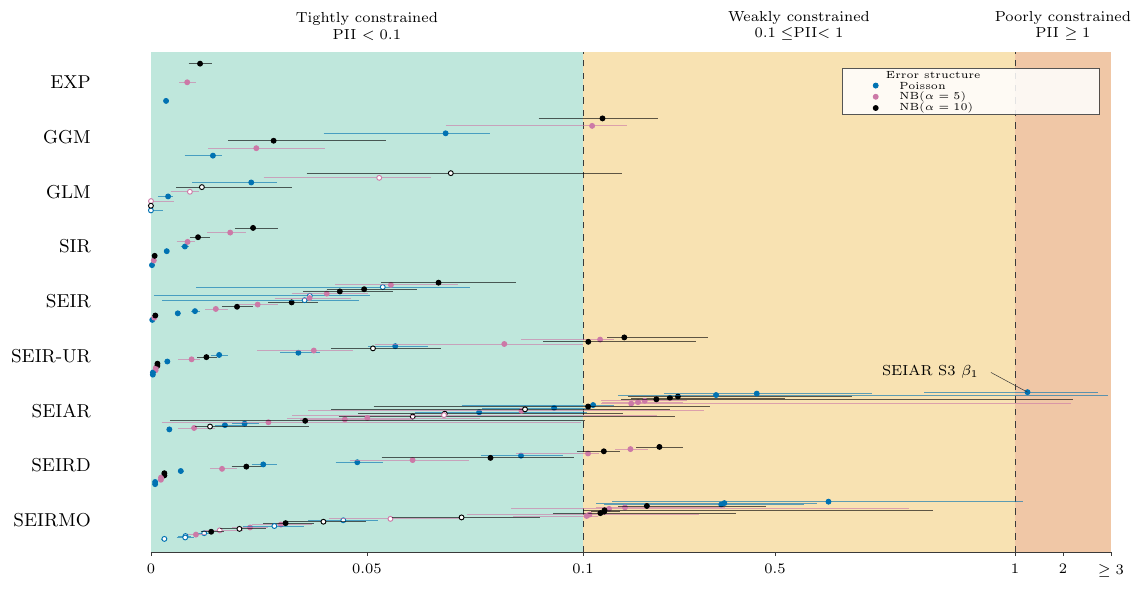}
\caption{Distribution of final-window PII values across models. Points show
the median PII for each model, scenario, parameter, and error-structure
combination; horizontal intervals show the central 95\% intervals across
replicates. Background bands mark the operational ranges: tightly constrained
($\mathrm{PII}<0.1$), weakly constrained
($0.1\leq\mathrm{PII}<1$), and poorly constrained
($\mathrm{PII}\geq1$). Colored circles indicate error structures; hollow
circles identify cases with PII$<0.1$ but empirical coverage below 90\%.}
\label{fig:pii_forest_summary}
\end{figure}

\subsection{Agreement between PII and NCIW classifications}

As a secondary benchmark, we compared PII with normalized confidence-interval
width (NCIW), an interval-width summary previously used to describe parameter
uncertainty~\cite{yoganandan2015lower,banerjee2020uncertainty}, here defined as
\begin{equation}
\mathrm{NCIW}=\frac{U-L}{2m},
\label{eq:nciw}
\end{equation}
where $m$ is the parameter estimate and $L$ and $U$ are the lower and upper
confidence-interval bounds. The same normalized interval-width form,
$(U-L)/(2m)$, is used as the component for each parameter in the
$CI_{\rm crit}$ criterion proposed by Lam et al.~\cite{lam2024evolving}.

For a fair comparison, the PII thresholds of 0.1 and 1 were matched to NCIW
thresholds of 0.115 and 1.423, respectively, using the geometric-mean
reference $m=\sqrt{LU}$. These matched NCIW thresholds were used only for
benchmarking and are not intended as general NCIW classification thresholds.

The two measures gave nearly identical final classifications
(Supplementary Table~\ref{tab:supp_pii_vs_nciw}). Of the 135 combinations,
134 (99.3\%) fell into the same constraint category. All 104 combinations
classified as tightly constrained by PII, including the 27 additionally
flagged as tightly constrained with low coverage, were also classified as
tightly constrained by NCIW. The only disagreement occurred for SEIAR
Scenario~3, $\beta_1$, under Poisson error: PII classified this parameter as
poorly constrained ($\mathrm{PII}=1.26$), whereas NCIW classified it as
weakly constrained ($\mathrm{NCIW}=1.11$).

At the final calibration window, PII and NCIW therefore provide nearly
identical classifications of parameter constraint, even though PII does not
depend on the parameter estimate. However, agreement between the two measures
does not imply adequate empirical coverage. Among the 104 combinations
classified as tightly constrained by both measures, 27 had empirical coverage
below 90\% and were therefore also flagged as tightly constrained with low
coverage. Thus, PII and NCIW provide closely aligned summaries of
confidence-interval width, while empirical coverage provides complementary
information on how often the confidence intervals contain the true parameter
value.

\subsection{Real-Data Application: 1918 San Francisco Influenza Outbreak}

In this section, we show a practical application of the PII using observed
epidemic data, where the true parameter values are unknown and parameter
constraint must be assessed directly from the fitted model. We consider daily
influenza incidence from the 1918 San Francisco epidemic~
\cite{chowell2007comparative,karami2026comparative} and recalibrate the models over progressively
longer calibration windows, $T=20,30,40,50,$ and $63$ days, under both Poisson
and negative-binomial observation models. This allows us to examine how
estimated parameter constraint changes as more of the epidemic trajectory is
included in the calibration data.

\paragraph{GLM.}
For the GLM, the final epidemic size $K$ was estimated within a narrow range
informed by the cumulative epidemic size (28,310--30,000), and the PII was
therefore summarized mainly for the growth-rate parameter $r$ and the
deceleration parameter $p$. Accordingly, the GLM PII results should be
interpreted given this externally informed constraint on $K$. Under both
observation models, as shown in
Figure~\ref{fig:GLM_realdata_SF1918_PII_r_p_combined}, parameter uncertainty
generally decreased as the calibration window increased, although the decrease
was not strictly monotonic. Under the negative-binomial observation model, the
PII for $r$ decreased from 0.19 at $T=20$ to 0.056 at $T=63$, while the PII
for $p$ decreased from 0.047 to 0.01. The Poisson observation model produced
narrower confidence intervals and lower PII values, with the PII for $r$
decreasing from 0.143 to 0.014 and that for $p$ from 0.035 to 0.002 over the
same interval. Thus, $p$ remained tightly constrained across all calibration
windows, whereas $r$ became tightly constrained once more of the epidemic
trajectory was available.

\begin{figure}[H]
\centering
\includegraphics[width=\textwidth]
{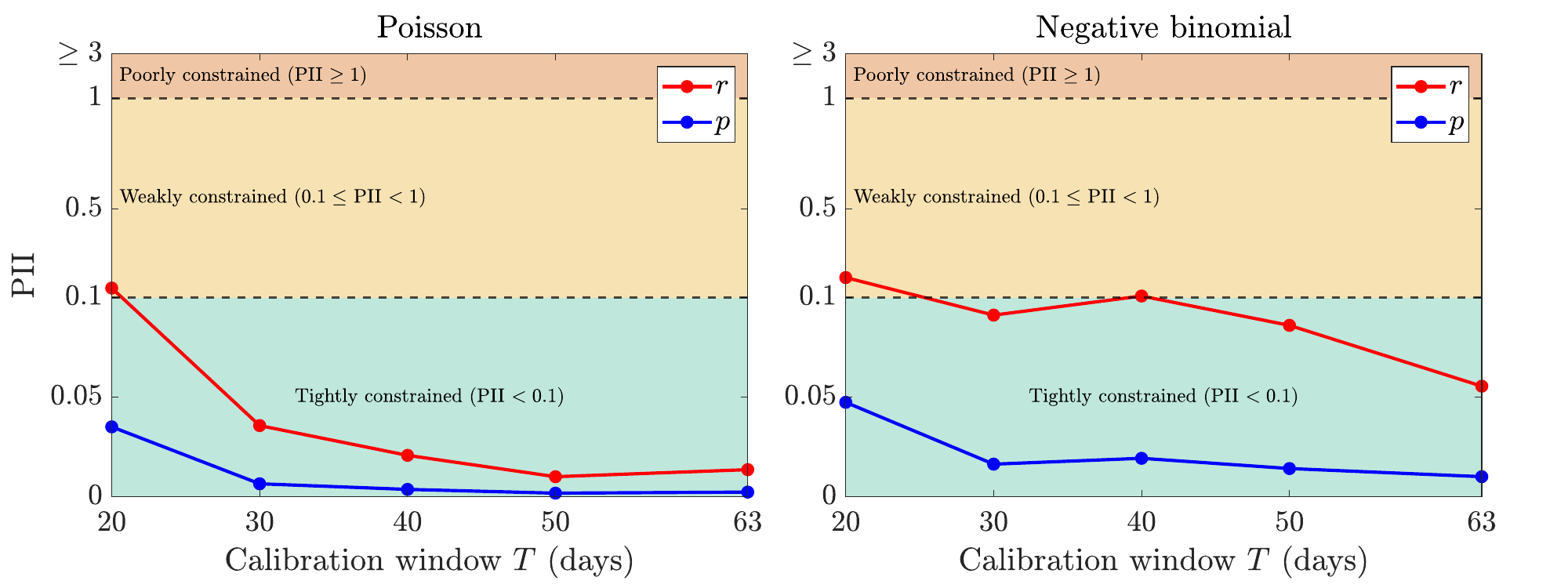}
\caption{Practical Identifiability Index (PII) for the GLM parameters $r$ and
$p$ across increasing calibration windows for the 1918 San Francisco
influenza data under Poisson (left) and negative-binomial (right) observation
models. Shaded regions indicate tightly constrained ($\mathrm{PII}<0.1$),
weakly constrained ($0.1\leq\mathrm{PII}<1$), and poorly constrained
($\mathrm{PII}\geq1$) parameters. The final epidemic size $K$ is not shown
because it was estimated within a narrow range informed by the full epidemic
wave.}
\label{fig:GLM_realdata_SF1918_PII_r_p_combined}
\end{figure}

\paragraph{SEIR-UR.}
We next considered the SEIR model with unreported infections, which accounts
for incomplete case reporting through the reporting fraction $\rho$. In this
model, only a fraction of new infections contributes to the observed
incidence, with $y(t)=\rho\kappa E(t)$. The transmission rate $\beta$,
incubation rate $\kappa$, reporting fraction $\rho$, and recovery rate
$\gamma$ were estimated, while the population size was fixed at
$N=550{,}000$. As shown in
Figure~\ref{fig:SEIRUR_realdata_SF1918_PII_combined}, the PII generally
decreased as more observations were included, although the two observation
models showed different patterns.

Under the Poisson model, all four parameters were weakly constrained at
$T=20$, whereas $\beta$, $\kappa$, and $\rho$ became tightly constrained as
the calibration window increased. The PII for $\rho$, for example, decreased
from 0.256 at $T=20$ to 0.01 at $T=63$. The recovery rate $\gamma$ showed a
different trend, becoming more tightly constrained over the intermediate
calibration windows before increasing again at the final calibration window.

The negative-binomial model produced broader confidence intervals and
generally larger PII values. The transmission rate $\beta$, incubation rate
$\kappa$, and reporting fraction $\rho$ remained predominantly weakly
constrained over the intermediate calibration windows, with $\rho$ becoming
tightly constrained at $T=63$. In contrast, uncertainty in the recovery rate
$\gamma$ remained substantial, and its lower confidence bound reached zero at
the final calibration window, resulting in $\mathrm{PII}=+\infty$. These
results show that adding more calibration data does not necessarily improve
parameter constraint uniformly across all parameters and that the observation
model used can substantially affect the resulting PII.

\begin{figure}[H]
\centering
\includegraphics[width=\textwidth]
{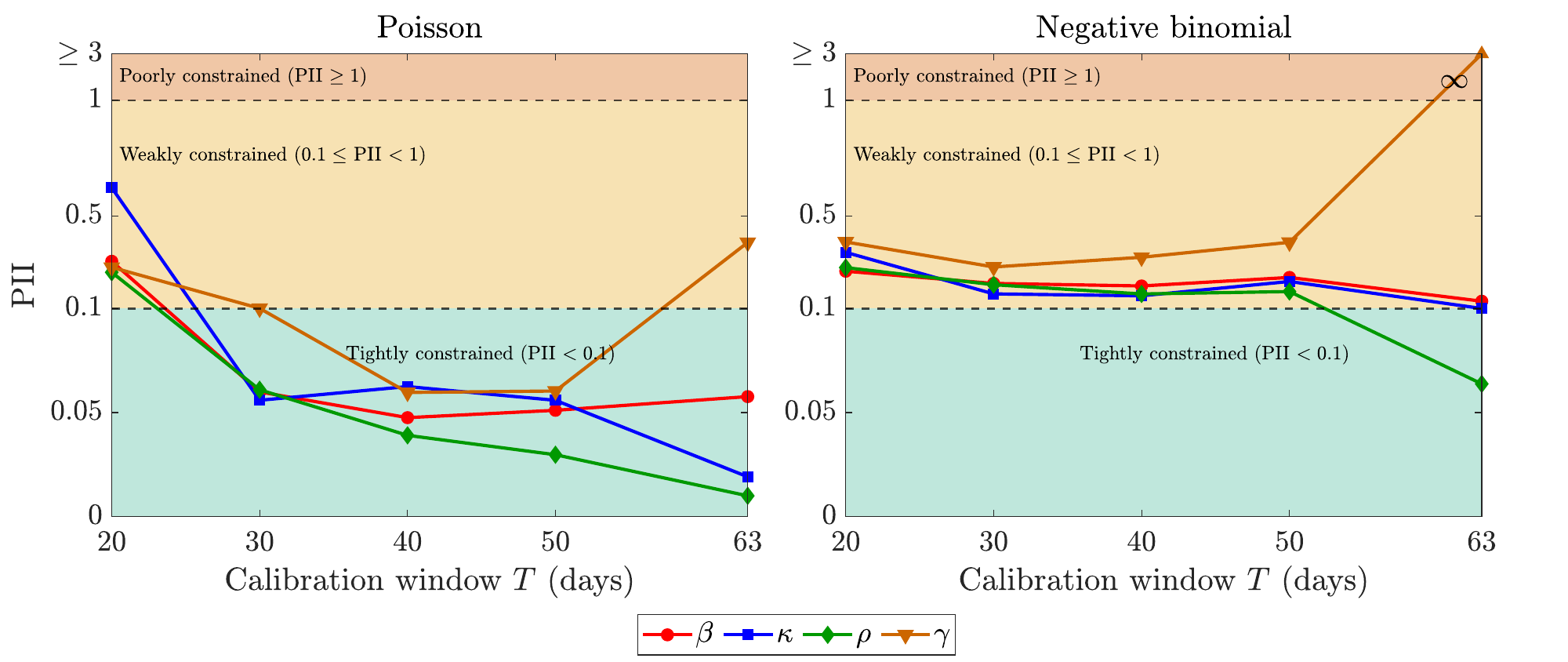}
\caption{Practical Identifiability Index (PII) for the SEIR model with
unreported infections (SEIR-UR) across increasing calibration windows for the
1918 San Francisco influenza data under Poisson (left) and negative-binomial
(right) observation models. PII values are shown for the transmission rate
$\beta$, incubation rate $\kappa$, reporting fraction $\rho$, and recovery
rate $\gamma$. Shaded regions indicate the PII classifications: tightly
constrained ($\mathrm{PII}<0.1$), weakly constrained
($0.1\leq\mathrm{PII}<1$), and poorly constrained
($\mathrm{PII}\geq1$). The infinite PII for $\gamma$ at the final
negative-binomial calibration window results from a confidence interval with
a lower bound of zero; for visualization, this value is displayed at the
upper plotting limit while the raw PII value remains $+\infty$.}
\label{fig:SEIRUR_realdata_SF1918_PII_combined}
\end{figure}
\section{Discussion}\label{sec:discussion}
\begingroup
\setlength{\parindent}{0pt}

This study introduced the Practical Identifiability Index (PII) as a
scale-invariant, easy-to-interpret summary of marginal parameter uncertainty
in dynamical models. The main contribution is that the PII provides a
standardized order-of-magnitude scale for comparing marginal parameter
uncertainty across parameters, models, error structures, and observation
designs, without requiring normalization by the parameter estimate.

By expressing confidence-interval width on an order-of-magnitude scale, the
PII provides a concise way to assess how tightly individual positive-valued
parameters are constrained by the available data under a specified model,
observation process, and analysis workflow. Across the simulation experiments,
several consistent patterns emerged: marginal uncertainty generally decreased
with increasing calibration-window length, increasing model complexity delayed
parameter constraint through parameter coupling, and higher observation noise
generally produced wider confidence intervals, although this pattern was not
uniform across all settings where parameters were estimated together.
Parameters associated with latent or indirectly observed processes also
remained weakly constrained for longer. These patterns show a practical use of
the PII: it provides a common scale for comparing which parameters have narrow
marginal intervals and which remain weakly constrained under the specified
model, data, and analysis workflow.

The comparison with NCIW further clarifies how the PII relates to existing
interval-width summaries. Using mathematically matched thresholds, PII and
NCIW produced the same final constraint classification in 134 of 135
model--scenario--parameter--error-structure combinations. All 104 combinations
classified as tightly constrained by PII were also classified as tightly
constrained by NCIW. Thus, the main value of the PII is not that it
systematically produces different classifications from normalized
interval-width measures, but that it provides nearly the same practical
classification on an easy-to-interpret multiplicative scale without depending
on the parameter estimate. The close agreement also shows that the PII is
consistent with established confidence-interval-width measures while retaining
its order-of-magnitude interpretation.

A key finding is that practical parameter constraint strongly depends on the
phase of the epidemic trajectory that is observed. For the simple growth
models without a peak, EXP and GGM, most parameters crossed the
one-order-of-magnitude boundary, $\mathrm{PII}<1$, using approximately
20--30 time units of data. In contrast, for the GLM, the growth rate $r$ and
deceleration parameter $p$ became tightly constrained before the peak, whereas
the final epidemic size $K$ generally required data near or beyond the peak,
crossing $\mathrm{PII}<1$ around $T=50$--60. Parameter uncertainty therefore
depends not only on the length of the calibration window, but also on whether
the observed trajectory includes the dynamical phase that informs each
parameter~\cite{sauer2021identifiability,saucedo2024comparative}. Longer
datasets are therefore not automatically more informative unless they include
the trajectory features that distinguish the parameters of interest.

The compartmental models further show how parameter coupling affects practical
parameter constraint. In the SIR model, the transmission rate $\beta$ crossed
$\mathrm{PII}<1$ very early when estimated alone, whereas estimating $\beta$
and $\gamma$ together required longer calibration windows because early
new-case data contain limited information for distinguishing transmission from
recovery. This effect became stronger in the SEIR model: when $\beta$,
$\kappa$, and $\gamma$ were estimated together, the window required to cross
$\mathrm{PII}<1$ depended on the error structure, occurring at $T=70$ under
Poisson and at $T=30$--50 under the negative-binomial settings. Increasing
model complexity therefore does not necessarily make all parameters poorly
constrained, but it spreads the available information across coupled
parameters and can delay the move to narrower marginal confidence intervals.

The extended epidemic models show additional challenges caused by latent and
partially observed processes. In the SEIRD model, the transmission and removal
rates crossed $\mathrm{PII}<1$ around $T=50$, whereas the infection fatality
fraction $\rho$ crossed earlier because new-death observations directly
informed this parameter. In the SEIR-UR model, the reporting fraction became
constrained relatively early, but the transmission and recovery rates required
longer windows closer to the epidemic peak. The SEIAR model was the most
challenging case: in the most complex estimation scenario, the asymptomatic
transmission rate $\beta_1$ remained poorly constrained under the Poisson
error structure even at $T=100$ and remained only weakly constrained
($0.1\leq\mathrm{PII}<1$) under overdispersed error structures. This shows a
fundamental difficulty in models with partially observed transmission
pathways, where parameters associated with unobserved compartments may remain
weakly constrained even when substantial new-case data are available~
\cite{tuncer2018structural}.

The multi-observable SEIR analysis further shows that practical parameter
constraint is not determined by model structure alone. Adding observations of
infectious and recovered individuals substantially reduced PII values relative
to calibration based only on new cases, especially for parameters associated
with latent progression and recovery. More informative observation
configurations can therefore partly offset short calibration windows and
parameter coupling. In applied settings, the PII can be used not only as a
diagnostic after model fitting but also as a guide for data collection:
parameters with persistently high PII can indicate where additional
observables, longer follow-up, external information, or model simplification
may be most valuable.

Several limitations should be noted. First, the models considered here include
simple growth models and commonly used compartmental epidemic models, but they
do not represent the full range of epidemiological structures. Models with age
structure, spatial coupling, behavioral feedback, network transmission, or
time-varying parameters may show different patterns of parameter constraint.
Second, the estimation scenarios do not cover every possible combination of
fixed and estimated parameters. Practical parameter constraint can depend
strongly on which parameters are estimated together, and parameters that are
well constrained in one scenario may become weakly constrained when additional
parameters are allowed to vary~\cite{roosa2019assessing}.

Third, the PII is based on the width of marginal confidence intervals and
therefore does not capture the full likelihood or posterior geometry. In
particular, it does not directly diagnose skewness, multimodality, flat
likelihood ridges, parameter correlations, structurally non-identifiable
parameter combinations, or estimator bias. A low PII should therefore be
interpreted as evidence of narrow marginal confidence intervals, not as proof
of complete identifiability or unbiased estimation. For detailed inference,
the PII should be interpreted together with complementary diagnostics such as
profile likelihoods, posterior distributions, sensitivity analysis, empirical
coverage, and structural identifiability analysis~\cite{raue2009structural}.
This point is especially important because, among the 104 final combinations
classified as tightly constrained by both PII and NCIW, 27 had empirical
coverage below 90\%, showing that narrow intervals do not necessarily imply
well-calibrated uncertainty. Accordingly, we recommend reporting PII together
with empirical coverage, profile likelihoods, posterior diagnostics, or other
checks of estimator bias whenever possible.

Fourth, numerical PII values may depend on inferential choices, including the
assumed observation model, bootstrap procedure, optimization bounds, and
confidence-interval construction. Because parameter estimation is performed
within finite parameter ranges, confidence intervals that approach or coincide
with these bounds may be limited by the imposed parameter bounds. The resulting
PII can therefore partly reflect the specified feasible range rather than
information from the data alone. Such bounds may be scientifically justified,
but their influence should still be considered when interpreting intervals
limited by these bounds.

When the lower confidence bound reaches zero, the definition
in~\eqref{eq:pii} yields $\mathrm{PII}=+\infty$, clearly indicating an
unbounded multiplicative interval ratio. Very small positive lower bounds
similarly produce very large finite PII values. Such cases should be
interpreted as warning signs of weak marginal parameter constraint rather than
as precise levels of uncertainty. Finally, our analysis was conducted under
idealized simulation conditions in which the data-generating process matched
the inference model. In real applications, model misspecification,
measurement error, reporting delays, changing surveillance intensity, and
unmodeled heterogeneity may further worsen practical non-identifiability~
\cite{murphy2024implementing}.

Overall, the PII provides a clear and easy-to-interpret diagnostic for
comparing marginal parameter uncertainty across models, parameters,
calibration windows, and observation designs. The tightly / weakly / poorly
constrained categories should be interpreted as practical, rule-of-thumb
benchmarks rather than universal definitions of practical identifiability.
The main value of the PII is not to replace established identifiability
methods, but to provide a standardized summary of marginal parameter
uncertainty that does not depend on the parameter estimate and can be reported
alongside these methods. In this way, the PII connects uncertainty
quantification, model comparison, and data-collection design in applied
dynamical modeling.

\endgroup
\section*{Code Availability}

All code and configuration files used in this study are publicly
available at \url{https://github.com/hkarami-GSU/PII/tree/main}.

\section*{Data accessibility}
The historical daily influenza incidence data for the 1918 San Francisco epidemic used in the real-data application are publicly available. Simulation data can be regenerated using the models, code, and configuration files provided in the public repository. Code and configuration files required to reproduce the simulations, calibration, bootstrap analyses, and figures are available at: \url{https://github.com/hkarami-GSU/PII/tree/main}.

\section*{Authors' contributions}
H.K. implemented the computational framework, performed the simulations, and generated the figures and tables. G.C. conceived the study, contributed to the methodological framing, interpreted the results, and drafted and revised the manuscript. All authors reviewed, edited, and approved the final manuscript.

\section*{Competing interests}
The authors declare no competing interests.

\section*{Funding}
This work was partially supported by National Science Foundation under grant number NSF ACED $\#2435886$.

\section*{Ethics}
The real-data application used publicly available historical epidemic data. This study did not involve direct interaction with human participants, animal subjects, or identifiable private information.

\section*{Use of AI and AI-assisted technologies}
During the preparation of this manuscript, the authors used AI-assisted technologies to support language editing, and code folder cleaning. The authors reviewed and edited all AI-assisted outputs and take full responsibility for the content of the manuscript.


\bibliographystyle{unsrt}
\bibliography{mybib}

\endgroup 
\beginsupplement
%

\definecolor{okGreen}{HTML}{009E73}
\definecolor{okOrange}{HTML}{E69F00}
\definecolor{okRed}{HTML}{D55E00}
\renewcommand{\clsID}{\cellcolor{okGreen!25}Tightly constrained}
\renewcommand{\clsIDlow}{\cellcolor{okGreen!8}\shortstack[l]{Tightly constrained\\with low coverage}}
\renewcommand{\clsWI}{\cellcolor{okOrange!30}Weakly constrained}
\renewcommand{\clsNI}{\cellcolor{okRed!35}Poorly constrained}
\renewcommand{\thetable}{S\arabic{table}}
\renewcommand{\tablename}{Supplementary Table}
\definecolor{okGreen}{HTML}{009E73}
\definecolor{okOrange}{HTML}{E69F00}
\definecolor{okRed}{HTML}{D55E00}

\title{Supplementary Material: Parameter uncertainty in dynamical models: a practical identifiability index}
\author[1,3]{Hamed Karami}
\author[1]{Alexandra Smirnova}
\author[2]{Sunmi Lee}
\author[2,3,*]{Gerardo Chowell}

\affil[1]{Department of Mathematics \& Statistics, Georgia State University, Atlanta, GA 30303, USA}
\affil[2]{Department of Applied Mathematics, Kyung Hee University, Yongin 17104, Korea}
\affil[3]{Department of Population Health Sciences, Georgia State University, Atlanta, GA 30303, USA}
\affil[*]{Corresponding author: \href{mailto:gchowell@gsu.edu}{gchowell@gsu.edu}}
\date{}

\maketitle

\section*{Software implementation and tutorial workflow}

We provide the code file
\texttt{plotPracticalIdentifiabilityResults.m} so users can explore different
practical identifiability settings using the methodology described above.
Users can select the model, error structure, calibration-window lengths, and
number of replicates and run their own analysis. The models currently available
are those considered in this study. Users can also use the
\textsc{QuantDiffForecast} toolbox to define their own model and apply it to
PII analysis.

In \texttt{plotPracticalIdentifiabilityResults.m}, the main options are listed
at the top of the file so that users can easily locate and adjust them as
needed. First, the user selects the model and estimation scenario by assigning
the corresponding options file handle to \texttt{options\_handle}. For example,
the exponential growth model under the negative-binomial setting can be
selected using

\begin{verbatim}
options_handle = @options_forecast_PII_EXPO_r_dist1_3;
\end{verbatim}

Other handles are available for GGM, GLM, SIR, SEIR, SEIRD, SEIR with
unreported infections, and SEIR with asymptomatic transmission, together with
the different parameter-estimation scenarios. These handles are listed at the
top of the code.

Next, the observation error structure is specified by

\begin{verbatim}
error_type = 'Poisson', 'Negbin5', or 'Negbin10';
\end{verbatim}

where \texttt{Negbin5} and \texttt{Negbin10} correspond to
negative-binomial data generation with dispersion values $\alpha=5$ and
$\alpha=10$, respectively. During fitting, $\alpha$ is estimated as a
nuisance parameter.

The calibration-window lengths are set through \texttt{windowsize1}. This can
be a single value, such as

\begin{verbatim}
windowsize1 = 20;
\end{verbatim}

or a vector of values, such as

\begin{verbatim}
windowsize1 = 20:10:50;
\end{verbatim}

to evaluate practical identifiability across multiple observation windows.

The number of simulation replicates is set by
\texttt{num\_replicates}. The parameter \texttt{run\_flag} controls whether
the computation is run or skipped when results have already been generated and
only plotting is needed. The parameter \texttt{plot\_type} lets the user
choose whether to generate PII summaries, confidence-interval plots for
parameter estimates, both types of figures, or no plots.

For users with access to a high-performance computing (HPC) cluster, we also
provide a cluster version of the code in the folder
\texttt{Final\_code\_PII\_cluster}. This version uses the same options files
and settings described above but replaces the local \texttt{parfor}
implementation with a single-task function, \texttt{main(taskID)}, designed
for SLURM job arrays. Each \texttt{taskID}, an integer from 1 to
\texttt{num\_windowsizes} $\times$ \texttt{num\_replicates}, is automatically
mapped to a unique (window size, replicate) pair.

For example, with

\begin{verbatim}
windowsize_array = 20:10:50;
num_replicates = 500;
\end{verbatim}

the total number of tasks is $4\times500=2000$, and the job can be submitted
as

\begin{verbatim}
sbatch --array=1-2000.
\end{verbatim}

An additional user setting, \texttt{factor1}, controls the data-generating
overdispersion value passed to the simulation wrapper. It is set to 1 for
Poisson, where it is not used by the Poisson error model, 5 for
\texttt{Negbin5}, and 10 for \texttt{Negbin10}. This setting does not fix
$\alpha$ during negative-binomial fitting; $\alpha$ remains estimated as a
nuisance parameter. All configuration options, including model selection,
window sizes, and number of replicates, are listed at the top of the cluster
code \texttt{main.m}.

\section*{Supplementary Tables}

\scriptsize
\setlength{\tabcolsep}{3pt}
\renewcommand{\arraystretch}{1.15}

\begin{longtable}{
p{1.3cm}
p{0.75cm}
p{2.4cm}
p{2.4cm}
p{3.0cm}
p{2.0cm}
p{2.5cm}}

\caption{Structural identifiability results for the models and estimation
scenarios considered in the simulation study.}
\label{tab:structural_identifiability}\\

\toprule
\textbf{Model} &
\textbf{Scen.} &
\textbf{Estimated} &
\textbf{Fixed} &
\textbf{Observation operator} &
\textbf{Structural status} &
\textbf{Reference} \\
\midrule
\endfirsthead

\toprule
\textbf{Model} &
\textbf{Scen.} &
\textbf{Estimated} &
\textbf{Fixed} &
\textbf{Observation operator} &
\textbf{Structural status} &
\textbf{Reference} \\
\midrule
\endhead

\midrule
\multicolumn{7}{r}{\textit{Continued on next page}}\\
\endfoot

\bottomrule
\endlastfoot

EXP
& 1
& $r$
& --
& $\dot C=rC$
& Global
& GGM with $p=1$; \cite{liyanage2025structural}. \\

\addlinespace

GGM
& 1
& $r,p$
& --
& $\dot C=rC^p$
& Global
& \cite{liyanage2025structural}. \\

\addlinespace

GLM
& 1
& $r,p,K$
& --
& $\dot C=rC^p(1-C/K)$
& Global
& \cite{liyanage2025structural}. \\

\addlinespace

SIR
& 1
& $\beta$
& $\gamma,N$
& $\dot C=\beta SI/N$
& --
& -- \\

SIR
& 2
& $\beta,\gamma$
& $N$
& $\dot C=\beta SI/N$
& --
& -- \\

\addlinespace

SEIR
& 1
& $\beta$
& $\kappa,\gamma,N$
& $\dot C=\kappa E$
& Global
& \cite{liyanage2026tutorial}. \\

SEIR
& 2
& $\beta,\gamma$
& $\kappa,N$
& $\dot C=\kappa E$
& Global
& \cite{liyanage2026tutorial}. \\

SEIR
& 3
& $\beta,\gamma,\kappa$
& $N$
& $\dot C=\kappa E$
& Global
& \cite{liyanage2026tutorial}. \\
\end{longtable}
\pagebreak
{\scriptsize
\setlength{\tabcolsep}{3pt}

\begin{longtable}{p{0.95in}p{0.75in}p{0.75in}p{0.75in}p{2.3in}}

\caption{Optimization bounds for model parameters and the nuisance dispersion
parameter in the simulation experiments and real-data analyses.}
\label{tab:supp_bounds}\\

\toprule
Model & Parameter & Lower bound & Upper bound & Notes \\
\midrule
\endfirsthead

\caption[]{Optimization bounds for model parameters and the nuisance dispersion
parameter (continued).}\\
\toprule
Model & Parameter & Lower bound & Upper bound & Notes \\
\midrule
\endhead

\bottomrule
\endfoot

\multicolumn{5}{l}{\textbf{Simulation experiments}} \\
\midrule

EXP
& $r$
& 0
& 1
& Estimated in Scenario~1. \\

GGM
& $r$
& 0
& 10
& Estimated in Scenario~1. \\

GGM
& $p$
& 0
& 1
& Estimated in Scenario~1. \\

GLM
& $r$
& 0.01
& 25
& Estimated in Scenario~1. \\

GLM
& $p$
& 0.01
& 1
& Estimated in Scenario~1. \\

GLM
& $K$
& 10
& 1,000,000
& Estimated in Scenario~1. \\

SIR
& $\beta$
& 0.001
& 10
& Estimated in Scenarios~1--2. \\
\end{longtable}
}
{\scriptsize
\setlength{\tabcolsep}{3pt}

\begin{longtable}{p{1.4in}p{1.2in}p{1.4in}}

\caption{Initial conditions used in the simulation experiments and real-data
analyses. All initial conditions are fixed during calibration.}
\label{tab:supp_initial_conditions}\\

\toprule
Model & Initial condition & Value \\
\midrule
\endfirsthead

\caption[]{Initial conditions used in the simulation experiments and real-data
analyses (continued).}\\
\toprule
Model & Initial condition & Value \\
\midrule
\endhead

\bottomrule
\endfoot

\multicolumn{3}{l}{\textbf{Simulation experiments}} \\
\midrule

EXP & $C_0$ & 5 \\
GGM & $C_0$ & 5 \\
GLM & $C_0$ & 6 \\

SIR & $S_0$ & $N-4=499,996$ \\
SIR & $I_0$ & 4 \\
SIR & $R_0$ & 0 \\
SIR & $C_0$ & 4 \\

SEIR and SEIRMO & $S_0$ & $N-4=499,996$ \\
SEIR and SEIRMO & $E_0$ & 0 \\
SEIR and SEIRMO & $I_0$ & 4 \\
SEIR and SEIRMO & $R_0$ & 0 \\
SEIR and SEIRMO & $C_0$ & 4 \\

SEIR-UR & $S_0$ & $N-5=499,995$ \\
SEIR-UR & $E_0$ & 0 \\
SEIR-UR & $I_0$ & 5 \\
SEIR-UR & $R_0$ & 0 \\
SEIR-UR & $C_0$ & 5 \\

SEIAR & $S_0$ & $N-4=499,996$ \\
SEIAR & $E_0$ & 0 \\
SEIAR & $I_0$ & 2 \\
SEIAR & $A_0$ & 2 \\
SEIAR & $R_0$ & 0 \\
SEIAR & $C_0$ & 4 \\

SEIRD & $S_0$ & $N-5=99,995$ \\
SEIRD & $E_0$ & 0 \\
SEIRD & $I_0$ & 5 \\
SEIRD & $R_0$ & 0 \\
SEIRD & $D_0$ & 5 \\

\midrule
\multicolumn{3}{l}{\textbf{Real-data analyses}} \\
\midrule

GLM & $C_0$ & 4 \\

SEIR-UR & $S_0$ & $N-4=549,996$ \\
SEIR-UR & $E_0$ & 0 \\
SEIR-UR & $I_0$ & 4 \\
SEIR-UR & $R_0$ & 0 \\
SEIR-UR & $C_0$ & 4 \\

\end{longtable}
}

\begin{footnotesize}
\setlength{\tabcolsep}{3pt}
\setlength{\LTcapwidth}{7.3in}
\renewcommand{\arraystretch}{1.05}
\begin{longtable}{llllr|rl c|rl c}
\caption{Detailed initial-versus-final-window comparison of parameter estimates,
confidence intervals, and PII values for every model, scenario, parameter, and error
structure. For each model--scenario--parameter--error combination, the table reports
the true parameter value, calibration-window length, 95\% confidence interval for the
estimated parameter, and median PII with central 95\% interval across replicates at the
initial and final calibration windows. PII summaries are shown in \textbf{bold} when
the median PII is below 1. An entry of $\infty$ indicates that zero lower confidence
endpoints occurred in enough replicates for the corresponding across-replicate PII
summary to be infinite.}
\label{tab:supp_pii_comprehensive}\\
\toprule
\textbf{Model} & \textbf{Sc.} & \textbf{Param.} & \textbf{Error} & \textbf{True} &
\multicolumn{3}{c|}{\textbf{Initial calibration}} &
\multicolumn{3}{c}{\textbf{Final calibration}} \\
 &  &  &  &  & $T$ & 95\% CI (param) & PII [95\% interval] & $T$ & 95\% CI (param) & PII [95\% interval] \\
\midrule
\endfirsthead
\multicolumn{11}{l}{\emph{(Supplementary Table~\ref{tab:supp_pii_comprehensive} continued)}}\\
\toprule
\textbf{Model} & \textbf{Sc.} & \textbf{Param.} & \textbf{Error} & \textbf{True} &
\multicolumn{3}{c|}{\textbf{Initial calibration}} &
\multicolumn{3}{c}{\textbf{Final calibration}} \\
 &  &  &  &  & $T$ & 95\% CI (param) & PII [95\% interval] & $T$ & 95\% CI (param) & PII [95\% interval] \\
\midrule
\endhead
\midrule
\multicolumn{11}{r}{\emph{(continued on next page)}}\\
\endfoot
\bottomrule
\endlastfoot
EXP & 1 & $r$ & Poisson & 0.14 & 20 & 0.140 [0.127, 0.151] & \textbf{0.07 [0.06, 0.09]} & 50 & 0.140 [0.139, 0.141] & \textbf{0.00 [0.00, 0.00]} \\
 &  &  & NB($\alpha{=}5$) & 0.14 & 20 & 0.136 [0.107, 0.160] & \textbf{0.17 [0.10, 0.30]} & 50 & 0.140 [0.139, 0.141] & \textbf{0.01 [0.01, 0.01]} \\
 &  &  & NB($\alpha{=}10$) & 0.14 & 20 & 0.128 [0.091, 0.161] & \textbf{0.24 [0.14, 0.50]} & 50 & 0.140 [0.138, 0.142] & \textbf{0.01 [0.01, 0.01]} \\
\midrule
GGM & 1 & $r$ & Poisson & 0.14 & 20 & 0.147 [0.123, 0.419] & \textbf{0.56 [0.40, 0.95]} & 50 & 0.140 [0.129, 0.152] & \textbf{0.07 [0.04, 0.08]} \\
 &  & $p$ & Poisson & 0.98 & 20 & 0.980 [0.603, 1.000] & \textbf{0.22 [0.14, 0.54]} & 50 & 0.980 [0.963, 0.997] & \textbf{0.01 [0.01, 0.02]} \\
 &  & $r$ & NB($\alpha{=}5$) & 0.14 & 20 & 0.149 [0.101, 1.296] & 1.13 [0.64, 1.58] & 50 & 0.138 [0.126, 0.166] & \textbf{0.12 [0.07, 0.19]} \\
 &  & $p$ & NB($\alpha{=}5$) & 0.98 & 20 & 1.000 [0.200, 1.000] & \textbf{0.70 [0.24, \(\infty\)]} & 50 & 0.983 [0.945, 1.000] & \textbf{0.02 [0.01, 0.04]} \\
 &  & $r$ & NB($\alpha{=}10$) & 0.14 & 20 & 0.146 [0.087, 1.649] & 1.35 [0.79, 1.74] & 50 & 0.134 [0.125, 0.174] & \textbf{0.14 [0.09, 0.26]} \\
\end{longtable}
\end{footnotesize}


\begin{footnotesize}
\setlength{\tabcolsep}{4pt}
\setlength{\LTcapwidth}{5.4in}
\renewcommand{\arraystretch}{1.1}
\begin{longtable}{lllrrrrl}
\caption{Cross-model synthesis of threshold windows, final-window Practical
Identifiability Index (PII) values, empirical coverage, and classification across all
models, scenarios, parameters, and error structures.}
\label{tab:supp_cross_model_synthesis}\\
\toprule
\textbf{Model} & \textbf{Sc.} & \textbf{Param.} & \textbf{Error} &
$T^{\star}_{\mathrm{PII}<1}$ & $T^{\star}_{\mathrm{cov}\geq 90\%}$ &
\textbf{Median PII [95\% interval]} ($T_{\max}$) & \textbf{Class} \\
\midrule
\endfirsthead
\multicolumn{8}{l}{\emph{(Supplementary Table~\ref{tab:supp_cross_model_synthesis} continued)}}\\
\toprule
\textbf{Model} & \textbf{Sc.} & \textbf{Param.} & \textbf{Error} &
$T^{\star}_{\mathrm{PII}<1}$ & $T^{\star}_{\mathrm{cov}\geq 90\%}$ &
\textbf{Median PII [95\% interval]} ($T_{\max}$) & \textbf{Class} \\
\midrule
\endhead
\midrule
\multicolumn{8}{r}{\emph{(continued on next page)}}\\
\endfoot
\bottomrule
\endlastfoot
EXP & 1 & $r$ & Poisson & 20 & 20 & \textbf{0.00 [0.00,0.00]} (cov.\ 94\%) & \clsID \\
 &  &  & NB($\alpha{=}5$) & 20 & 30 & \textbf{0.01 [0.01,0.01]} (cov.\ 92\%) & \clsID \\
 &  &  & NB($\alpha{=}10$) & 20 & 30 & \textbf{0.01 [0.01,0.01]} (cov.\ 91\%) & \clsID \\
\midrule
GGM & 1 & $r$ & Poisson & 20 & 20 & \textbf{0.07 [0.04,0.08]} (cov.\ 92\%) & \clsID \\
 &  & $p$ & Poisson & 20 & 20 & \textbf{0.01 [0.01,0.02]} (cov.\ 92\%) & \clsID \\
 &  & $r$ & NB($\alpha{=}5$) & 30 & 20 & \textbf{0.12 [0.07,0.19]} (cov.\ 97\%) & \clsWI \\
 &  & $p$ & NB($\alpha{=}5$) & 20 & 20 & \textbf{0.02 [0.01,0.04]} (cov.\ 97\%) & \clsID \\
 &  & $r$ & NB($\alpha{=}10$) & 30 & 20 & \textbf{0.14 [0.09,0.26]} (cov.\ 98\%) & \clsWI \\
\end{longtable}
\end{footnotesize}


{\scriptsize
\setlength{\tabcolsep}{4pt}
\renewcommand{\arraystretch}{1.05}
\begin{longtable}{llllrrr}
\caption{PII versus NCIW across models, scenarios, parameters, and error structures at the final calibration window.}
\label{tab:supp_pii_vs_nciw}\\
\toprule
\textbf{Model} & \textbf{Scenario} & \textbf{Parameter} & \textbf{Error structure} & \textbf{PII} & \textbf{NCIW} & \textbf{Coverage (\%)} \\
\midrule
\endfirsthead
\multicolumn{7}{l}{\emph{(Supplementary Table~\ref{tab:supp_pii_vs_nciw} continued)}}\\
\toprule
\textbf{Model} & \textbf{Scenario} & \textbf{Parameter} & \textbf{Error structure} & \textbf{PII} & \textbf{NCIW} & \textbf{Coverage (\%)} \\
\midrule
\endhead
\midrule
\multicolumn{7}{r}{\emph{(continued on next page)}}\\
\endfoot
\bottomrule
\endlastfoot
EXP & 1 & $r$ & Poisson & \cellcolor{okGreen!25}0.00351 & \cellcolor{okGreen!25}0.00404 & 94.4 \\
 &  &  & NB($\alpha{=}5$) & \cellcolor{okGreen!25}0.00839 & \cellcolor{okGreen!25}0.00966 & 92.2 \\
 &  &  & NB($\alpha{=}10$) & \cellcolor{okGreen!25}0.0114 & \cellcolor{okGreen!25}0.0131 & 91.2 \\
\midrule
GGM & 1 & $r$ & Poisson & \cellcolor{okGreen!25}0.0682 & \cellcolor{okGreen!25}0.0789 & 91.5 \\
 &  &  & NB($\alpha{=}5$) & \cellcolor{okOrange!30}0.119 & \cellcolor{okOrange!30}0.146 & 97.4 \\
 &  &  & NB($\alpha{=}10$) & \cellcolor{okOrange!30}0.141 & \cellcolor{okOrange!30}0.179 & 98.0 \\
\midrule
GGM & 1 & $p$ & Poisson & \cellcolor{okGreen!25}0.0144 & \cellcolor{okGreen!25}0.0165 & 92.4 \\
 &  &  & NB($\alpha{=}5$) & \cellcolor{okGreen!25}0.0244 & \cellcolor{okGreen!25}0.0278 & 97.4 \\
\end{longtable}
} 

\pagebreak
\section*{Supplementary Figures}

\section{EXP Model}

\subsection{Scenario~1: Estimated $\{r\}$}

\begin{figure}[H]
\centering
\includegraphics[width=\linewidth]{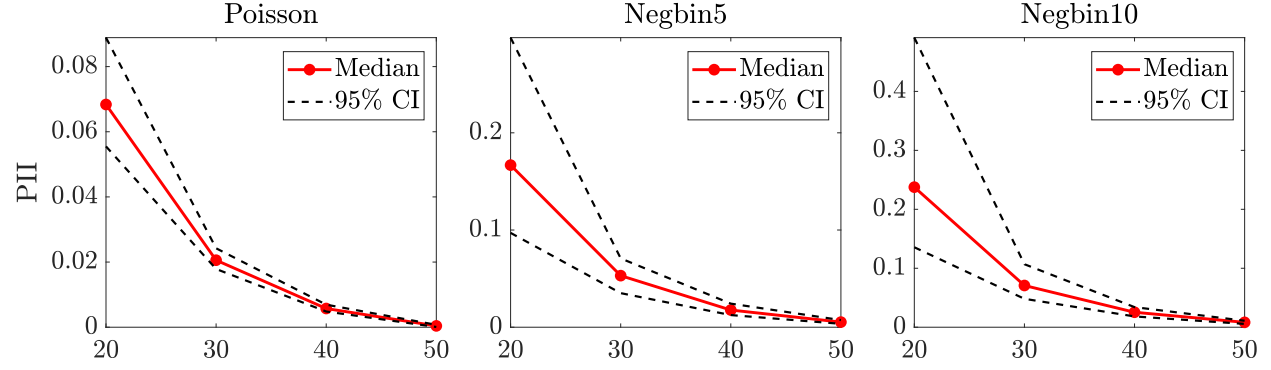}
\caption{Practical Identifiability Index (PII) for the exponential growth rate $r$ in the EXP model across calibration-window lengths $T=20, 30, 40,$ and $50$ under three error structures: Poisson, negative binomial with data-generating dispersion $\alpha=5$ (Negbin5), and negative binomial with data-generating dispersion $\alpha=10$ (Negbin10). Red lines show the median PII across replicates, and dashed black curves indicate the central 95\% interval.}
\label{fig:EXP_S1_PII_parameters}
\end{figure}

\begin{figure}[H]
\centering
\includegraphics[width=0.9\linewidth]{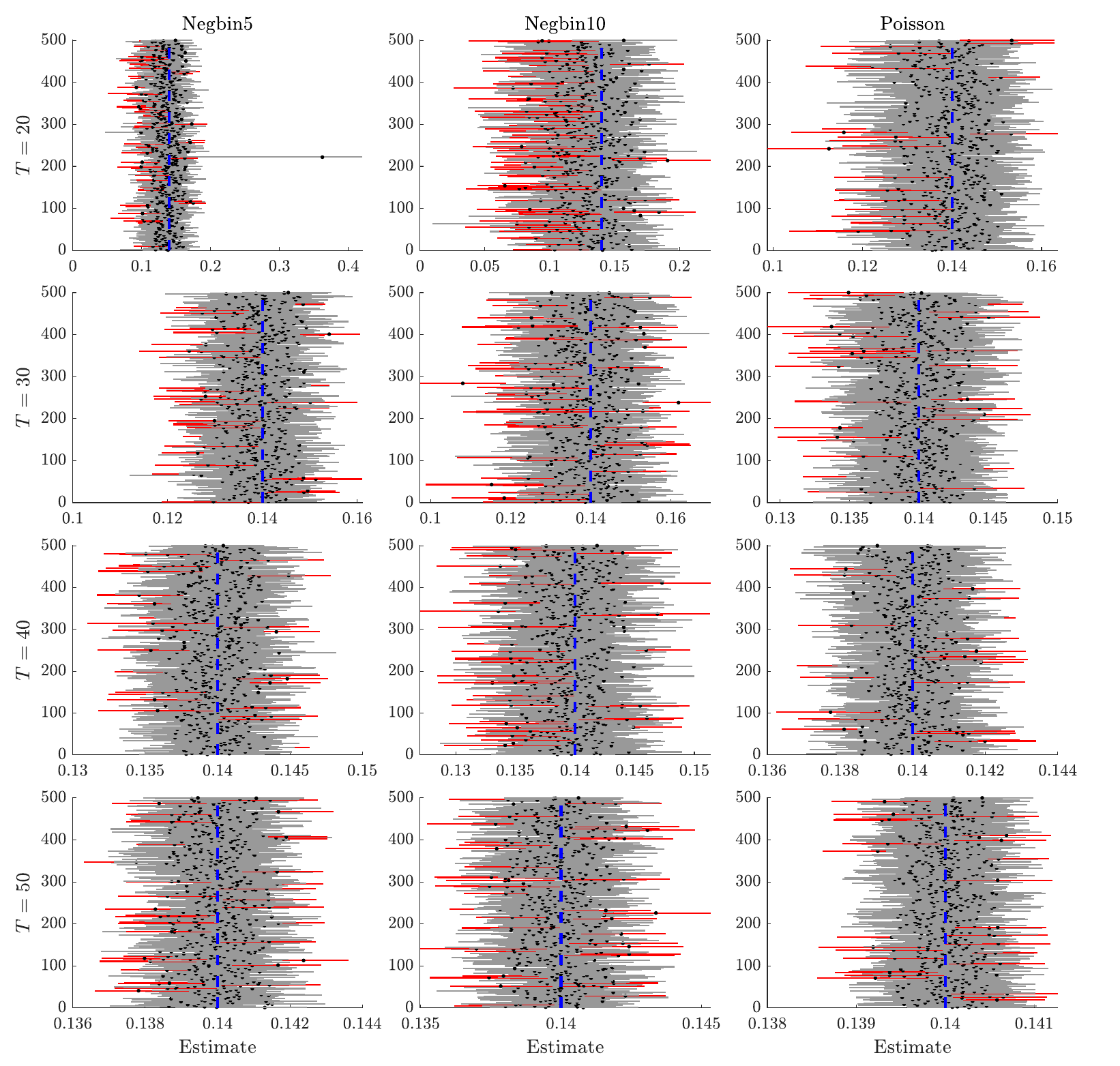}
\caption{
Parameter estimates and 95\% confidence intervals (CIs) for the exponential growth rate $r$ across 500 simulation replicates and calibration window lengths $T=20,30,40,50$, with the true value $r=0.14$ indicated by the vertical blue dashed line. Columns correspond to the error structures: negative binomial with data-generating dispersion parameter $\alpha=5$ (Negbin5), negative binomial with data-generating dispersion parameter $\alpha=10$ (Negbin10), and Poisson. Each horizontal line corresponds to a single simulation replicate, showing the bootstrap confidence interval obtained by resampling within that replicate, with the corresponding point estimate marked by a black dot at its center. Red intervals denote confidence intervals that do not contain the true value, whereas gray intervals denote those that do.
}
\label{fig:EXP_S1_CI_grid_param_1}
\end{figure}

\section{GGM Model}

\subsection{Scenario~1: Estimated $\{r,\, p\}$}

\begin{figure}[H]
\centering
\includegraphics[width=\linewidth]{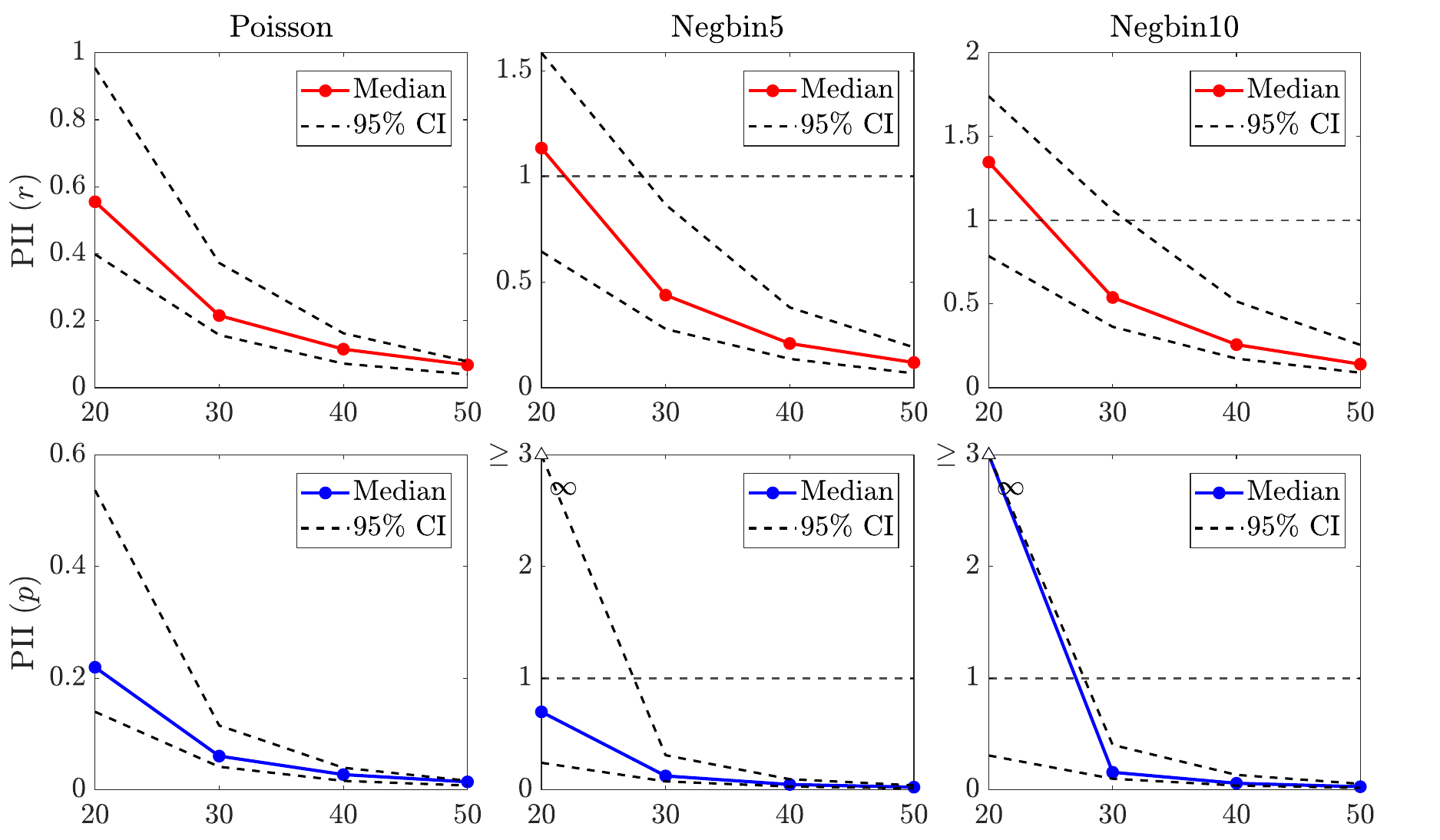}
\caption{Practical Identifiability Index (PII) for the growth rate $r$ and deceleration parameter $p$ in the GGM model across calibration-window lengths $T=20, 30, 40,$ and $50$ under three error structures: Poisson, negative binomial with data-generating dispersion $\alpha=5$ (Negbin5), and negative binomial with data-generating dispersion $\alpha=10$ (Negbin10). Red lines show the median PII across replicates, and dashed black curves indicate the central 95\% interval. For visualization only, within subplots containing PII values at or above 3, those values are displayed at an upper plotting limit of 3; uncapped values, including genuine infinite summaries, are retained for all numerical analyses. Upward triangles denote values displayed at the upper limit, and labels $\infty$ denote genuinely infinite PII summaries.}
\label{fig:GGM_S1_PII_parameters}
\end{figure}

\begin{figure}[H]
\centering
\includegraphics[width=0.9\linewidth]{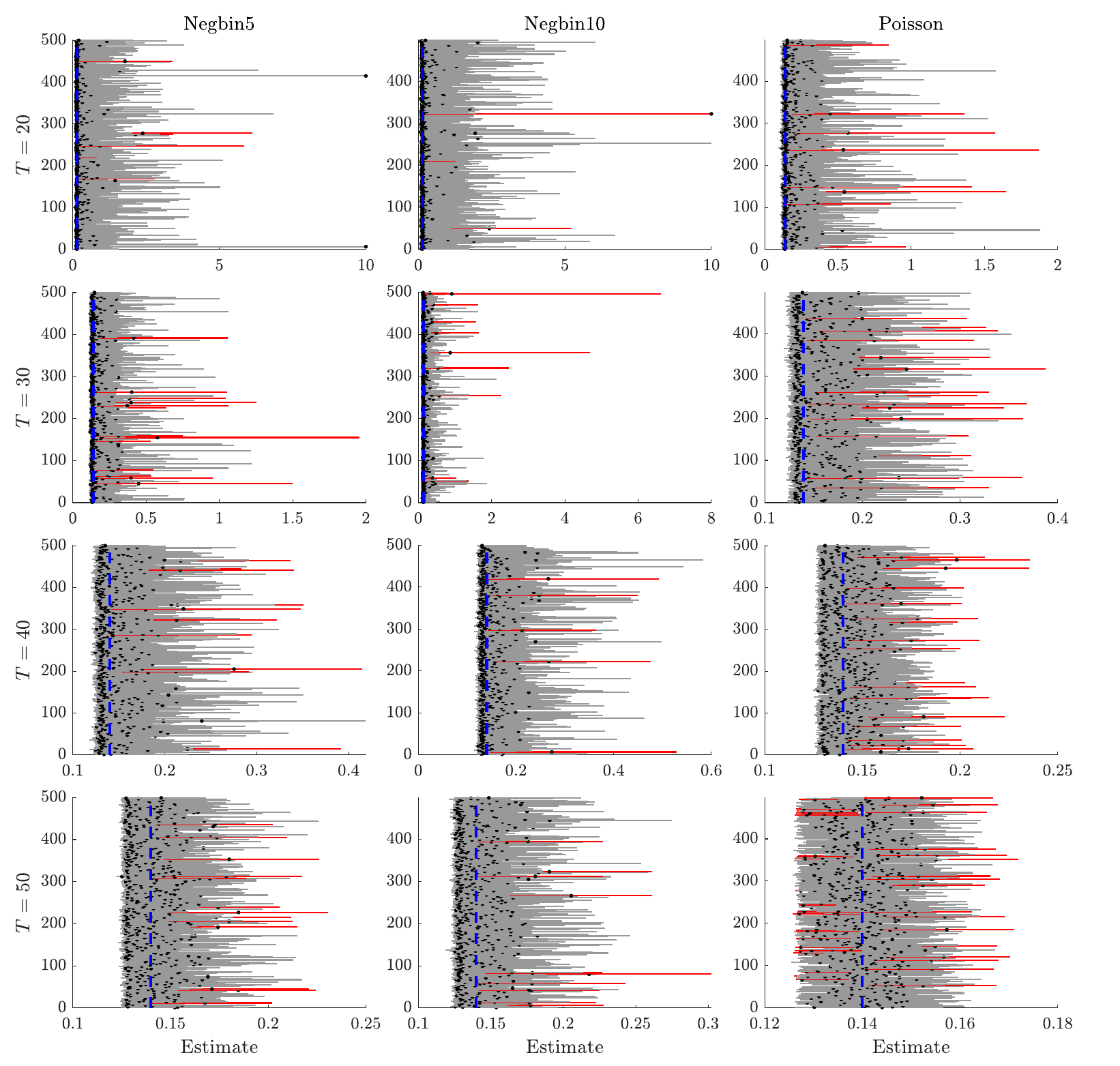}
\caption{
Parameter estimates and 95\% confidence intervals (CIs) for the growth rate $r$ across 500 simulation replicates and calibration window lengths $T=20,30,40,50$, with the true value $r=0.14$ indicated by the vertical blue dashed line. Columns correspond to the error structures: negative binomial with data-generating dispersion parameter $\alpha=5$ (Negbin5), negative binomial with data-generating dispersion parameter $\alpha=10$ (Negbin10), and Poisson. Each horizontal line corresponds to a single simulation replicate, showing the bootstrap confidence interval obtained by resampling within that replicate, with the corresponding point estimate marked by a black dot at its center. Red intervals denote confidence intervals that do not contain the true value, whereas gray intervals denote those that do.
}
\label{fig:GGM_S1_CI_grid_param_1}
\end{figure}

\begin{figure}[H]
\centering
\includegraphics[width=0.9\linewidth]{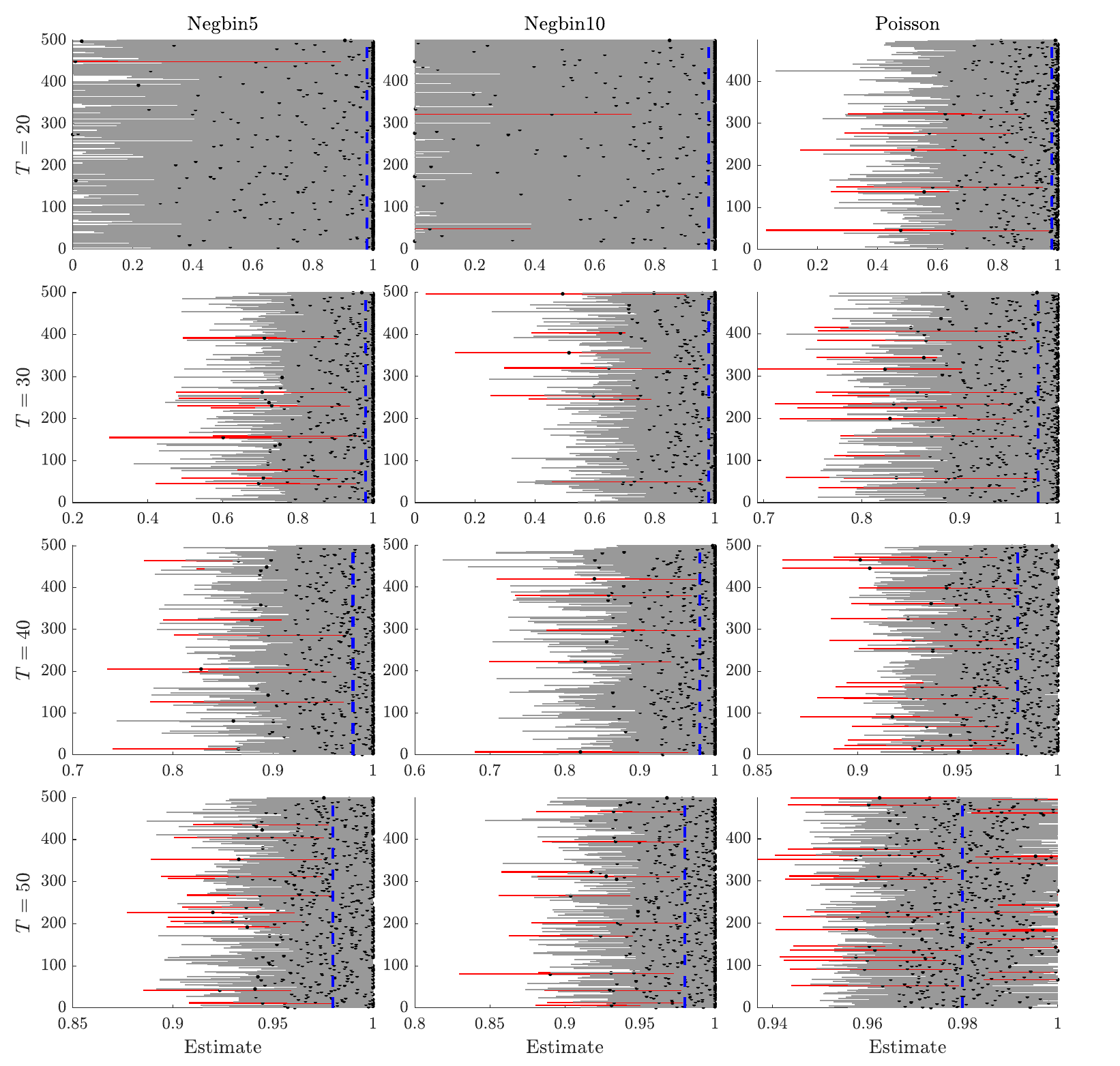}
\caption{
Parameter estimates and 95\% confidence intervals (CIs) for the deceleration parameter $p$ across 500 simulation replicates and calibration window lengths $T=20,30,40,50$, with the true value $p=0.98$ indicated by the vertical blue dashed line. Columns correspond to the error structures: negative binomial with data-generating dispersion parameter $\alpha=5$ (Negbin5), negative binomial with data-generating dispersion parameter $\alpha=10$ (Negbin10), and Poisson. Each horizontal line corresponds to a single simulation replicate, showing the bootstrap confidence interval obtained by resampling within that replicate, with the corresponding point estimate marked by a black dot at its center. Red intervals denote confidence intervals that do not contain the true value, whereas gray intervals denote those that do.
}
\label{fig:GGM_S1_CI_grid_param_2}
\end{figure}

\section{GLM Model}

\subsection{Scenario~1: Estimated $\{r,\, p,\, K\}$}

\begin{figure}[H]
\centering
\includegraphics[width=\linewidth]{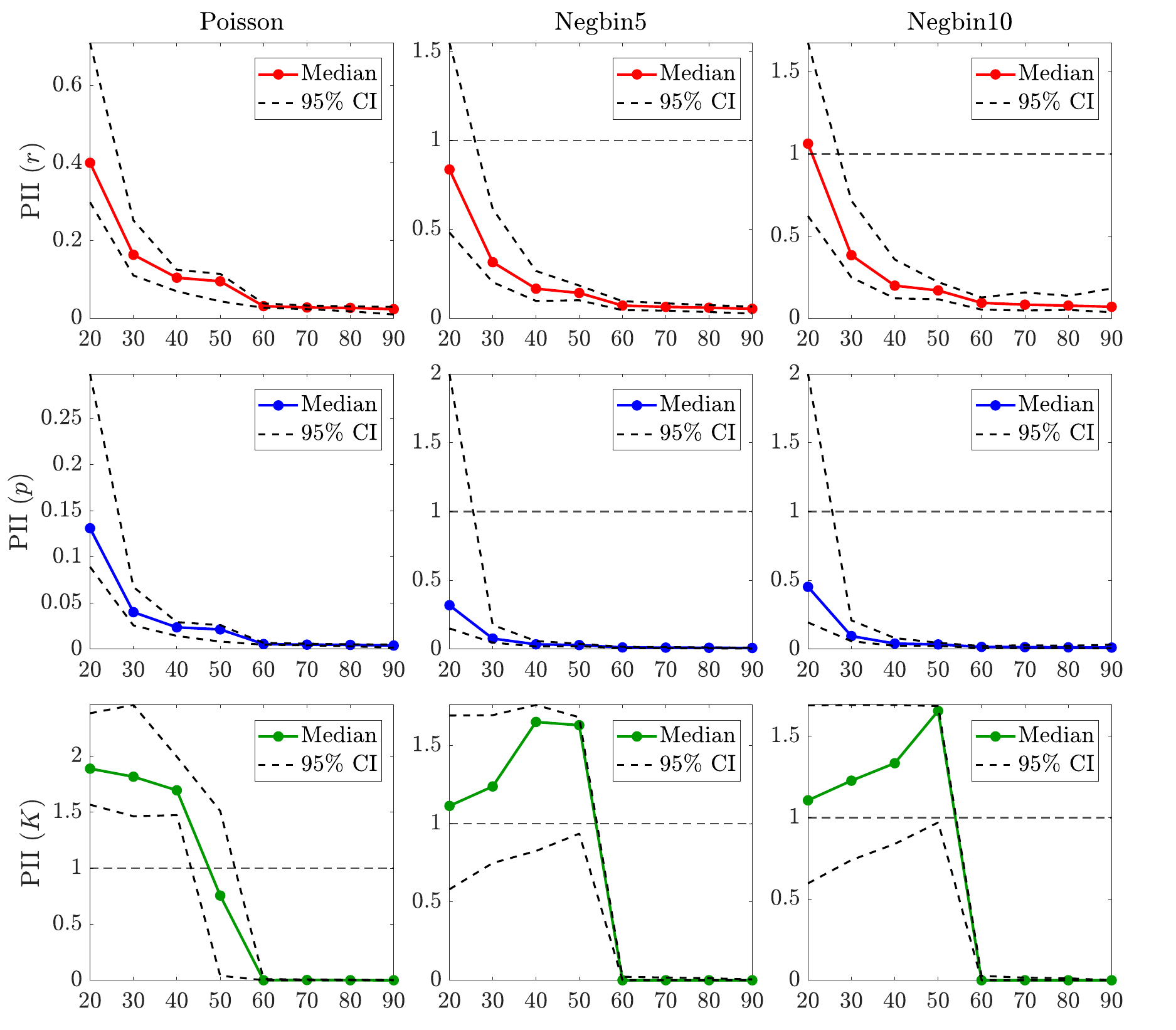}
\caption{Practical Identifiability Index (PII) for the growth rate $r$, deceleration parameter $p$, and carrying capacity $K$ in the GLM model across calibration-window lengths $T=20, 30, \ldots, 90$ under three error structures: Poisson, negative binomial with data-generating dispersion $\alpha=5$ (Negbin5), and negative binomial with data-generating dispersion $\alpha=10$ (Negbin10). Red lines show the median PII across replicates, and dashed black curves indicate the central 95\% interval.}
\label{fig:GLM_S1_PII_parameters}
\end{figure}

\begin{figure}[H]
\centering
\includegraphics[clip,trim=0 580 0 0,width=\linewidth]{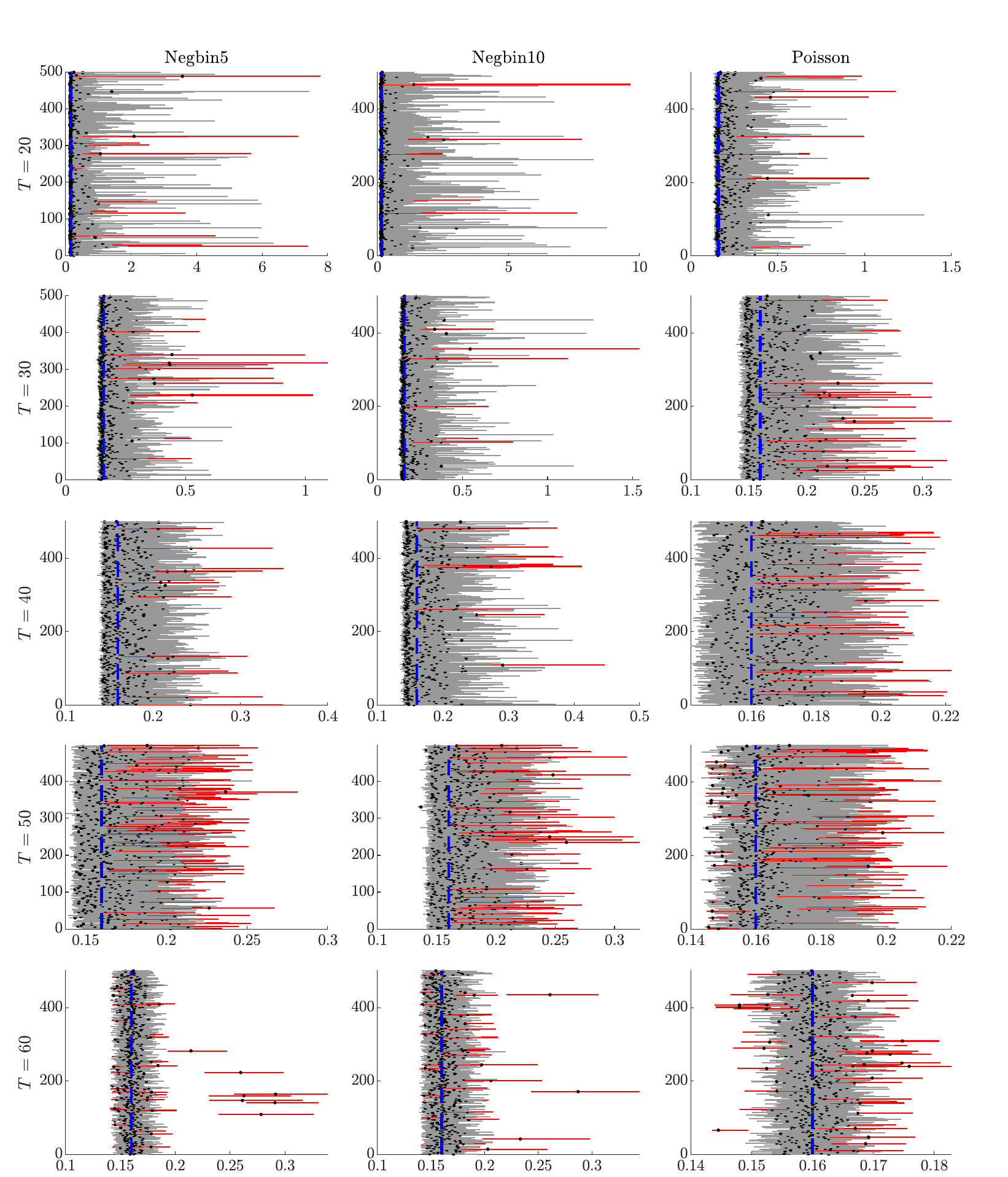}
\caption{
Parameter estimates and 95\% confidence intervals (CIs) for the growth rate $r$ across 500 simulation replicates and calibration window lengths $T=20,30,\ldots,90$, with the true value $r=0.16$ indicated by the vertical blue dashed line. Columns correspond to the error structures: negative binomial with data-generating dispersion parameter $\alpha=5$ (Negbin5), negative binomial with data-generating dispersion parameter $\alpha=10$ (Negbin10), and Poisson. Each horizontal line corresponds to a single simulation replicate, showing the bootstrap confidence interval obtained by resampling within that replicate, with the corresponding point estimate marked by a black dot at its center. Red intervals denote confidence intervals that do not contain the true value, whereas gray intervals denote those that do.
}
\label{fig:GLM_S1_CI_grid_param_1}
\end{figure}

\begin{figure}[H]
\centering
\includegraphics[clip,trim=0 0 0 920,width=\linewidth]{GLM_S1_CI_grid_param_1.pdf}
\caption*{Figure~\ref{fig:GLM_S1_CI_grid_param_1} (continued).}
\end{figure}

\begin{figure}[H]
\centering
\includegraphics[clip,trim=0 580 0 0,width=\linewidth]{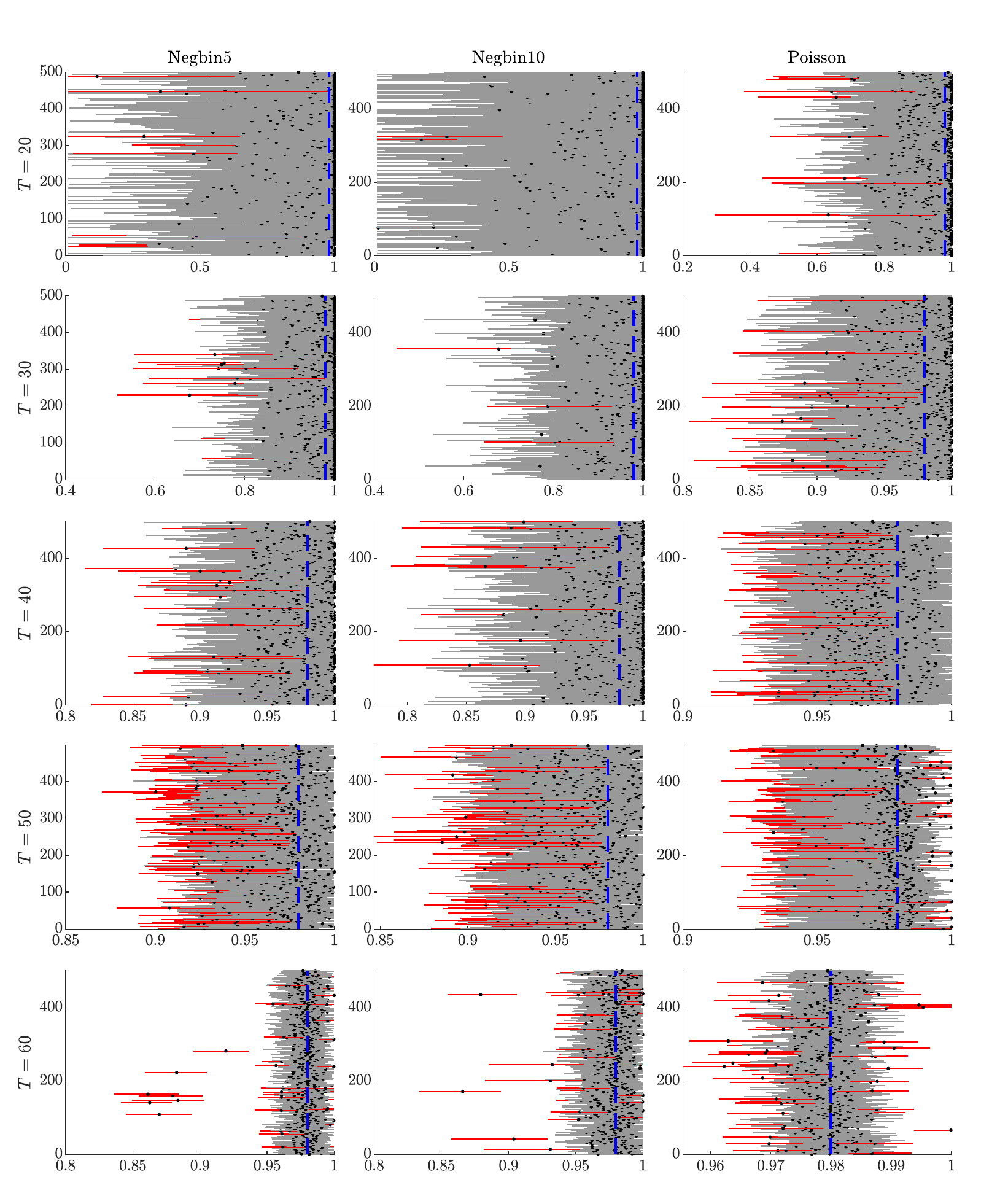}
\caption{
Parameter estimates and 95\% confidence intervals (CIs) for the deceleration parameter $p$ across 500 simulation replicates and calibration window lengths $T=20,30,\ldots,90$, with the true value $p=0.98$ indicated by the vertical blue dashed line. Columns correspond to the error structures: negative binomial with data-generating dispersion parameter $\alpha=5$ (Negbin5), negative binomial with data-generating dispersion parameter $\alpha=10$ (Negbin10), and Poisson. Each horizontal line corresponds to a single simulation replicate, showing the bootstrap confidence interval obtained by resampling within that replicate, with the corresponding point estimate marked by a black dot at its center. Red intervals denote confidence intervals that do not contain the true value, whereas gray intervals denote those that do.
}
\label{fig:GLM_S1_CI_grid_param_2}
\end{figure}

\begin{figure}[H]
\centering
\includegraphics[clip,trim=0 0 0 920,width=\linewidth]{GLM_S1_CI_grid_param_2.pdf}
\caption*{Figure~\ref{fig:GLM_S1_CI_grid_param_2} (continued).}
\end{figure}

\begin{figure}[H]
\centering
\includegraphics[clip,trim=0 570 0 0,width=\linewidth]{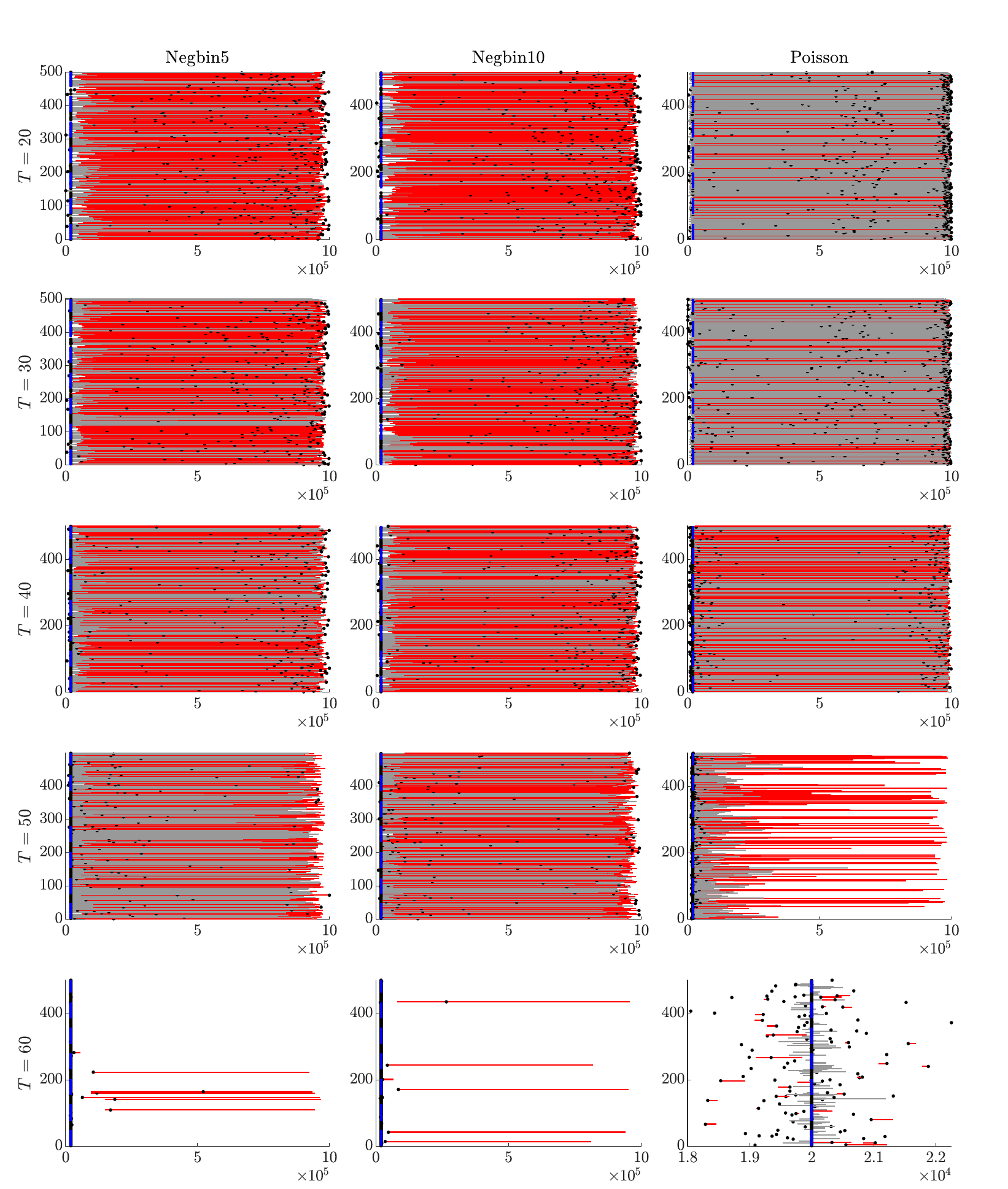}
\caption{
Parameter estimates and 95\% confidence intervals (CIs) for the carrying capacity $K$ across 500 simulation replicates and calibration window lengths $T=20,30,\ldots,90$, with the true value $K=20000$ indicated by the vertical blue dashed line. Columns correspond to the error structures: negative binomial with data-generating dispersion parameter $\alpha=5$ (Negbin5), negative binomial with data-generating dispersion parameter $\alpha=10$ (Negbin10), and Poisson. Each horizontal line corresponds to a single simulation replicate, showing the bootstrap confidence interval obtained by resampling within that replicate, with the corresponding point estimate marked by a black dot at its center. Red intervals denote confidence intervals that do not contain the true value, whereas gray intervals denote those that do.
}
\label{fig:GLM_S1_CI_grid_param_3}
\end{figure}

\begin{figure}[H]
\centering
\includegraphics[clip,trim=0 0 0 930,width=\linewidth]{GLM_S1_CI_grid_param_3.pdf}
\caption*{Figure~\ref{fig:GLM_S1_CI_grid_param_3} (continued).}
\end{figure}

\pagebreak
\section{SIR Model}

\subsection{Scenario~1: Estimated $\{\beta\}$; Fixed $\{\gamma,\, N\}$}

\begin{figure}[H]
\centering
\includegraphics[width=\linewidth]{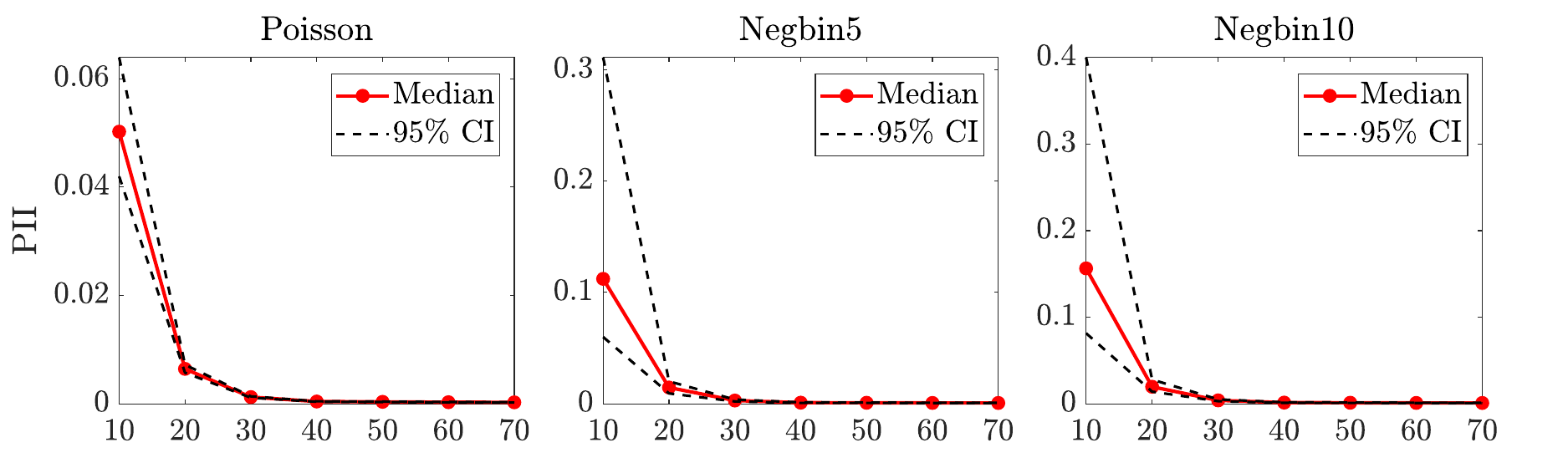}
\caption{Practical Identifiability Index (PII) for the transmission rate $\beta$ in the SIR model (Scenario~1) across calibration-window lengths $T=10, 20, \ldots, 70$ under three error structures: Poisson, negative binomial with data-generating dispersion $\alpha=5$ (Negbin5), and negative binomial with data-generating dispersion $\alpha=10$ (Negbin10). Red lines show the median PII across replicates, and dashed black curves indicate the central 95\% interval.}
\label{fig:SIR_S1_PII_parameters}
\end{figure}

\begin{figure}[H]
\centering
\includegraphics[clip,trim=0 580 0 0,width=\linewidth]{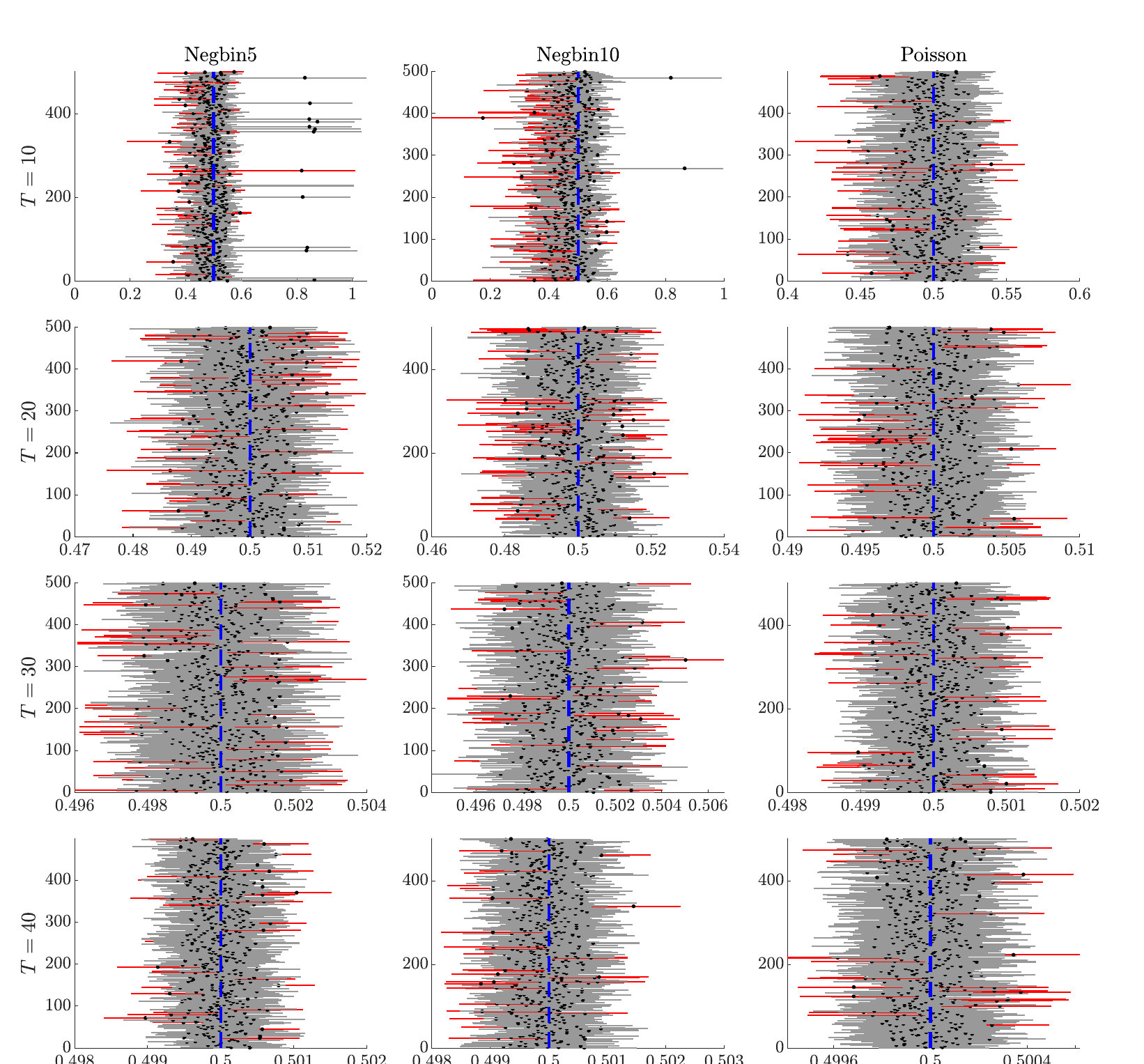}
\caption{
Parameter estimates and 95\% confidence intervals (CIs) for the transmission rate $\beta$ across 500 simulation replicates and calibration window lengths $T=10,20,\ldots,70$, with the true value $\beta=0.5$ indicated by the vertical blue dashed line. Columns correspond to the error structures: negative binomial with data-generating dispersion parameter $\alpha=5$ (Negbin5), negative binomial with data-generating dispersion parameter $\alpha=10$ (Negbin10), and Poisson. Each horizontal line corresponds to a single simulation replicate, showing the bootstrap confidence interval obtained by resampling within that replicate, with the corresponding point estimate marked by a black dot at its center. Red intervals denote confidence intervals that do not contain the true value, whereas gray intervals denote those that do.
}
\label{fig:SIR_S1_CI_grid_param_1}
\end{figure}

\begin{figure}[H]
\centering
\includegraphics[clip,trim=0 0 0 735,width=\linewidth]{SIR_S1_CI_grid_param_1.pdf}
\caption*{Figure~\ref{fig:SIR_S1_CI_grid_param_1} (continued).}
\end{figure}

\subsection{Scenario~2: Estimated $\{\beta,\, \gamma\}$; Fixed $\{N\}$}

\begin{figure}[H]
\centering
\includegraphics[width=\linewidth]{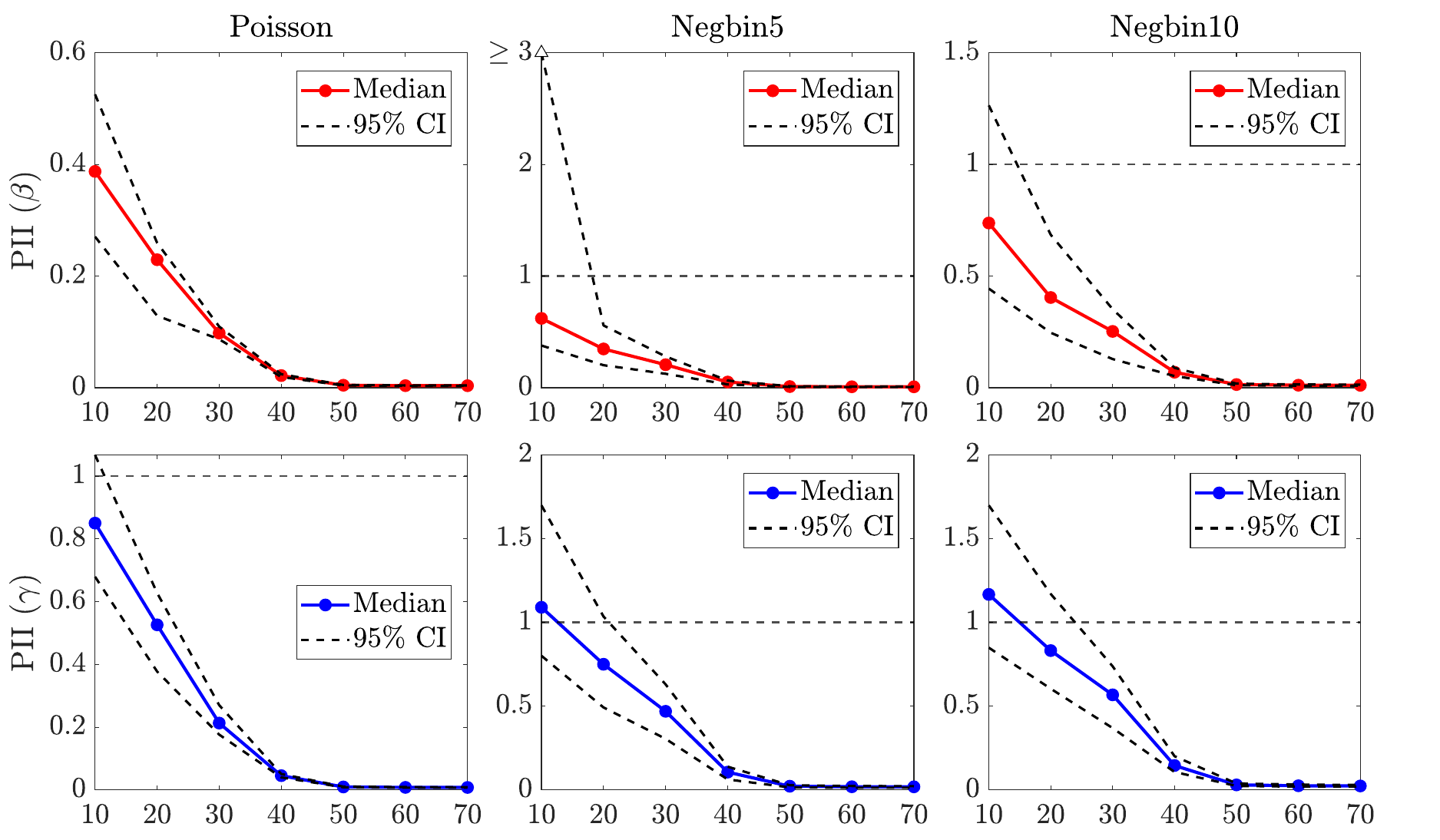}
\caption{Practical Identifiability Index (PII) for the transmission rate $\beta$ and recovery rate $\gamma$ in the SIR model (Scenario~2) across calibration-window lengths $T=10, 20, \ldots, 70$ under three error structures: Poisson, negative binomial with data-generating dispersion $\alpha=5$ (Negbin5), and negative binomial with data-generating dispersion $\alpha=10$ (Negbin10). Red lines show the median PII across replicates, and dashed black curves indicate the central 95\% interval. For visualization only, within subplots containing PII values at or above 3, those values are displayed at an upper plotting limit of 3; uncapped values are retained for all numerical analyses. Upward triangles denote values displayed at the upper limit.}
\label{fig:SIR_S2_PII_parameters}
\end{figure}

\begin{figure}[H]
\centering
\includegraphics[clip,trim=0 580 0 0,width=\linewidth]{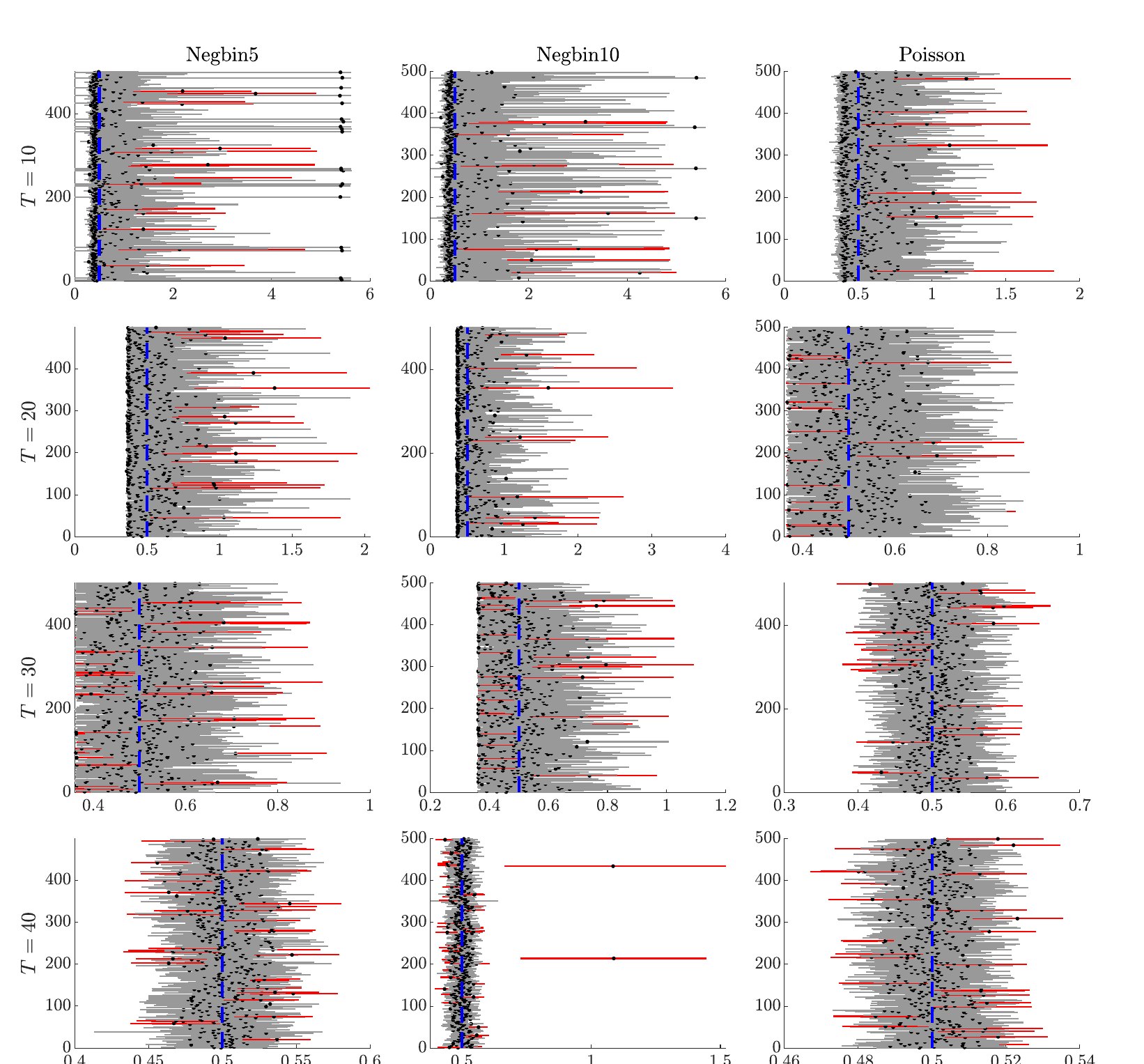}
\caption{
Parameter estimates and 95\% confidence intervals (CIs) for the transmission rate $\beta$ across 500 simulation replicates and calibration window lengths $T=10,20,\ldots,70$, with the true value $\beta=0.5$ indicated by the vertical blue dashed line. Columns correspond to the error structures: negative binomial with data-generating dispersion parameter $\alpha=5$ (Negbin5), negative binomial with data-generating dispersion parameter $\alpha=10$ (Negbin10), and Poisson. Each horizontal line corresponds to a single simulation replicate, showing the bootstrap confidence interval obtained by resampling within that replicate, with the corresponding point estimate marked by a black dot at its center. Red intervals denote confidence intervals that do not contain the true value, whereas gray intervals denote those that do.
}
\label{fig:SIR_S2_CI_grid_param_1}
\end{figure}

\begin{figure}[H]
\centering
\includegraphics[clip,trim=0 0 0 735,width=\linewidth]{SIR_S2_CI_grid_param_1.pdf}
\caption*{Figure~\ref{fig:SIR_S2_CI_grid_param_1} (continued).}
\end{figure}

\begin{figure}[H]
\centering
\includegraphics[clip,trim=0 580 0 0,width=\linewidth]{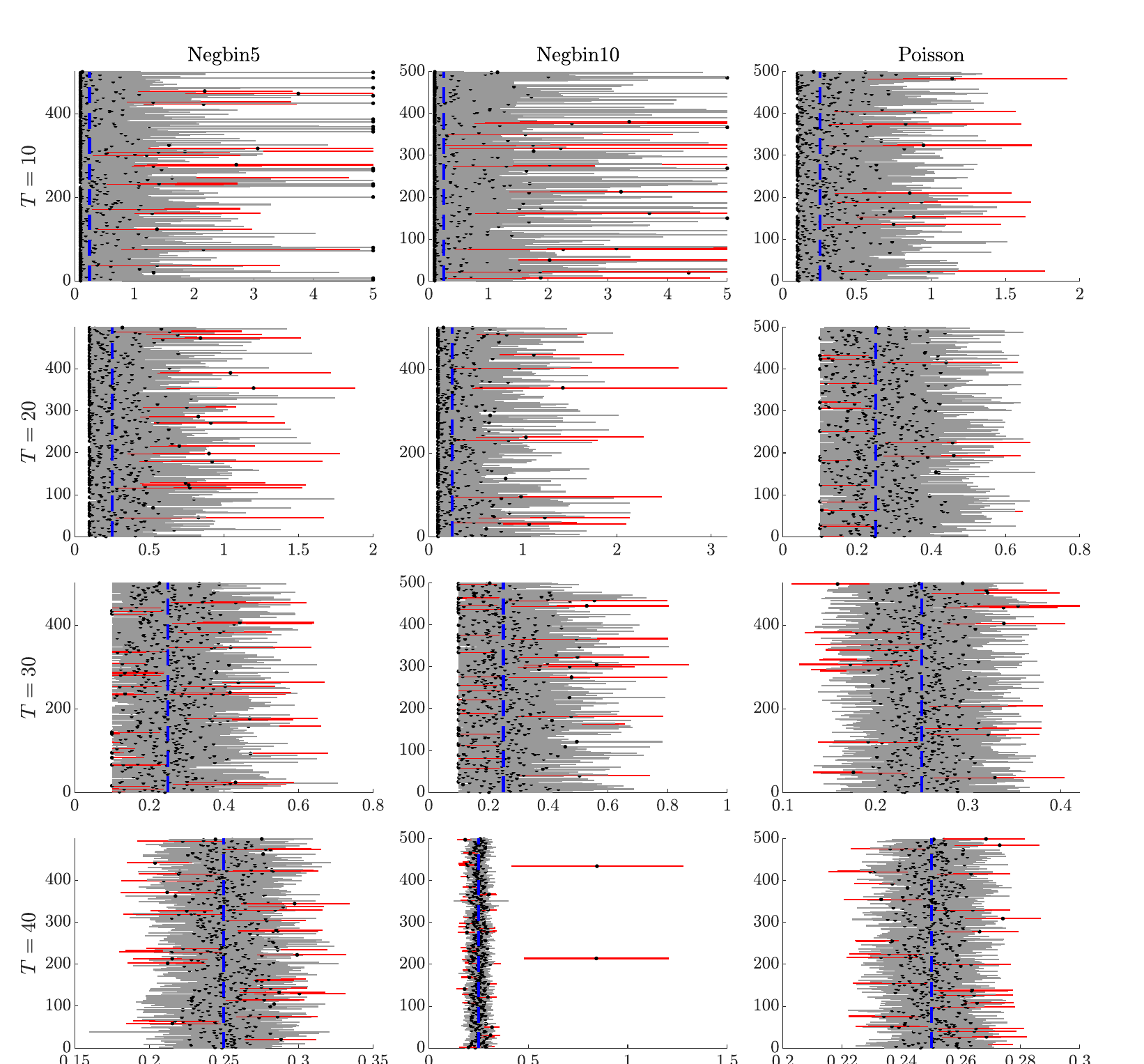}
\caption{
Parameter estimates and 95\% confidence intervals (CIs) for the recovery rate $\gamma$ across 500 simulation replicates and calibration window lengths $T=10,20,\ldots,70$, with the true value $\gamma=0.25$ indicated by the vertical blue dashed line. Columns correspond to the error structures: negative binomial with data-generating dispersion parameter $\alpha=5$ (Negbin5), negative binomial with data-generating dispersion parameter $\alpha=10$ (Negbin10), and Poisson. Each horizontal line corresponds to a single simulation replicate, showing the bootstrap confidence interval obtained by resampling within that replicate, with the corresponding point estimate marked by a black dot at its center. Red intervals denote confidence intervals that do not contain the true value, whereas gray intervals denote those that do.
}
\label{fig:SIR_S2_CI_grid_param_2}
\end{figure}

\begin{figure}[H]
\centering
\includegraphics[clip,trim=0 0 0 735,width=\linewidth]{SIR_S2_CI_grid_param_2.pdf}
\caption*{Figure~\ref{fig:SIR_S2_CI_grid_param_2} (continued).}
\end{figure}

\pagebreak
\section{SEIR Model}

\subsection{Scenario~1: Estimated $\{\beta\}$; Fixed $\{\kappa,\, \gamma,\, N\}$}

\begin{figure}[H]
\centering
\includegraphics[width=\linewidth]{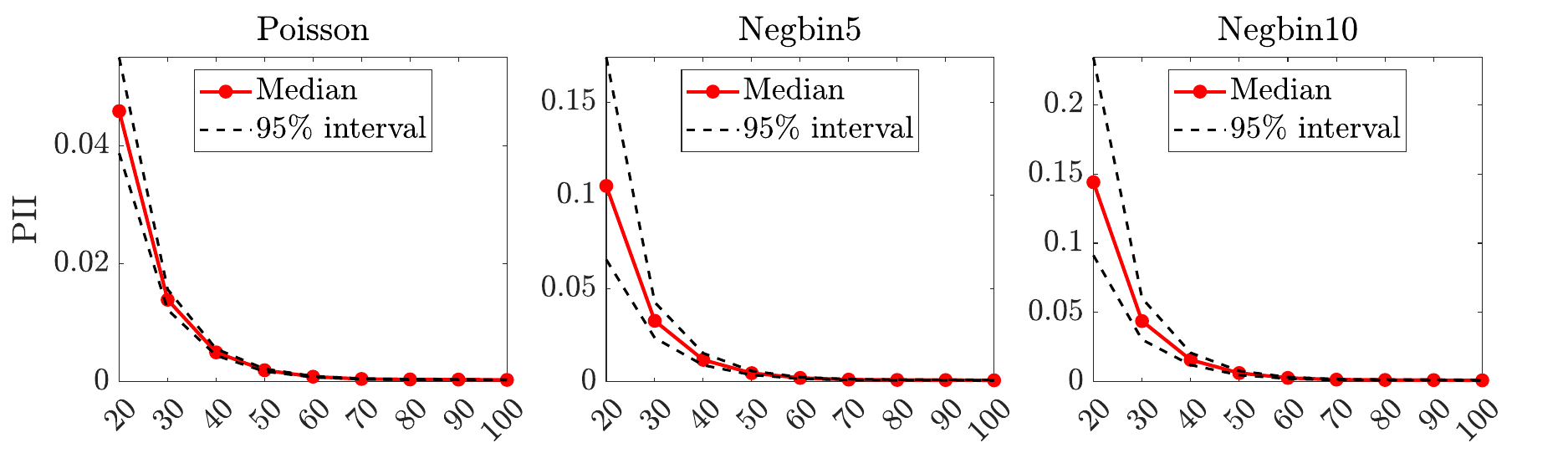}
\caption{Practical Identifiability Index (PII) for the transmission rate $\beta$ in the SEIR model (Scenario~1) across calibration-window lengths $T=20, 30, \ldots, 100$ under three error structures: Poisson, negative binomial with data-generating dispersion $\alpha=5$ (Negbin5), and negative binomial with data-generating dispersion $\alpha=10$ (Negbin10). Red lines show the median PII across replicates, and dashed black curves indicate the central 95\% interval.}
\label{fig:SEIR_S1_PII_parameters}
\end{figure}

\begin{figure}[H]
\centering
\includegraphics[clip,trim=0 760 0 0,width=\linewidth]{SEIR_S1_CI_grid_param_1.pdf}
\caption{
Parameter estimates and 95\% confidence intervals (CIs) for the transmission rate $\beta$ across 500 simulation replicates and calibration window lengths $T=20,30,\ldots,100$, with the true value $\beta=0.5$ indicated by the vertical blue dashed line. Columns correspond to the error structures: negative binomial with data-generating dispersion parameter $\alpha=5$ (Negbin5), negative binomial with data-generating dispersion parameter $\alpha=10$ (Negbin10), and Poisson. Each horizontal line corresponds to a single simulation replicate, showing the bootstrap confidence interval obtained by resampling within that replicate, with the corresponding point estimate marked by a black dot at its center. Red intervals denote confidence intervals that do not contain the true value, whereas gray intervals denote those that do.
}
\label{fig:SEIR_S1_CI_grid_param_1}
\end{figure}

\begin{figure}[H]
\centering
\includegraphics[clip,trim=0 0 0 920,width=\linewidth]{SEIR_S1_CI_grid_param_1.pdf}
\caption*{Figure~\ref{fig:SEIR_S1_CI_grid_param_1} (continued).}
\end{figure}

\subsection{Scenario~2: Estimated $\{\beta,\, \gamma\}$; Fixed $\{\kappa,\, N\}$}

\begin{figure}[H]
\centering
\includegraphics[width=\linewidth]{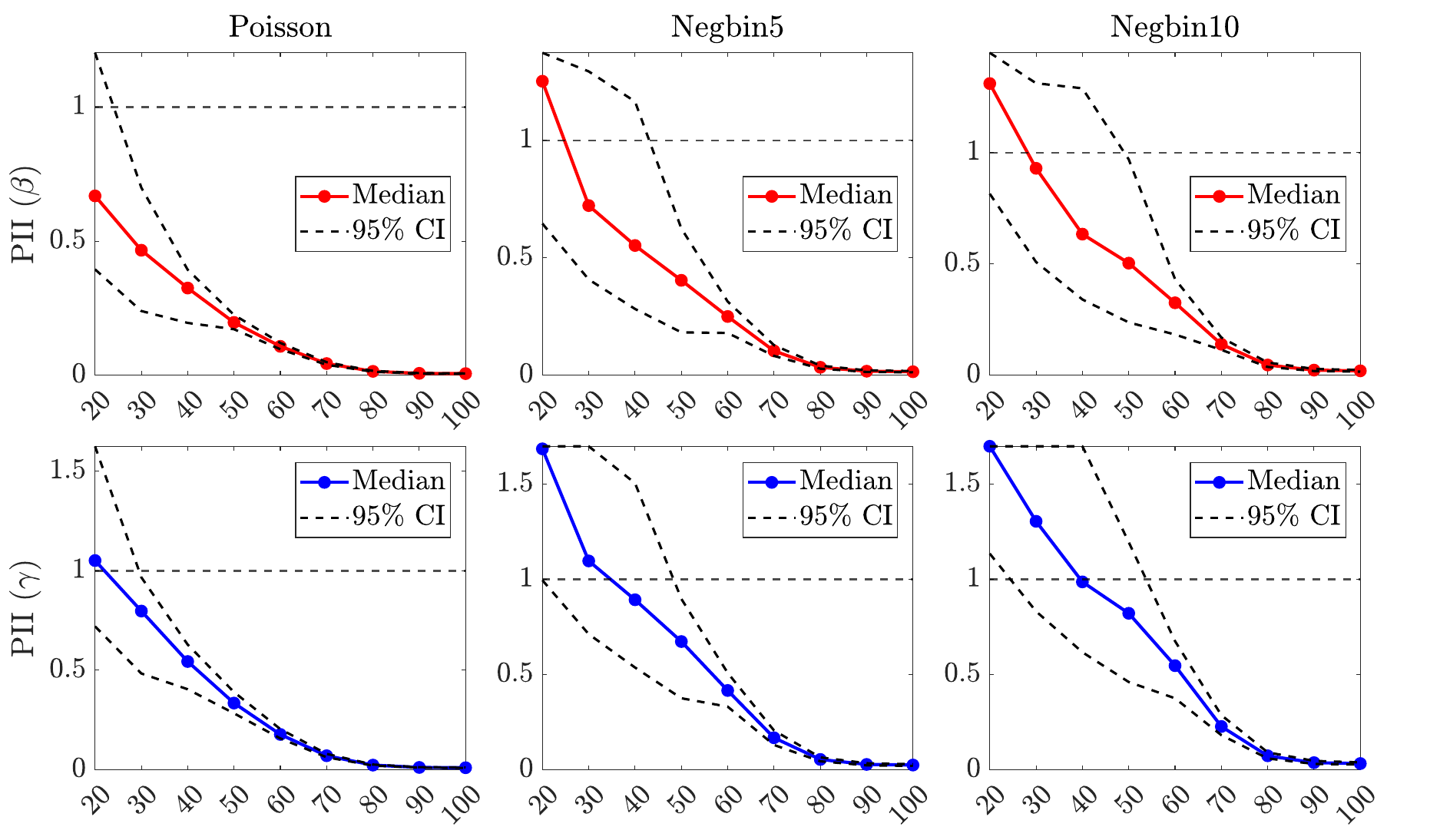}
\caption{Practical Identifiability Index (PII) for the transmission rate $\beta$ and recovery rate $\gamma$ in the SEIR model (Scenario~2) across calibration-window lengths $T=20, 30, \ldots, 100$ under three error structures: Poisson, negative binomial with data-generating dispersion $\alpha=5$ (Negbin5), and negative binomial with data-generating dispersion $\alpha=10$ (Negbin10). Red lines show the median PII across replicates, and dashed black curves indicate the central 95\% interval.}
\label{fig:SEIR_S2_PII_parameters}
\end{figure}

\begin{figure}[H]
\centering
\includegraphics[clip,trim=0 760 0 0,width=\linewidth]{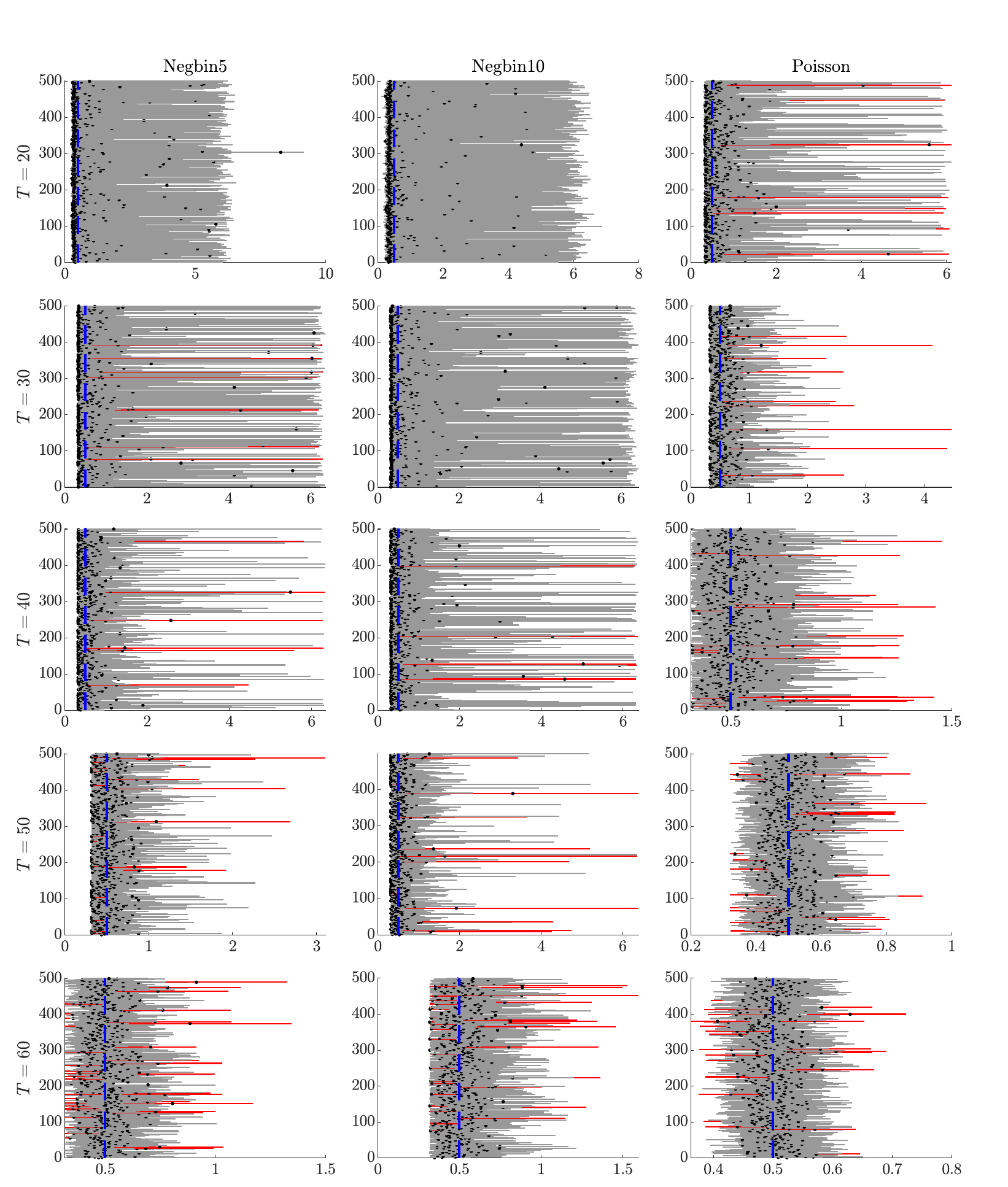}
\caption{
Parameter estimates and 95\% confidence intervals (CIs) for the transmission rate $\beta$ across 500 simulation replicates and calibration window lengths $T=20,30,\ldots,100$, with the true value $\beta=0.5$ indicated by the vertical blue dashed line. Columns correspond to the error structures: negative binomial with data-generating dispersion parameter $\alpha=5$ (Negbin5), negative binomial with data-generating dispersion parameter $\alpha=10$ (Negbin10), and Poisson. Each horizontal line corresponds to a single simulation replicate, showing the bootstrap confidence interval obtained by resampling within that replicate, with the corresponding point estimate marked by a black dot at its center. Red intervals denote confidence intervals that do not contain the true value, whereas gray intervals denote those that do.
}
\label{fig:SEIR_S2_CI_grid_param_1}
\end{figure}

\begin{figure}[H]
\centering
\includegraphics[clip,trim=0 0 0 920,width=\linewidth]{SEIR_S2_CI_grid_param_1.pdf}
\caption*{Figure~\ref{fig:SEIR_S2_CI_grid_param_1} (continued).}
\end{figure}

\begin{figure}[H]
\centering
\includegraphics[clip,trim=0 760 0 0,width=\linewidth]{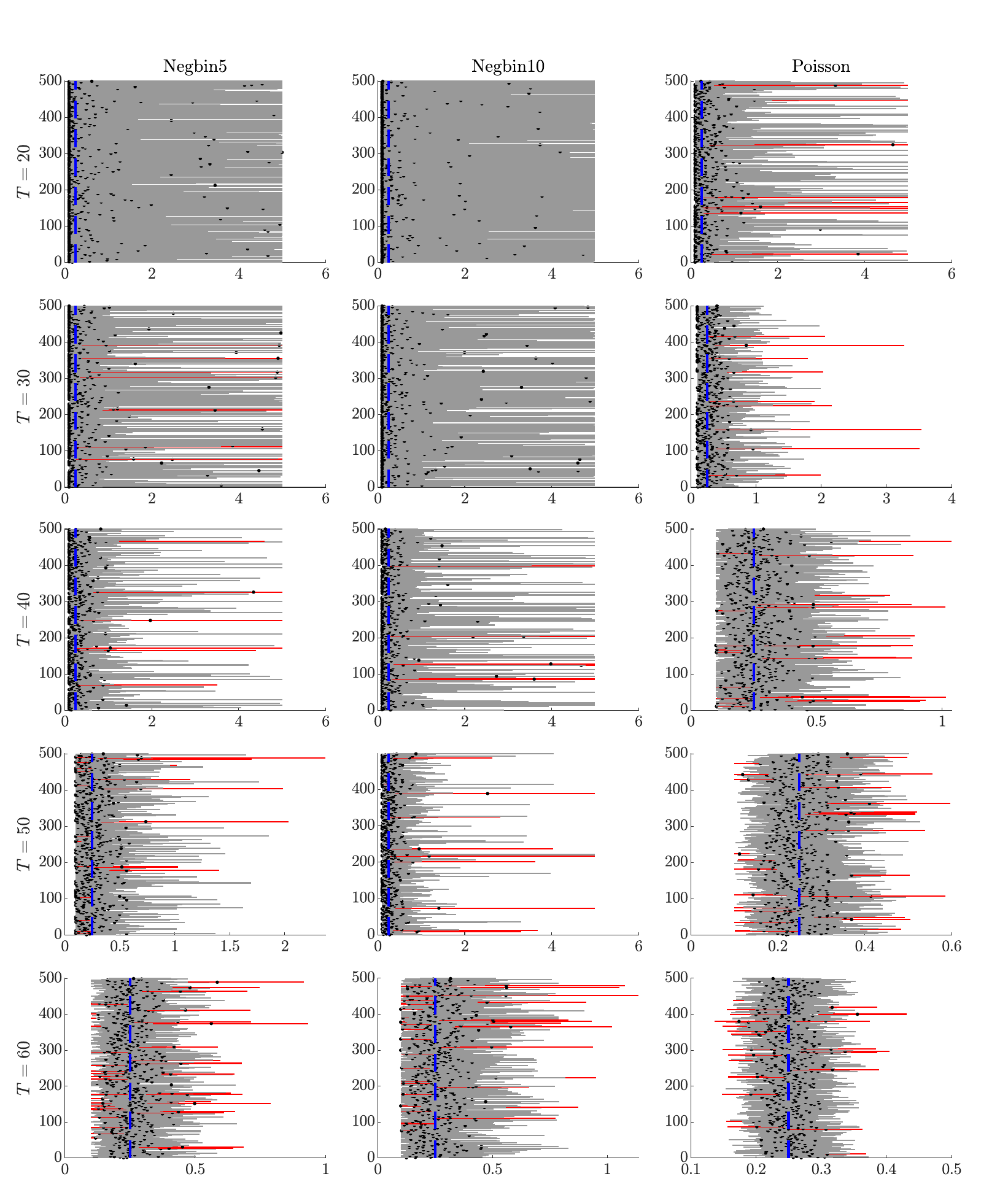}
\caption{
Parameter estimates and 95\% confidence intervals (CIs) for the recovery rate $\gamma$ across 500 simulation replicates and calibration window lengths $T=20,30,\ldots,100$, with the true value $\gamma=0.25$ indicated by the vertical blue dashed line. Columns correspond to the error structures: negative binomial with data-generating dispersion parameter $\alpha=5$ (Negbin5), negative binomial with data-generating dispersion parameter $\alpha=10$ (Negbin10), and Poisson. Each horizontal line corresponds to a single simulation replicate, showing the bootstrap confidence interval obtained by resampling within that replicate, with the corresponding point estimate marked by a black dot at its center. Red intervals denote confidence intervals that do not contain the true value, whereas gray intervals denote those that do.
}
\label{fig:SEIR_S2_CI_grid_param_3}
\end{figure}

\begin{figure}[H]
\centering
\includegraphics[clip,trim=0 0 0 920,width=\linewidth]{SEIR_S2_CI_grid_param_3.pdf}
\caption*{Figure~\ref{fig:SEIR_S2_CI_grid_param_3} (continued).}
\end{figure}

\subsection{Scenario~3: Estimated $\{\beta,\, \kappa,\, \gamma\}$; Fixed $\{N\}$}

\begin{figure}[H]
\centering
\includegraphics[width=\linewidth]{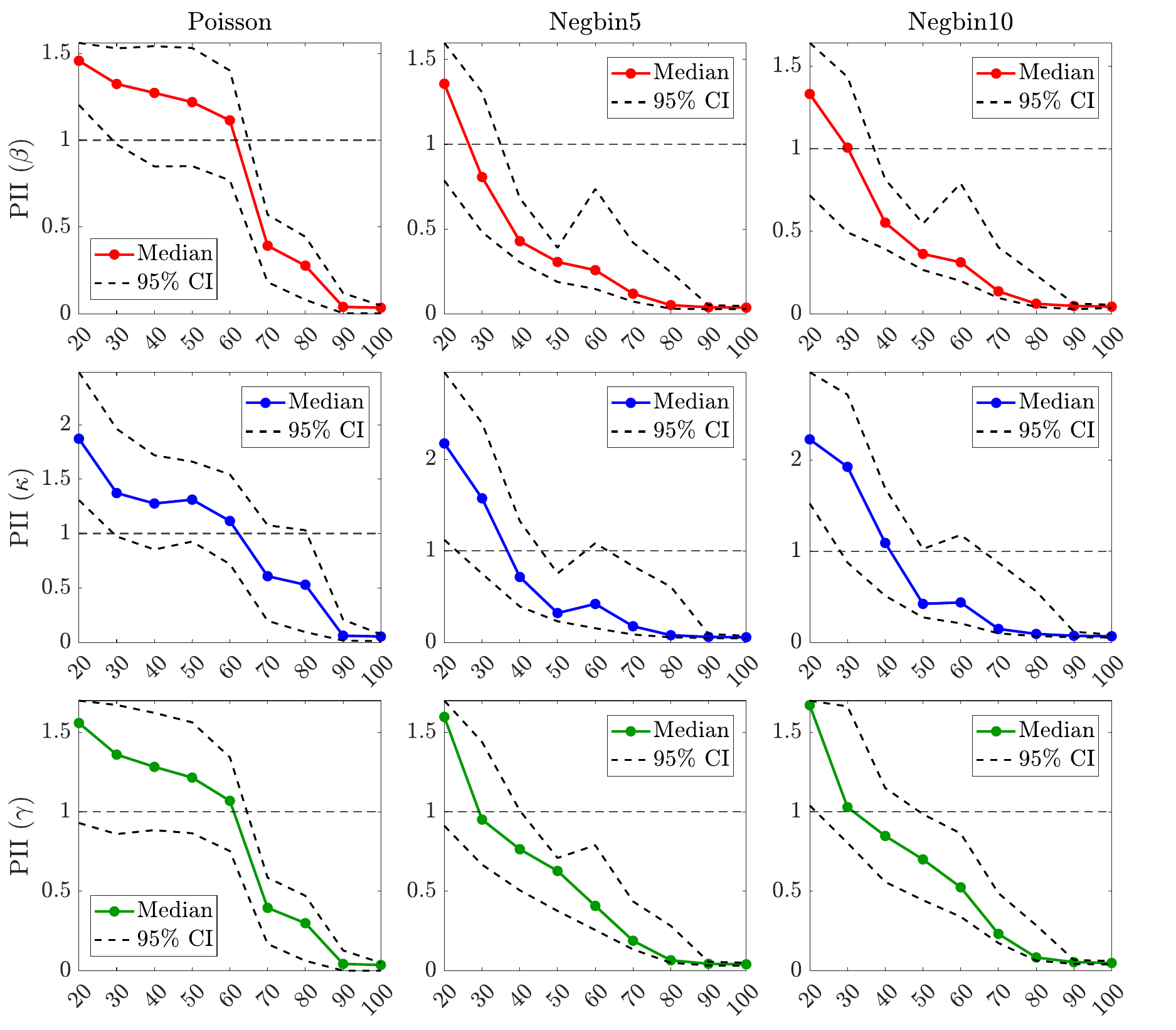}
\caption{Practical Identifiability Index (PII) for the transmission rate $\beta$, incubation rate $\kappa$, and recovery rate $\gamma$ in the SEIR model (Scenario~3) across calibration-window lengths $T=20, 30, \ldots, 100$ under three error structures: Poisson, negative binomial with data-generating dispersion $\alpha=5$ (Negbin5), and negative binomial with data-generating dispersion $\alpha=10$ (Negbin10). Red lines show the median PII across replicates, and dashed black curves indicate the central 95\% interval.}
\label{fig:SEIR_S3_PII_parameters}
\end{figure}

\begin{figure}[H]
\centering
\includegraphics[clip,trim=0 760 0 0,width=\linewidth]{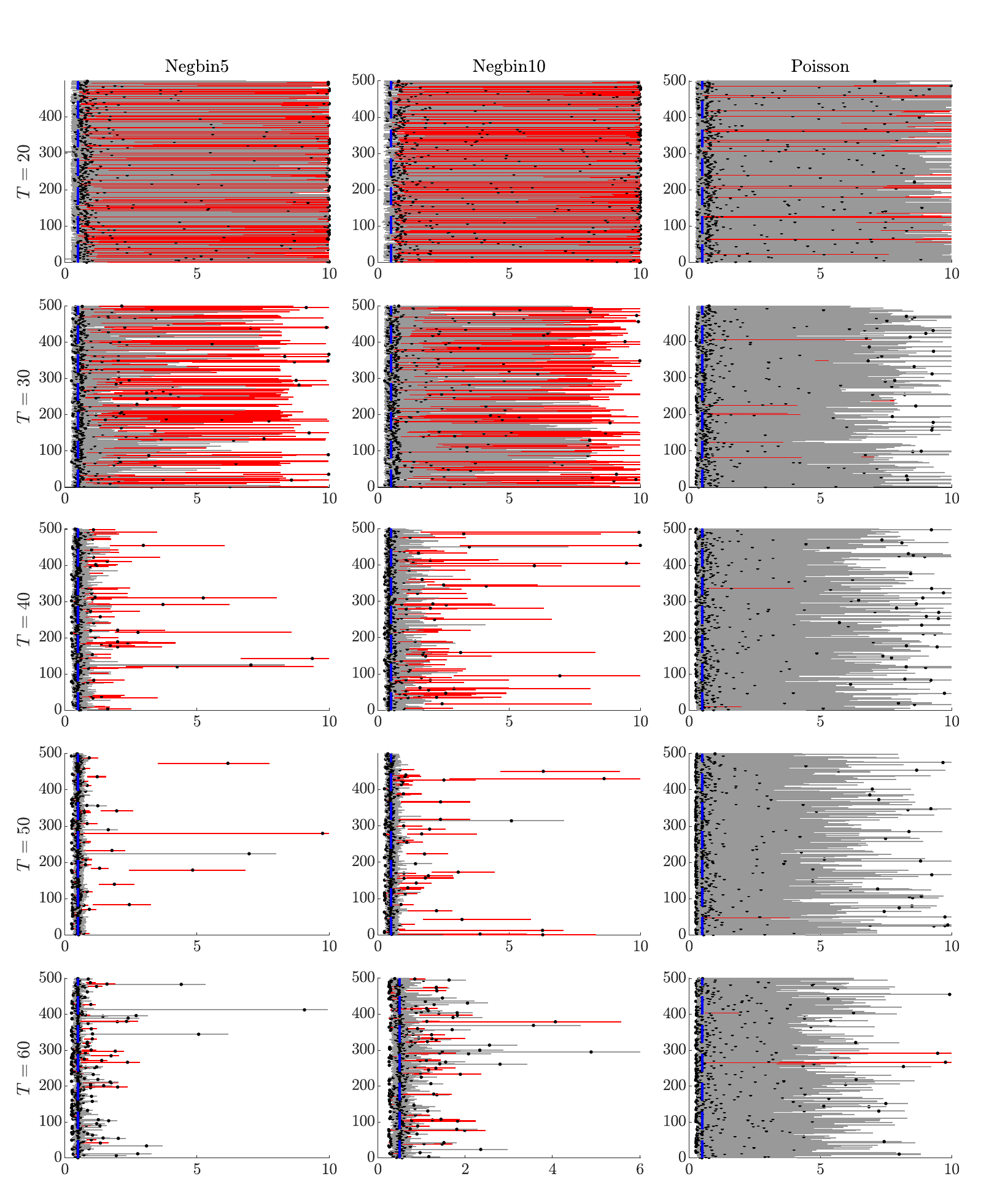}
\caption{
Parameter estimates and 95\% confidence intervals (CIs) for the transmission rate $\beta$ across 500 simulation replicates and calibration window lengths $T=20,30,\ldots,100$, with the true value $\beta=0.5$ indicated by the vertical blue dashed line. Columns correspond to the error structures: negative binomial with data-generating dispersion parameter $\alpha=5$ (Negbin5), negative binomial with data-generating dispersion parameter $\alpha=10$ (Negbin10), and Poisson. Each horizontal line corresponds to a single simulation replicate, showing the bootstrap confidence interval obtained by resampling within that replicate, with the corresponding point estimate marked by a black dot at its center. Red intervals denote confidence intervals that do not contain the true value, whereas gray intervals denote those that do.
}
\label{fig:SEIR_S3_CI_grid_param_1}
\end{figure}

\begin{figure}[H]
\centering
\includegraphics[clip,trim=0 0 0 920,width=\linewidth]{SEIR_S3_CI_grid_param_1.pdf}
\caption*{Figure~\ref{fig:SEIR_S3_CI_grid_param_1} (continued).}
\end{figure}

\begin{figure}[H]
\centering
\includegraphics[clip,trim=0 760 0 0,width=\linewidth]{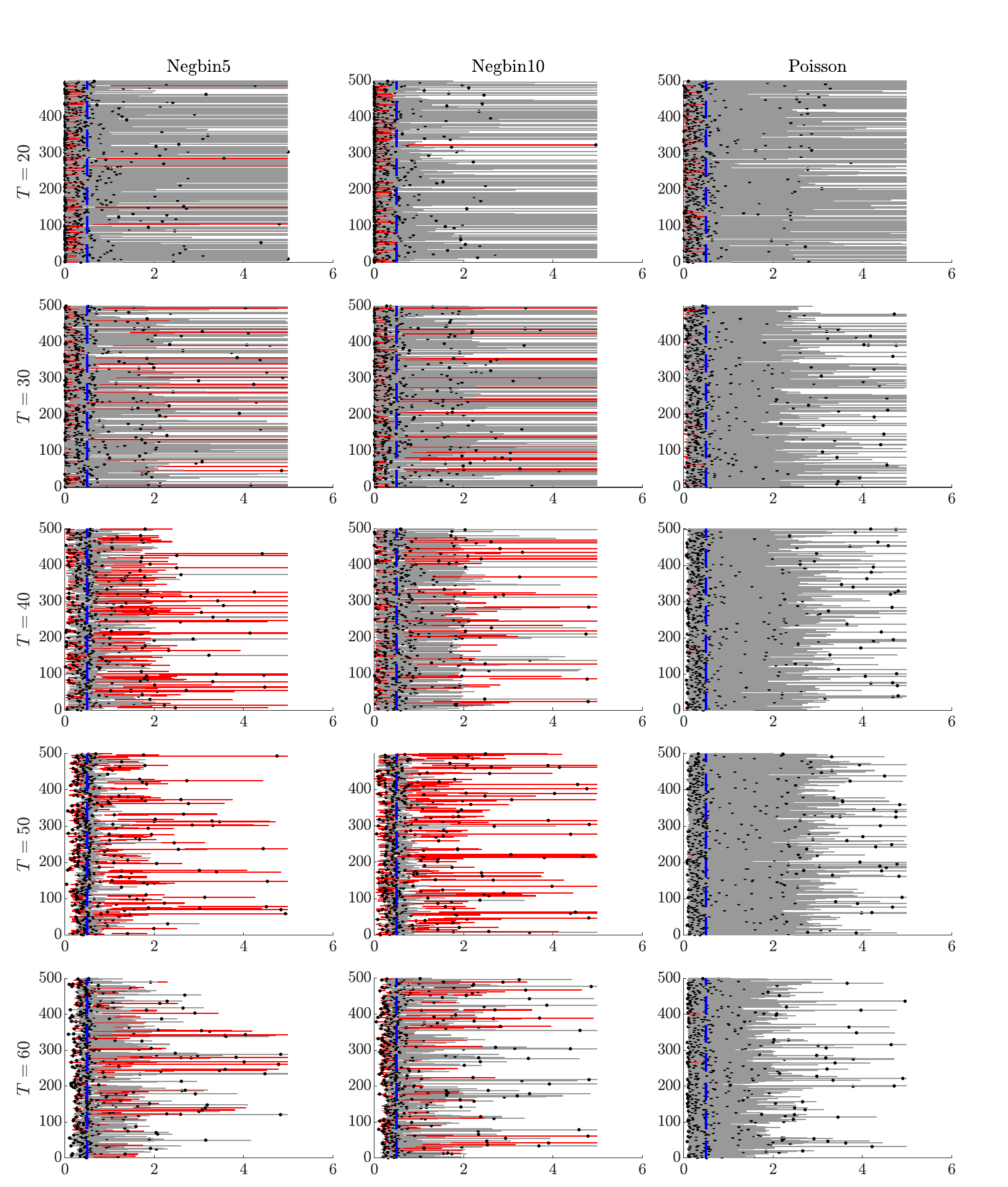}
\caption{
Parameter estimates and 95\% confidence intervals (CIs) for the incubation rate $\kappa$ across 500 simulation replicates and calibration window lengths $T=20,30,\ldots,100$, with the true value $\kappa=0.5$ indicated by the vertical blue dashed line. Columns correspond to the error structures: negative binomial with data-generating dispersion parameter $\alpha=5$ (Negbin5), negative binomial with data-generating dispersion parameter $\alpha=10$ (Negbin10), and Poisson. Each horizontal line corresponds to a single simulation replicate, showing the bootstrap confidence interval obtained by resampling within that replicate, with the corresponding point estimate marked by a black dot at its center. Red intervals denote confidence intervals that do not contain the true value, whereas gray intervals denote those that do.
}
\label{fig:SEIR_S3_CI_grid_param_2}
\end{figure}

\begin{figure}[H]
\centering
\includegraphics[clip,trim=0 0 0 920,width=\linewidth]{SEIR_S3_CI_grid_param_2.pdf}
\caption*{Figure~\ref{fig:SEIR_S3_CI_grid_param_2} (continued).}
\end{figure}

\begin{figure}[H]
\centering
\includegraphics[clip,trim=0 760 0 0,width=\linewidth]{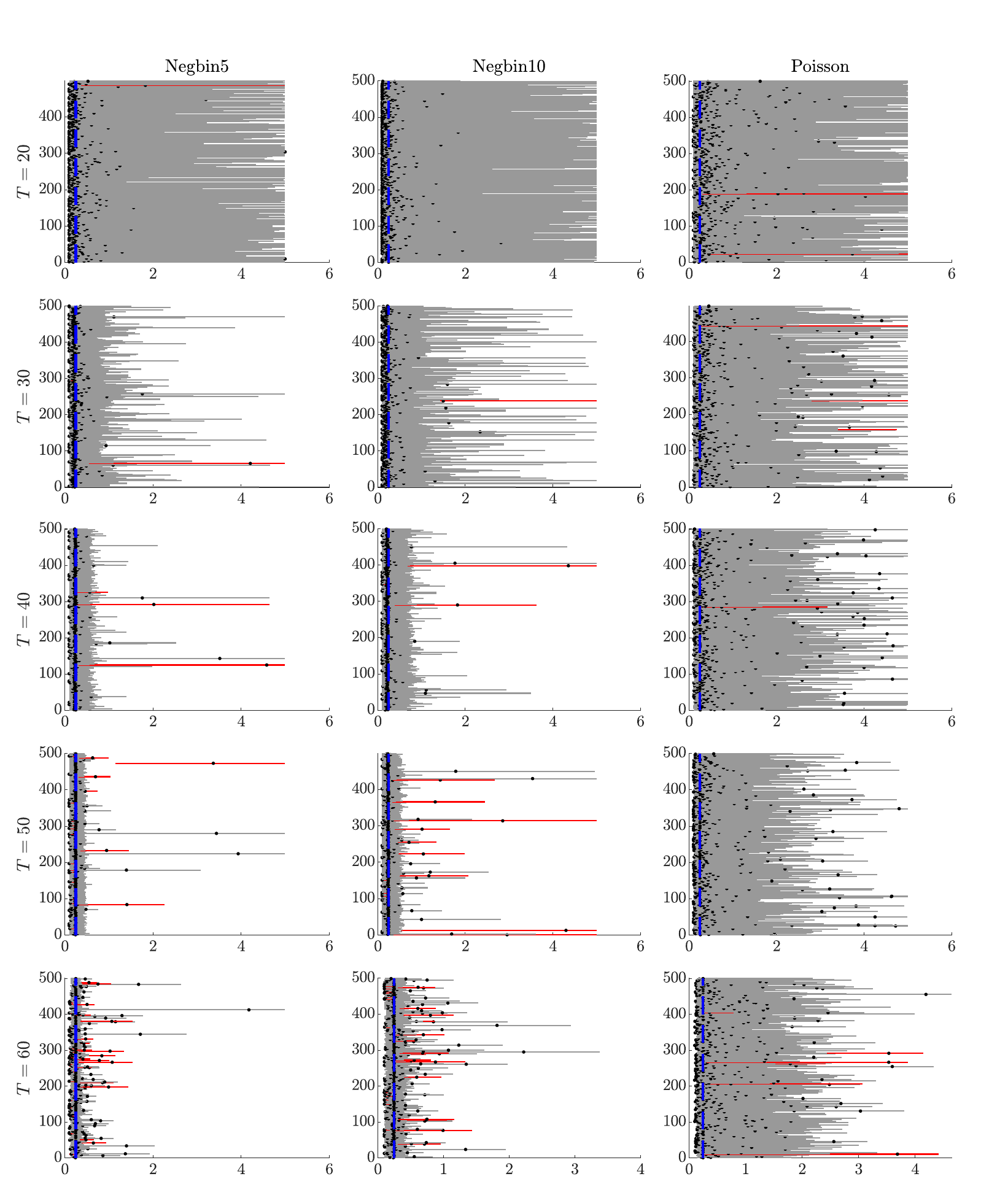}
\caption{
Parameter estimates and 95\% confidence intervals (CIs) for the recovery rate $\gamma$ across 500 simulation replicates and calibration window lengths $T=20,30,\ldots,100$, with the true value $\gamma=0.25$ indicated by the vertical blue dashed line. Columns correspond to the error structures: negative binomial with data-generating dispersion parameter $\alpha=5$ (Negbin5), negative binomial with data-generating dispersion parameter $\alpha=10$ (Negbin10), and Poisson. Each horizontal line corresponds to a single simulation replicate, showing the bootstrap confidence interval obtained by resampling within that replicate, with the corresponding point estimate marked by a black dot at its center. Red intervals denote confidence intervals that do not contain the true value, whereas gray intervals denote those that do.
}
\label{fig:SEIR_S3_CI_grid_param_3}
\end{figure}

\begin{figure}[H]
\centering
\includegraphics[clip,trim=0 0 0 920,width=\linewidth]{SEIR_S3_CI_grid_param_3.pdf}
\caption*{Figure~\ref{fig:SEIR_S3_CI_grid_param_3} (continued).}
\end{figure}

\section{SEIR-UR Model}

\subsection{Scenario~1: Estimated $\{\beta\}$; Fixed $\{\kappa,\, \rho,\, \gamma,\, N\}$}

\begin{figure}[H]
\centering
\includegraphics[width=\linewidth]{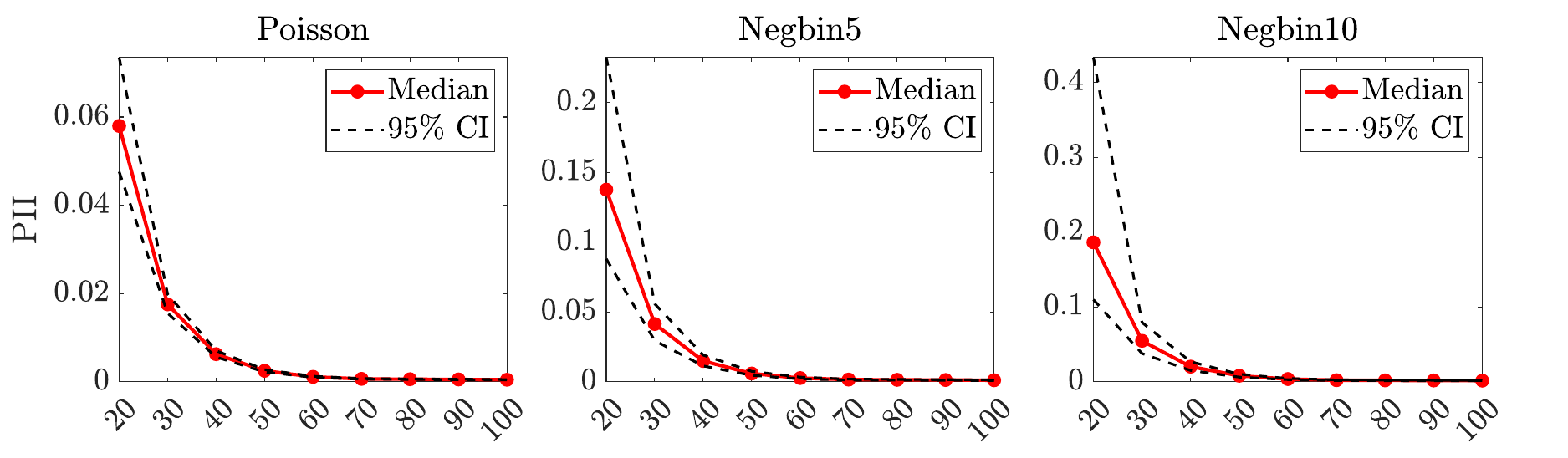}
\caption{Practical Identifiability Index (PII) for the transmission rate $\beta$ in the SEIR-UR model (Scenario~1) across calibration-window lengths $T=20, 30, \ldots, 100$ under three error structures: Poisson, negative binomial with data-generating dispersion $\alpha=5$ (Negbin5), and negative binomial with data-generating dispersion $\alpha=10$ (Negbin10). Red lines show the median PII across replicates, and dashed black curves indicate the central 95\% interval.}
\label{fig:SEIRunrep_S1_PII_parameters}
\end{figure}

\begin{figure}[H]
\centering
\includegraphics[clip,trim=0 760 0 0,width=\linewidth]{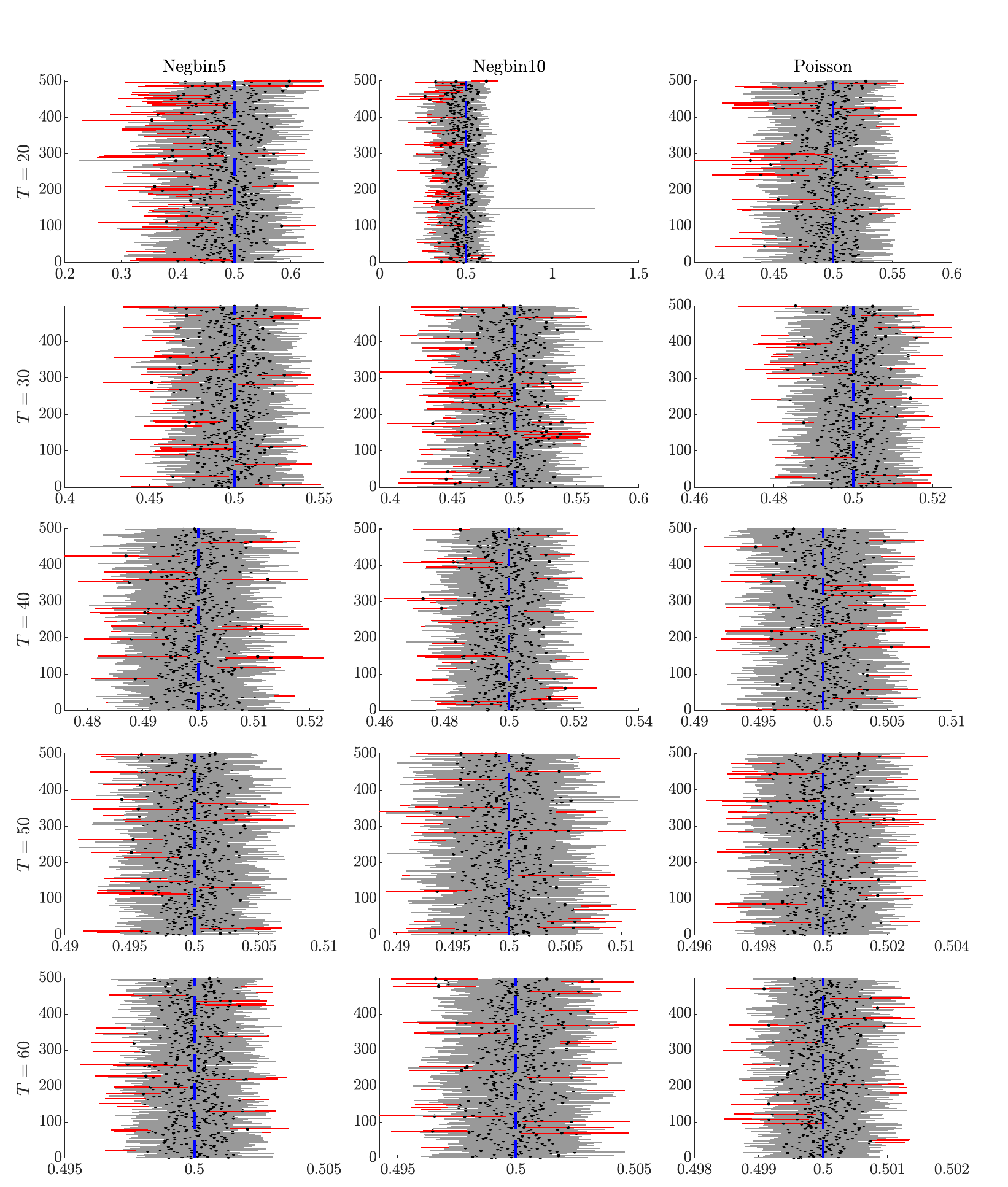}
\caption{
Parameter estimates and 95\% confidence intervals (CIs) for the transmission rate $\beta$ across 500 simulation replicates and calibration window lengths $T=20,30,\ldots,100$, with the true value $\beta=0.5$ indicated by the vertical blue dashed line. Columns correspond to the error structures: negative binomial with data-generating dispersion parameter $\alpha=5$ (Negbin5), negative binomial with data-generating dispersion parameter $\alpha=10$ (Negbin10), and Poisson. Each horizontal line corresponds to a single simulation replicate, showing the bootstrap confidence interval obtained by resampling within that replicate, with the corresponding point estimate marked by a black dot at its center. Red intervals denote confidence intervals that do not contain the true value, whereas gray intervals denote those that do.
}
\label{fig:SEIRunrep_S1_CI_grid_param_1}
\end{figure}

\begin{figure}[H]
\centering
\includegraphics[clip,trim=0 0 0 920,width=\linewidth]{SEIRunrep_S1_CI_grid_param_1.pdf}
\caption*{Figure~\ref{fig:SEIRunrep_S1_CI_grid_param_1} (continued).}
\end{figure}

\subsection{Scenario~2: Estimated $\{\beta,\, \rho\}$; Fixed $\{\kappa,\, \gamma,\, N\}$}

\begin{figure}[H]
\centering
\includegraphics[width=\linewidth]{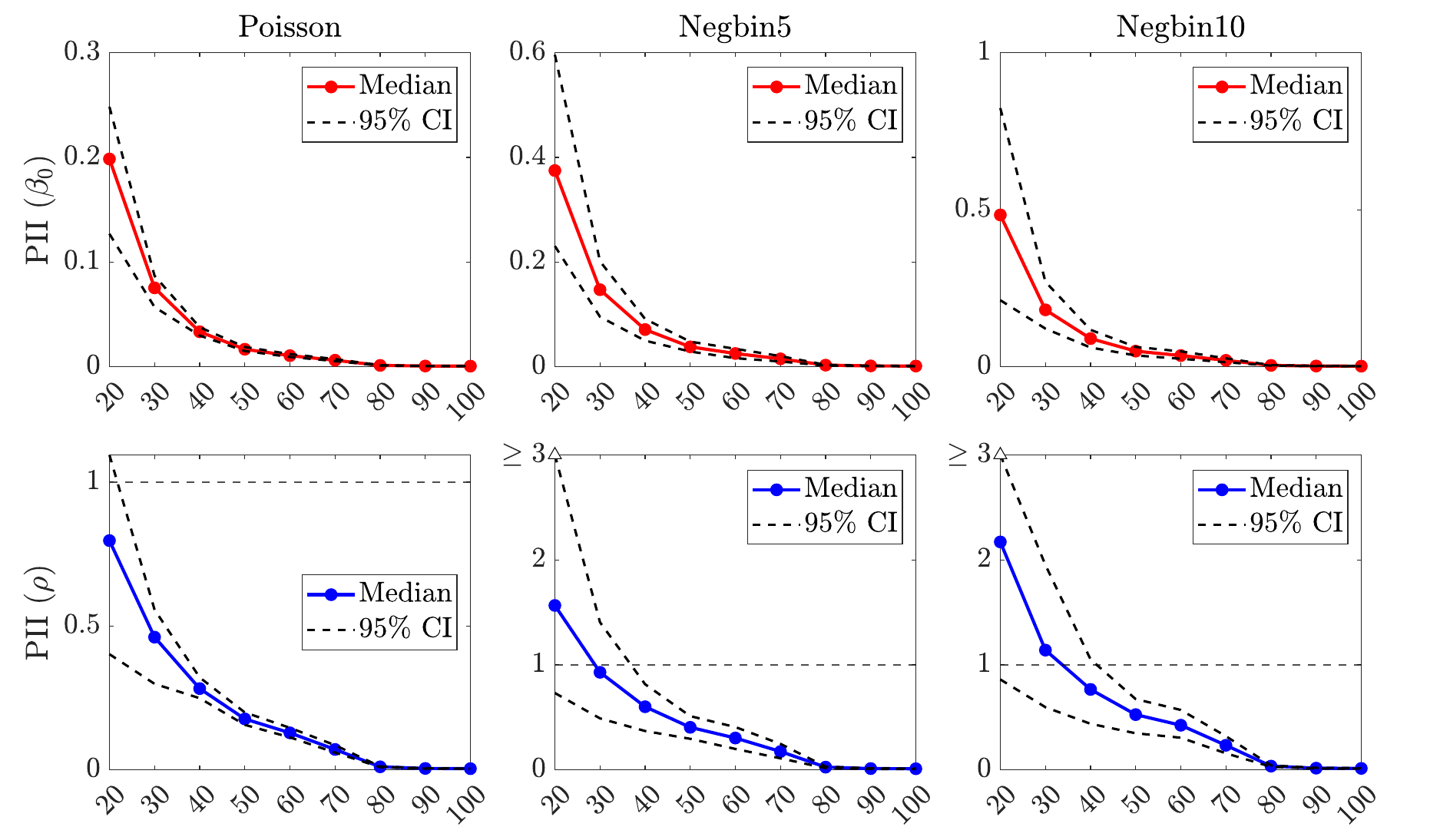}
\caption{Practical Identifiability Index (PII) for the transmission rate $\beta$ and reporting fraction $\rho$ in the SEIR-UR model (Scenario~2) across calibration-window lengths $T=20, 30, \ldots, 100$ under three error structures: Poisson, negative binomial with data-generating dispersion $\alpha=5$ (Negbin5), and negative binomial with data-generating dispersion $\alpha=10$ (Negbin10). Red lines show the median PII across replicates, and dashed black curves indicate the central 95\% interval. For visualization only, within subplots containing PII values at or above 3, those values are displayed at an upper plotting limit of 3; uncapped values are retained for all numerical analyses. Upward triangles denote values displayed at the upper limit.}
\label{fig:SEIRunrep_S2_PII_parameters}
\end{figure}

\begin{figure}[H]
\centering
\includegraphics[clip,trim=0 760 0 0,width=\linewidth]{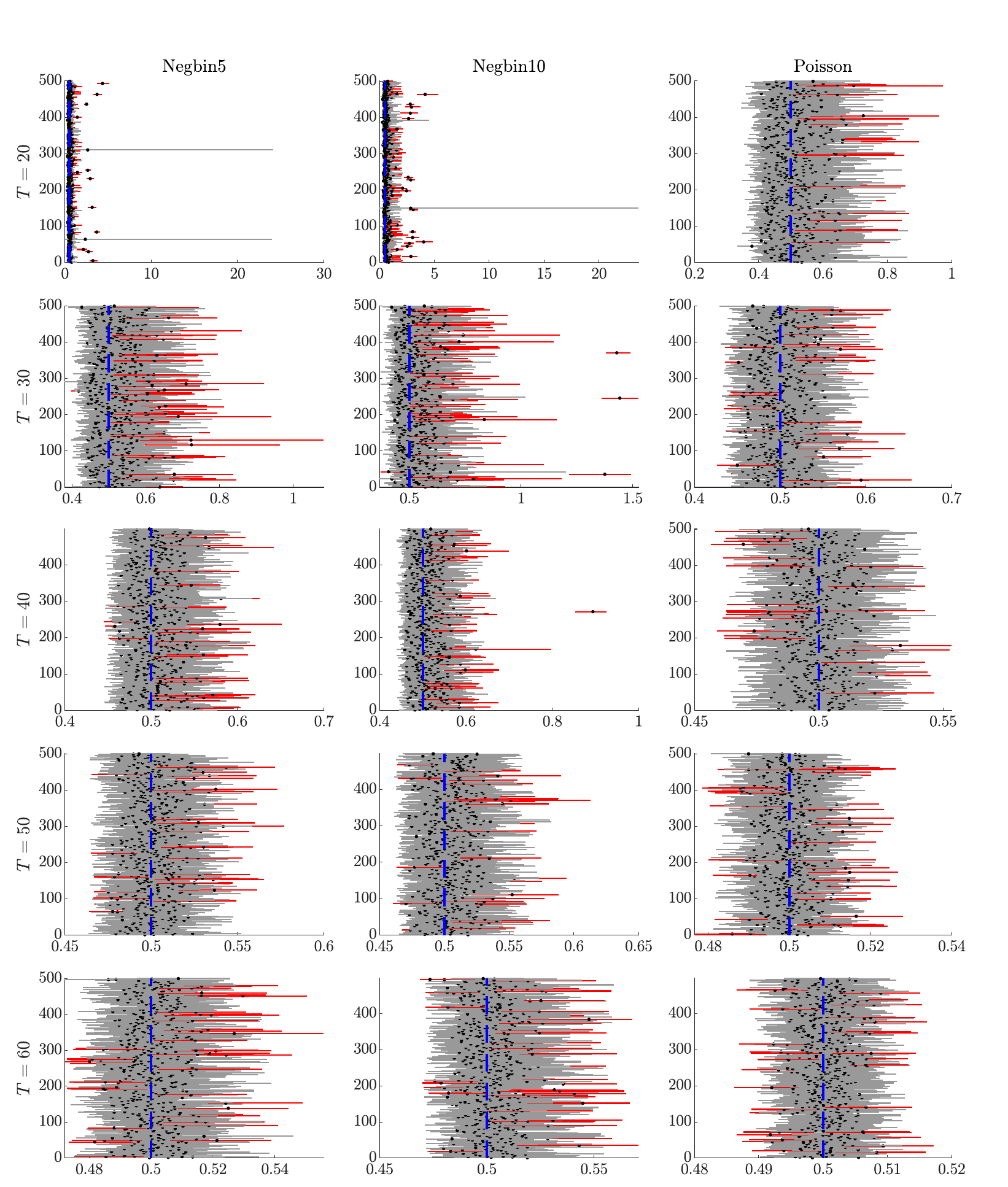}
\caption{
Parameter estimates and 95\% confidence intervals (CIs) for the transmission rate $\beta$ across 500 simulation replicates and calibration window lengths $T=20,30,\ldots,100$, with the true value $\beta=0.5$ indicated by the vertical blue dashed line. Columns correspond to the error structures: negative binomial with data-generating dispersion parameter $\alpha=5$ (Negbin5), negative binomial with data-generating dispersion parameter $\alpha=10$ (Negbin10), and Poisson. Each horizontal line corresponds to a single simulation replicate, showing the bootstrap confidence interval obtained by resampling within that replicate, with the corresponding point estimate marked by a black dot at its center. Red intervals denote confidence intervals that do not contain the true value, whereas gray intervals denote those that do.
}
\label{fig:SEIRunrep_S2_CI_grid_param_1}
\end{figure}

\begin{figure}[H]
\centering
\includegraphics[clip,trim=0 0 0 920,width=\linewidth]{SEIRunrep_S2_CI_grid_param_1.pdf}
\caption*{Figure~\ref{fig:SEIRunrep_S2_CI_grid_param_1} (continued).}
\end{figure}

\begin{figure}[H]
\centering
\includegraphics[clip,trim=0 760 0 0,width=\linewidth]{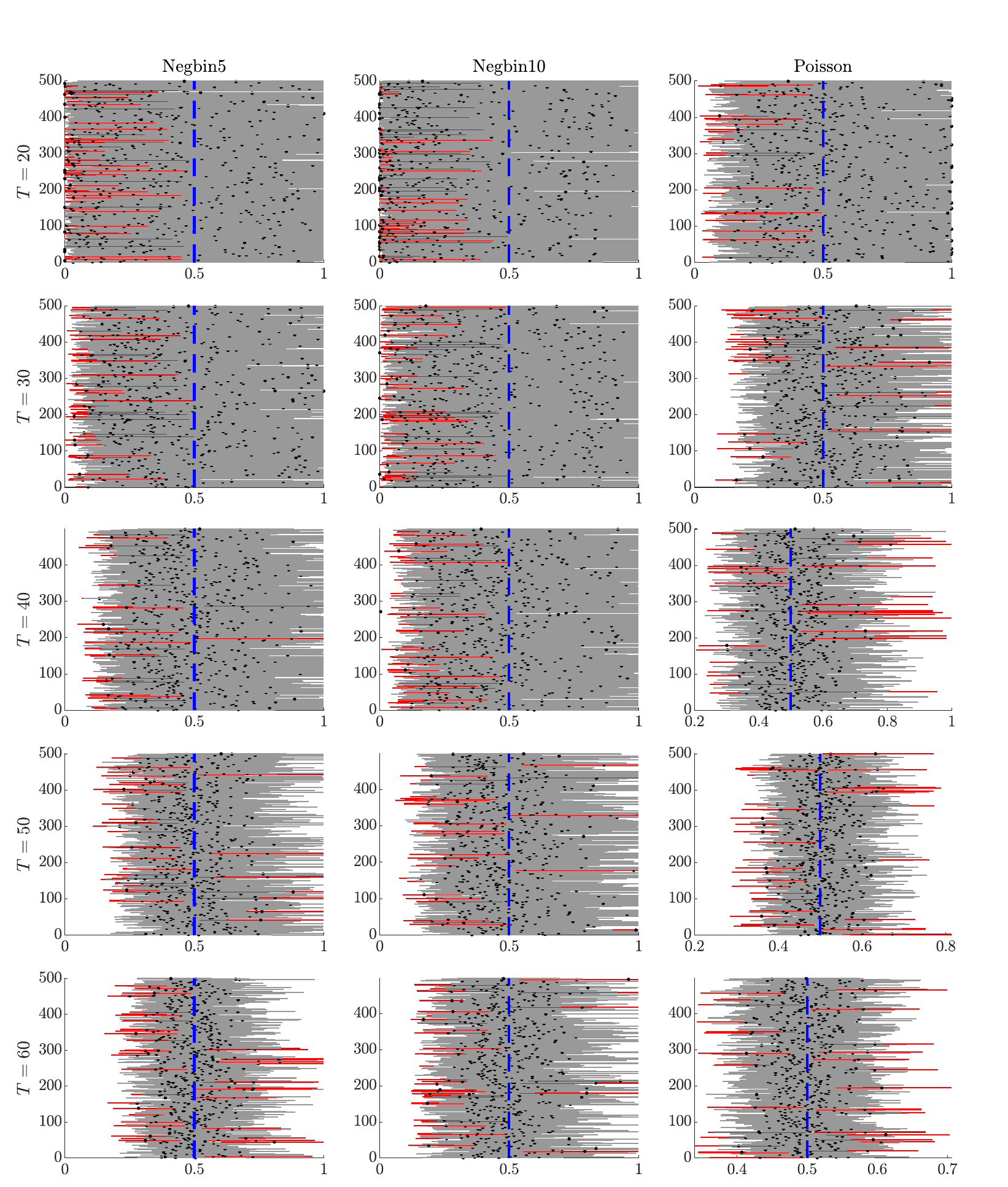}
\caption{
Parameter estimates and 95\% confidence intervals (CIs) for the reporting fraction $\rho$ across 500 simulation replicates and calibration window lengths $T=20,30,\ldots,100$, with the true value $\rho=0.5$ indicated by the vertical blue dashed line. Columns correspond to the error structures: negative binomial with data-generating dispersion parameter $\alpha=5$ (Negbin5), negative binomial with data-generating dispersion parameter $\alpha=10$ (Negbin10), and Poisson. Each horizontal line corresponds to a single simulation replicate, showing the bootstrap confidence interval obtained by resampling within that replicate, with the corresponding point estimate marked by a black dot at its center. Red intervals denote confidence intervals that do not contain the true value, whereas gray intervals denote those that do.
}
\label{fig:SEIRunrep_S2_CI_grid_param_3}
\end{figure}

\begin{figure}[H]
\centering
\includegraphics[clip,trim=0 0 0 920,width=\linewidth]{SEIRunrep_S2_CI_grid_param_3.pdf}
\caption*{Figure~\ref{fig:SEIRunrep_S2_CI_grid_param_3} (continued).}
\end{figure}

\subsection{Scenario~3: Estimated $\{\beta,\, \rho,\, \gamma\}$; Fixed $\{\kappa,\, N\}$}

\begin{figure}[H]
\centering
\includegraphics[width=\linewidth]{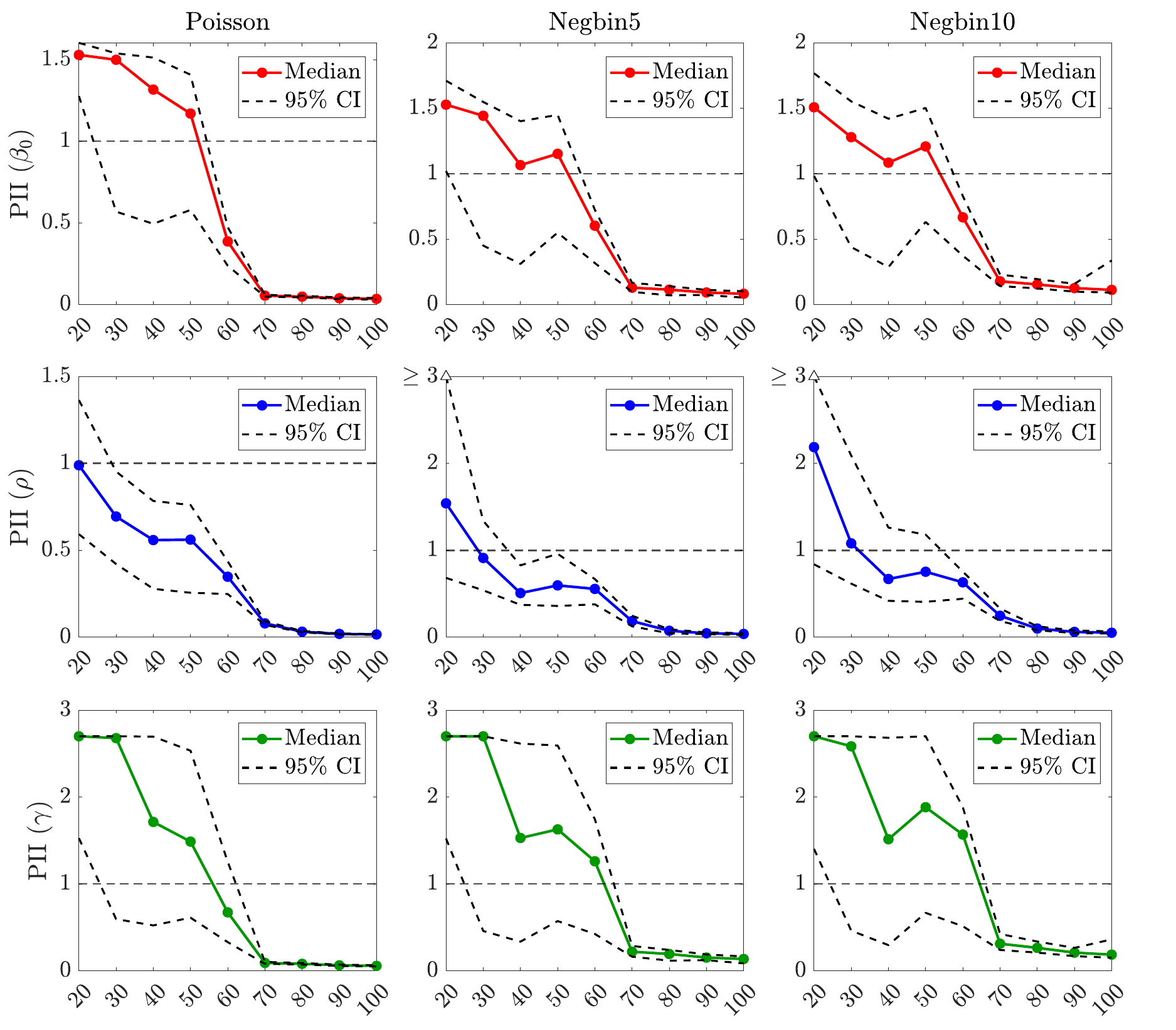}
\caption{Practical Identifiability Index (PII) for the transmission rate $\beta$, reporting fraction $\rho$, and recovery rate $\gamma$ in the SEIR-UR model (Scenario~3) across calibration-window lengths $T=20, 30, \ldots, 100$ under three error structures: Poisson, negative binomial with data-generating dispersion $\alpha=5$ (Negbin5), and negative binomial with data-generating dispersion $\alpha=10$ (Negbin10). Red lines show the median PII across replicates, and dashed black curves indicate the central 95\% interval. For visualization only, within subplots containing PII values at or above 3, those values are displayed at an upper plotting limit of 3; uncapped values are retained for all numerical analyses. Upward triangles denote values displayed at the upper limit.}
\label{fig:SEIRunrep_S3_PII_parameters}
\end{figure}

\begin{figure}[H]
\centering
\includegraphics[clip,trim=0 760 0 0,width=\linewidth]{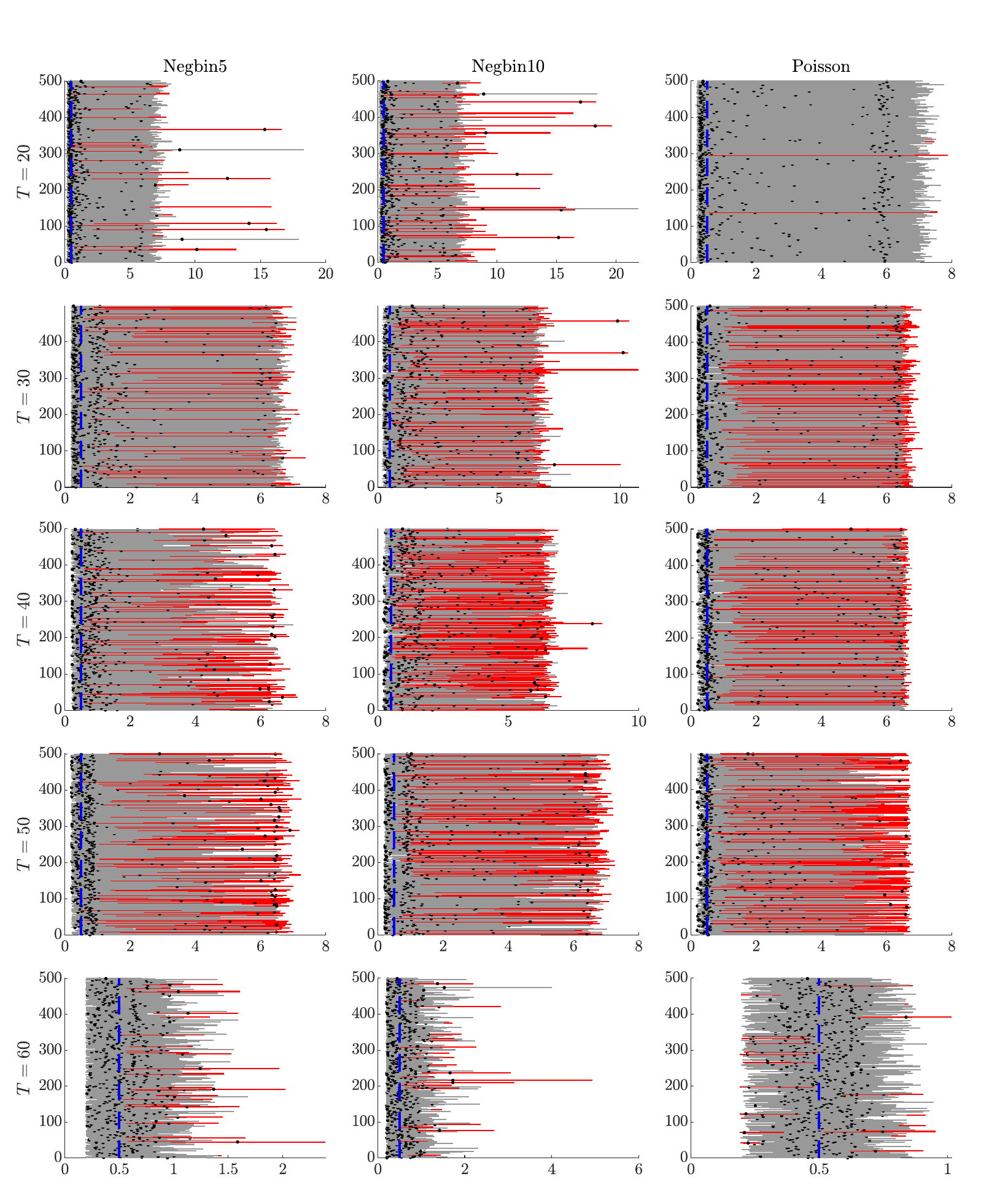}
\caption{
Parameter estimates and 95\% confidence intervals (CIs) for the transmission rate $\beta$ across 500 simulation replicates and calibration window lengths $T=20,30,\ldots,100$, with the true value $\beta=0.5$ indicated by the vertical blue dashed line. Columns correspond to the error structures: negative binomial with data-generating dispersion parameter $\alpha=5$ (Negbin5), negative binomial with data-generating dispersion parameter $\alpha=10$ (Negbin10), and Poisson. Each horizontal line corresponds to a single simulation replicate, showing the bootstrap confidence interval obtained by resampling within that replicate, with the corresponding point estimate marked by a black dot at its center. Red intervals denote confidence intervals that do not contain the true value, whereas gray intervals denote those that do.
}
\label{fig:SEIRunrep_S3_CI_grid_param_1}
\end{figure}

\begin{figure}[H]
\centering
\includegraphics[clip,trim=0 0 0 920,width=\linewidth]{SEIRunrep_S3_CI_grid_param_1.pdf}
\caption*{Figure~\ref{fig:SEIRunrep_S3_CI_grid_param_1} (continued).}
\end{figure}

\begin{figure}[H]
\centering
\includegraphics[clip,trim=0 760 0 0,width=\linewidth]{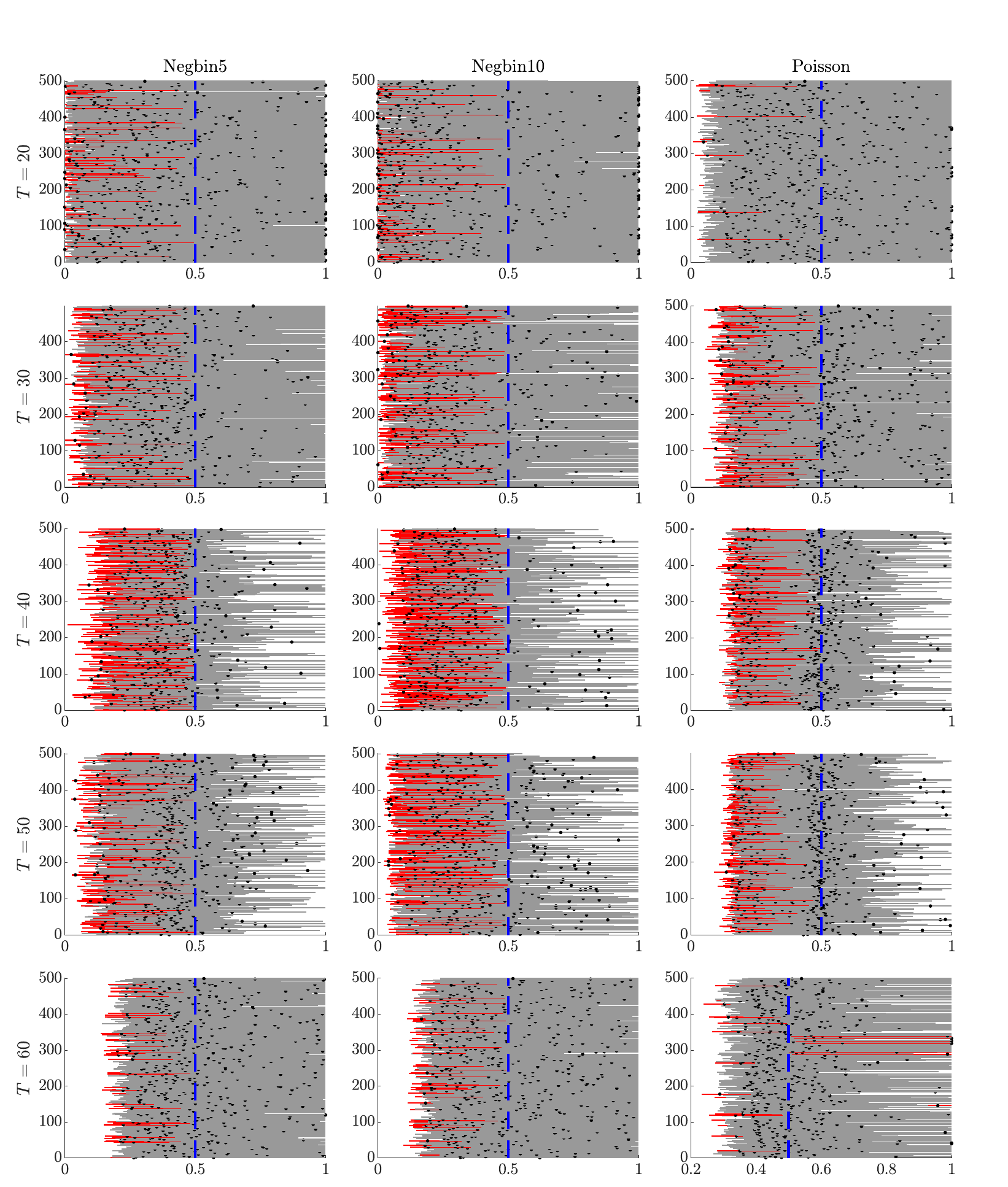}
\caption{
Parameter estimates and 95\% confidence intervals (CIs) for the reporting fraction $\rho$ across 500 simulation replicates and calibration window lengths $T=20,30,\ldots,100$, with the true value $\rho=0.5$ indicated by the vertical blue dashed line. Columns correspond to the error structures: negative binomial with data-generating dispersion parameter $\alpha=5$ (Negbin5), negative binomial with data-generating dispersion parameter $\alpha=10$ (Negbin10), and Poisson. Each horizontal line corresponds to a single simulation replicate, showing the bootstrap confidence interval obtained by resampling within that replicate, with the corresponding point estimate marked by a black dot at its center. Red intervals denote confidence intervals that do not contain the true value, whereas gray intervals denote those that do.
}
\label{fig:SEIRunrep_S3_CI_grid_param_3}
\end{figure}

\begin{figure}[H]
\centering
\includegraphics[clip,trim=0 0 0 920,width=\linewidth]{SEIRunrep_S3_CI_grid_param_3.pdf}
\caption*{Figure~\ref{fig:SEIRunrep_S3_CI_grid_param_3} (continued).}
\end{figure}

\begin{figure}[H]
\centering
\includegraphics[clip,trim=0 760 0 0,width=\linewidth]{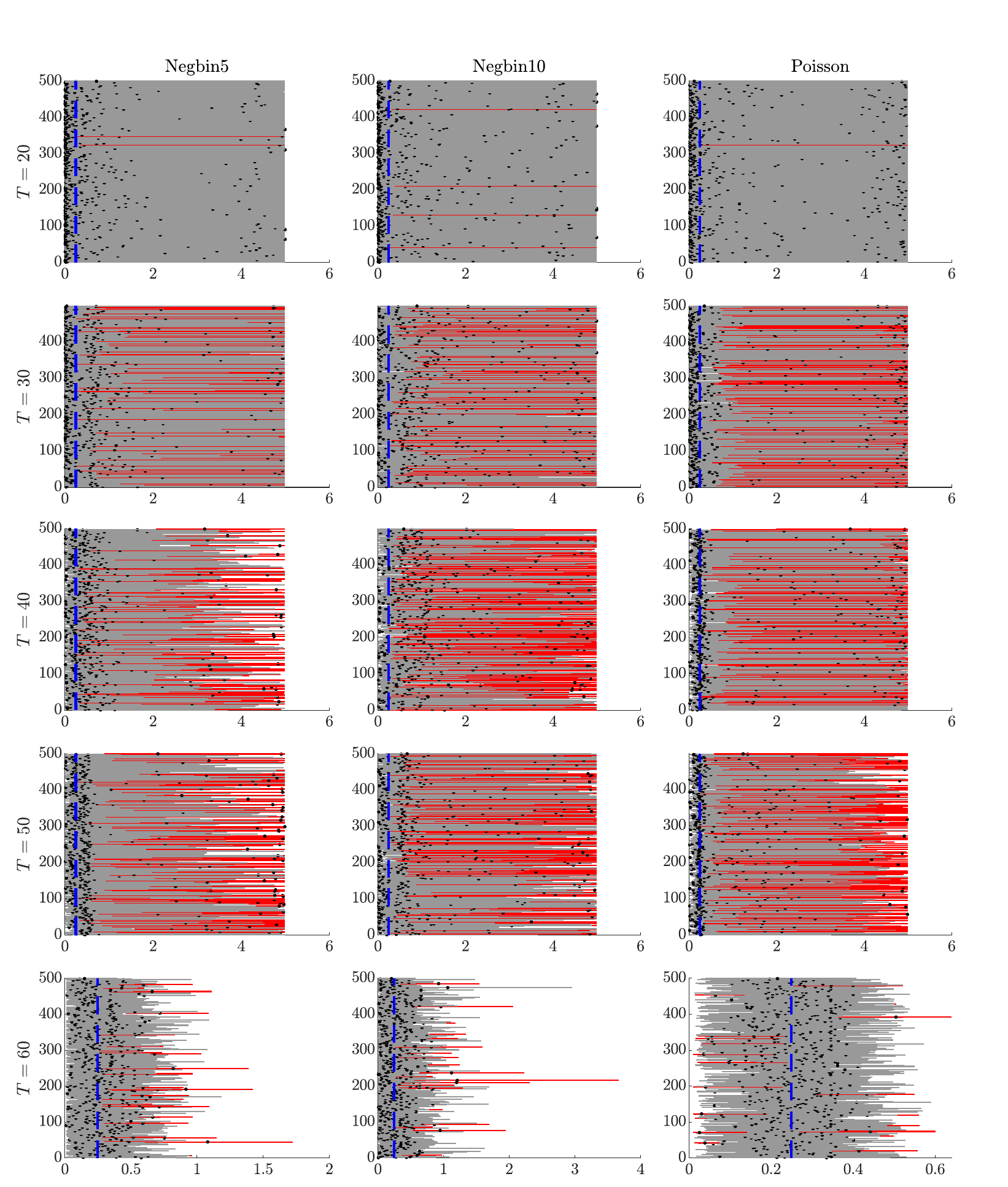}
\caption{
Parameter estimates and 95\% confidence intervals (CIs) for the recovery rate $\gamma$ across 500 simulation replicates and calibration window lengths $T=20,30,\ldots,100$, with the true value $\gamma=0.25$ indicated by the vertical blue dashed line. Columns correspond to the error structures: negative binomial with data-generating dispersion parameter $\alpha=5$ (Negbin5), negative binomial with data-generating dispersion parameter $\alpha=10$ (Negbin10), and Poisson. Each horizontal line corresponds to a single simulation replicate, showing the bootstrap confidence interval obtained by resampling within that replicate, with the corresponding point estimate marked by a black dot at its center. Red intervals denote confidence intervals that do not contain the true value, whereas gray intervals denote those that do.
}
\label{fig:SEIRunrep_S3_CI_grid_param_4}
\end{figure}

\begin{figure}[H]
\centering
\includegraphics[clip,trim=0 0 0 920,width=\linewidth]{SEIRunrep_S3_CI_grid_param_4.pdf}
\caption*{Figure~\ref{fig:SEIRunrep_S3_CI_grid_param_4} (continued).}
\end{figure}

\section{SEIAR Model}

\subsection{Scenario~1: Estimated $\{\beta_0,\, \beta_1\}$; Fixed $\{\kappa,\, \rho,\, \gamma,\, N\}$}

\begin{figure}[H]
\centering
\includegraphics[width=\linewidth]{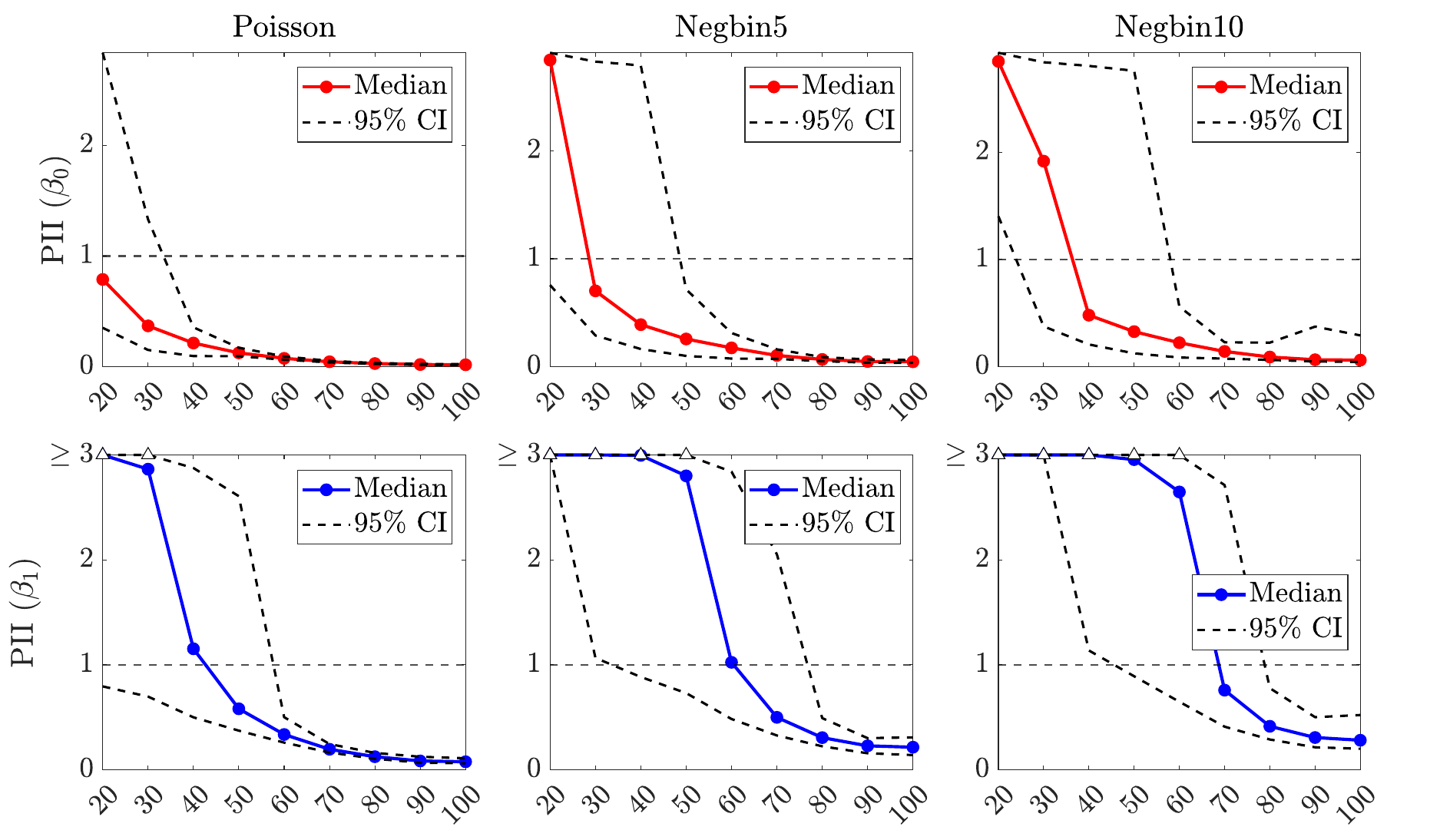}
\caption{Practical Identifiability Index (PII) for the symptomatic transmission rate $\beta_0$ and asymptomatic transmission rate $\beta_1$ in the SEIAR model (Scenario~1) across calibration-window lengths $T=20, 30, \ldots, 100$ under three error structures: Poisson, negative binomial with data-generating dispersion $\alpha=5$ (Negbin5), and negative binomial with data-generating dispersion $\alpha=10$ (Negbin10). Red lines show the median PII across replicates, and dashed black curves indicate the central 95\% interval. For visualization only, within subplots containing PII values at or above 3, those values are displayed at an upper plotting limit of 3; uncapped values are retained for all numerical analyses. Upward triangles denote values displayed at the upper limit.}
\label{fig:SEIRasymp_S1_PII_parameters}
\end{figure}

\begin{figure}[H]
\centering
\includegraphics[clip,trim=0 760 0 0,width=\linewidth]{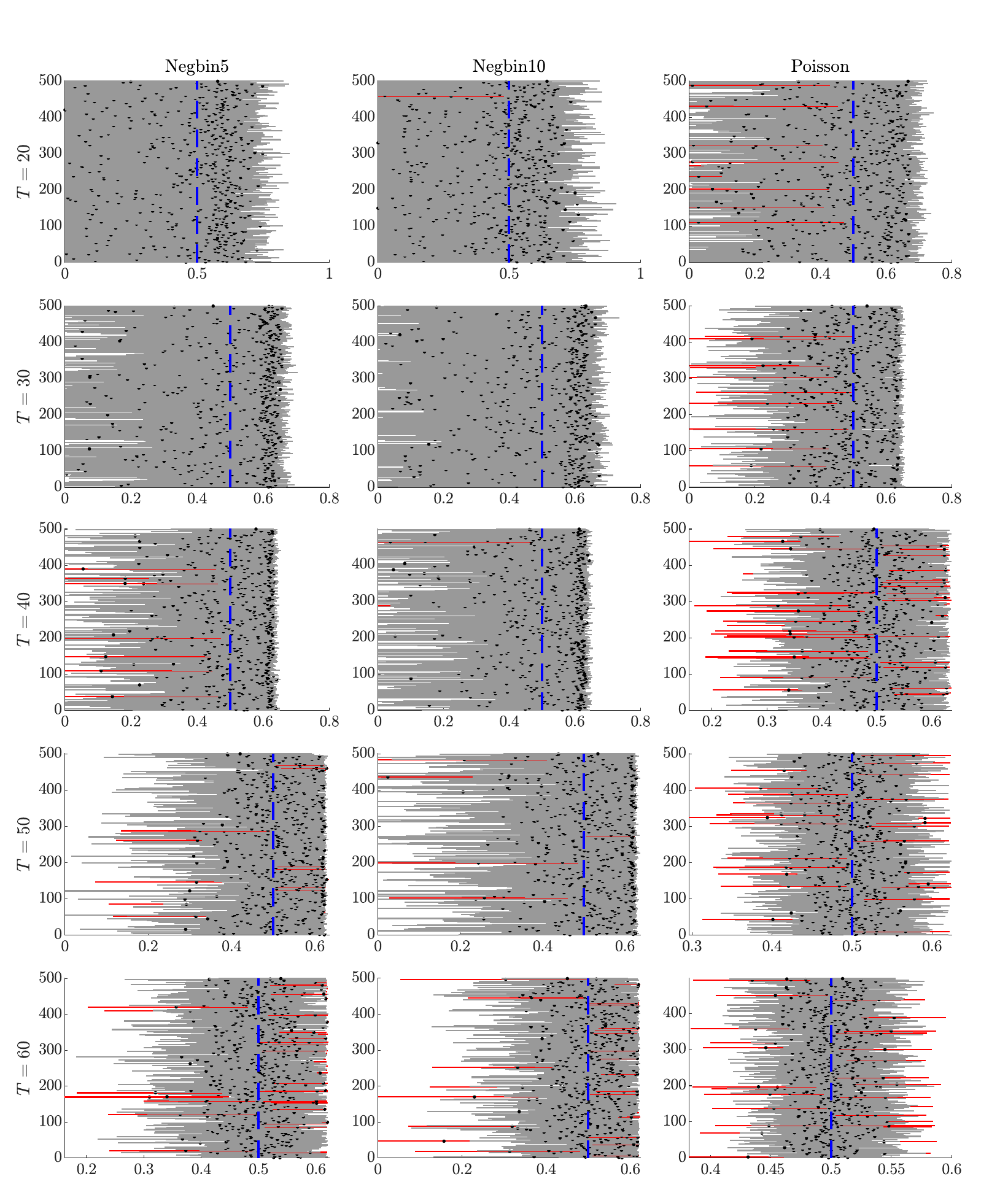}
\caption{
Parameter estimates and 95\% confidence intervals (CIs) for the symptomatic transmission rate $\beta_0$ across 500 simulation replicates and calibration window lengths $T=20,30,\ldots,100$, with the true value $\beta_0=0.5$ indicated by the vertical blue dashed line. Columns correspond to the error structures: negative binomial with data-generating dispersion parameter $\alpha=5$ (Negbin5), negative binomial with data-generating dispersion parameter $\alpha=10$ (Negbin10), and Poisson. Each horizontal line corresponds to a single simulation replicate, showing the bootstrap confidence interval obtained by resampling within that replicate, with the corresponding point estimate marked by a black dot at its center. Red intervals denote confidence intervals that do not contain the true value, whereas gray intervals denote those that do.
}
\label{fig:SEIRasymp_S1_CI_grid_param_1}
\end{figure}

\begin{figure}[H]
\centering
\includegraphics[clip,trim=0 0 0 920,width=\linewidth]{SEIRasymp_S1_CI_grid_param_1.pdf}
\caption*{Figure~\ref{fig:SEIRasymp_S1_CI_grid_param_1} (continued).}
\end{figure}

\begin{figure}[H]
\centering
\includegraphics[clip,trim=0 760 0 0,width=\linewidth]{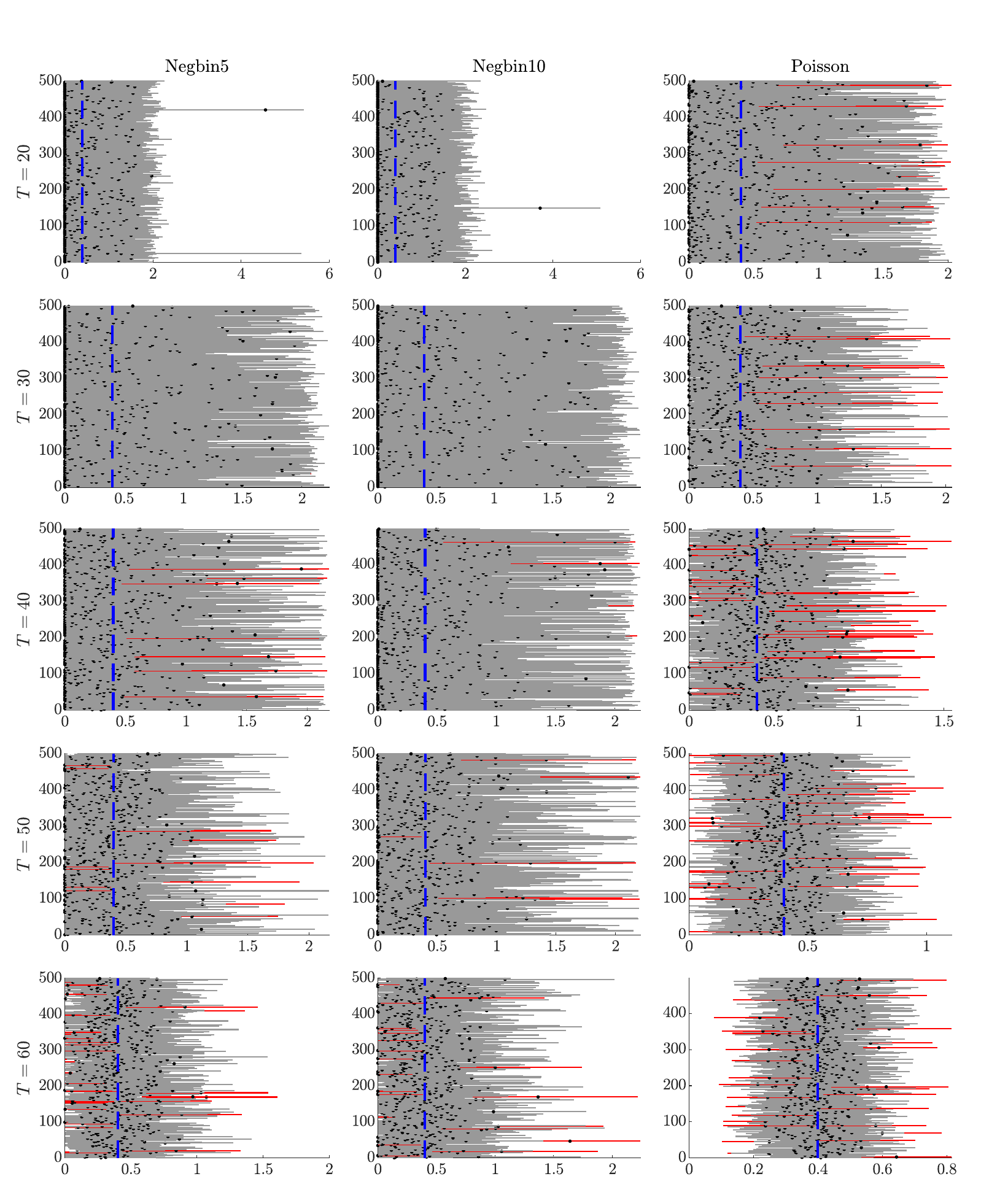}
\caption{
Parameter estimates and 95\% confidence intervals (CIs) for the asymptomatic transmission rate $\beta_1$ across 500 simulation replicates and calibration window lengths $T=20,30,\ldots,100$, with the true value $\beta_1=0.4$ indicated by the vertical blue dashed line. Columns correspond to the error structures: negative binomial with data-generating dispersion parameter $\alpha=5$ (Negbin5), negative binomial with data-generating dispersion parameter $\alpha=10$ (Negbin10), and Poisson. Each horizontal line corresponds to a single simulation replicate, showing the bootstrap confidence interval obtained by resampling within that replicate, with the corresponding point estimate marked by a black dot at its center. Red intervals denote confidence intervals that do not contain the true value, whereas gray intervals denote those that do.
}
\label{fig:SEIRasymp_S1_CI_grid_param_2}
\end{figure}

\begin{figure}[H]
\centering
\includegraphics[clip,trim=0 0 0 920,width=\linewidth]{SEIRasymp_S1_CI_grid_param_2.pdf}
\caption*{Figure~\ref{fig:SEIRasymp_S1_CI_grid_param_2} (continued).}
\end{figure}

\subsection{Scenario~2: Estimated $\{\beta_0,\, \beta_1,\, \rho\}$; Fixed $\{\kappa,\, \gamma,\, N\}$}

\begin{figure}[H]
\centering
\includegraphics[width=\linewidth]{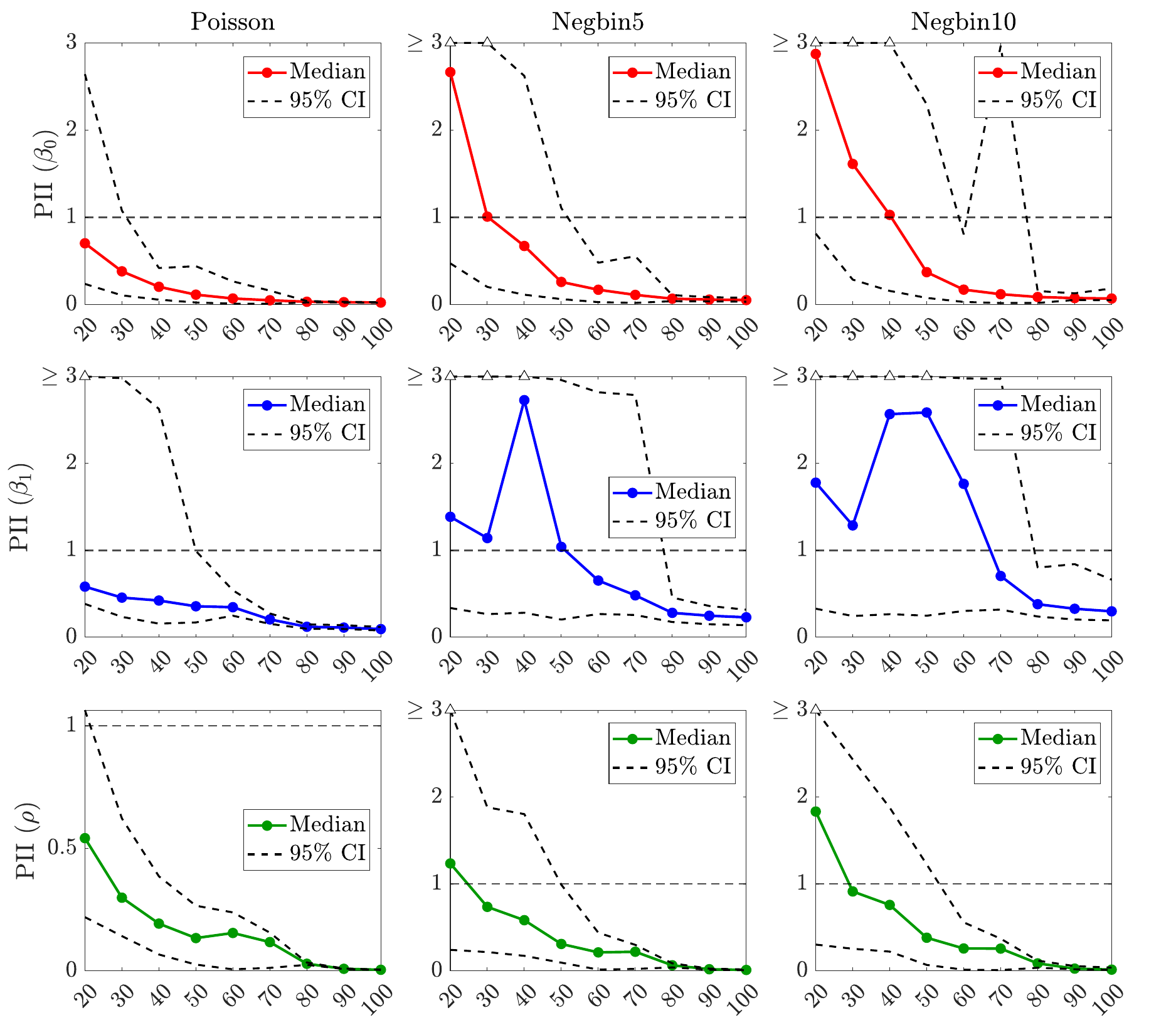}
\caption{Practical Identifiability Index (PII) for the symptomatic transmission rate $\beta_0$, asymptomatic transmission rate $\beta_1$, and symptomatic fraction $\rho$ in the SEIAR model (Scenario~2) across calibration-window lengths $T=20, 30, \ldots, 100$ under three error structures: Poisson, negative binomial with data-generating dispersion $\alpha=5$ (Negbin5), and negative binomial with data-generating dispersion $\alpha=10$ (Negbin10). Red lines show the median PII across replicates, and dashed black curves indicate the central 95\% interval. For visualization only, within subplots containing PII values at or above 3, those values are displayed at an upper plotting limit of 3; uncapped values are retained for all numerical analyses. Upward triangles denote values displayed at the upper limit.}
\label{fig:SEIRasymp_S2_PII_parameters}
\end{figure}

\begin{figure}[H]
\centering
\includegraphics[clip,trim=0 760 0 0,width=\linewidth]{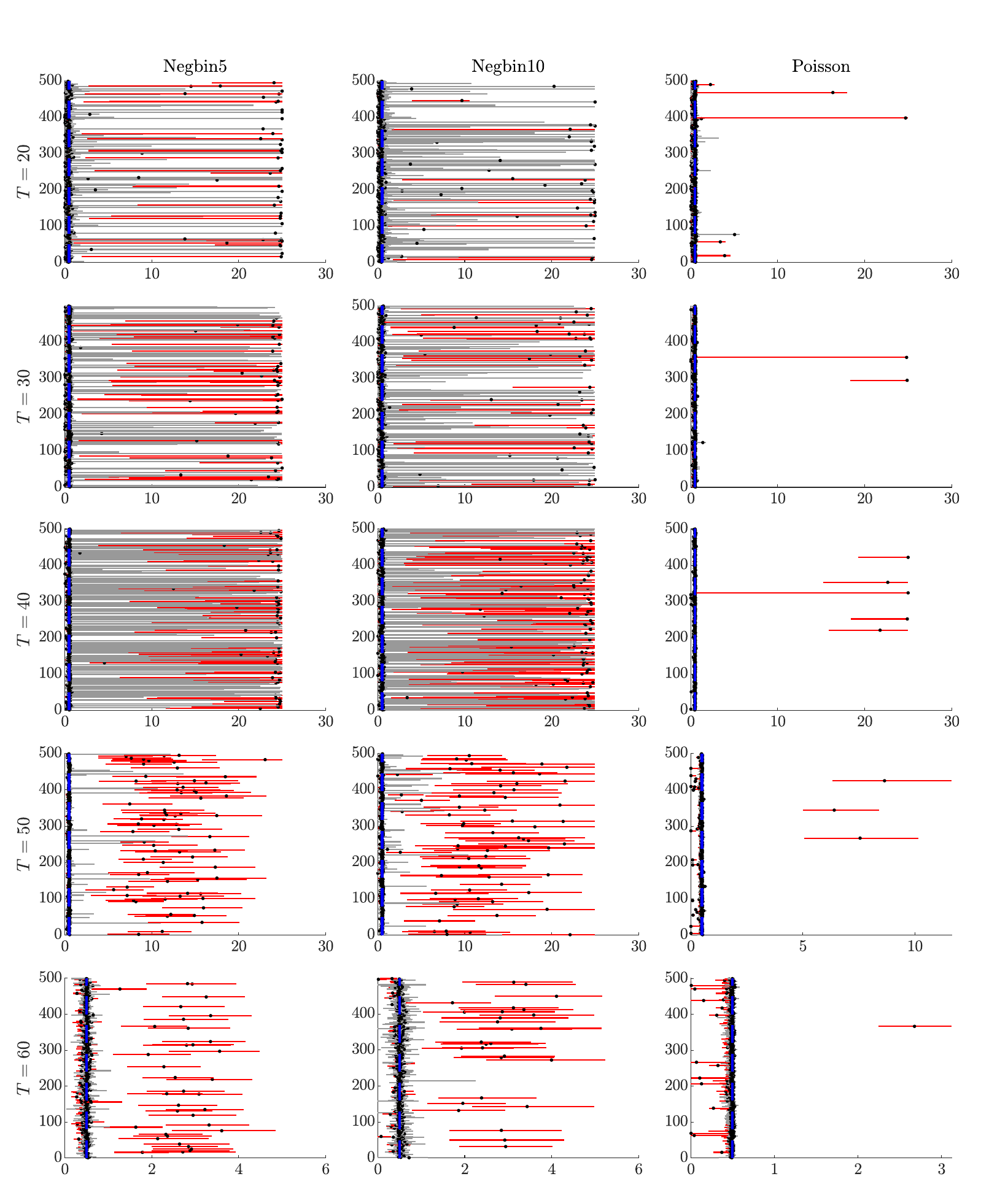}
\caption{
Parameter estimates and 95\% confidence intervals (CIs) for the symptomatic transmission rate $\beta_0$ across 500 simulation replicates and calibration window lengths $T=20,30,\ldots,100$, with the true value $\beta_0=0.5$ indicated by the vertical blue dashed line. Columns correspond to the error structures: negative binomial with data-generating dispersion parameter $\alpha=5$ (Negbin5), negative binomial with data-generating dispersion parameter $\alpha=10$ (Negbin10), and Poisson. Each horizontal line corresponds to a single simulation replicate, showing the bootstrap confidence interval obtained by resampling within that replicate, with the corresponding point estimate marked by a black dot at its center. Red intervals denote confidence intervals that do not contain the true value, whereas gray intervals denote those that do.
}
\label{fig:SEIRasymp_S2_CI_grid_param_1}
\end{figure}

\begin{figure}[H]
\centering
\includegraphics[clip,trim=0 0 0 920,width=\linewidth]{SEIRasymp_S2_CI_grid_param_1.pdf}
\caption*{Figure~\ref{fig:SEIRasymp_S2_CI_grid_param_1} (continued).}
\end{figure}

\begin{figure}[H]
\centering
\includegraphics[clip,trim=0 760 0 0,width=\linewidth]{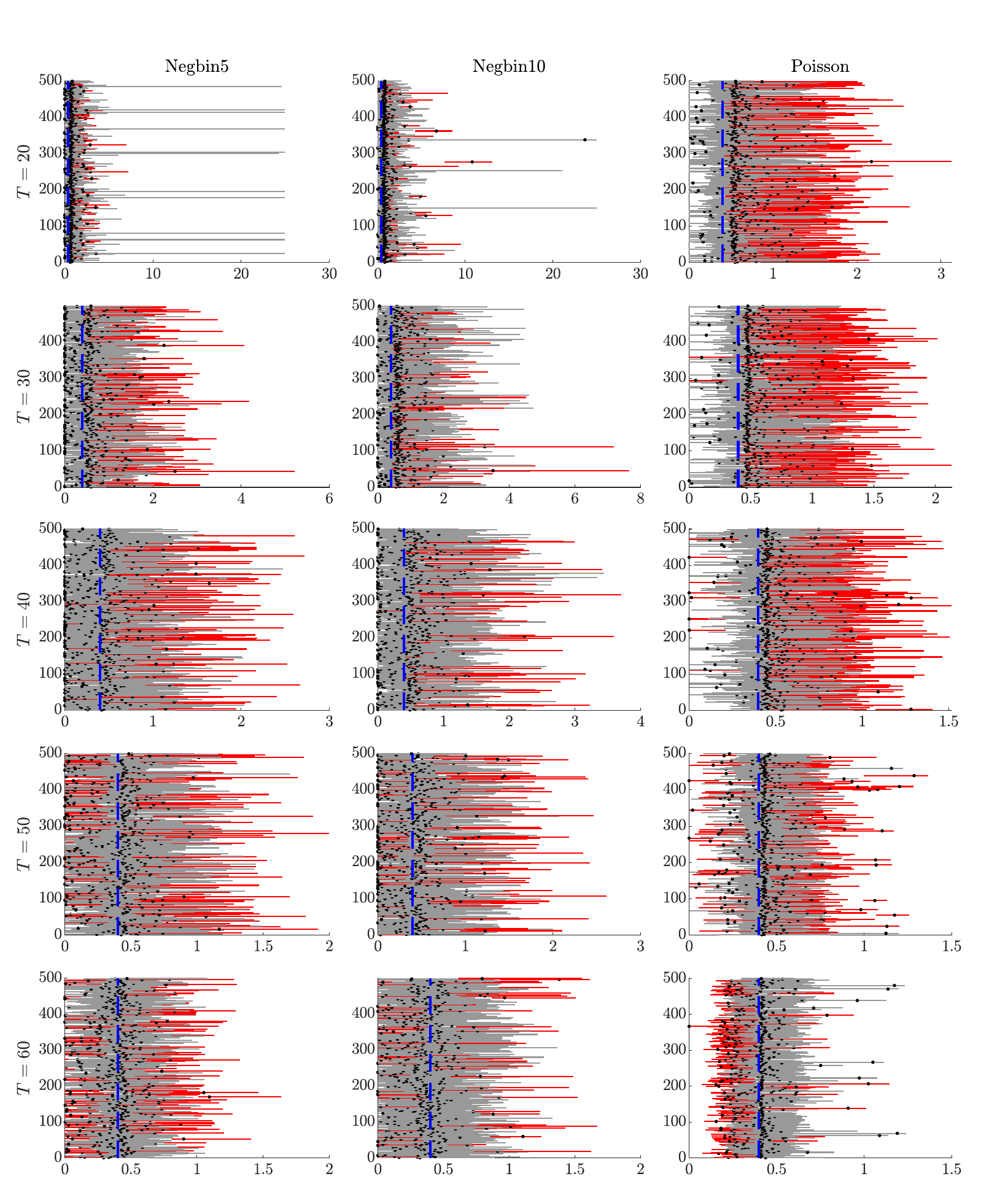}
\caption{
Parameter estimates and 95\% confidence intervals (CIs) for the asymptomatic transmission rate $\beta_1$ across 500 simulation replicates and calibration window lengths $T=20,30,\ldots,100$, with the true value $\beta_1=0.4$ indicated by the vertical blue dashed line. Columns correspond to the error structures: negative binomial with data-generating dispersion parameter $\alpha=5$ (Negbin5), negative binomial with data-generating dispersion parameter $\alpha=10$ (Negbin10), and Poisson. Each horizontal line corresponds to a single simulation replicate, showing the bootstrap confidence interval obtained by resampling within that replicate, with the corresponding point estimate marked by a black dot at its center. Red intervals denote confidence intervals that do not contain the true value, whereas gray intervals denote those that do.
}
\label{fig:SEIRasymp_S2_CI_grid_param_2}
\end{figure}

\begin{figure}[H]
\centering
\includegraphics[clip,trim=0 0 0 920,width=\linewidth]{SEIRasymp_S2_CI_grid_param_2.pdf}
\caption*{Figure~\ref{fig:SEIRasymp_S2_CI_grid_param_2} (continued).}
\end{figure}

\begin{figure}[H]
\centering
\includegraphics[clip,trim=0 760 0 0,width=\linewidth]{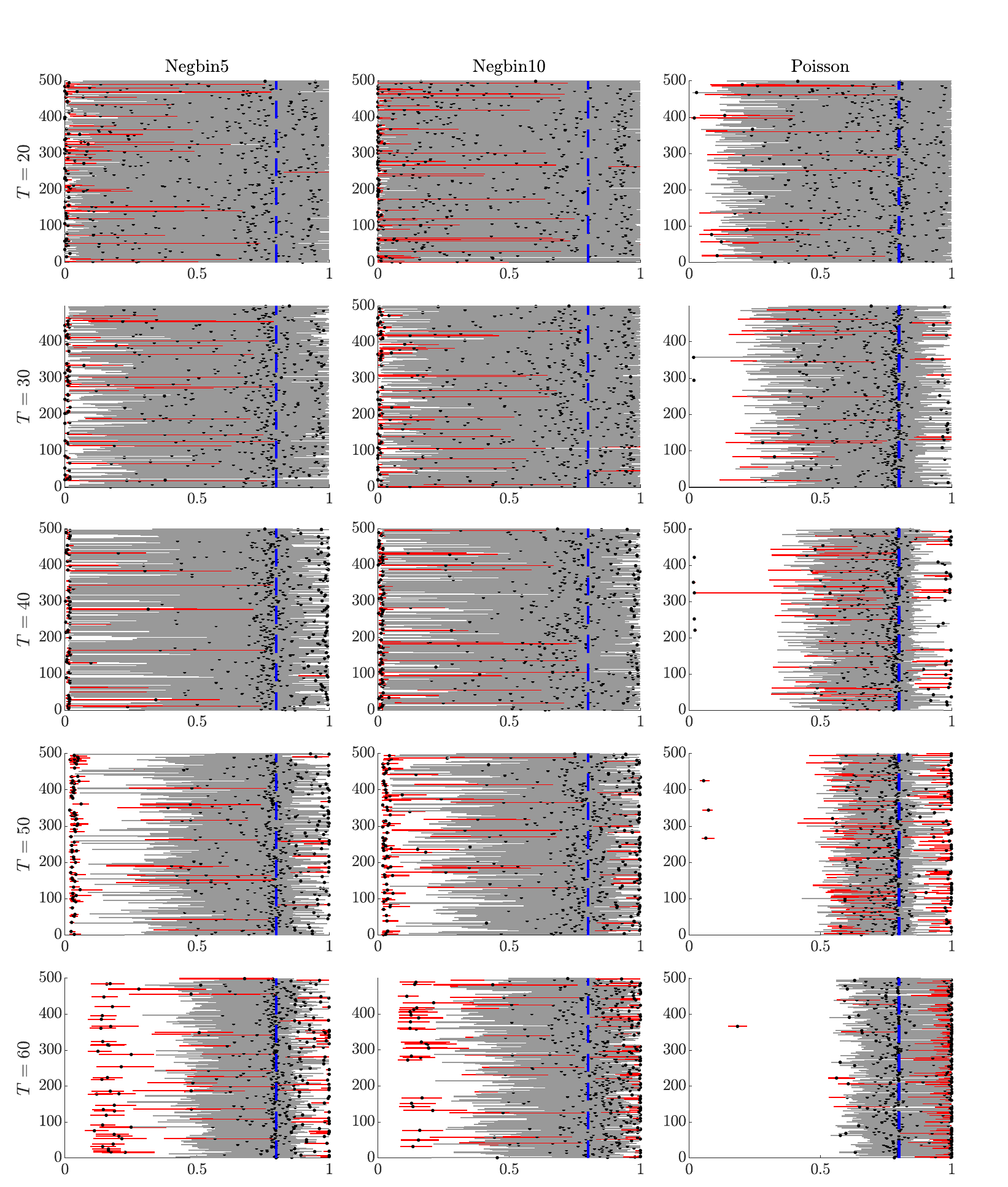}
\caption{
Parameter estimates and 95\% confidence intervals (CIs) for the symptomatic fraction $\rho$ across 500 simulation replicates and calibration window lengths $T=20,30,\ldots,100$, with the true value $\rho=0.8$ indicated by the vertical blue dashed line. Columns correspond to the error structures: negative binomial with data-generating dispersion parameter $\alpha=5$ (Negbin5), negative binomial with data-generating dispersion parameter $\alpha=10$ (Negbin10), and Poisson. Each horizontal line corresponds to a single simulation replicate, showing the bootstrap confidence interval obtained by resampling within that replicate, with the corresponding point estimate marked by a black dot at its center. Red intervals denote confidence intervals that do not contain the true value, whereas gray intervals denote those that do.
}
\label{fig:SEIRasymp_S2_CI_grid_param_4}
\end{figure}

\begin{figure}[H]
\centering
\includegraphics[clip,trim=0 0 0 920,width=\linewidth]{SEIRasymp_S2_CI_grid_param_4.pdf}
\caption*{Figure~\ref{fig:SEIRasymp_S2_CI_grid_param_4} (continued).}
\end{figure}

\subsection{Scenario~3: Estimated $\{\beta_0,\, \beta_1,\, \rho,\, \gamma\}$; Fixed $\{\kappa,\, N\}$}

\begin{figure}[H]
\centering
\includegraphics[width=\linewidth]{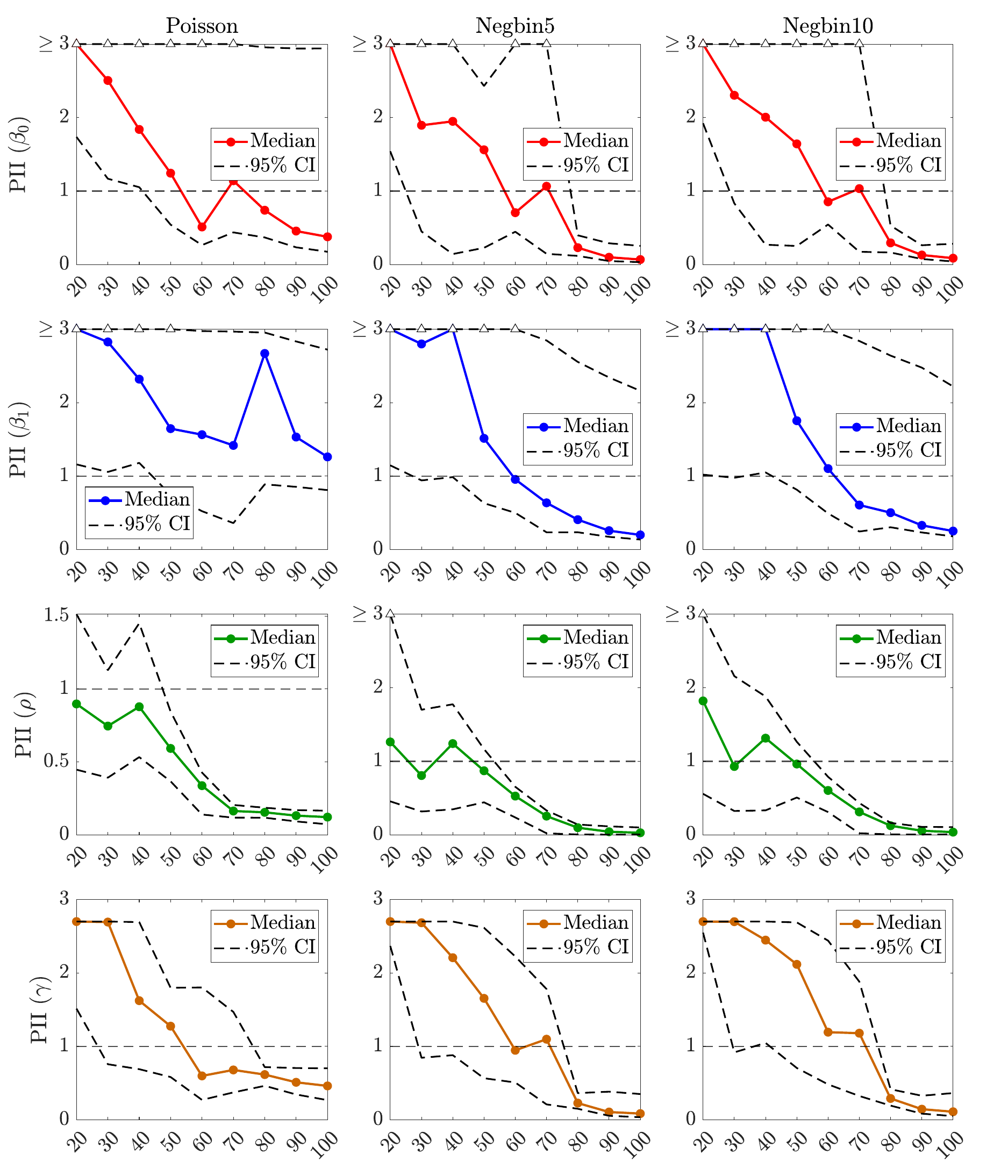}
\caption{Practical Identifiability Index (PII) for the symptomatic transmission rate $\beta_0$, asymptomatic transmission rate $\beta_1$, symptomatic fraction $\rho$, and recovery rate $\gamma$ in the SEIAR model (Scenario~3) across calibration-window lengths $T=20, 30, \ldots, 100$ under three error structures: Poisson, negative binomial with data-generating dispersion $\alpha=5$ (Negbin5), and negative binomial with data-generating dispersion $\alpha=10$ (Negbin10). Red lines show the median PII across replicates, and dashed black curves indicate the central 95\% interval. For visualization only, within subplots containing PII values at or above 3, those values are displayed at an upper plotting limit of 3; uncapped values are retained for all numerical analyses. Upward triangles denote values displayed at the upper limit.}
\label{fig:SEIRasymp_S3_PII_parameters}
\end{figure}

\begin{figure}[H]
\centering
\includegraphics[clip,trim=0 760 0 0,width=\linewidth]{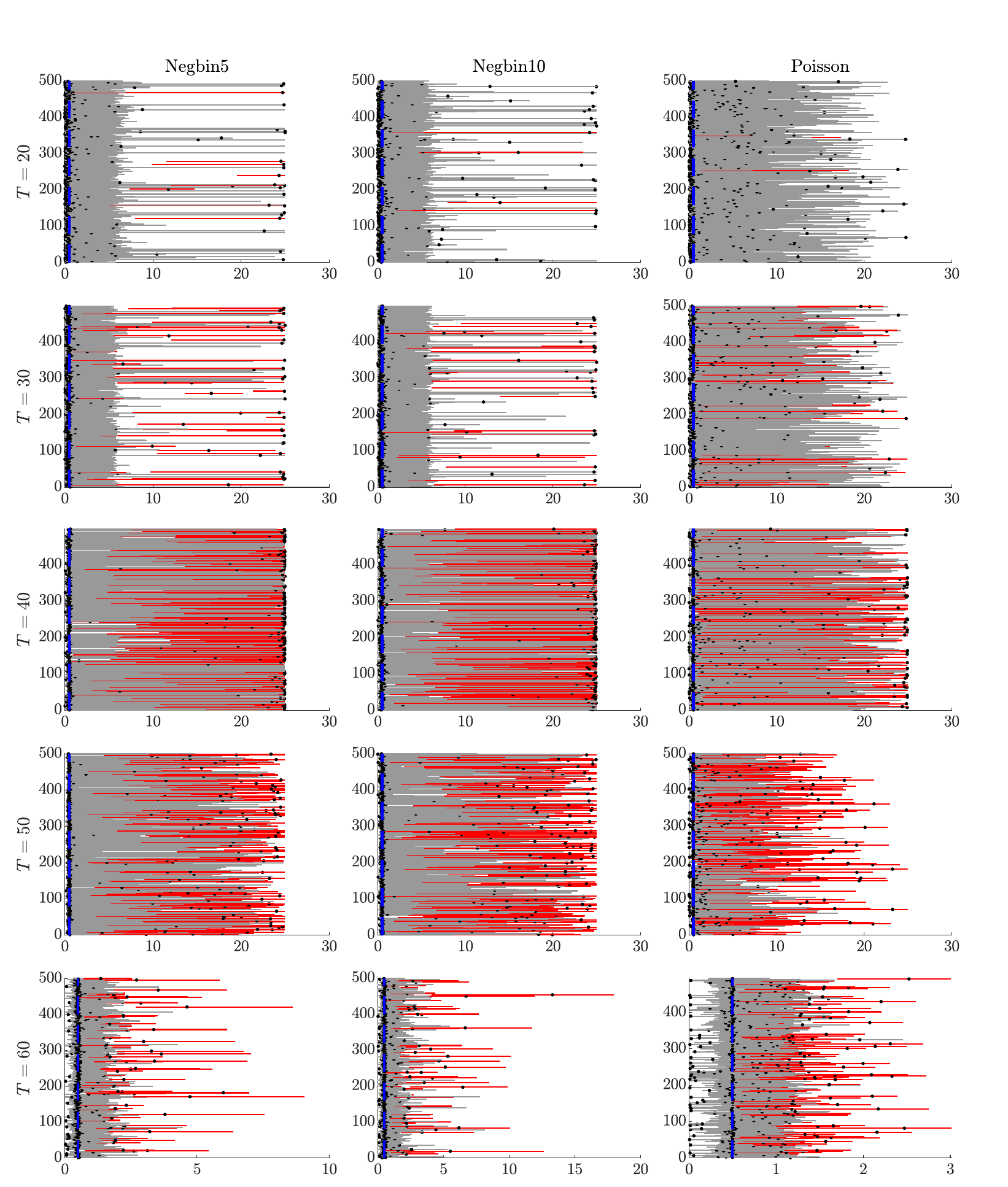}
\caption{
Parameter estimates and 95\% confidence intervals (CIs) for the symptomatic transmission rate $\beta_0$ across 500 simulation replicates and calibration window lengths $T=20,30,\ldots,100$, with the true value $\beta_0=0.5$ indicated by the vertical blue dashed line. Columns correspond to the error structures: negative binomial with data-generating dispersion parameter $\alpha=5$ (Negbin5), negative binomial with data-generating dispersion parameter $\alpha=10$ (Negbin10), and Poisson. Each horizontal line corresponds to a single simulation replicate, showing the bootstrap confidence interval obtained by resampling within that replicate, with the corresponding point estimate marked by a black dot at its center. Red intervals denote confidence intervals that do not contain the true value, whereas gray intervals denote those that do.
}
\label{fig:SEIRasymp_S3_CI_grid_param_1}
\end{figure}

\begin{figure}[H]
\centering
\includegraphics[clip,trim=0 0 0 920,width=\linewidth]{SEIRasymp_S3_CI_grid_param_1.pdf}
\caption*{Figure~\ref{fig:SEIRasymp_S3_CI_grid_param_1} (continued).}
\end{figure}

\begin{figure}[H]
\centering
\includegraphics[clip,trim=0 760 0 0,width=\linewidth]{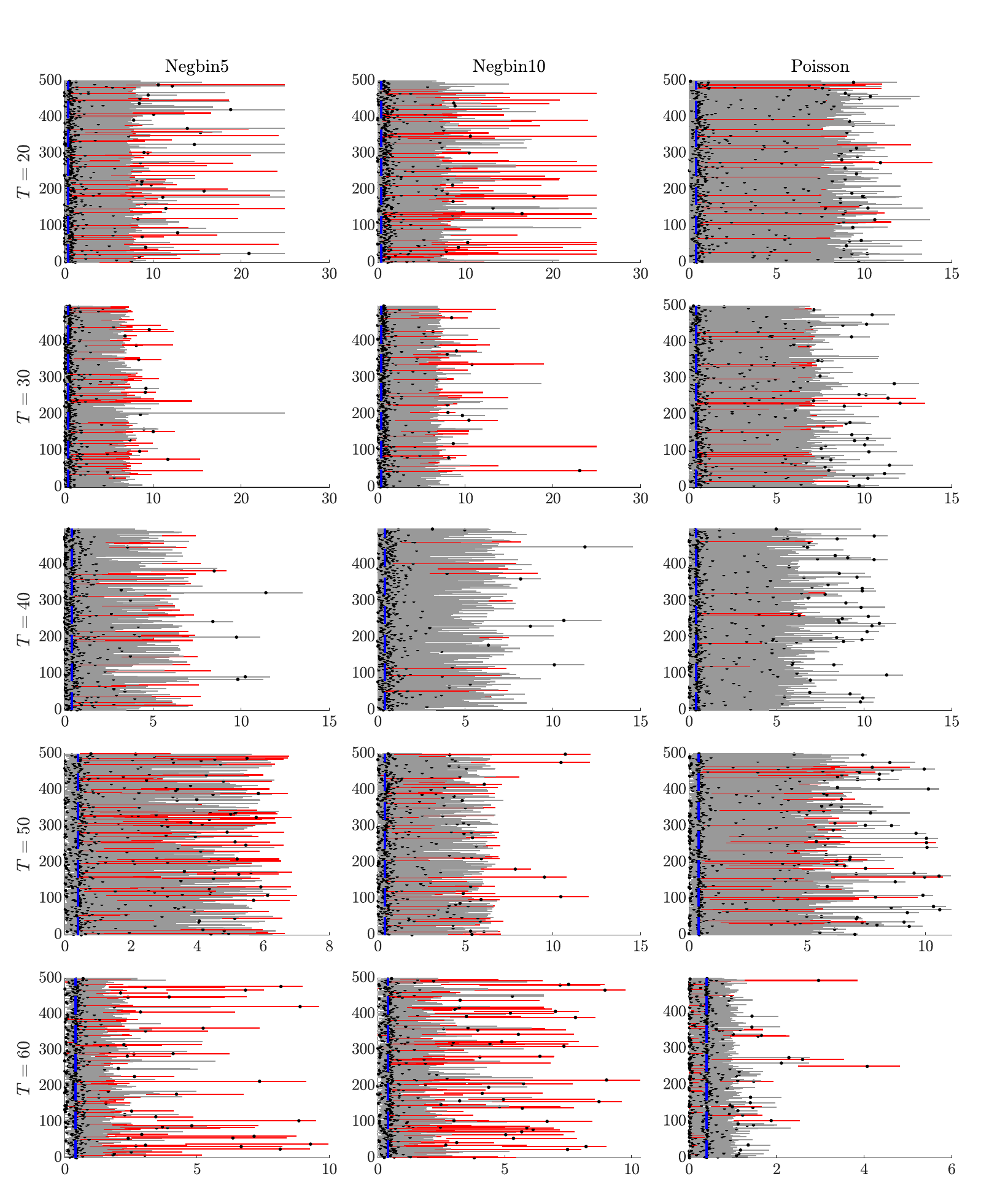}
\caption{
Parameter estimates and 95\% confidence intervals (CIs) for the asymptomatic transmission rate $\beta_1$ across 500 simulation replicates and calibration window lengths $T=20,30,\ldots,100$, with the true value $\beta_1=0.4$ indicated by the vertical blue dashed line. Columns correspond to the error structures: negative binomial with data-generating dispersion parameter $\alpha=5$ (Negbin5), negative binomial with data-generating dispersion parameter $\alpha=10$ (Negbin10), and Poisson. Each horizontal line corresponds to a single simulation replicate, showing the bootstrap confidence interval obtained by resampling within that replicate, with the corresponding point estimate marked by a black dot at its center. Red intervals denote confidence intervals that do not contain the true value, whereas gray intervals denote those that do.
}
\label{fig:SEIRasymp_S3_CI_grid_param_2}
\end{figure}

\begin{figure}[H]
\centering
\includegraphics[clip,trim=0 0 0 920,width=\linewidth]{SEIRasymp_S3_CI_grid_param_2.pdf}
\caption*{Figure~\ref{fig:SEIRasymp_S3_CI_grid_param_2} (continued).}
\end{figure}

\begin{figure}[H]
\centering
\includegraphics[clip,trim=0 760 0 0,width=\linewidth]{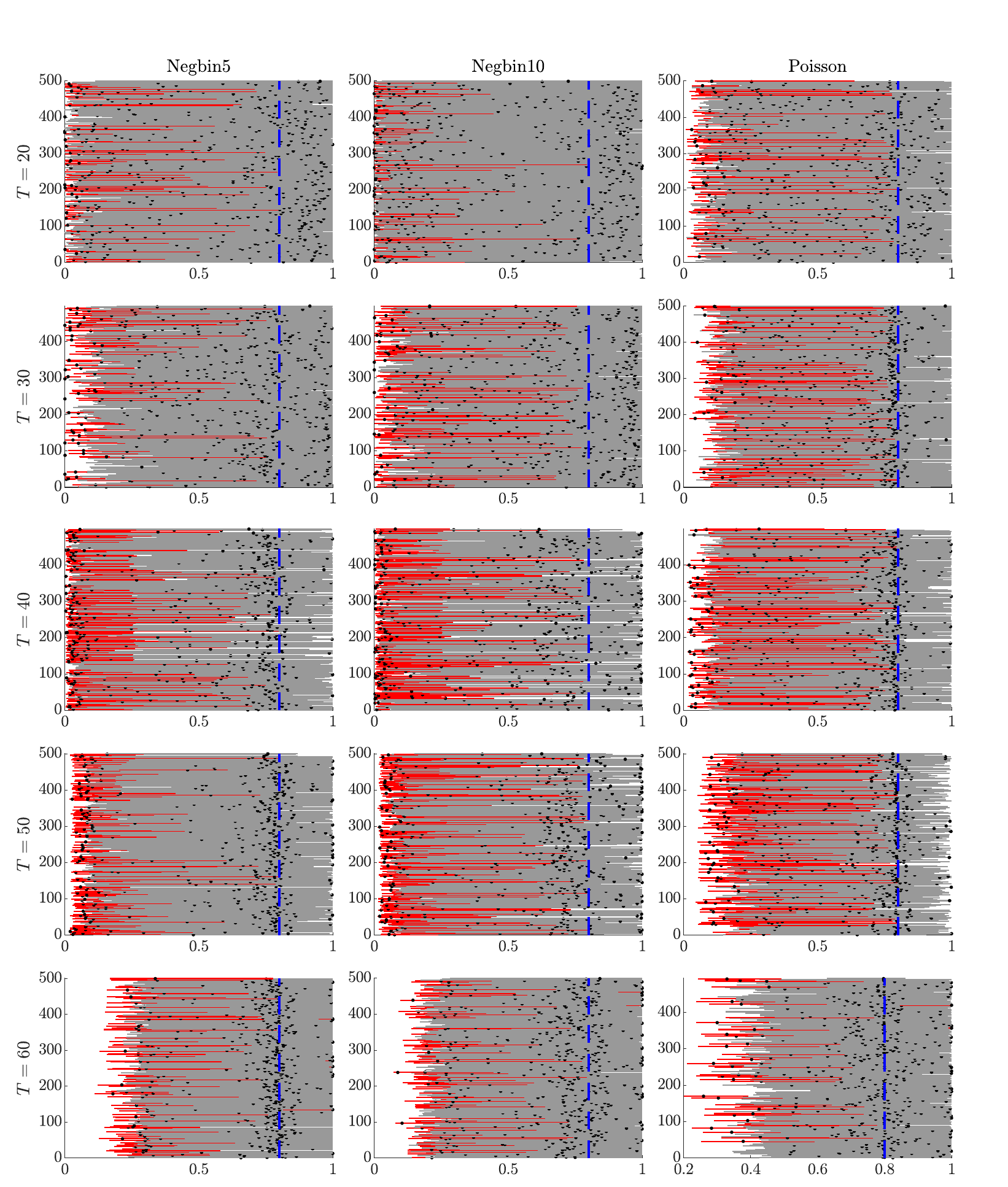}
\caption{
Parameter estimates and 95\% confidence intervals (CIs) for the symptomatic fraction $\rho$ across 500 simulation replicates and calibration window lengths $T=20,30,\ldots,100$, with the true value $\rho=0.8$ indicated by the vertical blue dashed line. Columns correspond to the error structures: negative binomial with data-generating dispersion parameter $\alpha=5$ (Negbin5), negative binomial with data-generating dispersion parameter $\alpha=10$ (Negbin10), and Poisson. Each horizontal line corresponds to a single simulation replicate, showing the bootstrap confidence interval obtained by resampling within that replicate, with the corresponding point estimate marked by a black dot at its center. Red intervals denote confidence intervals that do not contain the true value, whereas gray intervals denote those that do.
}
\label{fig:SEIRasymp_S3_CI_grid_param_4}
\end{figure}

\begin{figure}[H]
\centering
\includegraphics[clip,trim=0 0 0 920,width=\linewidth]{SEIRasymp_S3_CI_grid_param_4.pdf}
\caption*{Figure~\ref{fig:SEIRasymp_S3_CI_grid_param_4} (continued).}
\end{figure}

\begin{figure}[H]
\centering
\includegraphics[clip,trim=0 760 0 0,width=\linewidth]{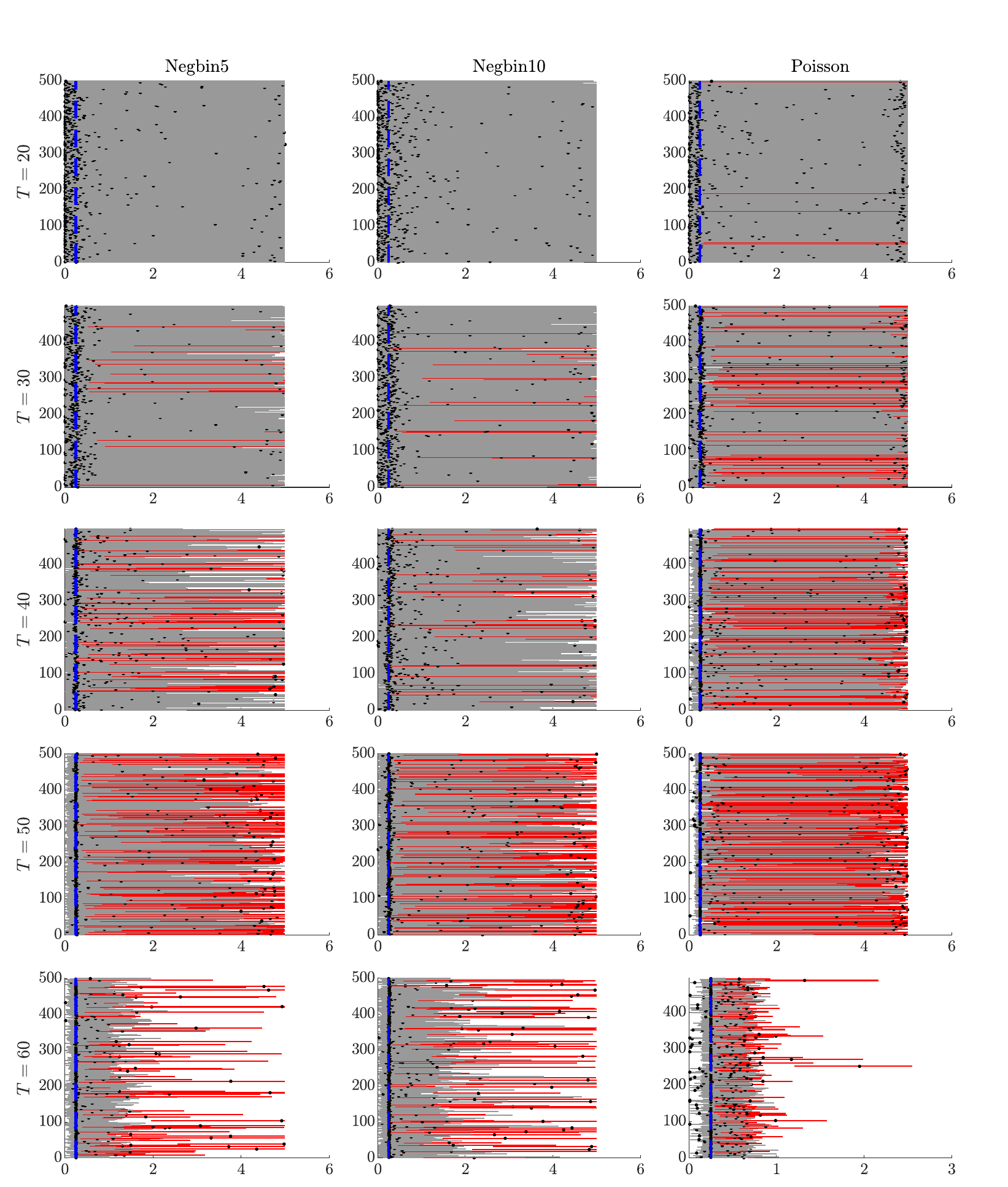}
\caption{
Parameter estimates and 95\% confidence intervals (CIs) for the recovery rate $\gamma$ across 500 simulation replicates and calibration window lengths $T=20,30,\ldots,100$, with the true value $\gamma=0.25$ indicated by the vertical blue dashed line. Columns correspond to the error structures: negative binomial with data-generating dispersion parameter $\alpha=5$ (Negbin5), negative binomial with data-generating dispersion parameter $\alpha=10$ (Negbin10), and Poisson. Each horizontal line corresponds to a single simulation replicate, showing the bootstrap confidence interval obtained by resampling within that replicate, with the corresponding point estimate marked by a black dot at its center. Red intervals denote confidence intervals that do not contain the true value, whereas gray intervals denote those that do.
}
\label{fig:SEIRasymp_S3_CI_grid_param_5}
\end{figure}

\begin{figure}[H]
\centering
\includegraphics[clip,trim=0 0 0 920,width=\linewidth]{SEIRasymp_S3_CI_grid_param_5.pdf}
\caption*{Figure~\ref{fig:SEIRasymp_S3_CI_grid_param_5} (continued).}
\end{figure}

\section{SEIRD Model}

\subsection{Scenario~1: Estimated $\{\beta\}$; Fixed $\{\kappa,\, \rho,\, \gamma,\, N\}$}

\begin{figure}[H]
\centering
\includegraphics[width=\linewidth]{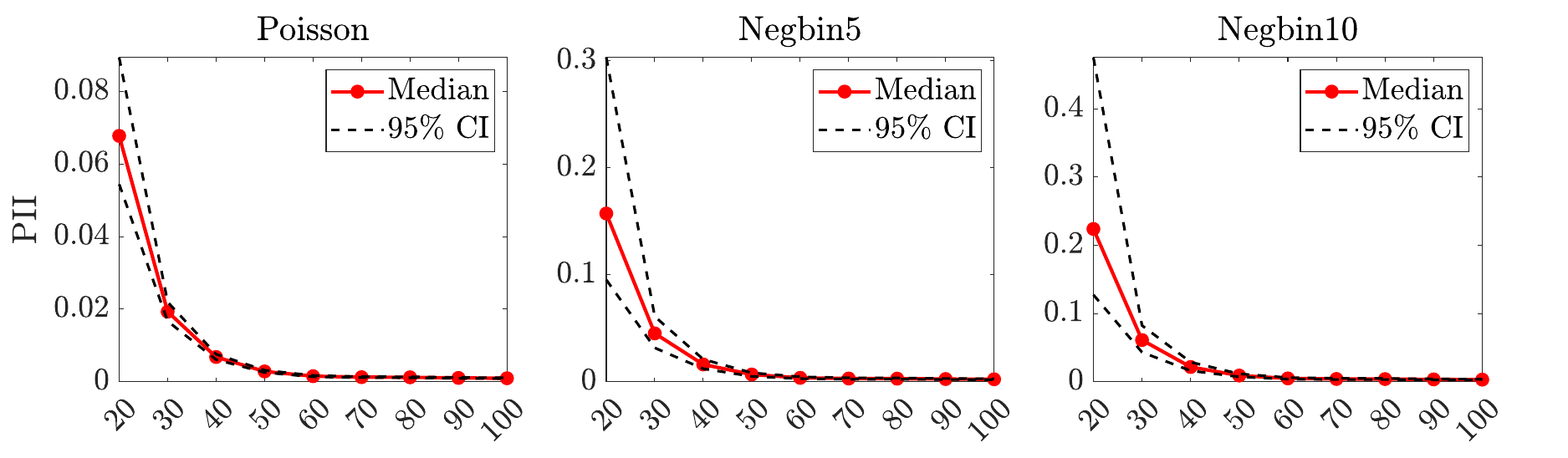}
\caption{Practical Identifiability Index (PII) for the transmission rate $\beta$ in the SEIRD model (Scenario~1) across calibration-window lengths $T=20, 30, \ldots, 100$ under three error structures: Poisson, negative binomial with data-generating dispersion $\alpha=5$ (Negbin5), and negative binomial with data-generating dispersion $\alpha=10$ (Negbin10). Red lines show the median PII across replicates, and dashed black curves indicate the central 95\% interval.}
\label{fig:SEIRD_S1_PII_parameters}
\end{figure}

\begin{figure}[H]
\centering
\includegraphics[clip,trim=0 760 0 0,width=\linewidth]{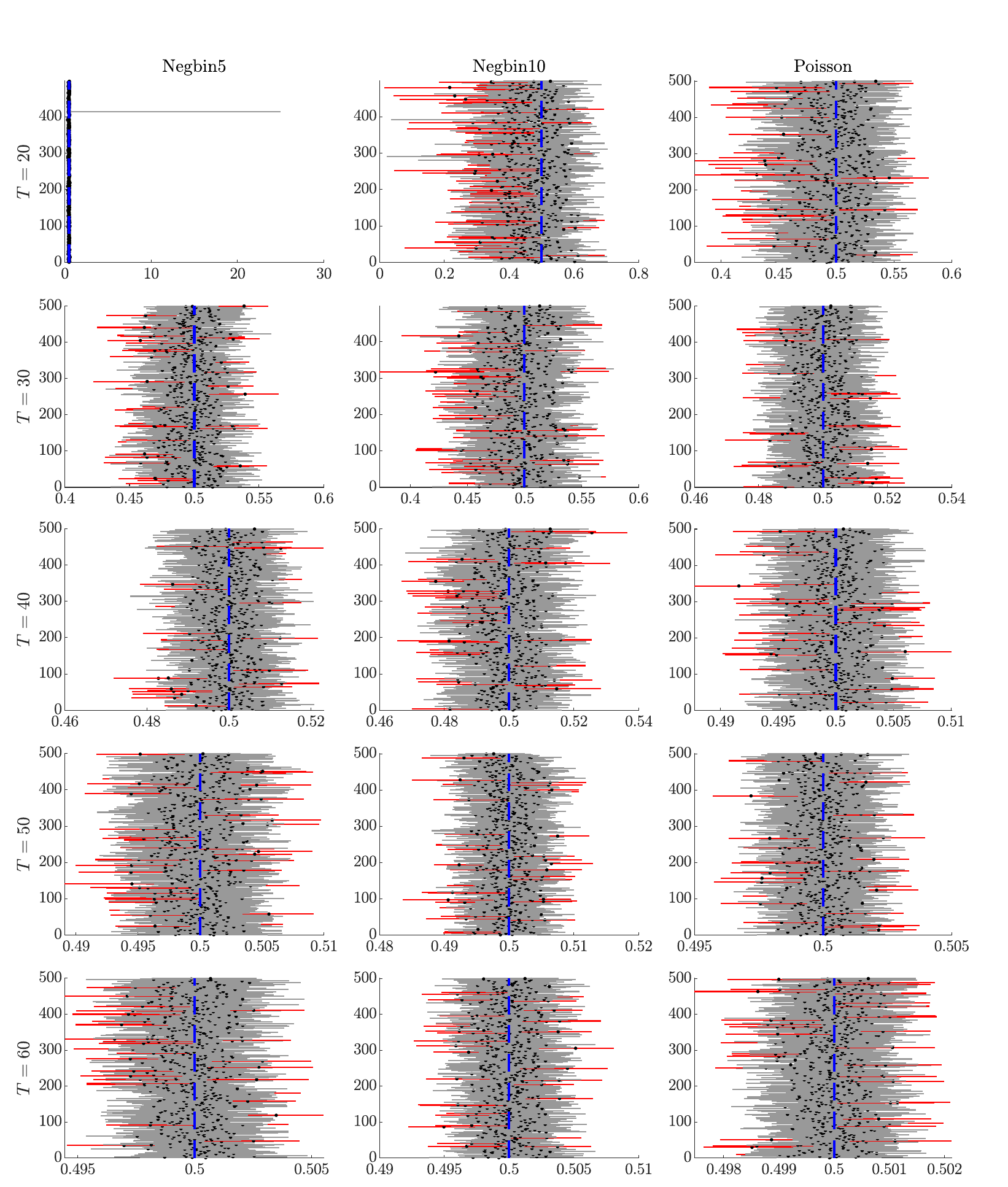}
\caption{
Parameter estimates and 95\% confidence intervals (CIs) for the transmission rate $\beta$ across 500 simulation replicates and calibration window lengths $T=20,30,\ldots,100$, with the true value $\beta=0.5$ indicated by the vertical blue dashed line. Columns correspond to the error structures: negative binomial with data-generating dispersion parameter $\alpha=5$ (Negbin5), negative binomial with data-generating dispersion parameter $\alpha=10$ (Negbin10), and Poisson. Each horizontal line corresponds to a single simulation replicate, showing the bootstrap confidence interval obtained by resampling within that replicate, with the corresponding point estimate marked by a black dot at its center. Red intervals denote confidence intervals that do not contain the true value, whereas gray intervals denote those that do.
}
\label{fig:SEIRD_S1_CI_grid_param_1}
\end{figure}

\begin{figure}[H]
\centering
\includegraphics[clip,trim=0 0 0 920,width=\linewidth]{SEIRD_S1_CI_grid_param_1.pdf}
\caption*{Figure~\ref{fig:SEIRD_S1_CI_grid_param_1} (continued).}
\end{figure}

\subsection{Scenario~2: Estimated $\{\beta,\, \rho\}$; Fixed $\{\kappa,\, \gamma,\, N\}$}

\begin{figure}[H]
\centering
\includegraphics[width=\linewidth]{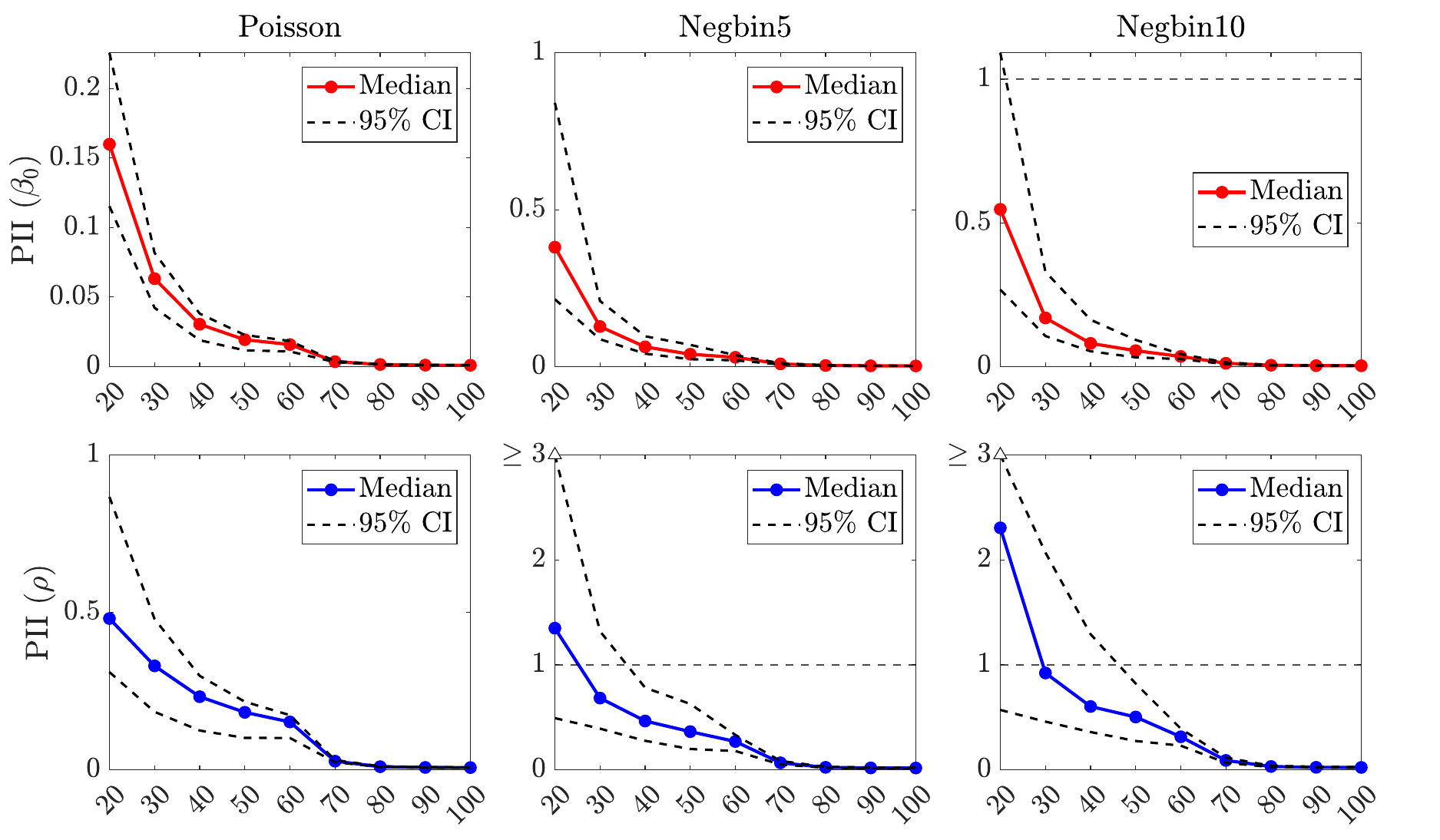}
\caption{Practical Identifiability Index (PII) for the transmission rate $\beta$ and infection fatality fraction $\rho$ in the SEIRD model (Scenario~2) across calibration-window lengths $T=20, 30, \ldots, 100$ under three error structures: Poisson, negative binomial with data-generating dispersion $\alpha=5$ (Negbin5), and negative binomial with data-generating dispersion $\alpha=10$ (Negbin10). Red lines show the median PII across replicates, and dashed black curves indicate the central 95\% interval. For visualization only, within subplots containing PII values at or above 3, those values are displayed at an upper plotting limit of 3; uncapped values are retained for all numerical analyses. Upward triangles denote values displayed at the upper limit.}
\label{fig:SEIRD_S2_PII_parameters}
\end{figure}

\begin{figure}[H]
\centering
\includegraphics[clip,trim=0 760 0 0,width=\linewidth]{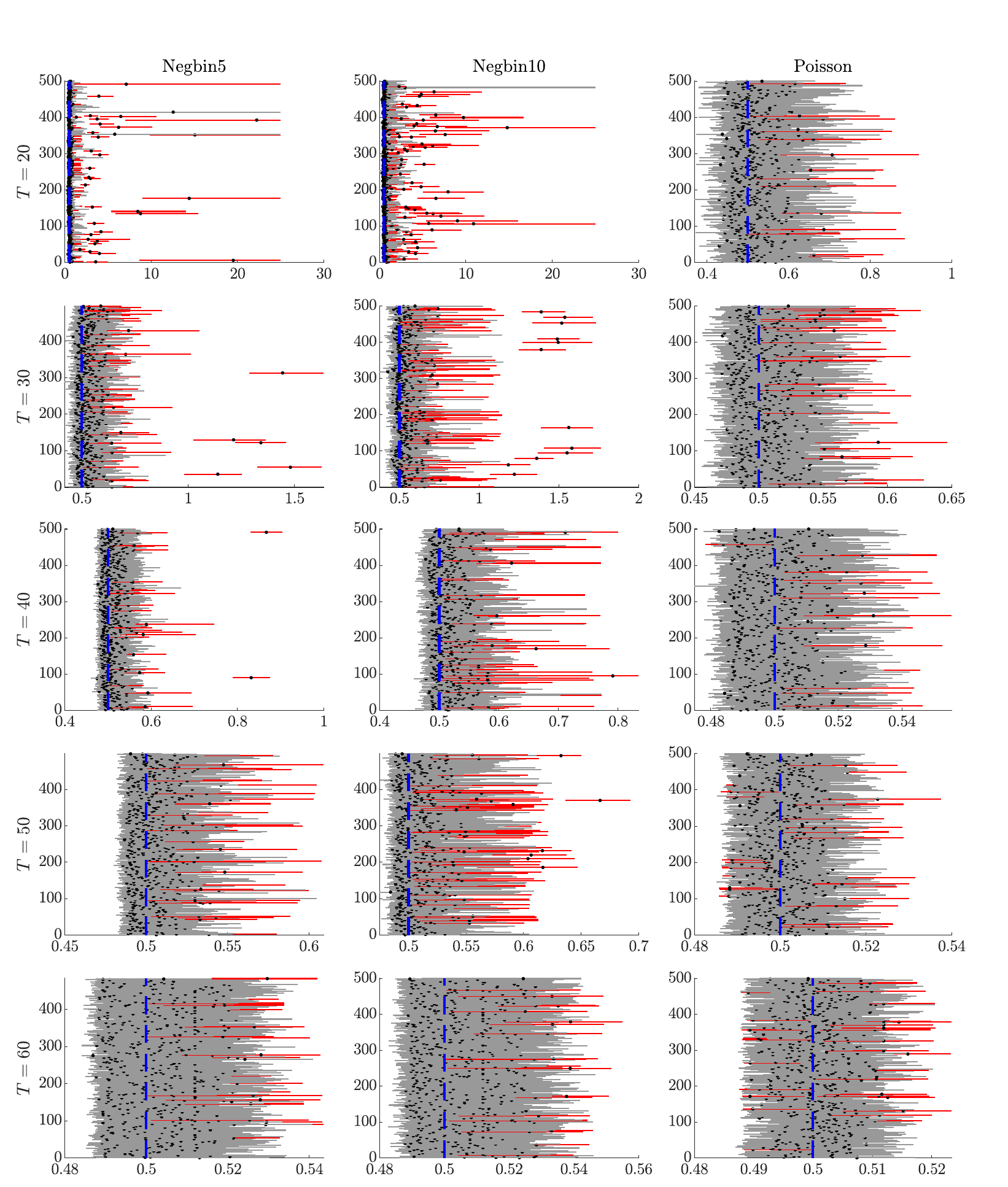}
\caption{
Parameter estimates and 95\% confidence intervals (CIs) for the transmission rate $\beta$ across 500 simulation replicates and calibration window lengths $T=20,30,\ldots,100$, with the true value $\beta=0.5$ indicated by the vertical blue dashed line. Columns correspond to the error structures: negative binomial with data-generating dispersion parameter $\alpha=5$ (Negbin5), negative binomial with data-generating dispersion parameter $\alpha=10$ (Negbin10), and Poisson. Each horizontal line corresponds to a single simulation replicate, showing the bootstrap confidence interval obtained by resampling within that replicate, with the corresponding point estimate marked by a black dot at its center. Red intervals denote confidence intervals that do not contain the true value, whereas gray intervals denote those that do.
}
\label{fig:SEIRD_S2_CI_grid_param_1}
\end{figure}

\begin{figure}[H]
\centering
\includegraphics[clip,trim=0 0 0 920,width=\linewidth]{SEIRD_S2_CI_grid_param_1.pdf}
\caption*{Figure~\ref{fig:SEIRD_S2_CI_grid_param_1} (continued).}
\end{figure}

\begin{figure}[H]
\centering
\includegraphics[clip,trim=0 760 0 0,width=\linewidth]{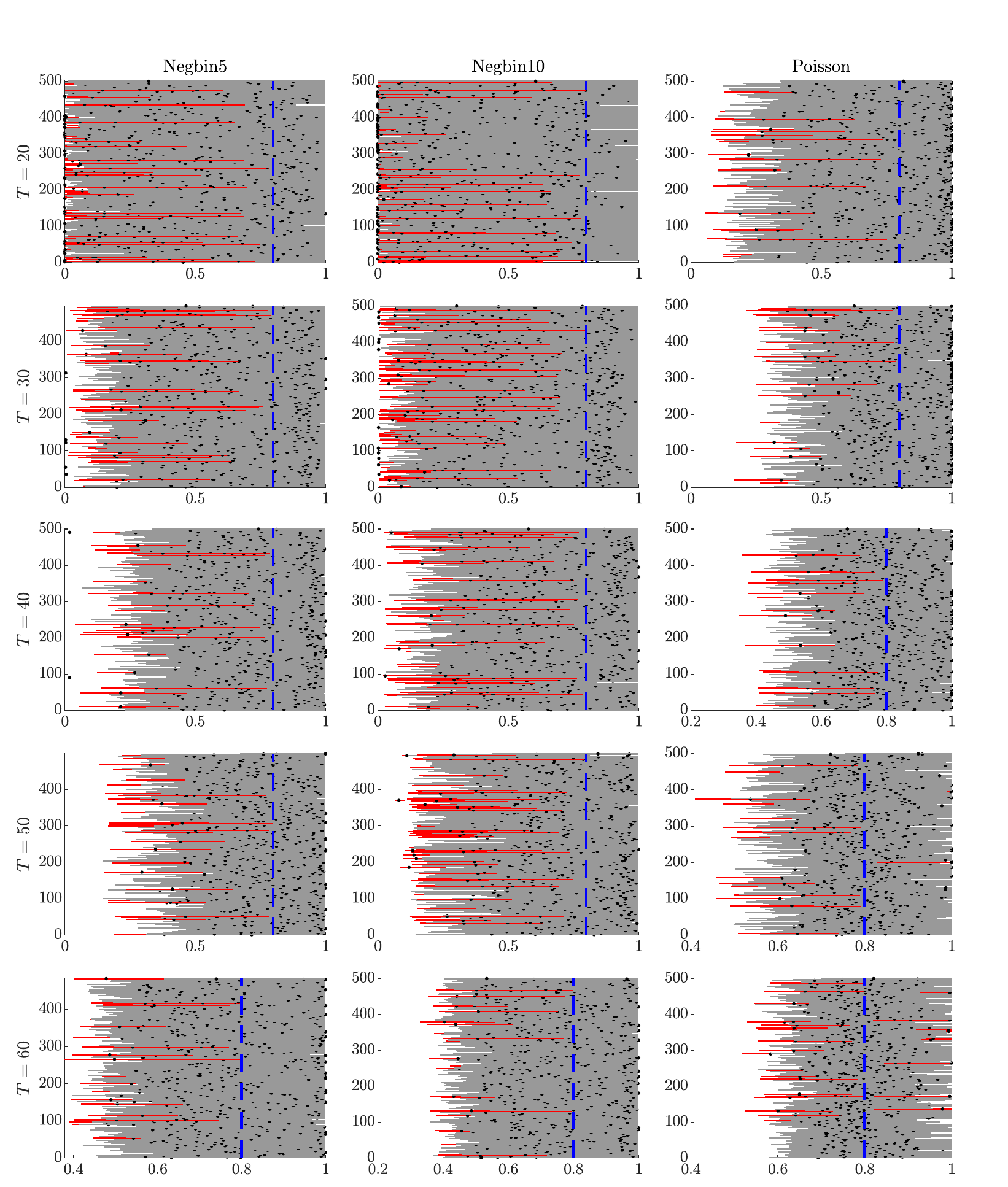}
\caption{
Parameter estimates and 95\% confidence intervals (CIs) for the infection fatality fraction $\rho$ across 500 simulation replicates and calibration window lengths $T=20,30,\ldots,100$, with the true value $\rho=0.8$ indicated by the vertical blue dashed line. Columns correspond to the error structures: negative binomial with data-generating dispersion parameter $\alpha=5$ (Negbin5), negative binomial with data-generating dispersion parameter $\alpha=10$ (Negbin10), and Poisson. Each horizontal line corresponds to a single simulation replicate, showing the bootstrap confidence interval obtained by resampling within that replicate, with the corresponding point estimate marked by a black dot at its center. Red intervals denote confidence intervals that do not contain the true value, whereas gray intervals denote those that do.
}
\label{fig:SEIRD_S2_CI_grid_param_3}
\end{figure}

\begin{figure}[H]
\centering
\includegraphics[clip,trim=0 0 0 920,width=\linewidth]{SEIRD_S2_CI_grid_param_3.pdf}
\caption*{Figure~\ref{fig:SEIRD_S2_CI_grid_param_3} (continued).}
\end{figure}

\subsection{Scenario~3: Estimated $\{\beta,\, \rho,\, \gamma\}$; Fixed $\{\kappa,\, N\}$}

\begin{figure}[H]
\centering
\includegraphics[width=\linewidth]{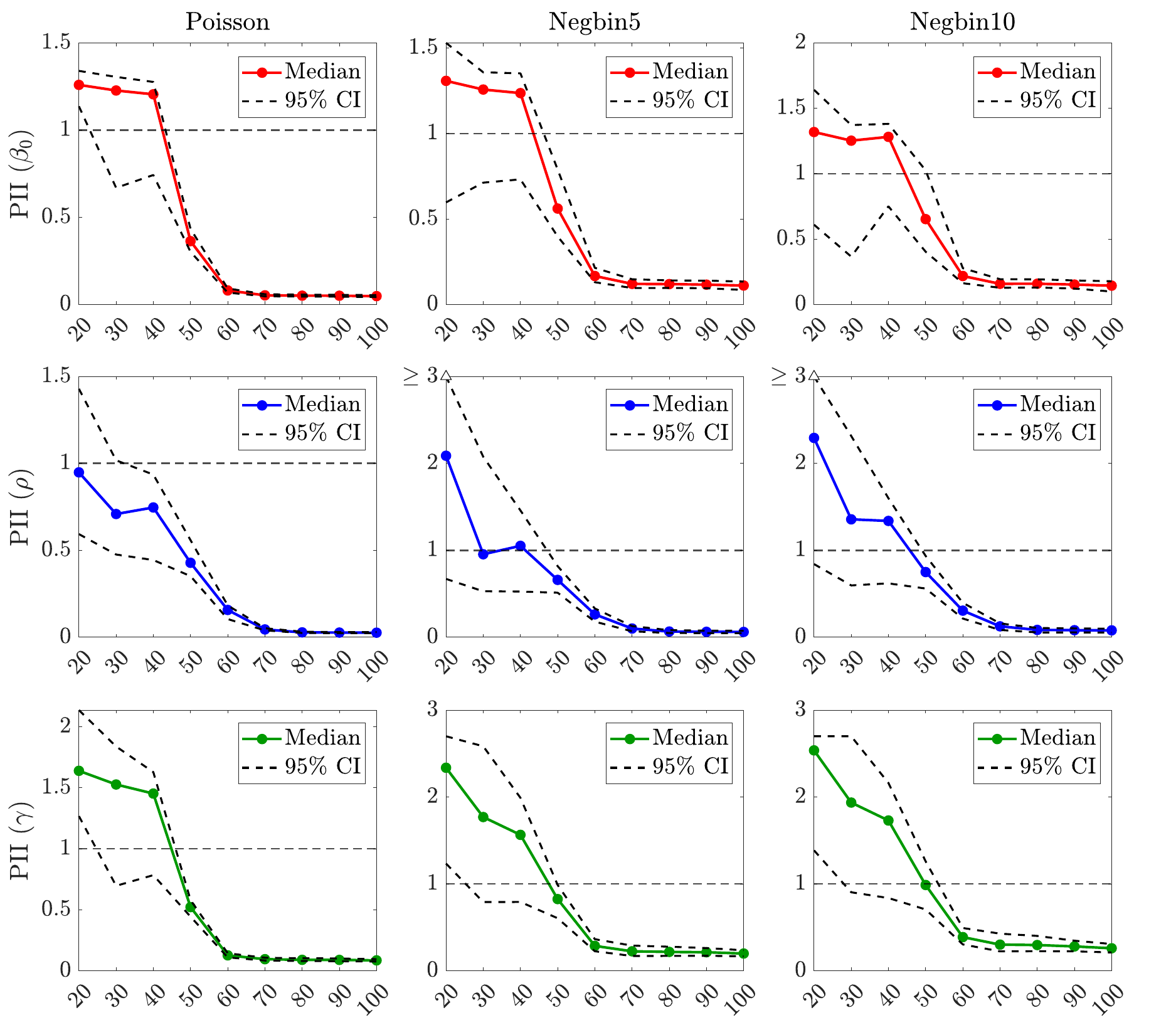}
\caption{Practical Identifiability Index (PII) for the transmission rate $\beta$, infection fatality fraction $\rho$, and removal rate $\gamma$ in the SEIRD model (Scenario~3) across calibration-window lengths $T=20, 30, \ldots, 100$ under three error structures: Poisson, negative binomial with data-generating dispersion $\alpha=5$ (Negbin5), and negative binomial with data-generating dispersion $\alpha=10$ (Negbin10). Red lines show the median PII across replicates, and dashed black curves indicate the central 95\% interval. For visualization only, within subplots containing PII values at or above 3, those values are displayed at an upper plotting limit of 3; uncapped values are retained for all numerical analyses. Upward triangles denote values displayed at the upper limit.}
\label{fig:SEIRD_S3_PII_parameters}
\end{figure}

\begin{figure}[H]
\centering
\includegraphics[clip,trim=0 760 0 0,width=\linewidth]{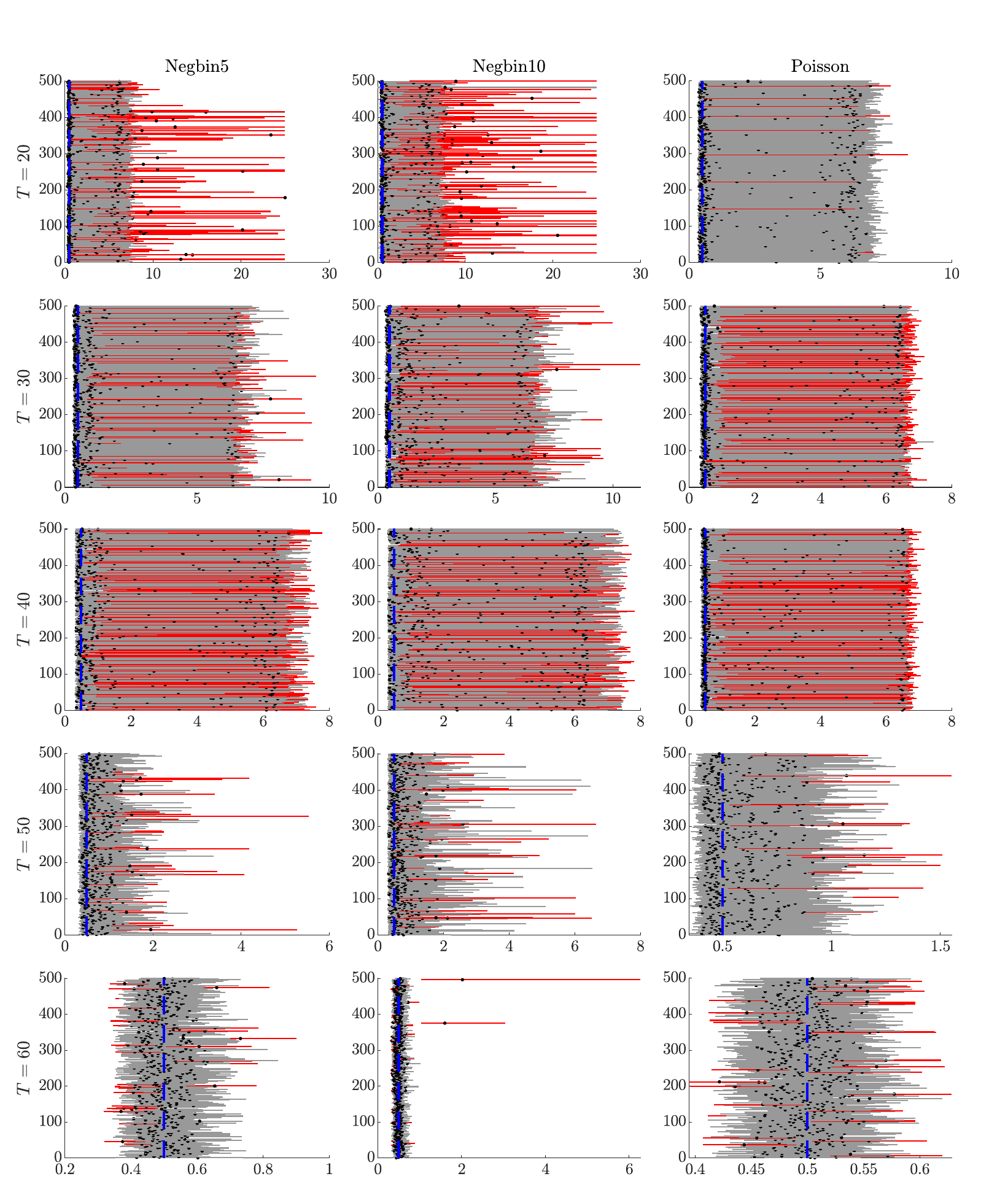}
\caption{
Parameter estimates and 95\% confidence intervals (CIs) for the transmission rate $\beta$ across 500 simulation replicates and calibration window lengths $T=20,30,\ldots,100$, with the true value $\beta=0.5$ indicated by the vertical blue dashed line. Columns correspond to the error structures: negative binomial with data-generating dispersion parameter $\alpha=5$ (Negbin5), negative binomial with data-generating dispersion parameter $\alpha=10$ (Negbin10), and Poisson. Each horizontal line corresponds to a single simulation replicate, showing the bootstrap confidence interval obtained by resampling within that replicate, with the corresponding point estimate marked by a black dot at its center. Red intervals denote confidence intervals that do not contain the true value, whereas gray intervals denote those that do.
}
\label{fig:SEIRD_S3_CI_grid_param_1}
\end{figure}

\begin{figure}[H]
\centering
\includegraphics[clip,trim=0 0 0 920,width=\linewidth]{SEIRD_S3_CI_grid_param_1.pdf}
\caption*{Figure~\ref{fig:SEIRD_S3_CI_grid_param_1} (continued).}
\end{figure}

\begin{figure}[H]
\centering
\includegraphics[clip,trim=0 760 0 0,width=\linewidth]{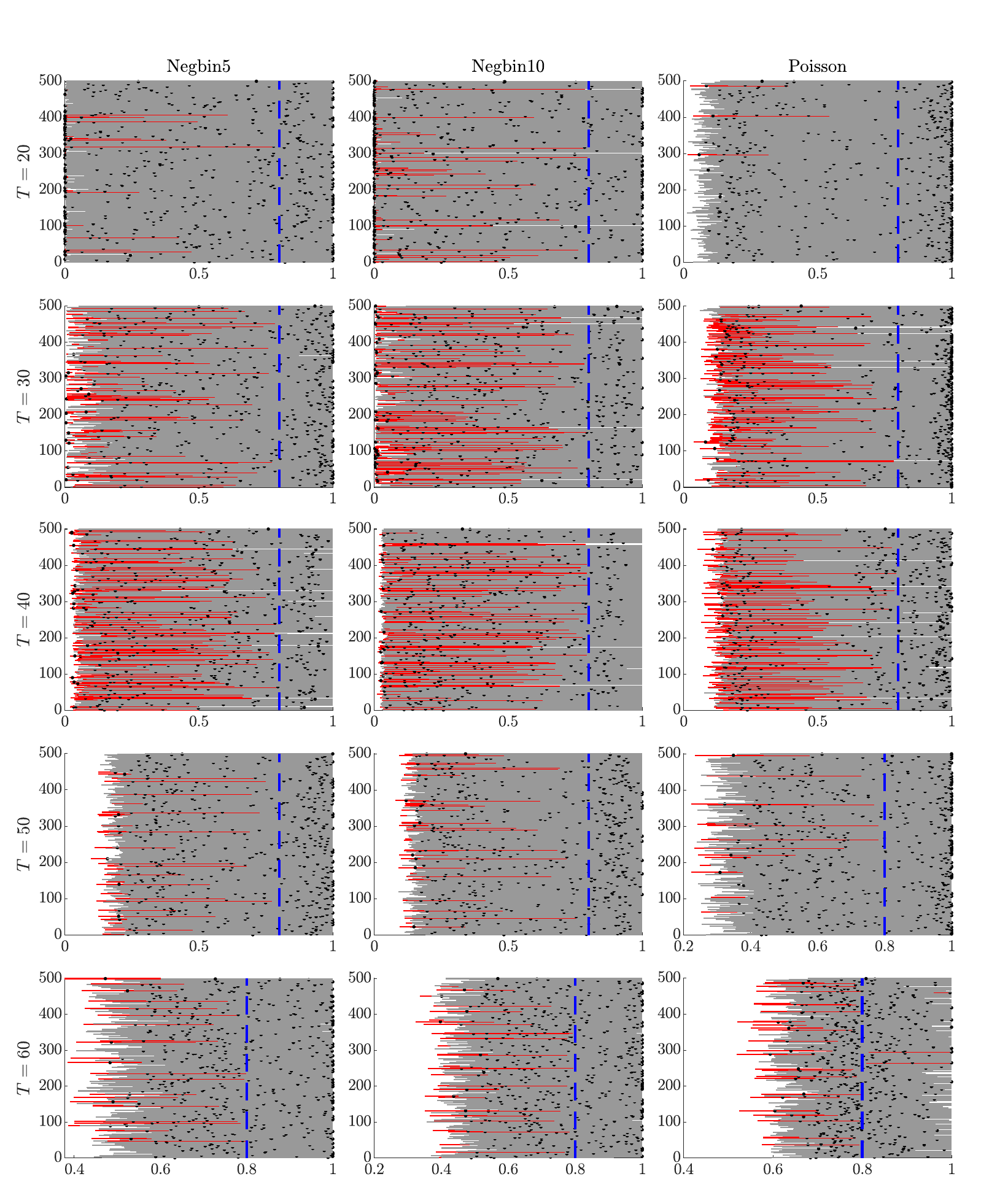}
\caption{
Parameter estimates and 95\% confidence intervals (CIs) for the infection fatality fraction $\rho$ across 500 simulation replicates and calibration window lengths $T=20,30,\ldots,100$, with the true value $\rho=0.8$ indicated by the vertical blue dashed line. Columns correspond to the error structures: negative binomial with data-generating dispersion parameter $\alpha=5$ (Negbin5), negative binomial with data-generating dispersion parameter $\alpha=10$ (Negbin10), and Poisson. Each horizontal line corresponds to a single simulation replicate, showing the bootstrap confidence interval obtained by resampling within that replicate, with the corresponding point estimate marked by a black dot at its center. Red intervals denote confidence intervals that do not contain the true value, whereas gray intervals denote those that do.
}
\label{fig:SEIRD_S3_CI_grid_param_3}
\end{figure}

\begin{figure}[H]
\centering
\includegraphics[clip,trim=0 0 0 920,width=\linewidth]{SEIRD_S3_CI_grid_param_3.pdf}
\caption*{Figure~\ref{fig:SEIRD_S3_CI_grid_param_3} (continued).}
\end{figure}

\begin{figure}[H]
\centering
\includegraphics[clip,trim=0 760 0 0,width=\linewidth]{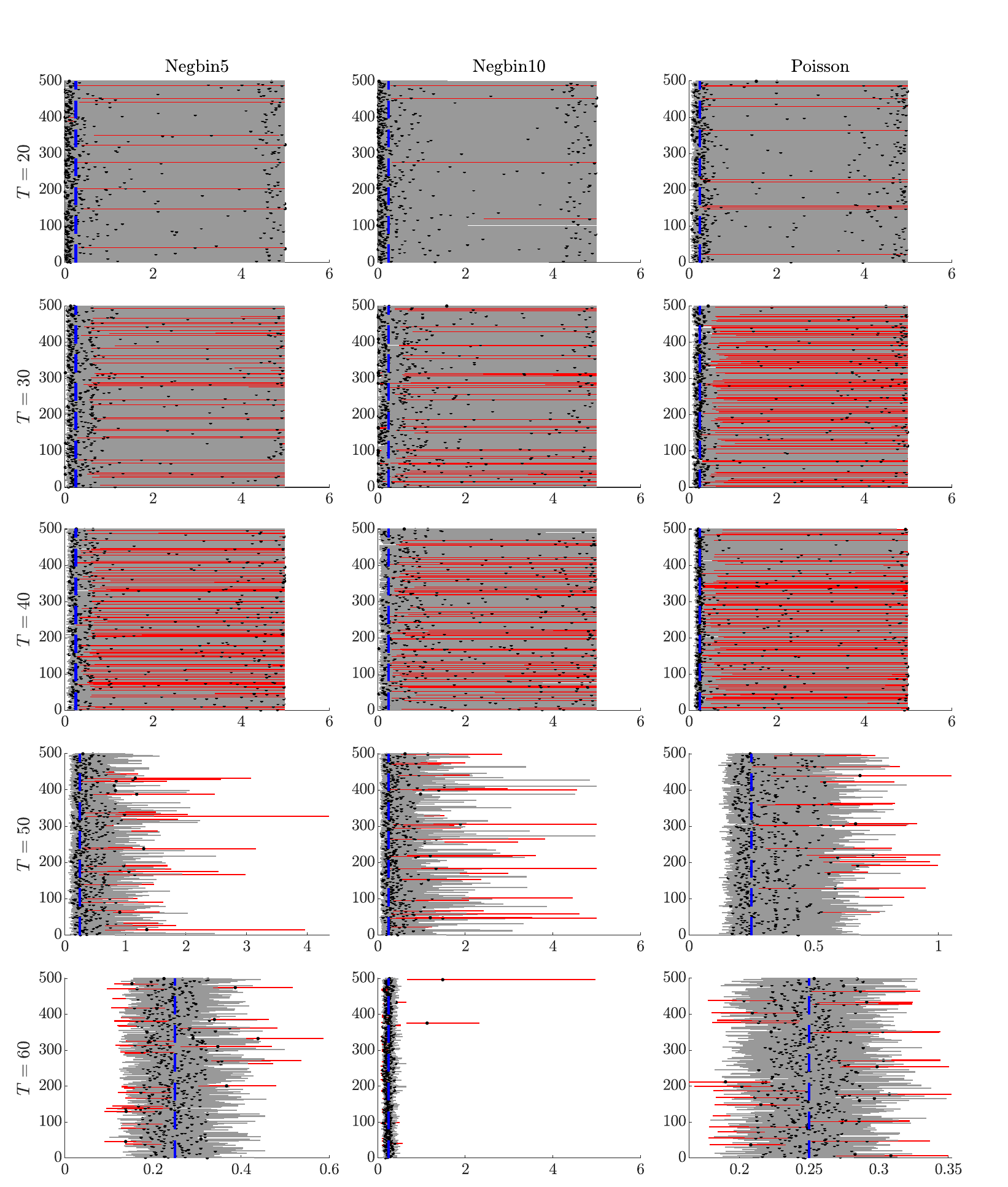}
\caption{
Parameter estimates and 95\% confidence intervals (CIs) for the removal rate $\gamma$ across 500 simulation replicates and calibration window lengths $T=20,30,\ldots,100$, with the true value $\gamma=0.25$ indicated by the vertical blue dashed line. Columns correspond to the error structures: negative binomial with data-generating dispersion parameter $\alpha=5$ (Negbin5), negative binomial with data-generating dispersion parameter $\alpha=10$ (Negbin10), and Poisson. Each horizontal line corresponds to a single simulation replicate, showing the bootstrap confidence interval obtained by resampling within that replicate, with the corresponding point estimate marked by a black dot at its center. Red intervals denote confidence intervals that do not contain the true value, whereas gray intervals denote those that do.
}
\label{fig:SEIRD_S3_CI_grid_param_4}
\end{figure}

\begin{figure}[H]
\centering
\includegraphics[clip,trim=0 0 0 920,width=\linewidth]{SEIRD_S3_CI_grid_param_4.pdf}
\caption*{Figure~\ref{fig:SEIRD_S3_CI_grid_param_4} (continued).}
\end{figure}

\pagebreak
\section{SEIR Model: Multi-Observable Scenarios}

To examine the effect of additional data streams on practical identifiability, we consider the SEIR model under Scenario~3 ($\beta,\,\kappa,\,\gamma$ estimated; $N$ fixed) with three observation configurations:
(i)~incidence only ($dC/dt$),
(ii)~incidence and prevalence ($dC/dt,\,I$), and
(iii)~incidence, prevalence, and recovered counts ($dC/dt,\,I,\,R$).
All other settings (true parameters, error structure, and bootstrap
parameters) remain identical to the standard SEIR Scenario~3 experiments.
The SEIRMO analysis uses calibration-window lengths
$T=30,35,\ldots,70$.

\subsection{Single observable: $\{dC/dt\}$}

\begin{figure}[H]
\centering
\includegraphics[width=\linewidth]{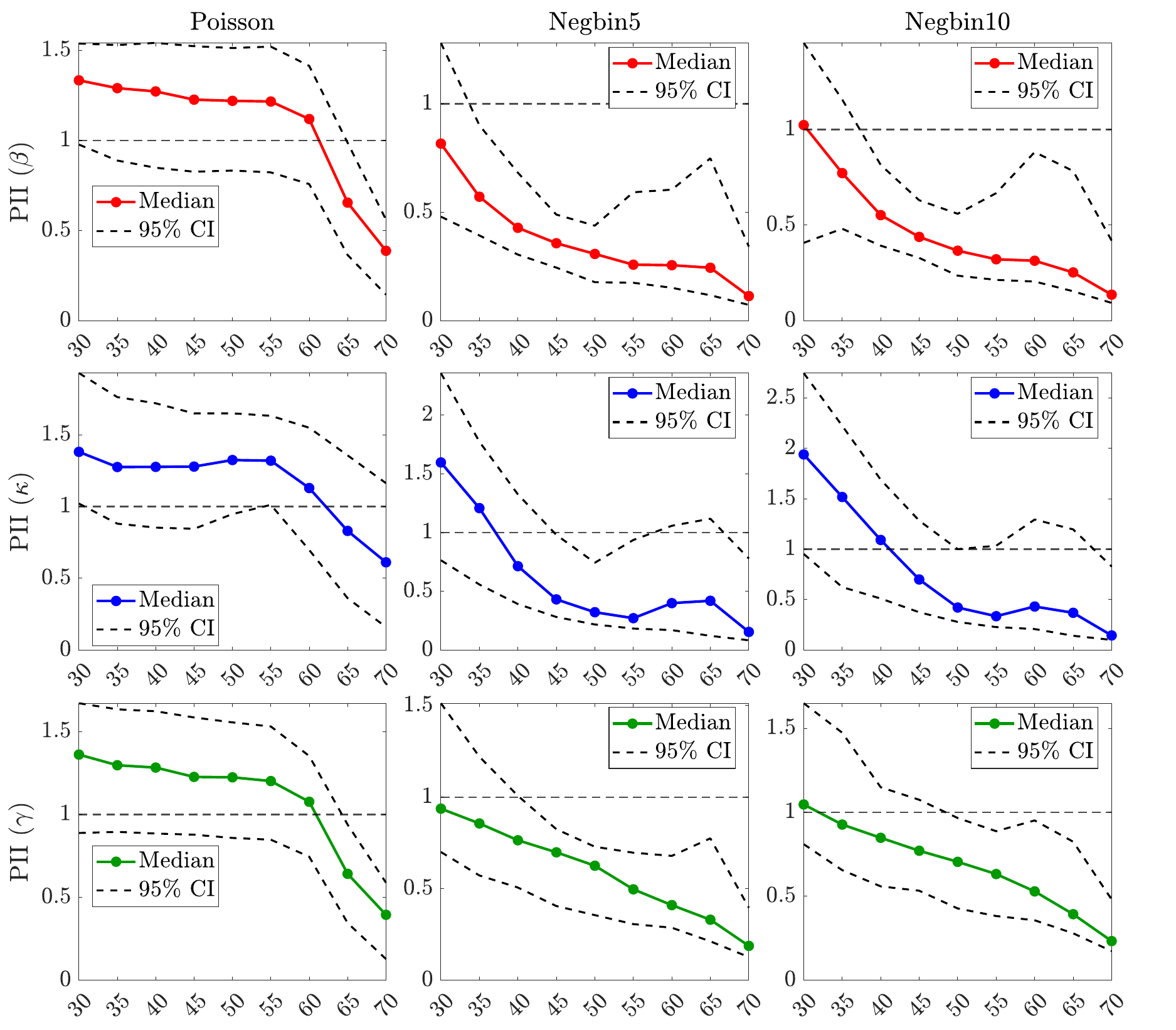}
\caption{Practical Identifiability Index (PII) for $\beta$, $\kappa$, and $\gamma$ in the SEIR model using a single observable ($dC/dt$) across calibration-window lengths $T=30, 35, \ldots, 70$ under three error structures: Poisson, negative binomial with data-generating dispersion $\alpha=5$ (Negbin5), and negative binomial with data-generating dispersion $\alpha=10$ (Negbin10). Red lines show the median PII across replicates, and dashed black curves indicate the central 95\% interval.}
\label{fig:SEIR_vars_dCdt_PII}
\end{figure}

\begin{figure}[H]
\centering
\includegraphics[clip,trim=0 760 0 0,width=\linewidth]{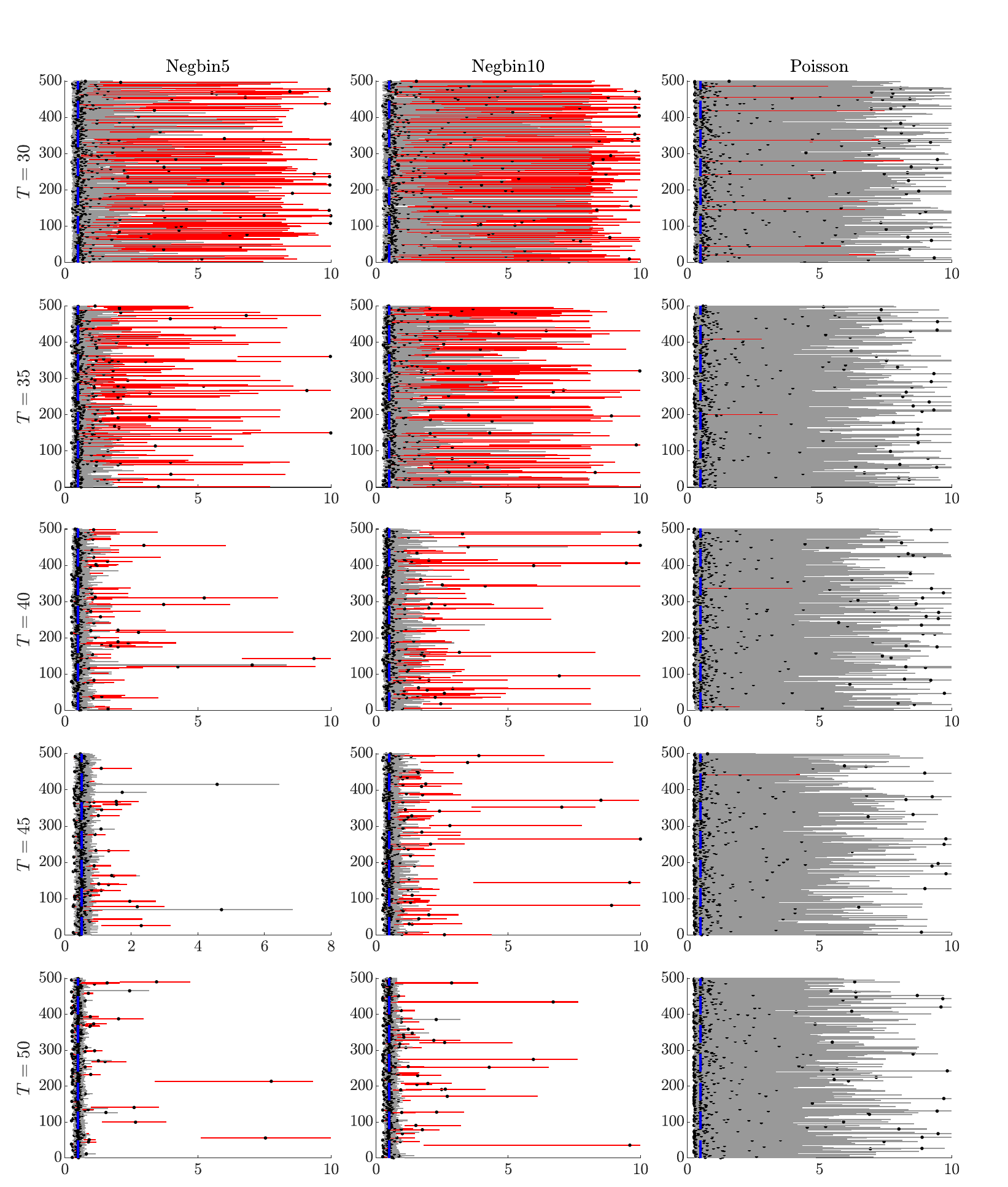}
\caption{Parameter estimates and 95\% CIs for $\beta$ (single observable $dC/dt$) across 500 replicates and calibration window lengths $T=30,35,\ldots,70$, with the true value $\beta=0.5$ indicated by the vertical blue dashed line. Red intervals do not contain the true value; gray intervals do.}
\label{fig:SEIR_vars_dCdt_CI_param_1}
\end{figure}

\begin{figure}[H]
\centering
\includegraphics[clip,trim=0 0 0 920,width=\linewidth]{SEIR_vars_dCdt_CI_grid_param_1.pdf}
\caption*{Figure~\ref{fig:SEIR_vars_dCdt_CI_param_1} (continued).}
\end{figure}

\begin{figure}[H]
\centering
\includegraphics[clip,trim=0 760 0 0,width=\linewidth]{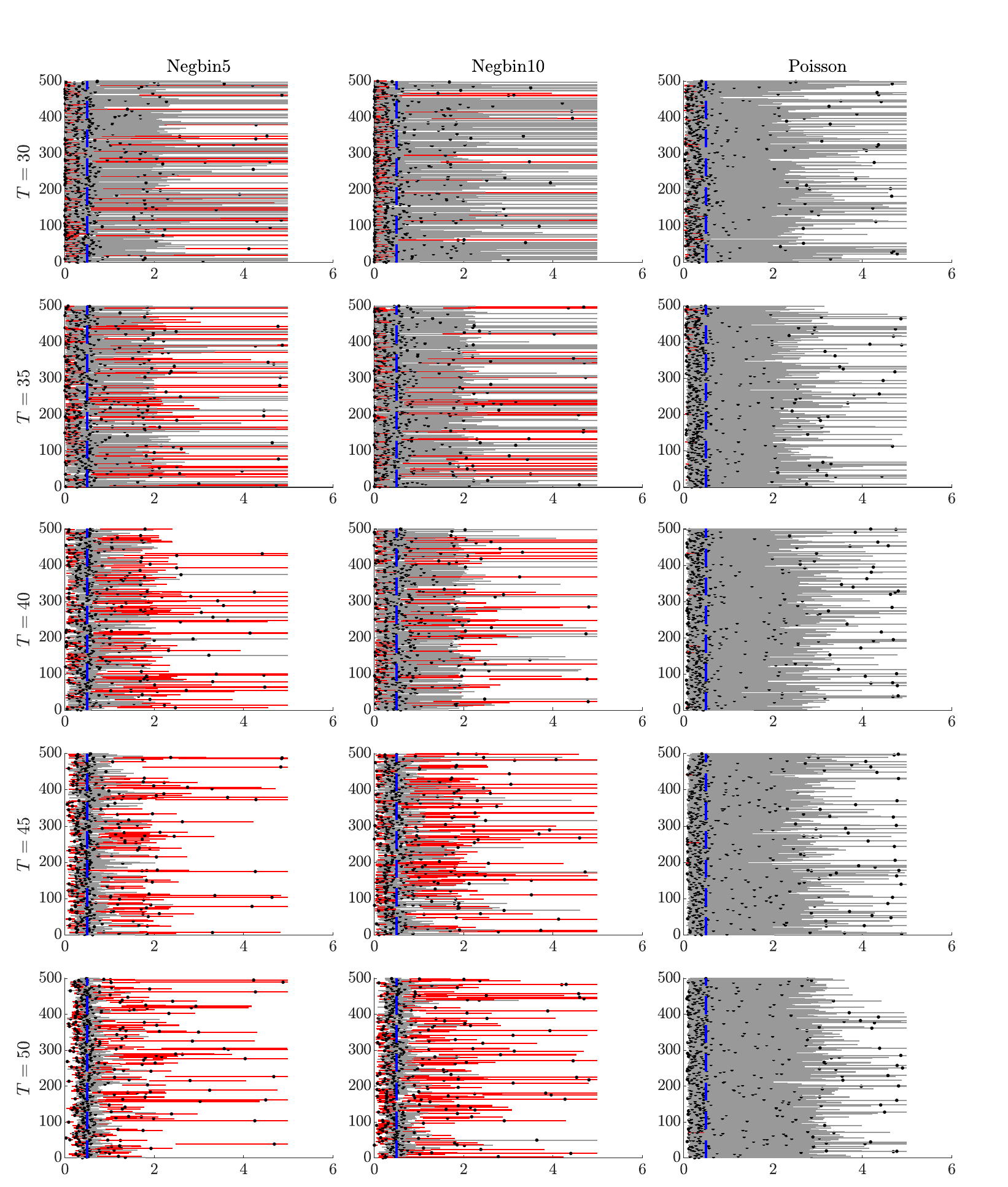}
\caption{Parameter estimates and 95\% CIs for $\kappa$ (single observable $dC/dt$) across 500 replicates and calibration window lengths $T=30,35,\ldots,70$, with the true value $\kappa=0.5$ indicated by the vertical blue dashed line.}
\label{fig:SEIR_vars_dCdt_CI_param_2}
\end{figure}

\begin{figure}[H]
\centering
\includegraphics[clip,trim=0 0 0 920,width=\linewidth]{SEIR_vars_dCdt_CI_grid_param_2.pdf}
\caption*{Figure~\ref{fig:SEIR_vars_dCdt_CI_param_2} (continued).}
\end{figure}

\begin{figure}[H]
\centering
\includegraphics[clip,trim=0 760 0 0,width=\linewidth]{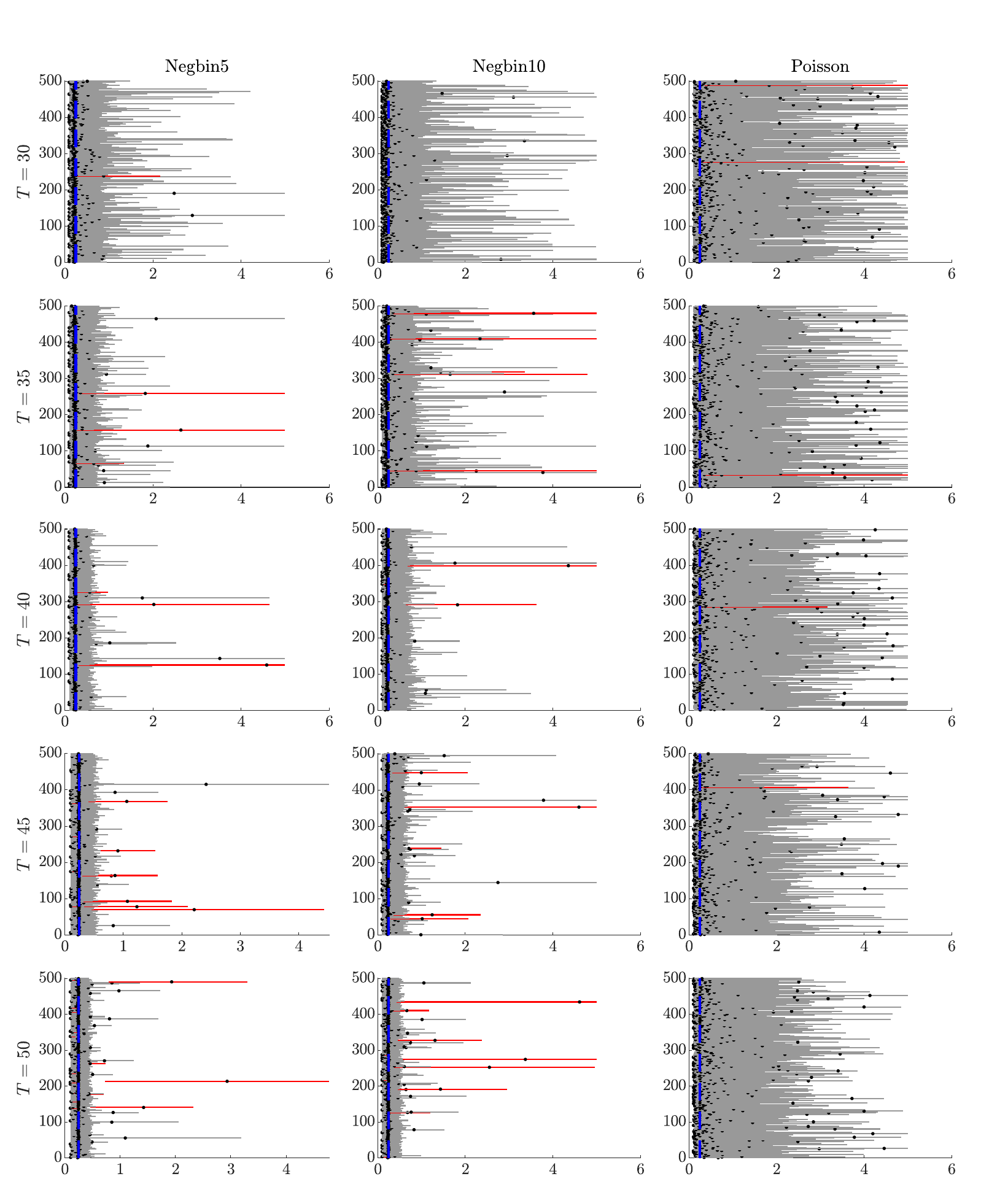}
\caption{Parameter estimates and 95\% CIs for $\gamma$ (single observable $dC/dt$) across 500 replicates and calibration window lengths $T=30,35,\ldots,70$, with the true value $\gamma=0.25$ indicated by the vertical blue dashed line.}
\label{fig:SEIR_vars_dCdt_CI_param_3}
\end{figure}

\begin{figure}[H]
\centering
\includegraphics[clip,trim=0 0 0 920,width=\linewidth]{SEIR_vars_dCdt_CI_grid_param_3.pdf}
\caption*{Figure~\ref{fig:SEIR_vars_dCdt_CI_param_3} (continued).}
\end{figure}

\subsection{Two observables: $\{dC/dt,\, I\}$}

\begin{figure}[H]
\centering
\includegraphics[width=\linewidth]{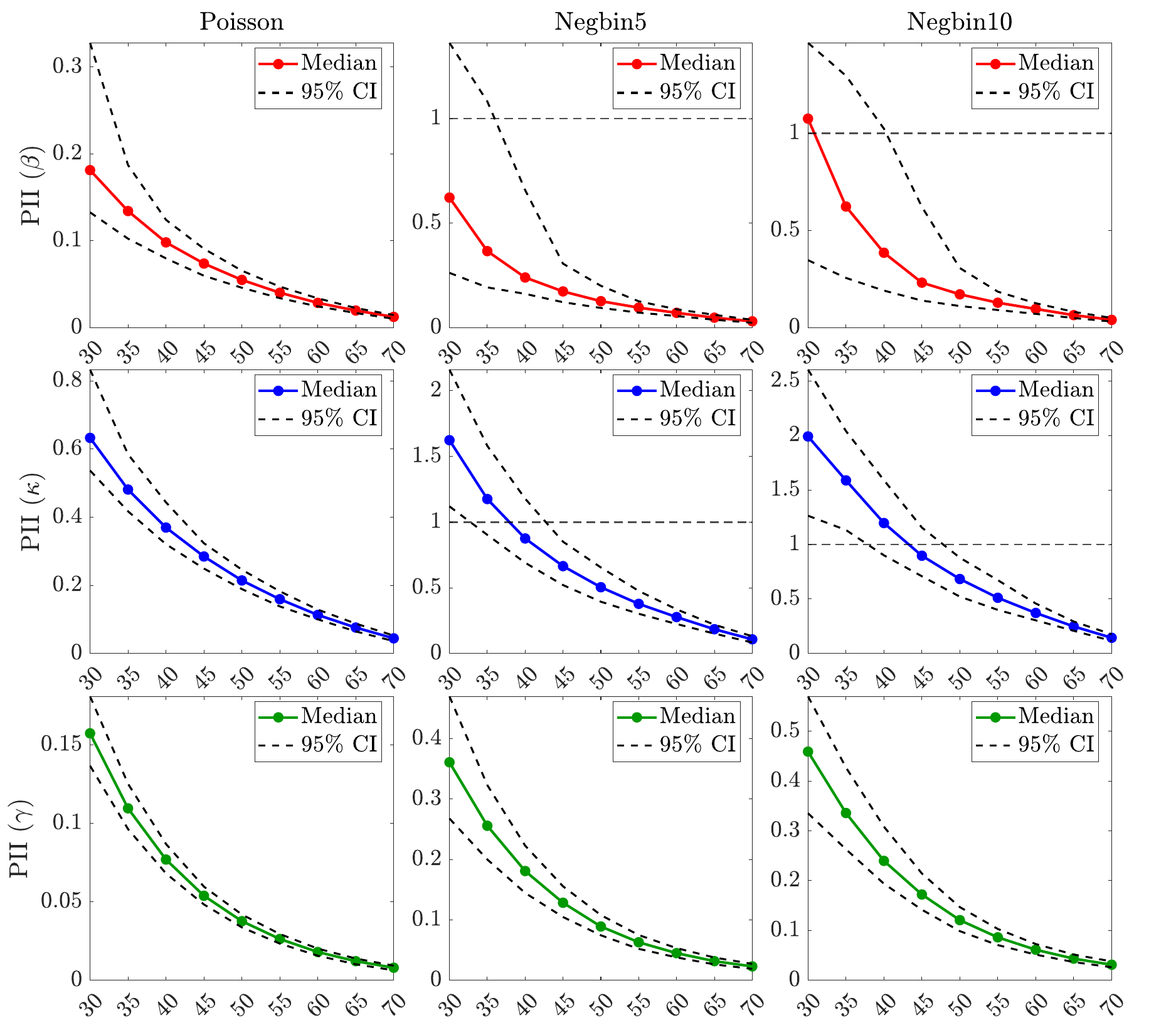}
\caption{Practical Identifiability Index (PII) for $\beta$, $\kappa$, and $\gamma$ in the SEIR model using two observables ($dC/dt$ and $I$) across calibration-window lengths $T=30, 35, \ldots, 70$ under three error structures: Poisson, negative binomial with data-generating dispersion $\alpha=5$ (Negbin5), and negative binomial with data-generating dispersion $\alpha=10$ (Negbin10). Red lines show the median PII across replicates, and dashed black curves indicate the central 95\% interval.}
\label{fig:SEIR_vars_2vars_PII}
\end{figure}

\begin{figure}[H]
\centering
\includegraphics[clip,trim=0 760 0 0,width=\linewidth]{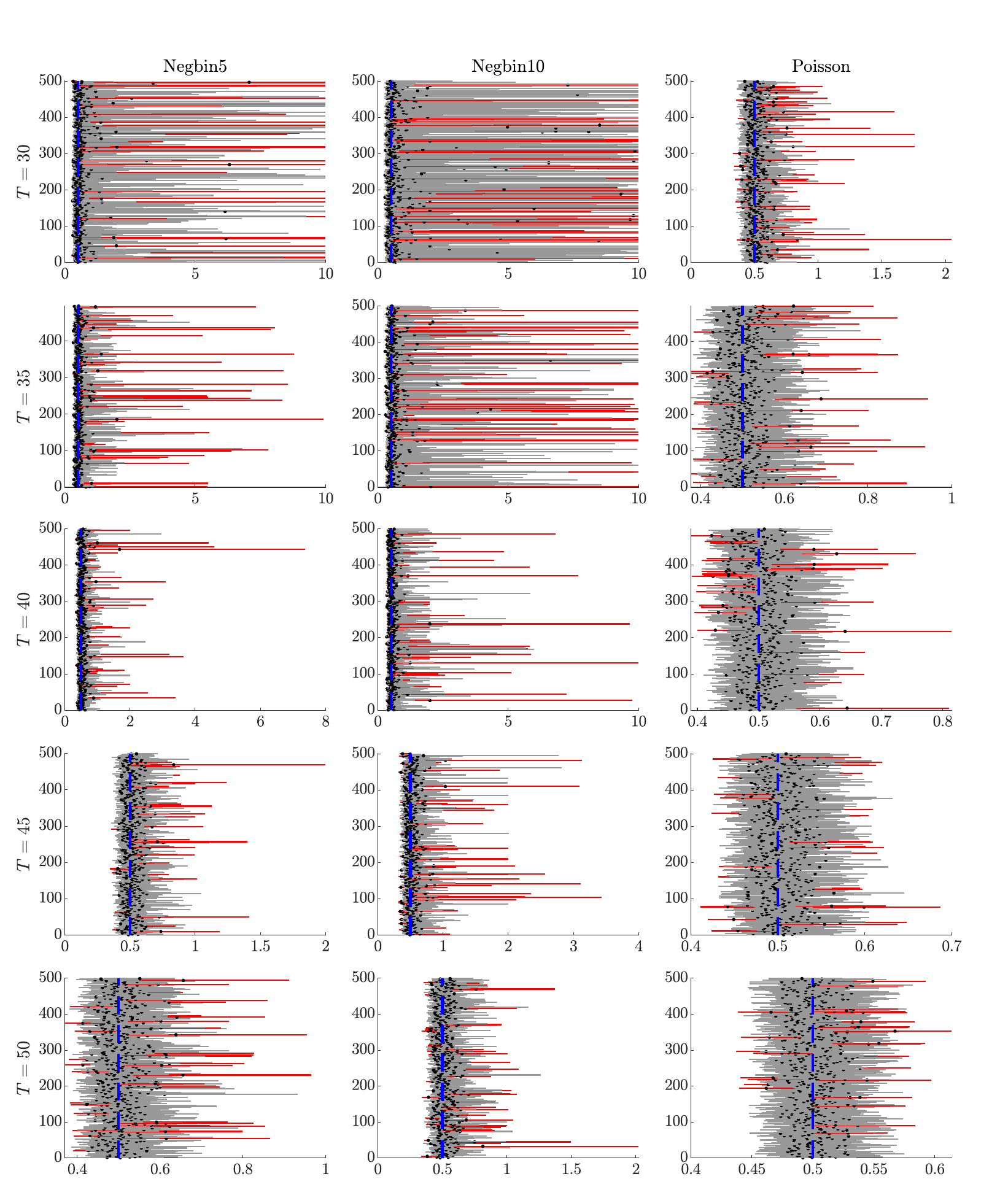}
\caption{Parameter estimates and 95\% CIs for $\beta$ (two observables: $dC/dt,\,I$) across 500 replicates and calibration window lengths $T=30,35,\ldots,70$, with the true value $\beta=0.5$ indicated by the vertical blue dashed line. Red intervals do not contain the true value; gray intervals do.}
\label{fig:SEIR_vars_2vars_CI_param_1}
\end{figure}

\begin{figure}[H]
\centering
\includegraphics[clip,trim=0 0 0 920,width=\linewidth]{SEIR_vars_2vars_CI_grid_param_1.pdf}
\caption*{Figure~\ref{fig:SEIR_vars_2vars_CI_param_1} (continued).}
\end{figure}

\begin{figure}[H]
\centering
\includegraphics[clip,trim=0 760 0 0,width=\linewidth]{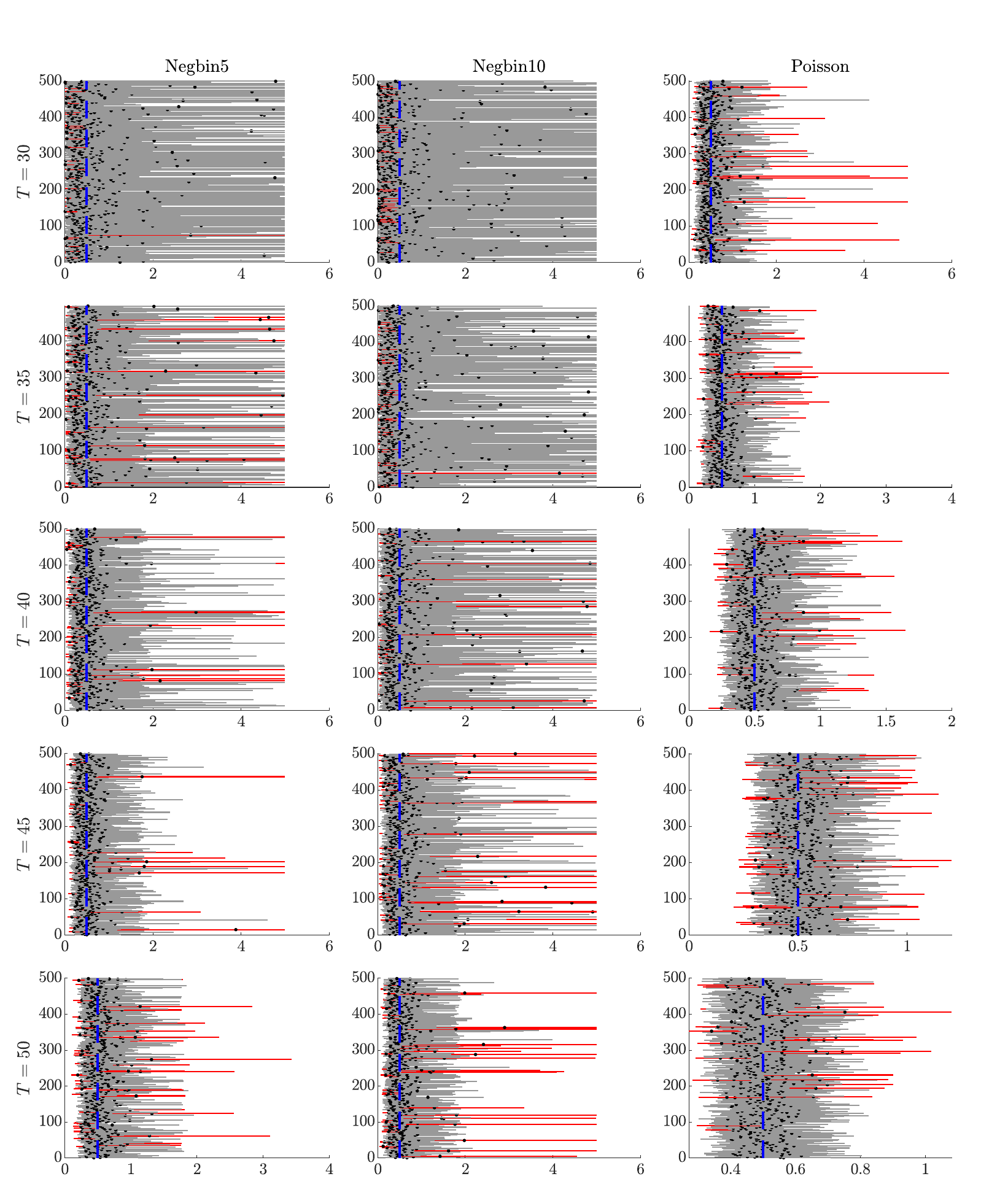}
\caption{Parameter estimates and 95\% CIs for $\kappa$ (two observables: $dC/dt,\,I$) across 500 replicates and calibration window lengths $T=30,35,\ldots,70$, with the true value $\kappa=0.5$ indicated by the vertical blue dashed line.}
\label{fig:SEIR_vars_2vars_CI_param_2}
\end{figure}

\begin{figure}[H]
\centering
\includegraphics[clip,trim=0 0 0 920,width=\linewidth]{SEIR_vars_2vars_CI_grid_param_2.pdf}
\caption*{Figure~\ref{fig:SEIR_vars_2vars_CI_param_2} (continued).}
\end{figure}

\begin{figure}[H]
\centering
\includegraphics[clip,trim=0 760 0 0,width=\linewidth]{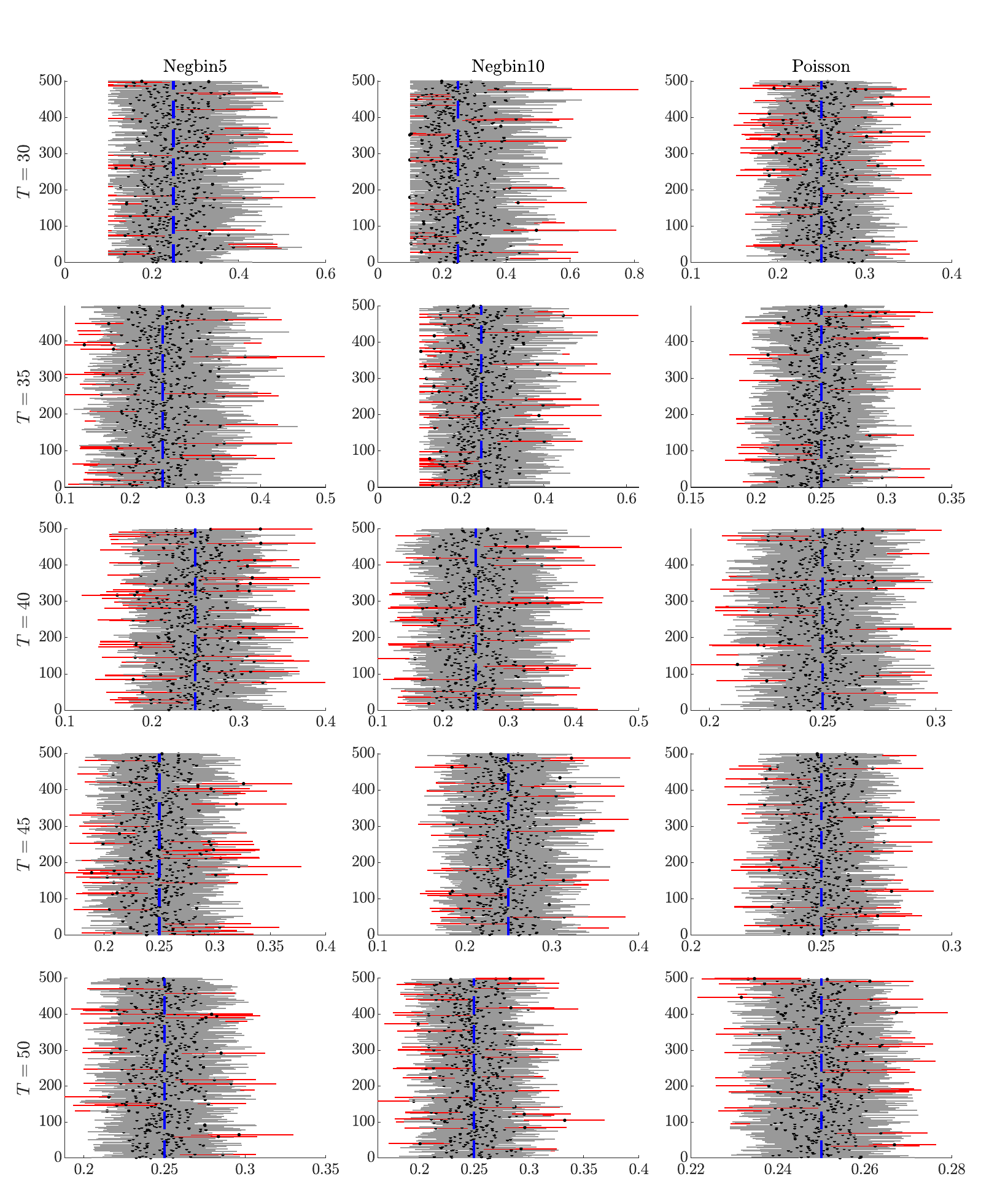}
\caption{Parameter estimates and 95\% CIs for $\gamma$ (two observables: $dC/dt,\,I$) across 500 replicates and calibration window lengths $T=30,35,\ldots,70$, with the true value $\gamma=0.25$ indicated by the vertical blue dashed line.}
\label{fig:SEIR_vars_2vars_CI_param_3}
\end{figure}

\begin{figure}[H]
\centering
\includegraphics[clip,trim=0 0 0 920,width=\linewidth]{SEIR_vars_2vars_CI_grid_param_3.pdf}
\caption*{Figure~\ref{fig:SEIR_vars_2vars_CI_param_3} (continued).}
\end{figure}

\subsection{Three observables: $\{dC/dt,\, I,\, R\}$}

\begin{figure}[H]
\centering
\includegraphics[width=\linewidth]{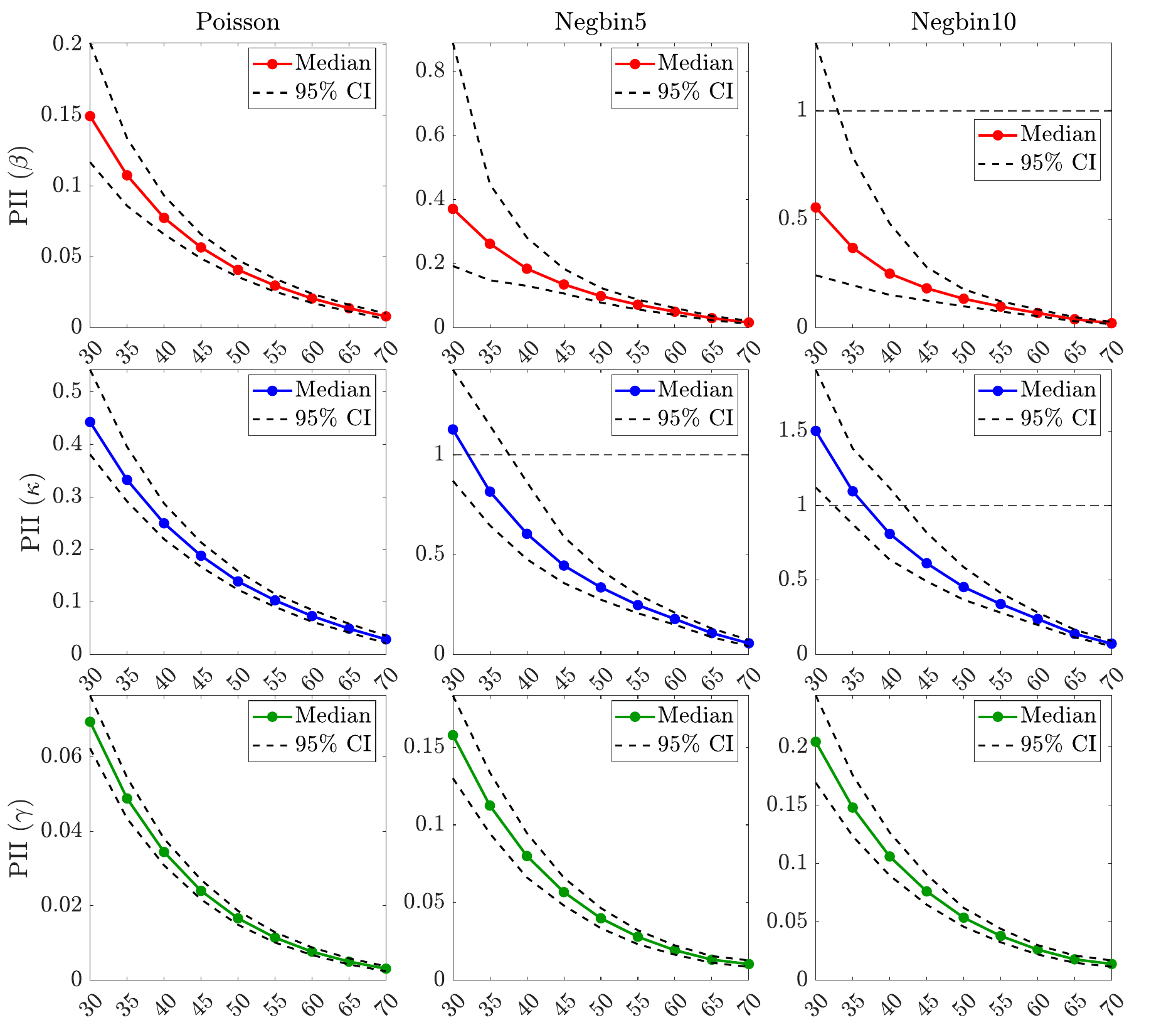}
\caption{Practical Identifiability Index (PII) for $\beta$, $\kappa$, and $\gamma$ in the SEIR model using three observables ($dC/dt$, $I$, and $R$) across calibration-window lengths $T=30, 35, \ldots, 70$ under three error structures: Poisson, negative binomial with data-generating dispersion $\alpha=5$ (Negbin5), and negative binomial with data-generating dispersion $\alpha=10$ (Negbin10). Red lines show the median PII across replicates, and dashed black curves indicate the central 95\% interval.}
\label{fig:SEIR_vars_3vars_PII}
\end{figure}

\begin{figure}[H]
\centering
\includegraphics[clip,trim=0 760 0 0,width=\linewidth]{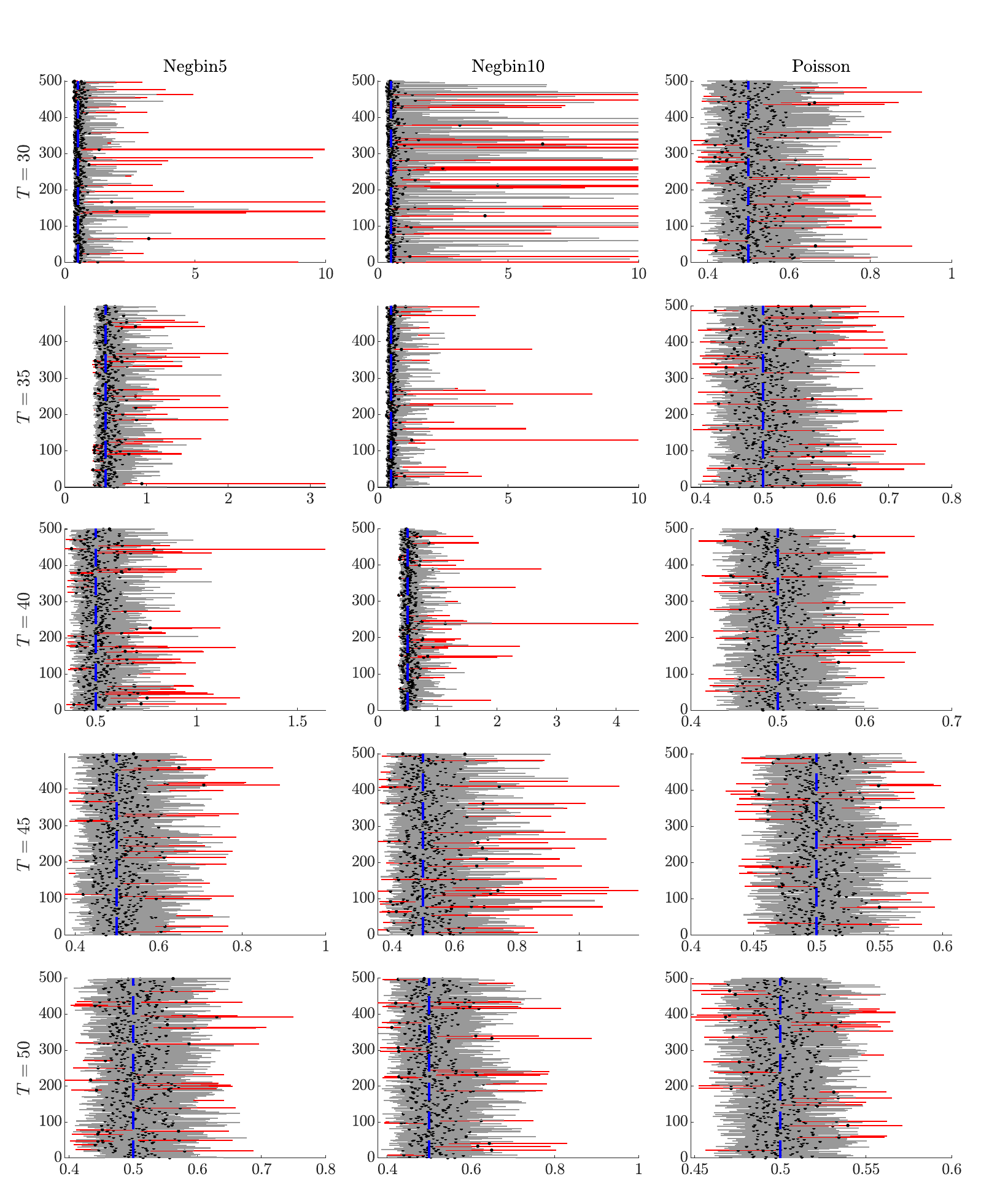}
\caption{Parameter estimates and 95\% CIs for $\beta$ (three observables: $dC/dt,\,I,\,R$) across 500 replicates and calibration window lengths $T=30,35,\ldots,70$, with the true value $\beta=0.5$ indicated by the vertical blue dashed line. Red intervals do not contain the true value; gray intervals do.}
\label{fig:SEIR_vars_3vars_CI_param_1}
\end{figure}

\begin{figure}[H]
\centering
\includegraphics[clip,trim=0 0 0 920,width=\linewidth]{SEIR_vars_3vars_CI_grid_param_1.pdf}
\caption*{Figure~\ref{fig:SEIR_vars_3vars_CI_param_1} (continued).}
\end{figure}

\begin{figure}[H]
\centering
\includegraphics[clip,trim=0 760 0 0,width=\linewidth]{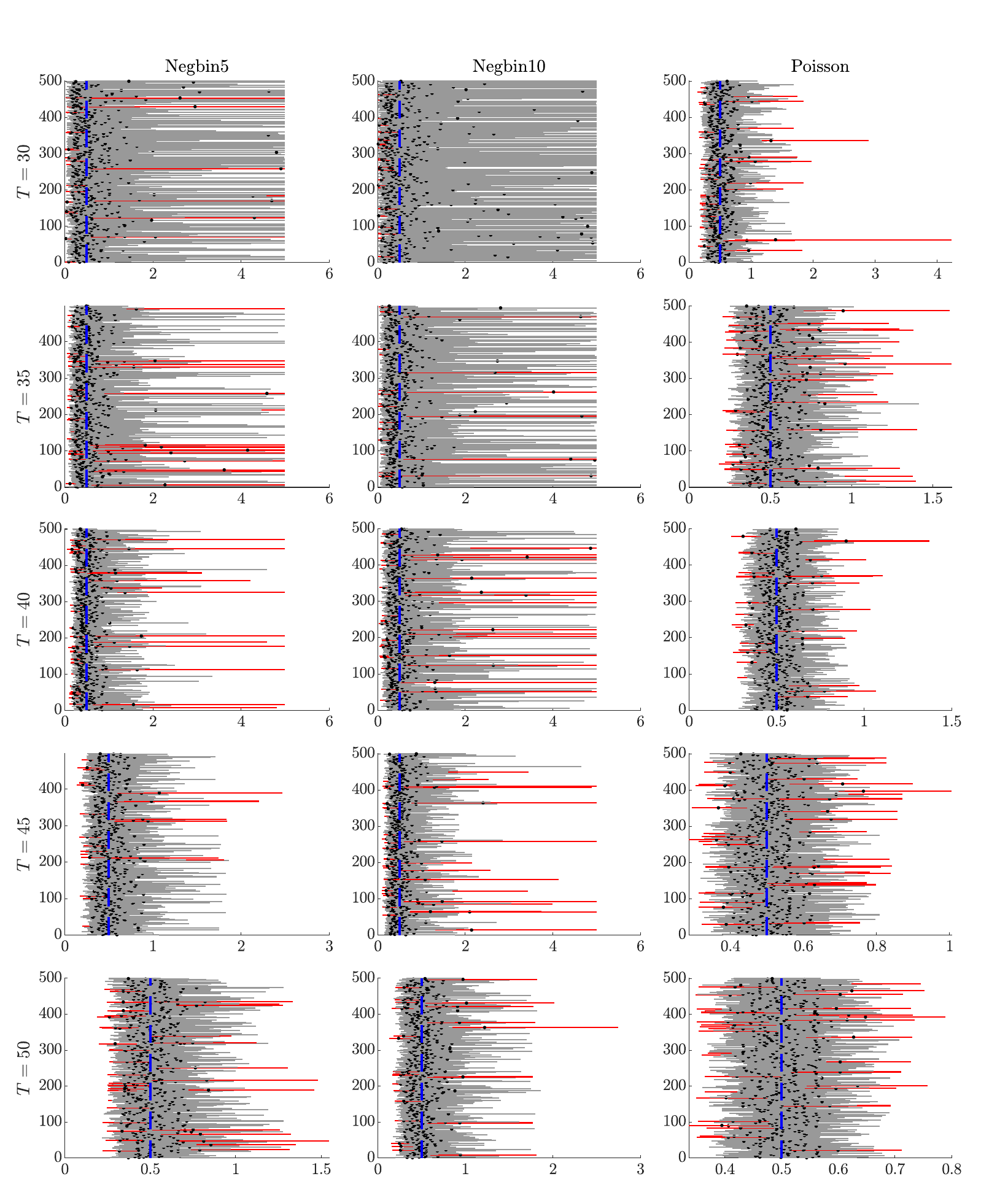}
\caption{Parameter estimates and 95\% CIs for $\kappa$ (three observables: $dC/dt,\,I,\,R$) across 500 replicates and calibration window lengths $T=30,35,\ldots,70$, with the true value $\kappa=0.5$ indicated by the vertical blue dashed line.}
\label{fig:SEIR_vars_3vars_CI_param_2}
\end{figure}

\begin{figure}[H]
\centering
\includegraphics[clip,trim=0 0 0 920,width=\linewidth]{SEIR_vars_3vars_CI_grid_param_2.pdf}
\caption*{Figure~\ref{fig:SEIR_vars_3vars_CI_param_2} (continued).}
\end{figure}

\begin{figure}[H]
\centering
\includegraphics[clip,trim=0 760 0 0,width=\linewidth]{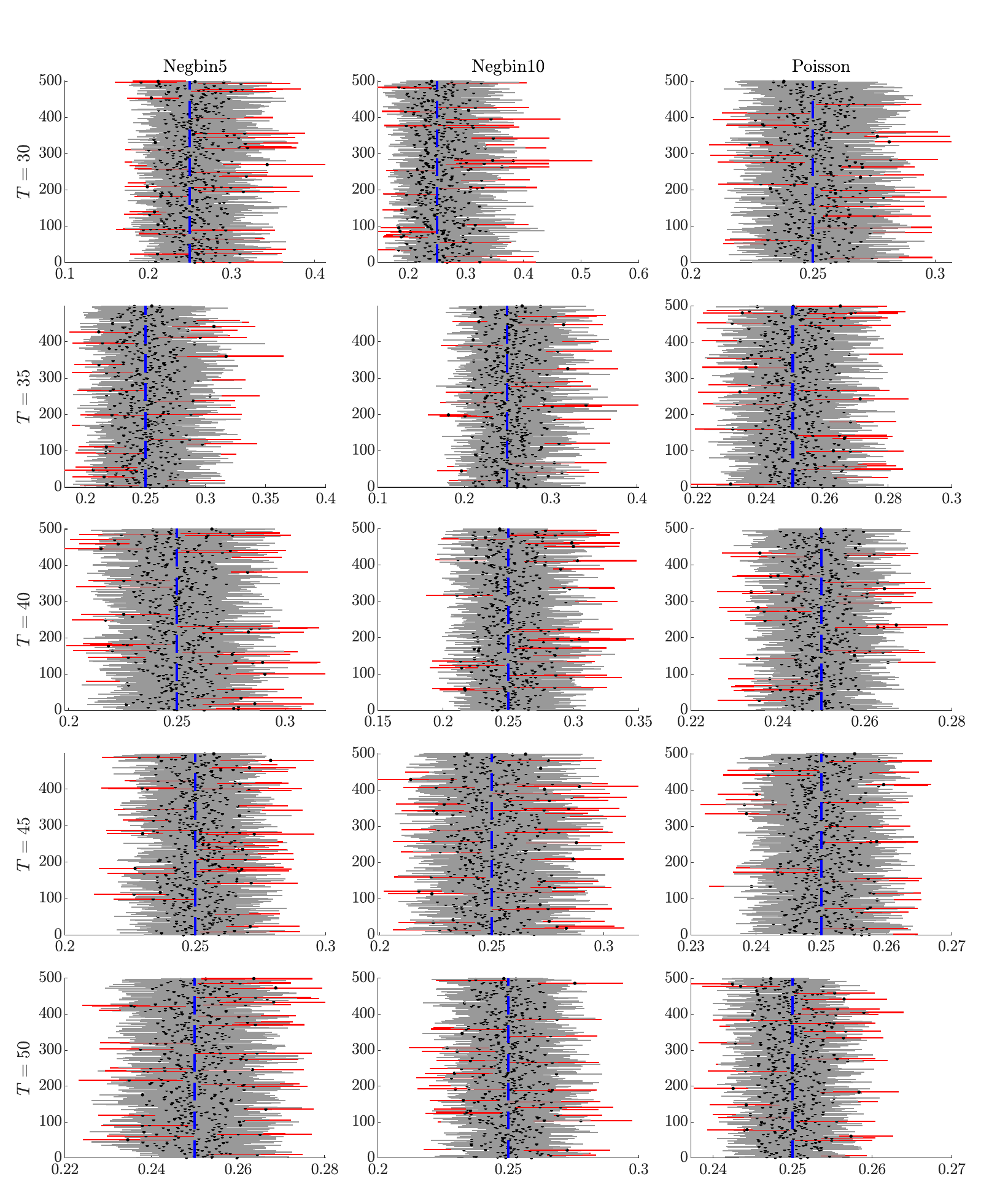}
\caption{Parameter estimates and 95\% CIs for $\gamma$ (three observables: $dC/dt,\,I,\,R$) across 500 replicates and calibration window lengths $T=30,35,\ldots,70$, with the true value $\gamma=0.25$ indicated by the vertical blue dashed line.}
\label{fig:SEIR_vars_3vars_CI_param_3}
\end{figure}

\begin{figure}[H]
\centering
\includegraphics[clip,trim=0 0 0 920,width=\linewidth]{SEIR_vars_3vars_CI_grid_param_3.pdf}
\caption*{Figure~\ref{fig:SEIR_vars_3vars_CI_param_3} (continued).}
\end{figure}


\end{document}